\documentclass{article}
\usepackage[utf8]{inputenc}
\usepackage{graphicx}
\usepackage{subfigure}

\usepackage{amssymb}
\usepackage{amsthm}
\usepackage{url}
\usepackage{amsmath}
\usepackage{subfigure}
\usepackage{textcomp}
\usepackage[affil-it]{authblk}
\usepackage[margin=1.5cm]{geometry}
\usepackage{lscape}
\usepackage{xcolor}

\graphicspath{{figs/}}

\title{On memfractance of plants and fungi}
\author[1,*]{Alexander E. Beasley}
\author[3]{Mohammed-Salah Abdelouahab}
\author[2]{Ren\'{e} Lozi}
\author[1]{Anna L. Powell}
\author[1]{Andrew Adamatzky}

\affil[1]{Unconventional Computing Laboratory, UWE, Bristol, UK}
\affil[*]{Corresponding author: Alexander Beasley, alex.beasley@uwe.ac.uk}
\affil[2]{Laboratory of Mathematics and their interactions, University Centre Abdelhafid Boussouf, Mila 43000, Algeria}
\affil[3]{Universit\'{e} C\^{o}te d’Azur, CNRS, LJAD, Nice, France}
\date{\today}

\begin{document}

\maketitle

\begin{abstract}
\noindent
The key feature of a memristor is that the resistance is a function of its previous resistance, thereby the behaviour of the device is influenced by changing the way in which potential is applied across it. Ultimately, information can be encoded on memristors, which can then be used to implement a number of circuit topologies.%~\cite{borghetti2010memristive, linn2012beyond, kvatinsky2014magic, borghetti2009hybrid, ho2009nonvolatile}. 
Biological substrates have already been shown to exhibit some memristive properties.%~\cite{martinsen2010memristance, kosta2011human, volkov2014memristors, paper:apples_memristor, gale2015slime, del2019two,chiolerio2020resistance}. 
It is, therefore, logical that all biological media will follow this trend to some degree. In this paper we demonstrate that a range of yet untested specimens exhibit memristive properties, including mediums such as water and dampened wood shavings on which we can cultivate biological specimens. We propose that memristance is not a binary property $\{0,1\}$,  but rather a continuum on the scale [0,1]. The results imply that there is great potential for hybrid electronic systems that combine traditional electronic typologies with naturally occurring specimens.

\vspace{5mm}


\end{abstract}

\section{Introduction}

\begin{figure}[!bp]
    \centering
    \includegraphics[width=0.8\textwidth]{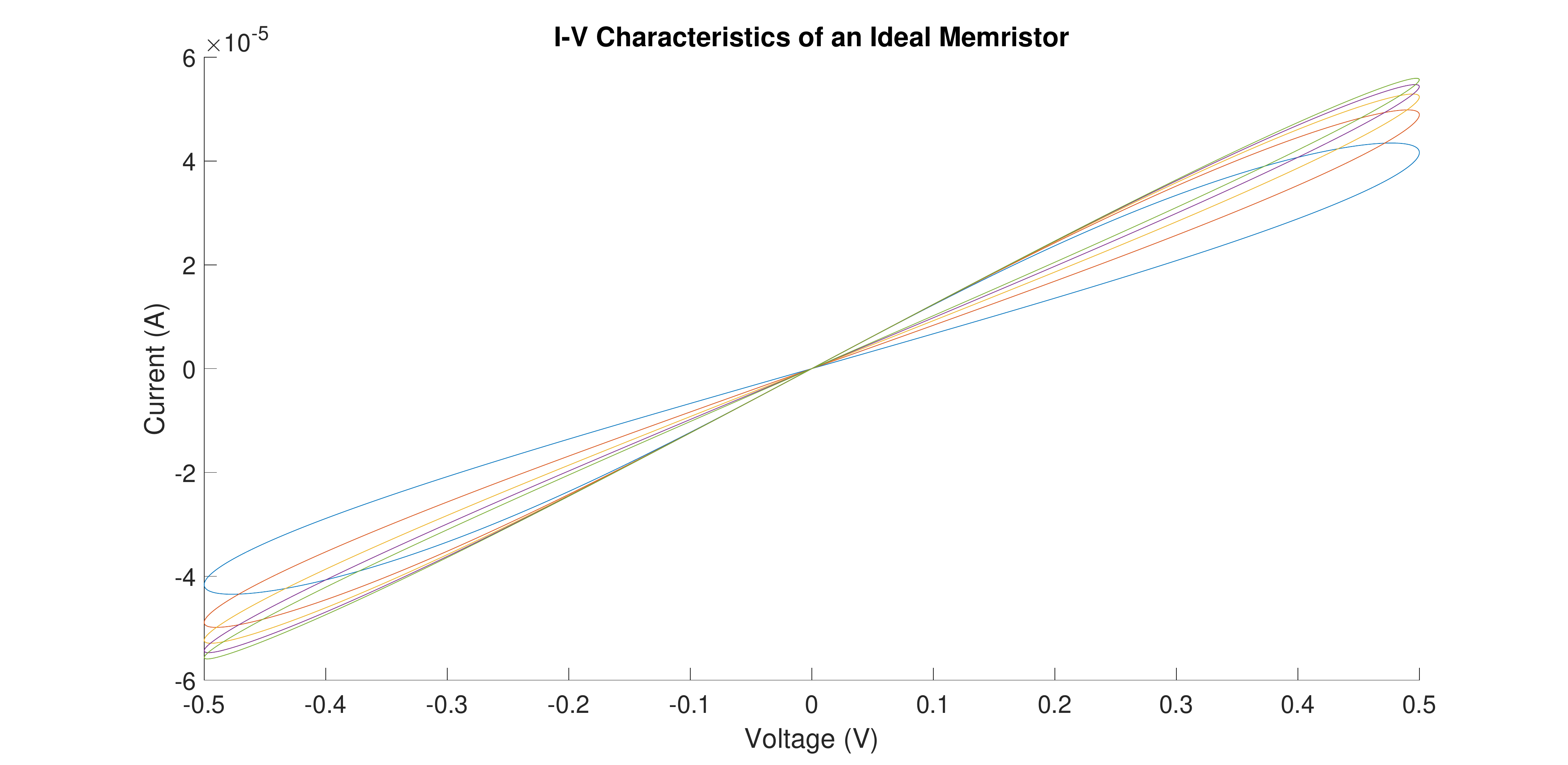}
    \caption{I-V characteristics from a model of an ideal memristor~\cite{web:matlab_memristor_model}.}
    \label{fig:ideal_memristor}
\end{figure}

Originally proposed by Leon Chua in 1971~\cite{paper:memristor}, the memristor poses an fourth basic circuit element, whose characteristics differ from that of $R$, $L$ and $C$ elements. The model of an optimal memristor (Fig.~\ref{fig:ideal_memristor}) shows a number of key features: (1) lobes in the positive and negative half of the cycle, and (2) a `pinch' (or crossing) point at 0V.

Memristance has been seen in nano-scale devices, where solid-state electronic and ionic transport are coupled under an external bias voltage~\cite{paper:missing_memristor}. Strukov~\emph{et al.} posit that the hysteric I-V characteristics observed in thin-film, two-terminal devices can be understood as memristive. However, this is observed behaviour of devices that already have other, large signal behaviours.\par

Finding a true memristor is by no means an easy task, however, a number of studies have turned to nature to provide the answer, with varying success. 
Memristive properties of organic polymers have been studied since 2005~\cite{erokhin2005hybrid} in experiments with hybrid electronic devices based on polyaniline-polyethylenoxide junction~\cite{erokhin2005hybrid}.
Memristive properties of living creatures and their organs and fluids have been demonstrated in 
skin~\cite{martinsen2010memristance},
blood~\cite{kosta2011human},
plants~\cite{volkov2014memristors} (including fruits~\cite{paper:apples_memristor}),
slime mould~\cite{gale2015slime},
tubulin microtubules~\cite{del2019two,chiolerio2020resistance}.

From a more global point of view, mem-fractance which involves fractional calculus, is a general paradigm for unifying and enlarging the family of memristive, mem-capacitive and mem-inductive elements.\par

This paper presents a study of the I-V characteristics of a number of specimens of plants, fungi, and cultivation mediums. Why choose specimens from nature? Previous work has demonstrated significant potential for the use of naturally occurring substances as memristors. Taking these studies as a basis, it is proposed that any substance taken from nature --- that has once been living --- will exhibit the same memristive properties. \par

Why we are looking for memristive properties? A memristor is a material implication~\cite{borghetti2010memristive,kvatinsky2013memristor}. Therefore, memristors can be used for constructing other logical circuits, statefull logic operations~\cite{borghetti2010memristive}, logic operations in passive crossbar arrays of memristos~\cite{linn2012beyond}, memory aided logic circuits~\cite{kvatinsky2014magic}, self-programmable logic circuits~\cite{borghetti2009hybrid}, and, indeed, memory devices~\cite{ho2009nonvolatile}. If the substances show memristive properties then we can implement a large variety of memory and computing devices embedded directly into hybrid electronic circuits that utilise naturally occurring resources.\par

The rest of this paper is organised as follows. Section~\ref{sec:experimentation} details the experimental set up used to examine the I-V characteristics of fungal fruit bodies. Section~\ref{sec:results} presents the results from the experimentation. A mathematical modelling onion mem-fractance is presented in Sect.~\ref{sec:model}. A discussion of the results is given in Sect.~\ref{sec:discussions} and finally conclusions are given in Sect.~\ref{sec:discussions}.

\section{Experimental method}
\label{sec:experimentation}

A number of subjects were identified for the purposes of testing the I-V characteristics of biological medium. Samples fall under the following categories: fruiting bodies, flora, fungi and water. In addition, a number of control samples were also subject to test (dry wood shavings and de-ionised water).\par

Fruits and vegetables used in experiments are
%apples (Oaklands British Apples, variety Cameo, Kent, Tesco Stores Ltd.), 
large garlic (origin Spain, Tesco Stores Ltd.), 
aubergine (origin Spain, Aldi UK), 
onion (origin UK, Aldi UK), 
potato (origin UK, Aldi UK), 
banana (Aldi UK), 
cucumber (origin Spain, Aldi UK), 
mango (origin Peru, Aldi UK), and
bell pepper (origin Spain, Aldi UK). 
Plants used in experiments are \emph{Echeveria pulidonis} and \emph{Senecio ficoides}. Fungi used in experiments grey oyster fungi \emph{Pleurotus ostreatus} (Ann Miller's Speciality Mushrooms Ltd, UK)  cultivated on wood shavings.

\begin{figure}[!tbp]
    \centering
    \includegraphics[width=0.6\textwidth]{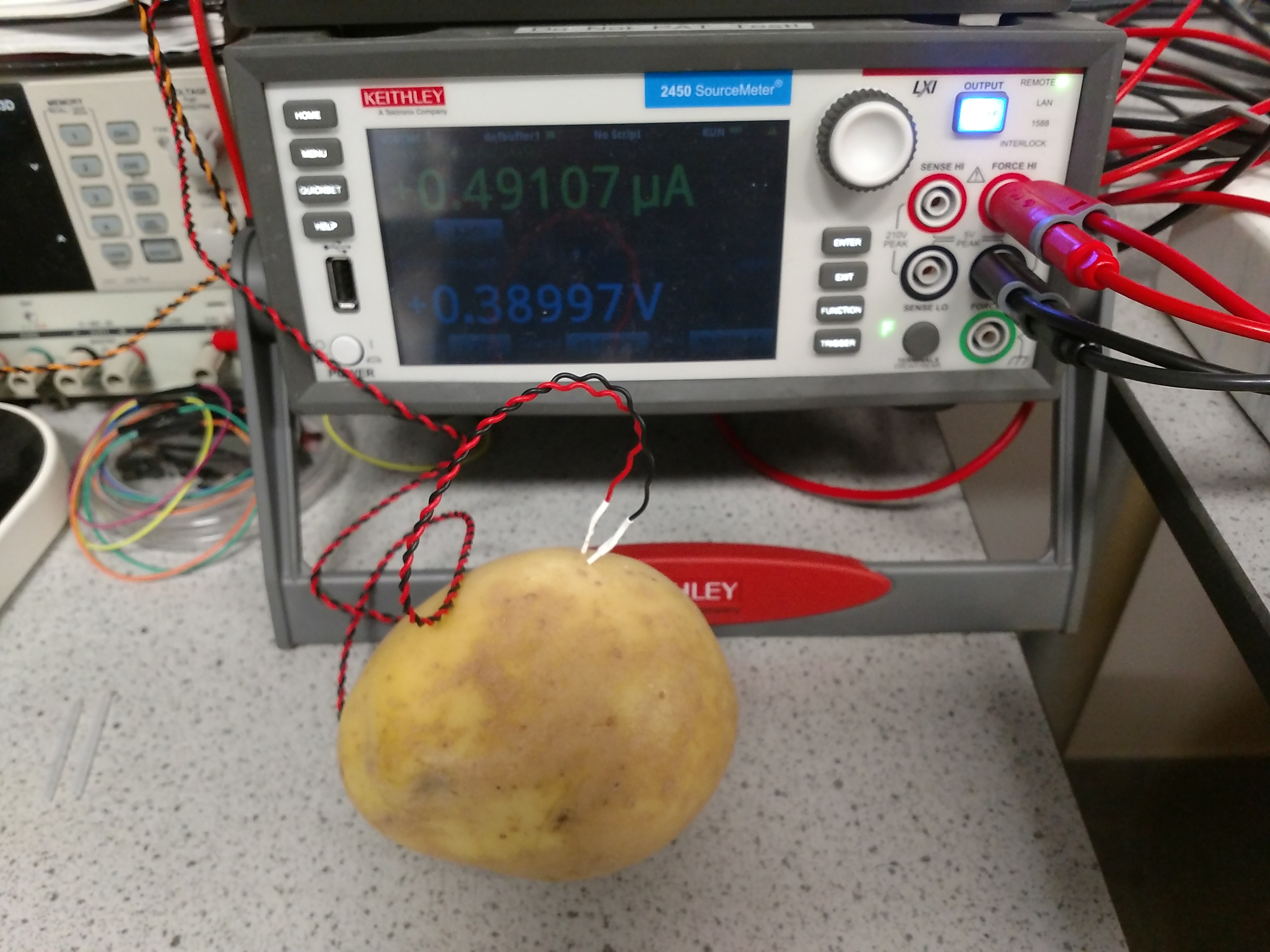}
    \caption{Sample specimen under test using Keithley SMU.}
    \label{fig:test_set_up}
\end{figure}

Iridium-coated stainless steel sub-dermal needles with twisted cables (Spes Medica SRL, Italy)  were inserted approximately 10~mm apart in each of the samples under test, such as in the example in Fig.~\ref{fig:test_set_up}. I-V sweeps were performed on the each of the samples using a Keithley Source Measure Unit (SMU) 2450 (Keithley Instruments, USA) under a range of conditions: 
    \begin{itemize}
        \item {Cyclic voltammetry performed from -0V5 to 0V5 and -1V to 1V.} The voltage limits of the cyclic voltammetry are limited as to not exceed the electrolysis of water.
        \item {Delay between consecutive voltage settings: 0.01s, 0.1s and 1s.}
    \end{itemize}
%In addition, samples of mycelium were also tested under a range of lighting conditions as they are a photosensitive medium: 
%    \begin{itemize}
%        \item {Darkened condition - substrate covered}
%        \item {Ambient lab lighting - 965\,Lux}
%        \item {Intense lighting - 1500\,Lux}
%    \end{itemize}
The composition of tap water is measured to be the following: 
conductivity 573 micro Siemens (measured at 22.7\textdegree C with Eutch Instruments, Model CONDG+),
pH 6.86 (measured at 22.5\textdegree C with VWR pH 10), pH 7.0  standardised at 25\textdegree C,
total dissolved solids 393 ppm,
salinity: 0.3 ppm.
 All conditions were repeated a number of times and the resulting I-V curves processed using MATLAB.\par

\section{Results}
\label{sec:results}

A number of key findings have been drawn from the tests that support the assertion that any biological object exhibits memristance. This section present a subset of the results from I-V characterisation of the specimens. All raw data plots for I-V characteristics may be found in the appendices (section~\ref{sec:Appendix}).\par

\subsection{Vegetables and plants}

\begin{figure}[!tbp]
    \centering
    \subfigure[]{\includegraphics[width=0.45\textwidth]{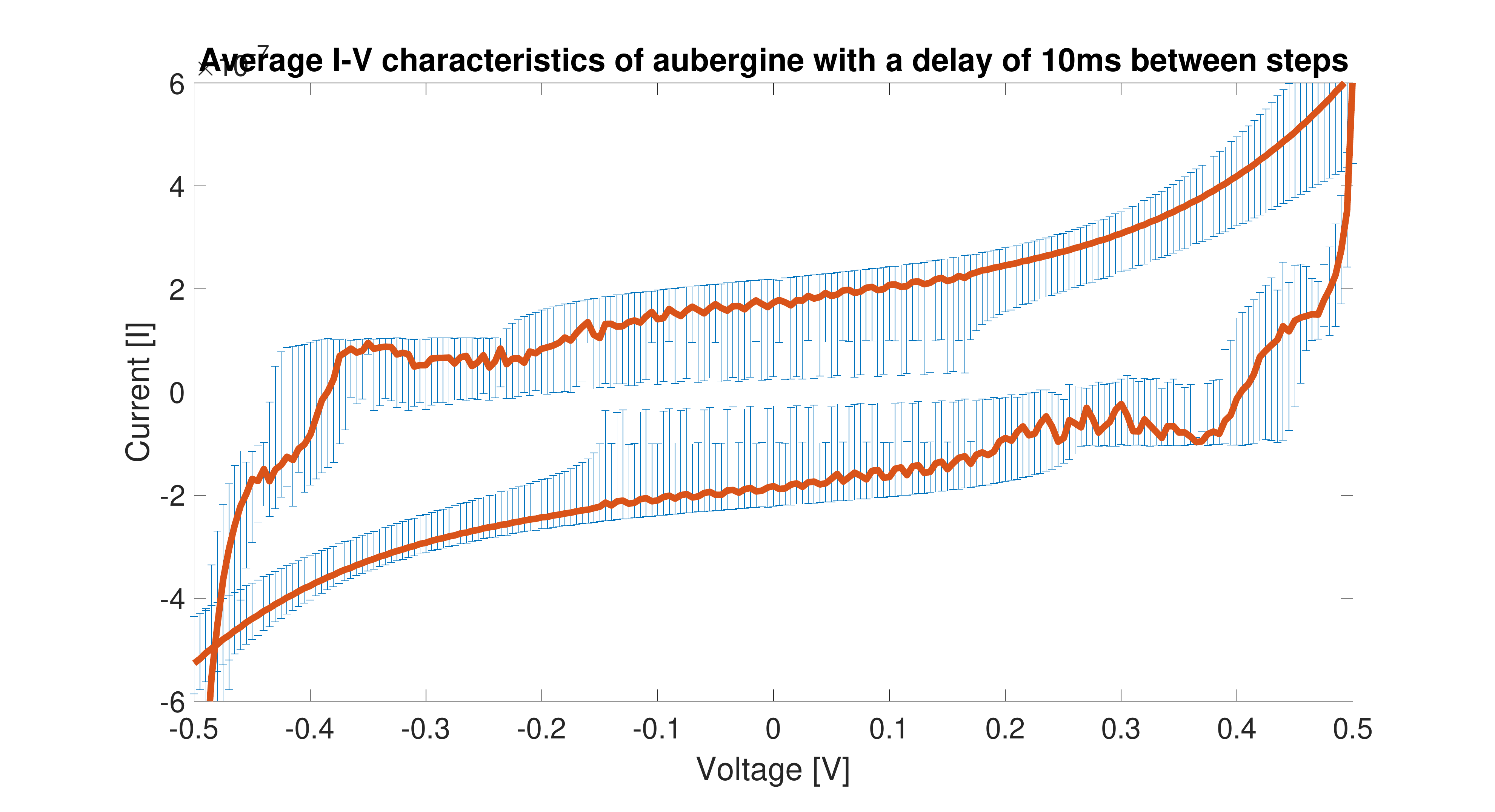}}
    \subfigure[]{\includegraphics[width=0.45\textwidth]{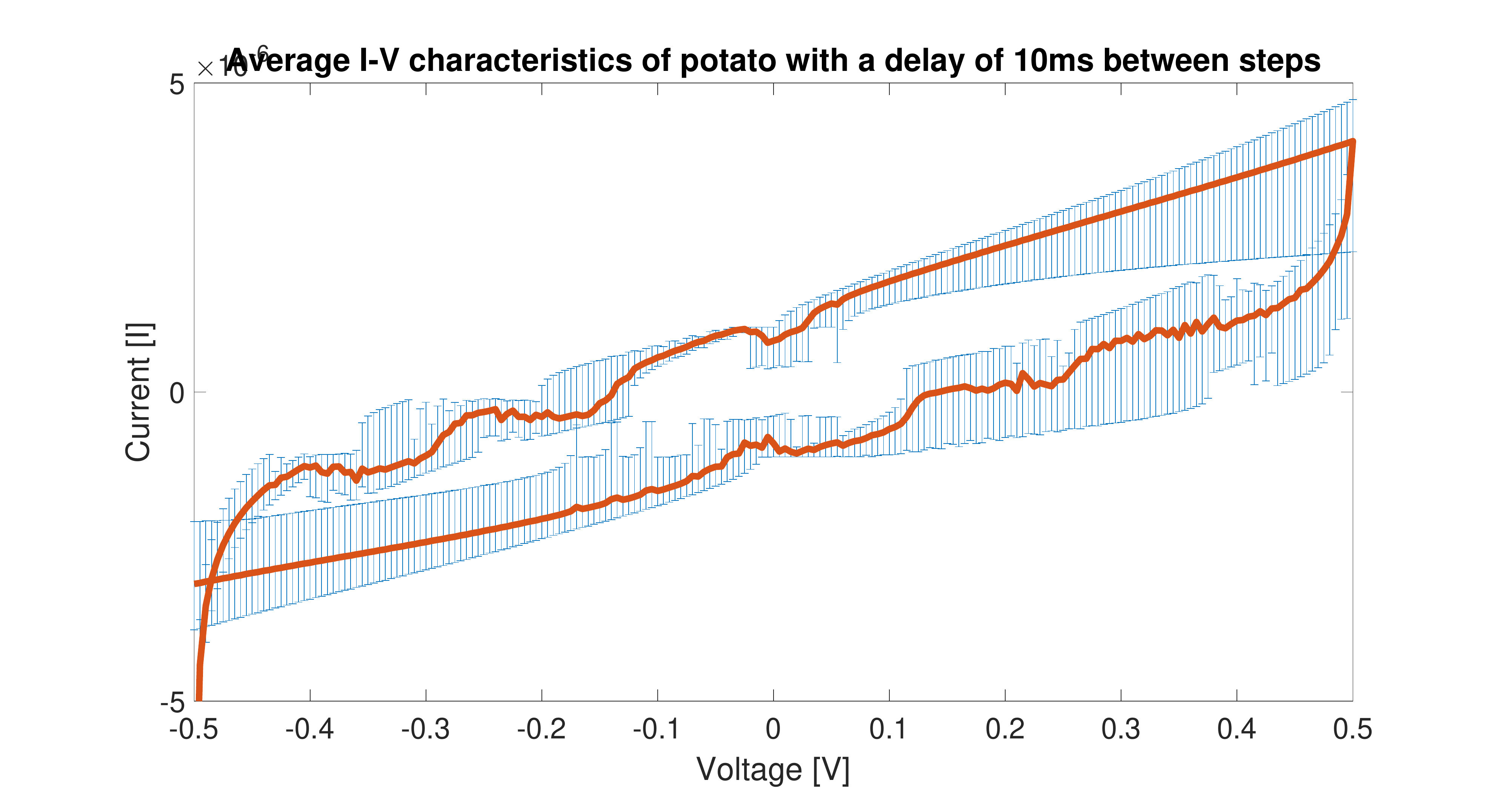}}
    \subfigure[]{\includegraphics[width=0.45\textwidth]{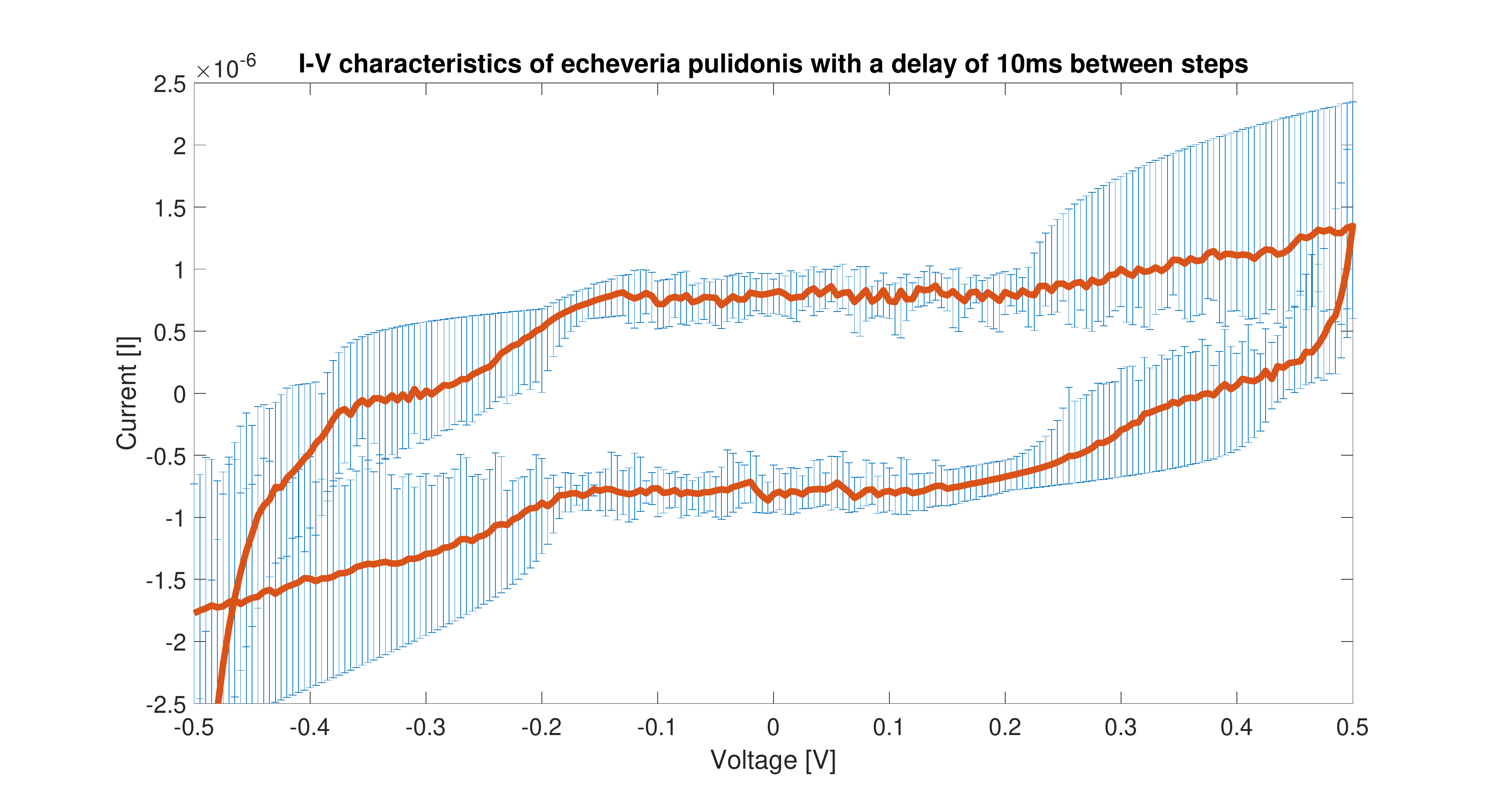}}
    \subfigure[]{\includegraphics[width=0.45\textwidth]{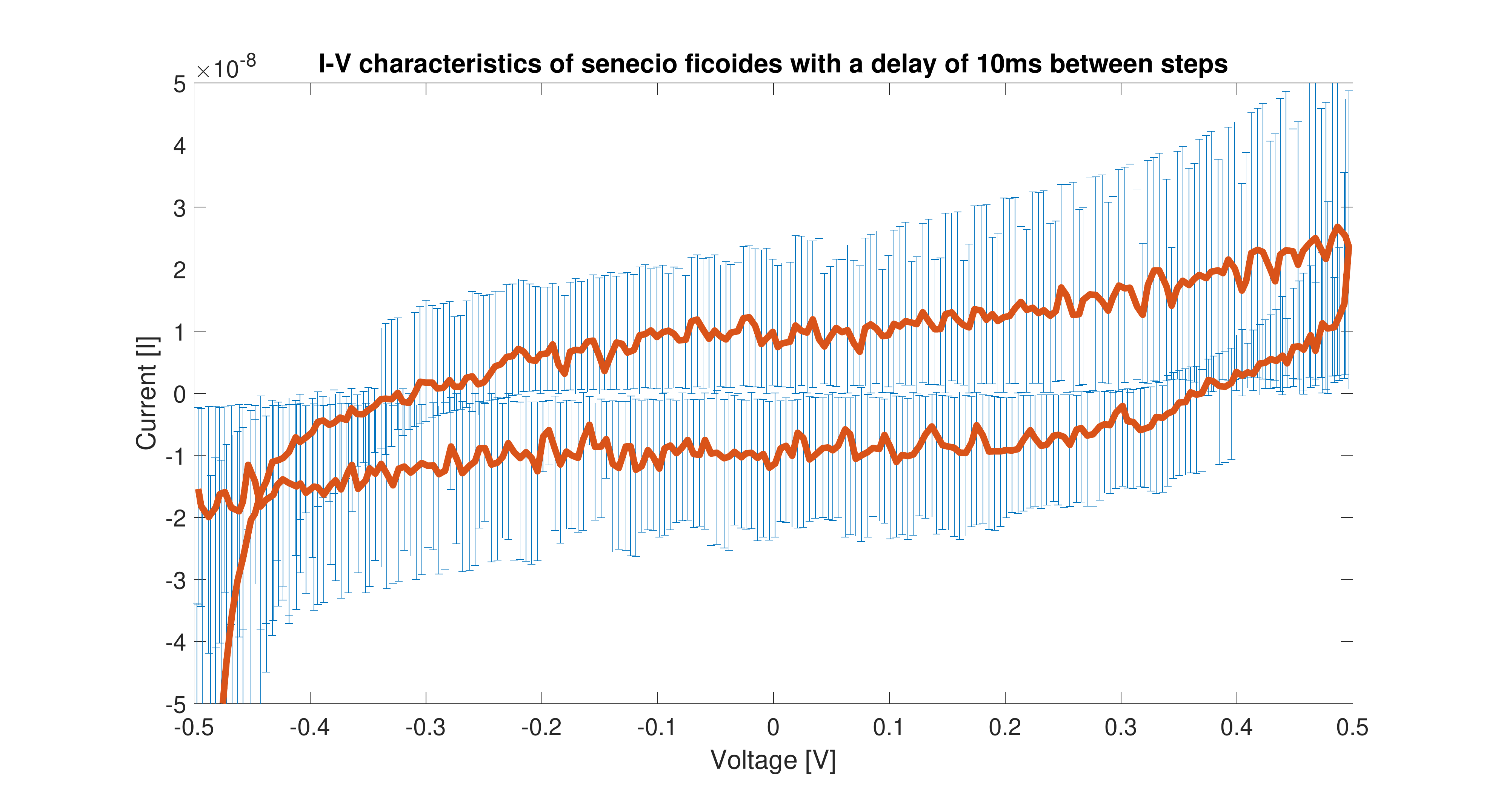}}
    \caption{I-V characteristics of specimens, error bars shown. Cyclic voltammetry performed over -0.5V to 0.5V, with a step delay of 10ms. (a) aubergine, (b) potato, (c) \emph{Echeveria pulidonis}, (d) \emph{Senecio ficoides}. }
    \label{fig:example_plots}
\end{figure}

From the species tested in this study, it is seen that the I-V sweeps that the resistance of the subject is varied depending not only on the previously applied voltage but also on the frequency with which the voltage is changed. As a general rule, the faster the voltage is changed the greater the divergence in conducted current between the positive and negative phases of the cyclic voltammetry. Additionally, increasing the frequency of the voltage will yield a larger conducted current. Figure~\ref{fig:example_plots} shows the average I-V response for a selection of the test specimens.\par 

Test substrates were subjected to cyclic voltammetry over two voltage ranges. Naturally, for the larger voltage range, the maximum conducted current is also far greater. The greater the applied voltage, the closer the test subject is to a breakdown voltage where larger current flow can be expected, similar behaviour to a p-n-p junction in a traditional silicon semi-conductor.\par

\subsection{Mycelium}

%\begin{figure}[!tbp]
%    \centering
%    \subfigure[]{\includegraphics[width=\textwidth]{concat_average_mycelium_1Vpp.pdf}}
%    \subfigure[]{\includegraphics[width=\textwidth]{concat_average_mycelium_2Vpp.pdf}}
%    \caption{Average mycelium I-V characteristics under all lighting conditions. (a) cyclic voltammetry performed from -0.5V to 0.5V. (b) cyclic voltammetry performed from -1V to 1V. }
%    \label{fig:mycelium_ave}
%\end{figure}

\begin{figure}
    \centering
    \subfigure[]{\includegraphics[width=0.49\textwidth]{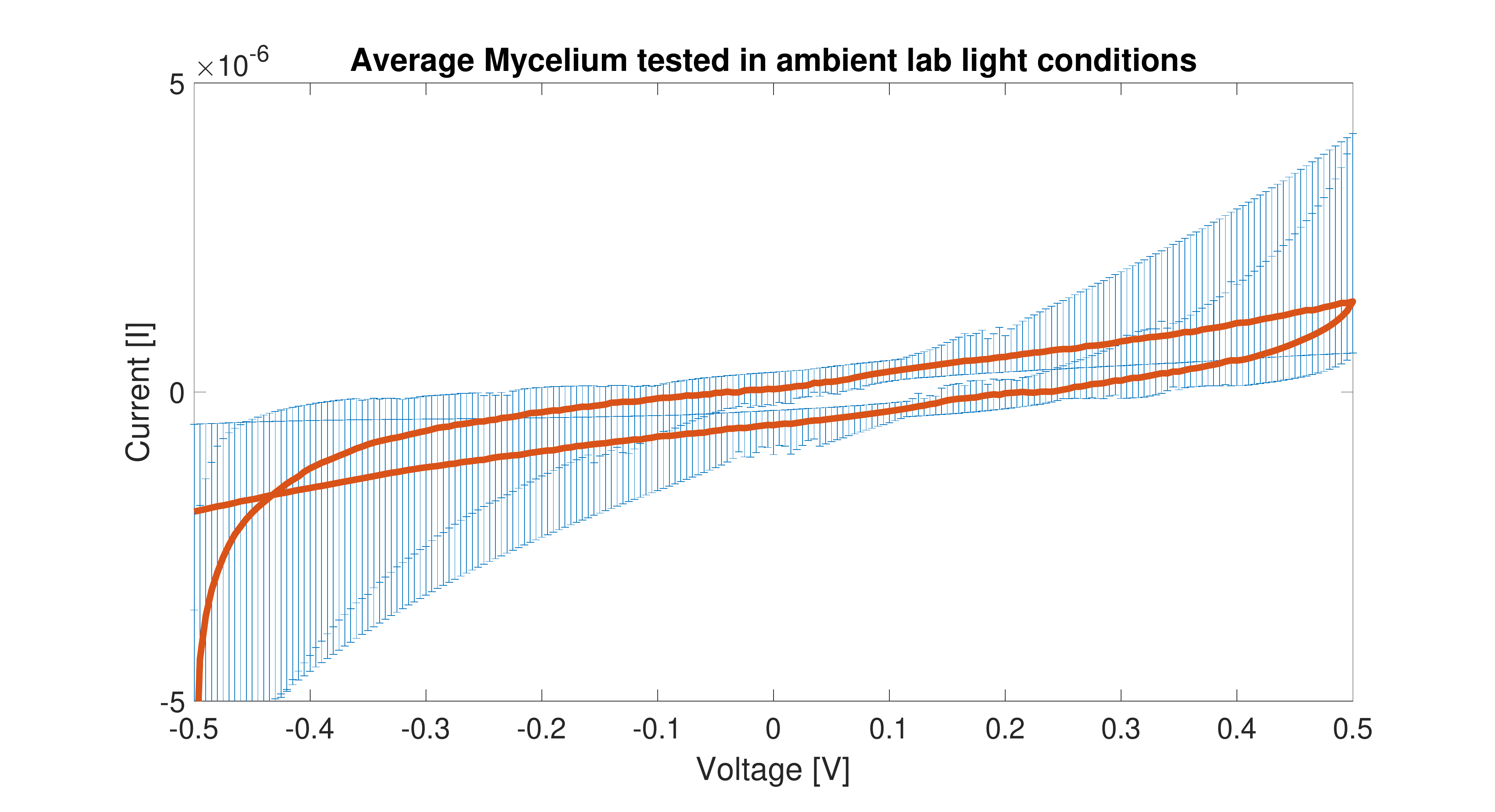}}
    \subfigure[]{\includegraphics[width=0.49\textwidth]{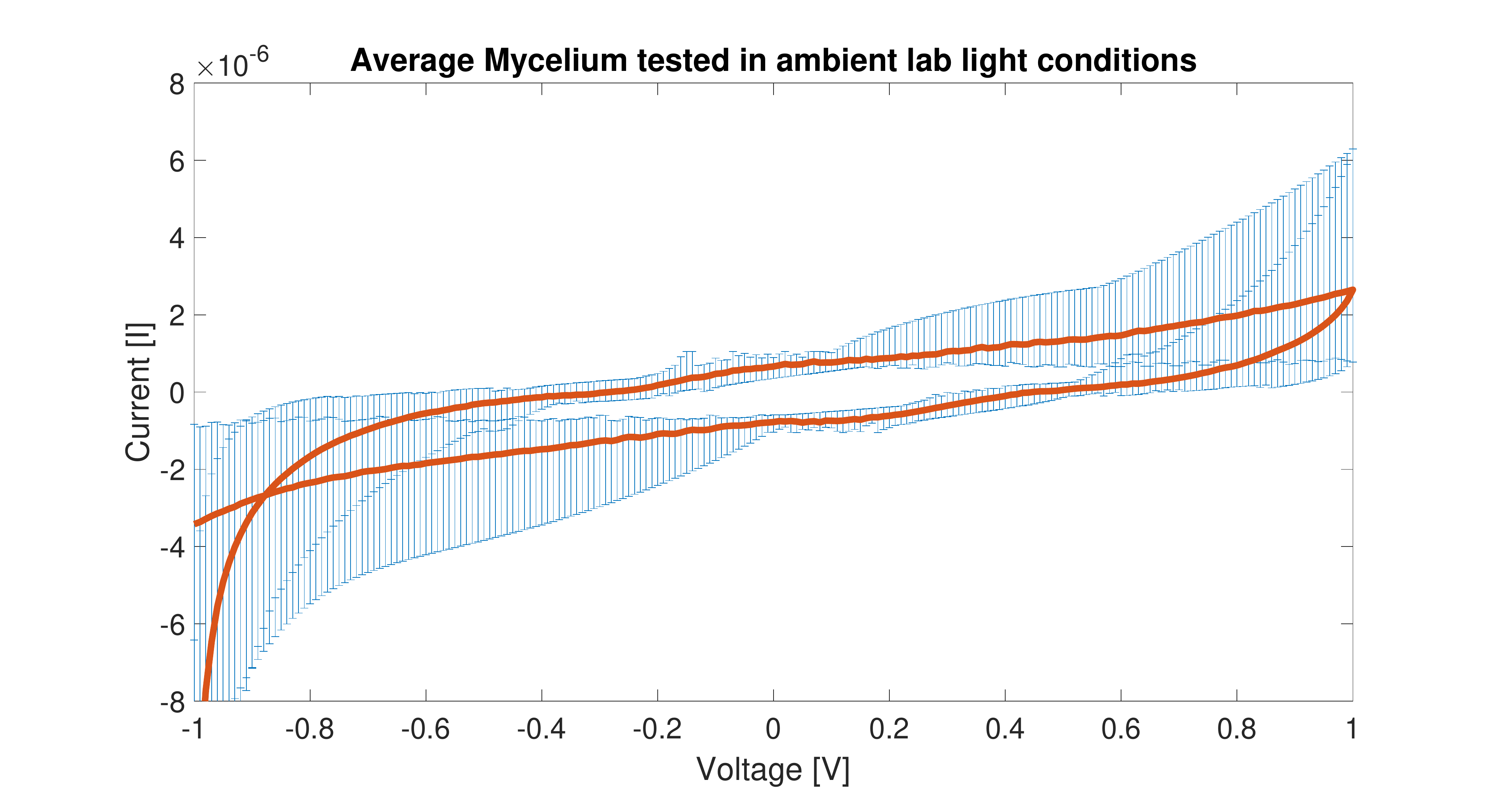}}
    \caption{Average mycelium I-V characteristics. (a) cyclic voltammetry performed from -0.5V to 0.5V. (b) cyclic voltammetry performed from -1V to 1V.}
    \label{fig:mycelium_ave}
\end{figure}

\begin{figure}
    \centering
    \subfigure[]{\includegraphics[width=0.49\textwidth]{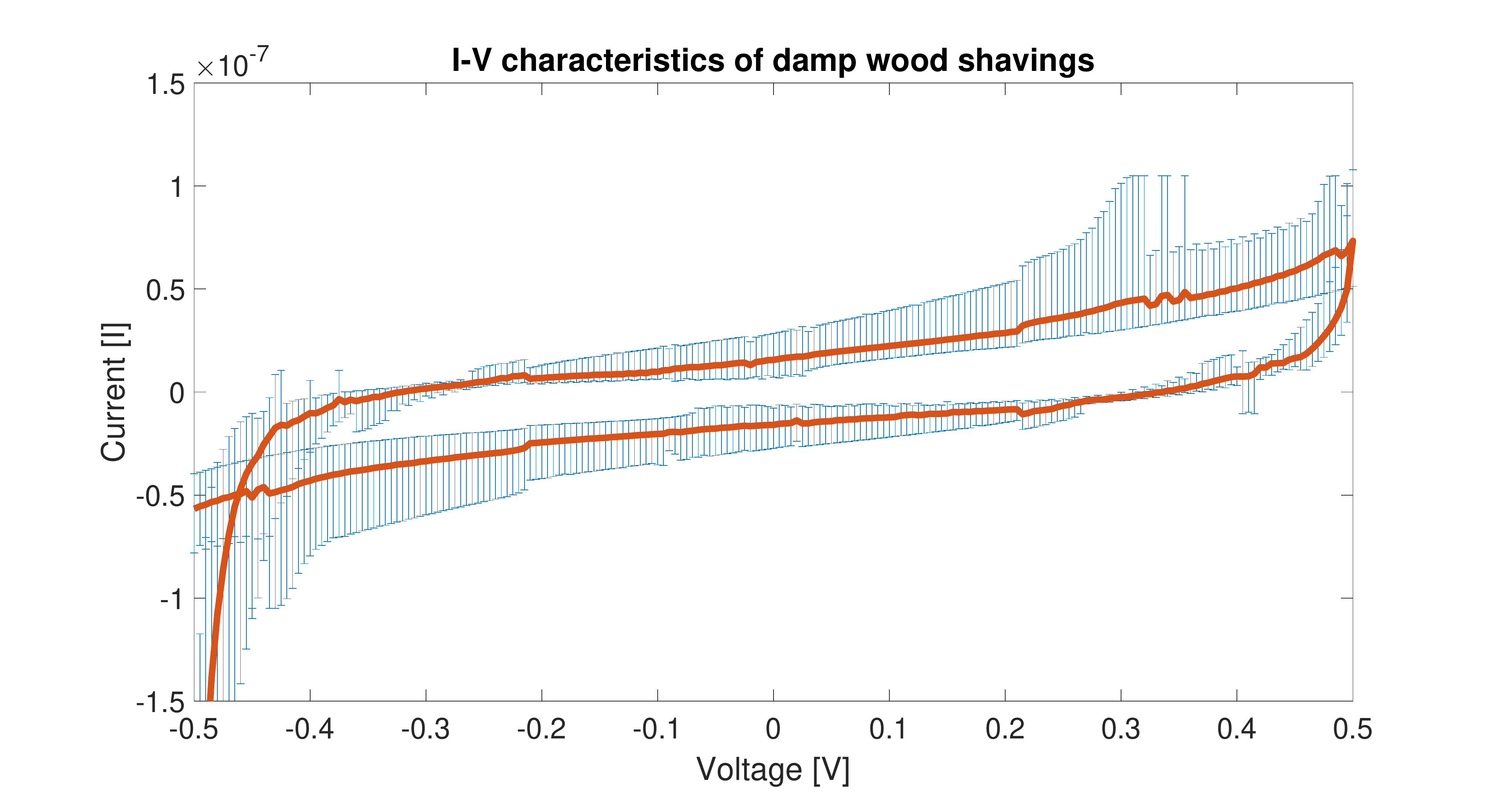}}
    \subfigure[]{\includegraphics[width=0.49\textwidth]{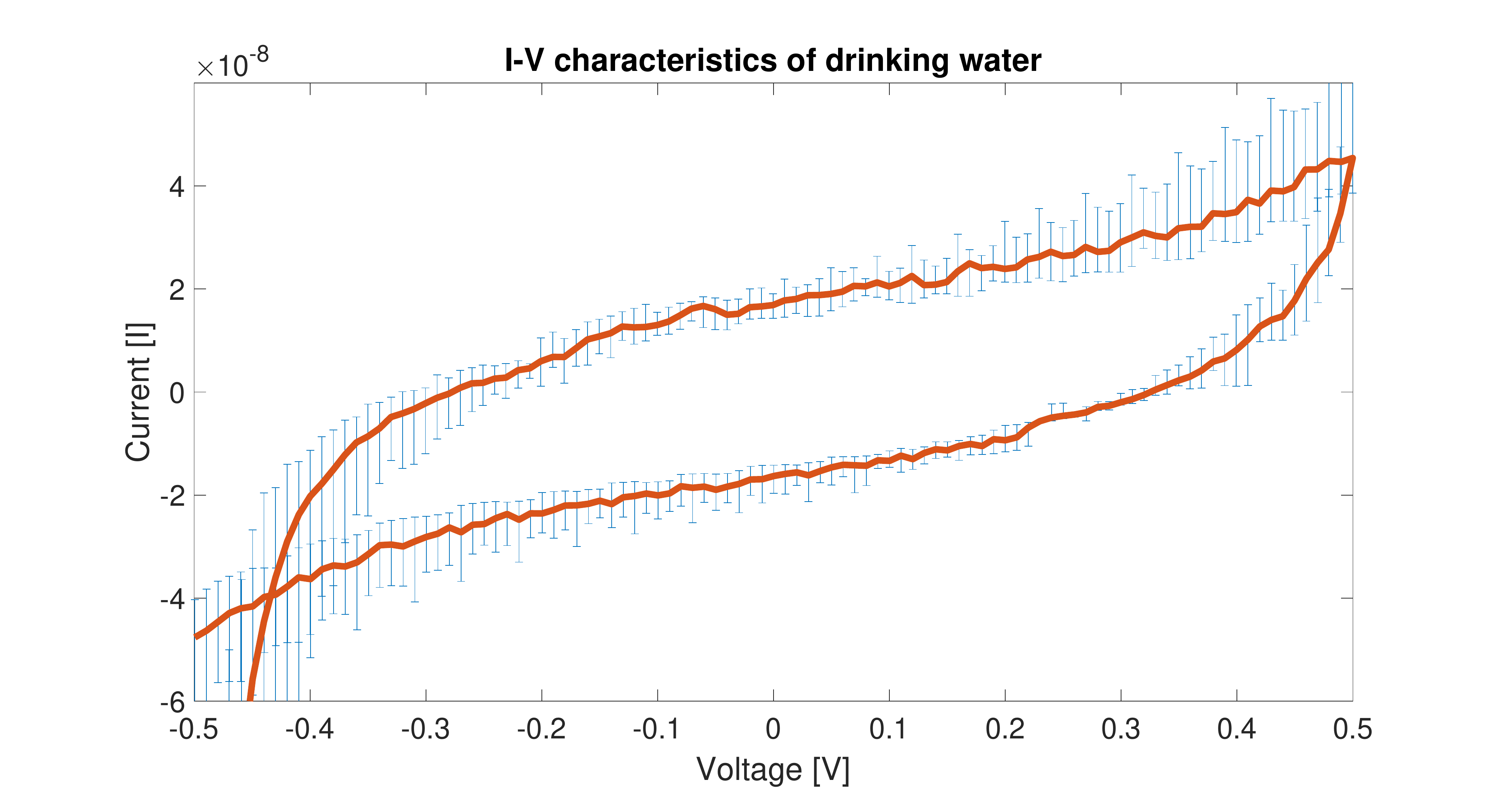}}
    \subfigure[]{\includegraphics[width=0.49\textwidth]{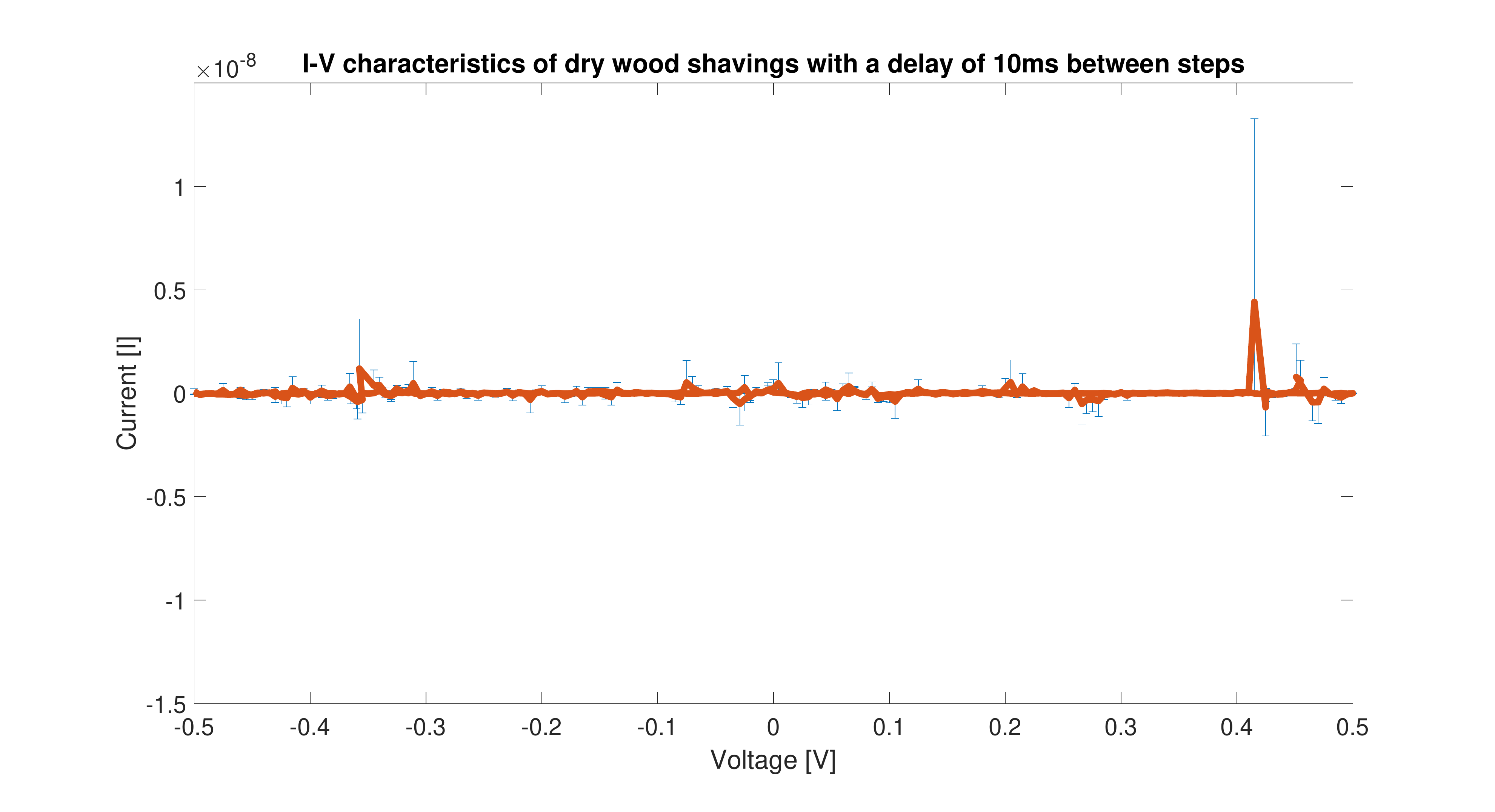}
    \label{fig:dry_shavings}}
    \caption{Average I-V characterisation of control mediums. Cyclic voltammetry performed over -0.5V to 0.5V. (a) damp wood shavings. (b) Bristol tap water sample. (c) dry wood shavings}
    \label{fig:controls}
\end{figure}

This study also covered the use of mycelium as a memristor, Fig.~\ref{fig:mycelium_ave}. The mycelium exhibits the same memristive properties as the other fruiting bodies and flora. However, the mycelium is cultivated on dampened wood shavings. Therefore, cyclic voltammetry was conducted on both dampened wood shavings and tap water, Fig.~\ref{fig:controls}, to explore the potential memristance of the growth medium. The I-V sweeps demonstrate that the control samples also exhibit memristive properties, albeit with a lower conducted current than the mycelium. This is something that is not seen in dry wood shavings (Fig.~\ref{fig:dry_shavings}). The dry shavings respond in a similar way as an open circuit for the test set-up. It is therefore concluded that the addition of water to the growth medium provides a transport mechanism that allows the conduction of current. Tap water has a number of impurities dissolved in it that act as charge carriers, combining this with the wood shavings in a thin layer increases the conducted current compared to the tap water alone in volume. \par    

\subsection{Spiking}
\label{sec:spiking}

The cyclic voltammetry of the subject matter illustrates that periods of `spiking' (oscillations) occur in their I-V characteristics. The spiking behaviour is important as it is a classic component of devices that exhibit memristive properties~\cite{gale2014emergent,prezioso2016spiking,serrano2013proposal}.
By way of example, MATLAB was used to detect the occurrence of the spikes in I-V traces from some of the samples that were tested and the results are shown below.

Figure~\ref{fig:spikes_aubergine} shows the spiking density from the aubergine and Fig.~\ref{fig:spikes_echeveria} shows the spiking density from the plant \emph{Echeveria pulidonis}. It is clear that spiking tends to occur over sections of the I-V curve, for a number of periods of the oscillation (also shown in figures of I-V sweeps from Sect.~\ref{sec:results}). However, there are instances where individual spikes can occur over the waveform. These are characterised by having a larger voltage interval from other occurrences of spikes. The number of spikes, or length of an oscillation period, will vary from between different samples. The important note is that these spikes are exhibited, thereby reinforcing the the resistance of the substance is a function of the previous voltage state and frequency of the voltage swing. \par

\begin{figure}[!tbp]
    \centering
    \subfigure[]{\includegraphics[width=0.45\textwidth]{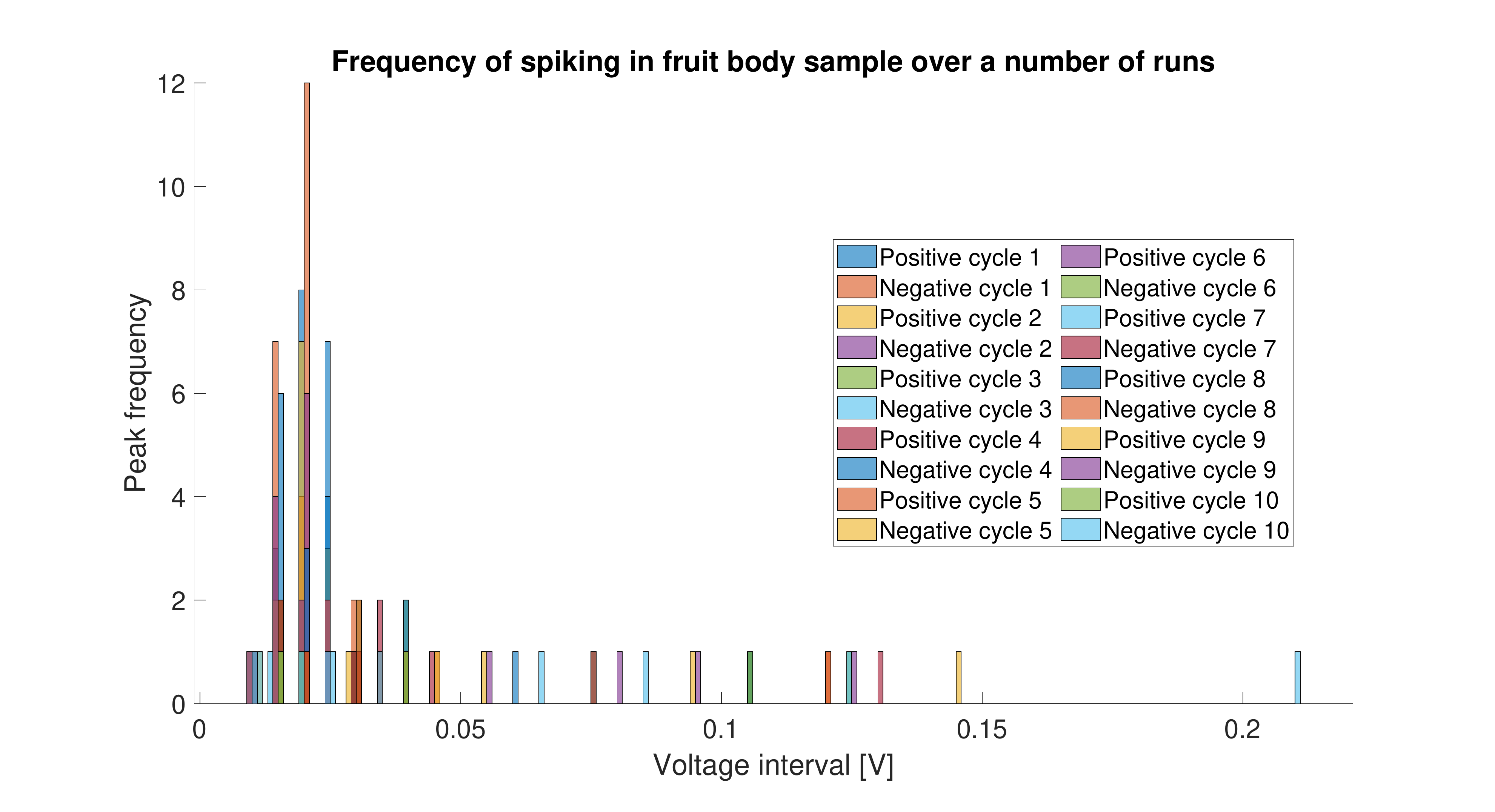}}
    \subfigure[]{\includegraphics[width=0.45\textwidth]{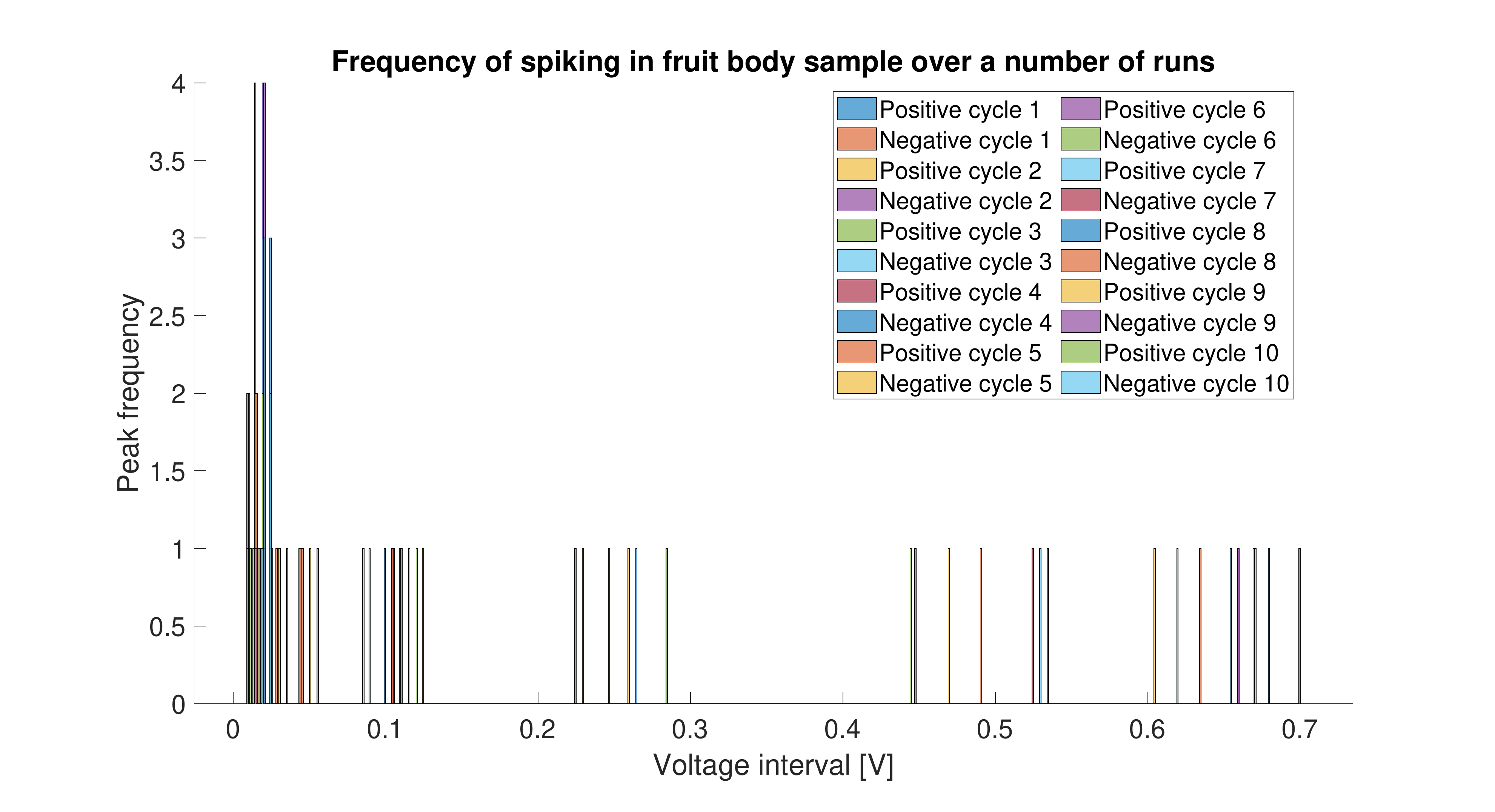}}
    \subfigure[]{\includegraphics[width=0.45\textwidth]{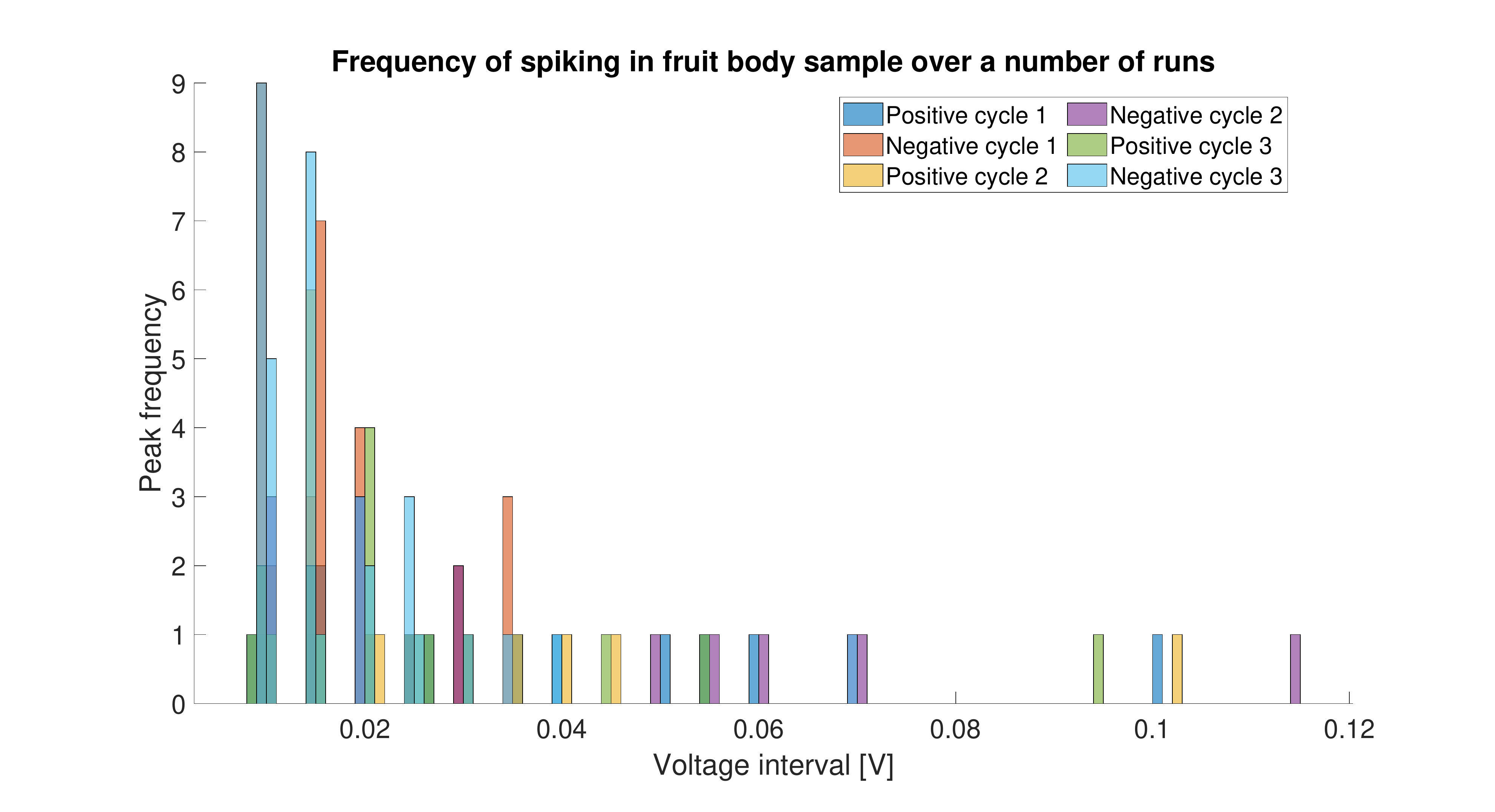}}
    \subfigure[]{\includegraphics[width=0.45\textwidth]{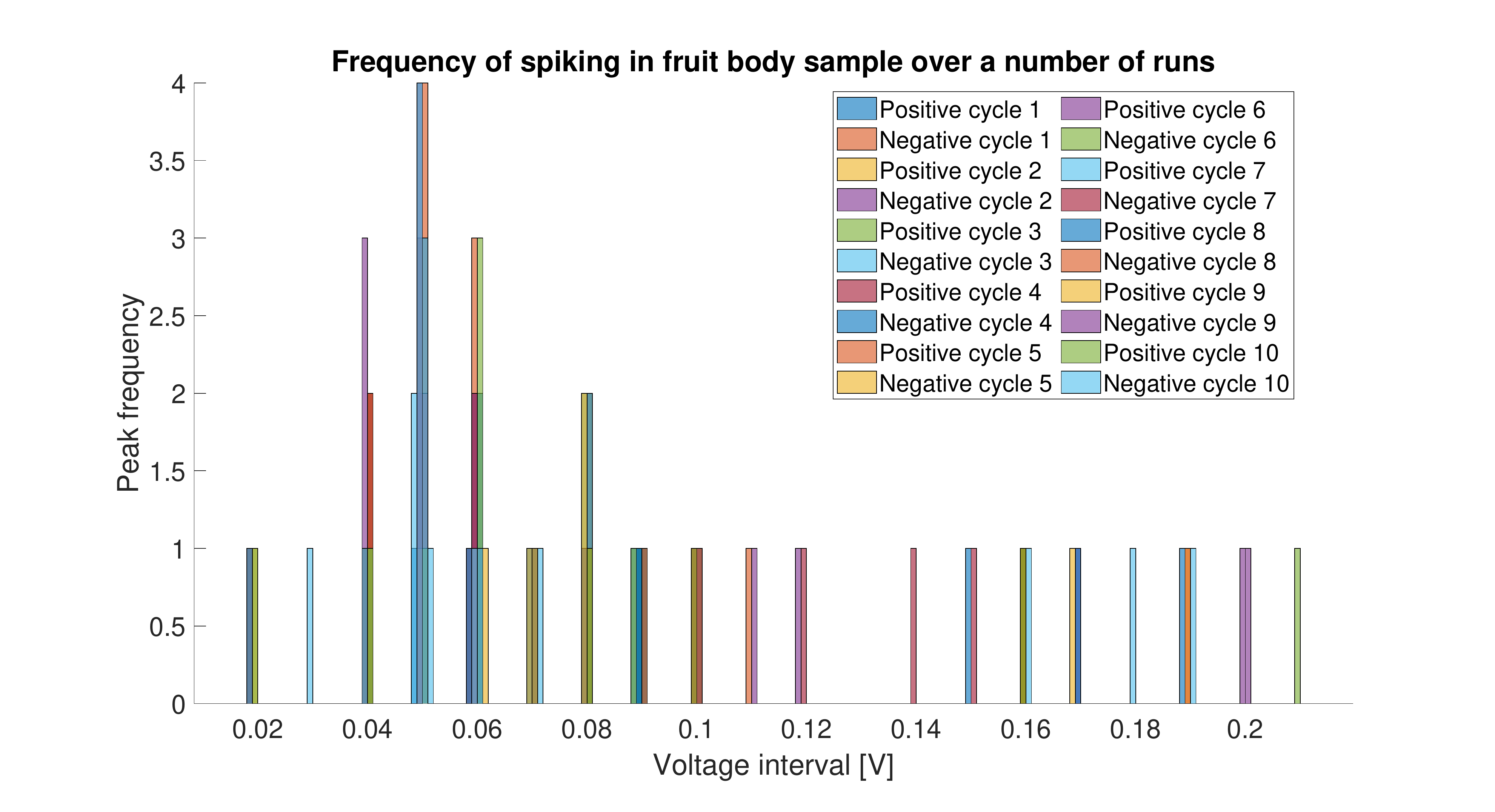}}
    \subfigure[]{\includegraphics[width=0.45\textwidth]{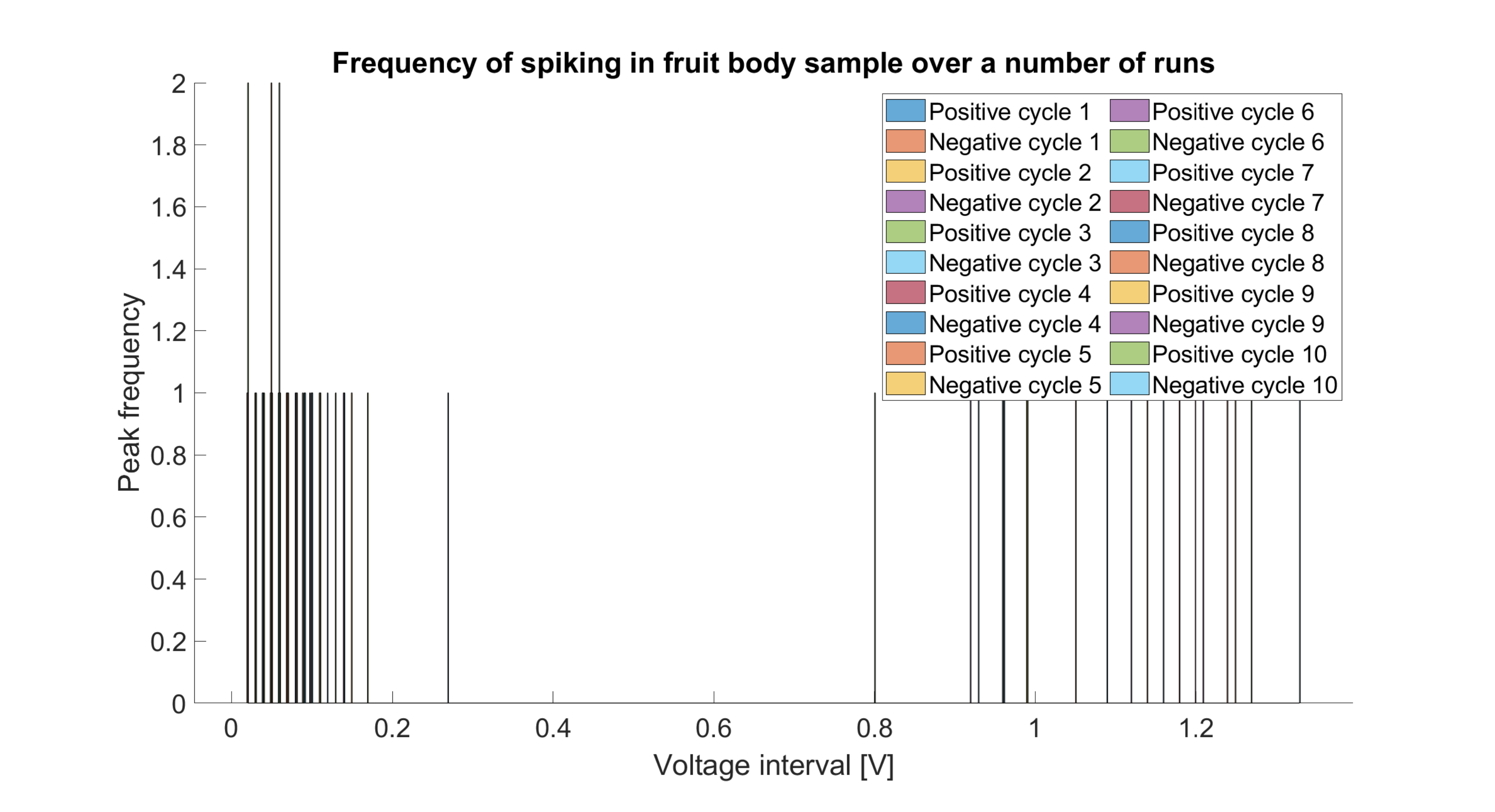}}
    \subfigure[]{\includegraphics[width=0.45\textwidth]{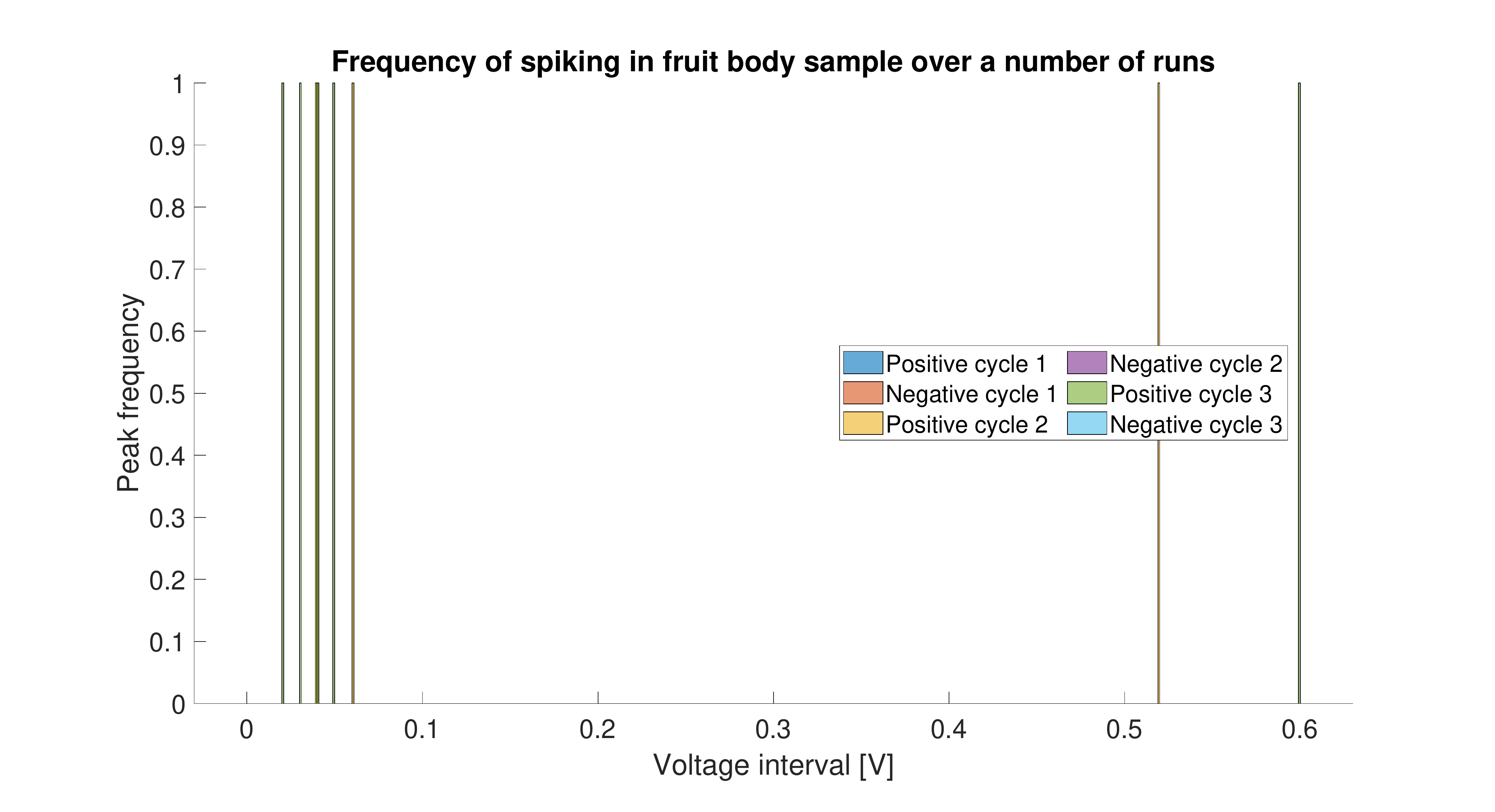}}
    \caption{Frequency of voltage interval between spikes for cyclic voltammetry of aubergine under the following conditions: (a) -0.5V to 0.5V, sample delay 10ms, (b) -0.5V to 0.5V, sample delay 100ms, (c) -0.5V to 0.5V, sample delay 1000ms, (d) -1V to 1V, sample delay 10ms, (e) -1V to 1V, sample delay 100ms, (f) -1V to 1V, sample delay 1000ms. }
    \label{fig:spikes_aubergine}
\end{figure}

\begin{figure}[!tbp]
    \centering
    \subfigure[]{\includegraphics[width=0.45\textwidth]{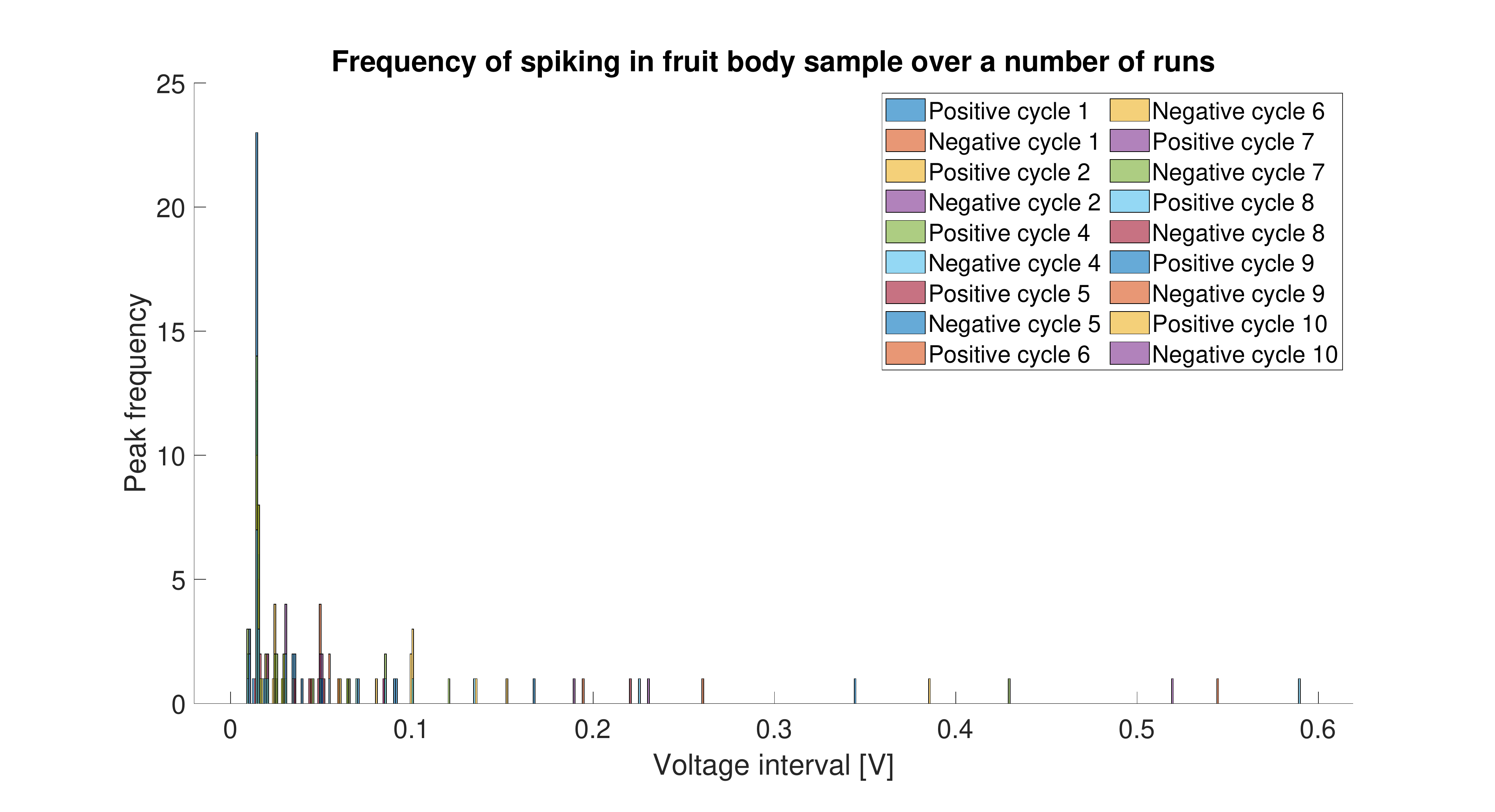}}
    \subfigure[]{\includegraphics[width=0.45\textwidth]{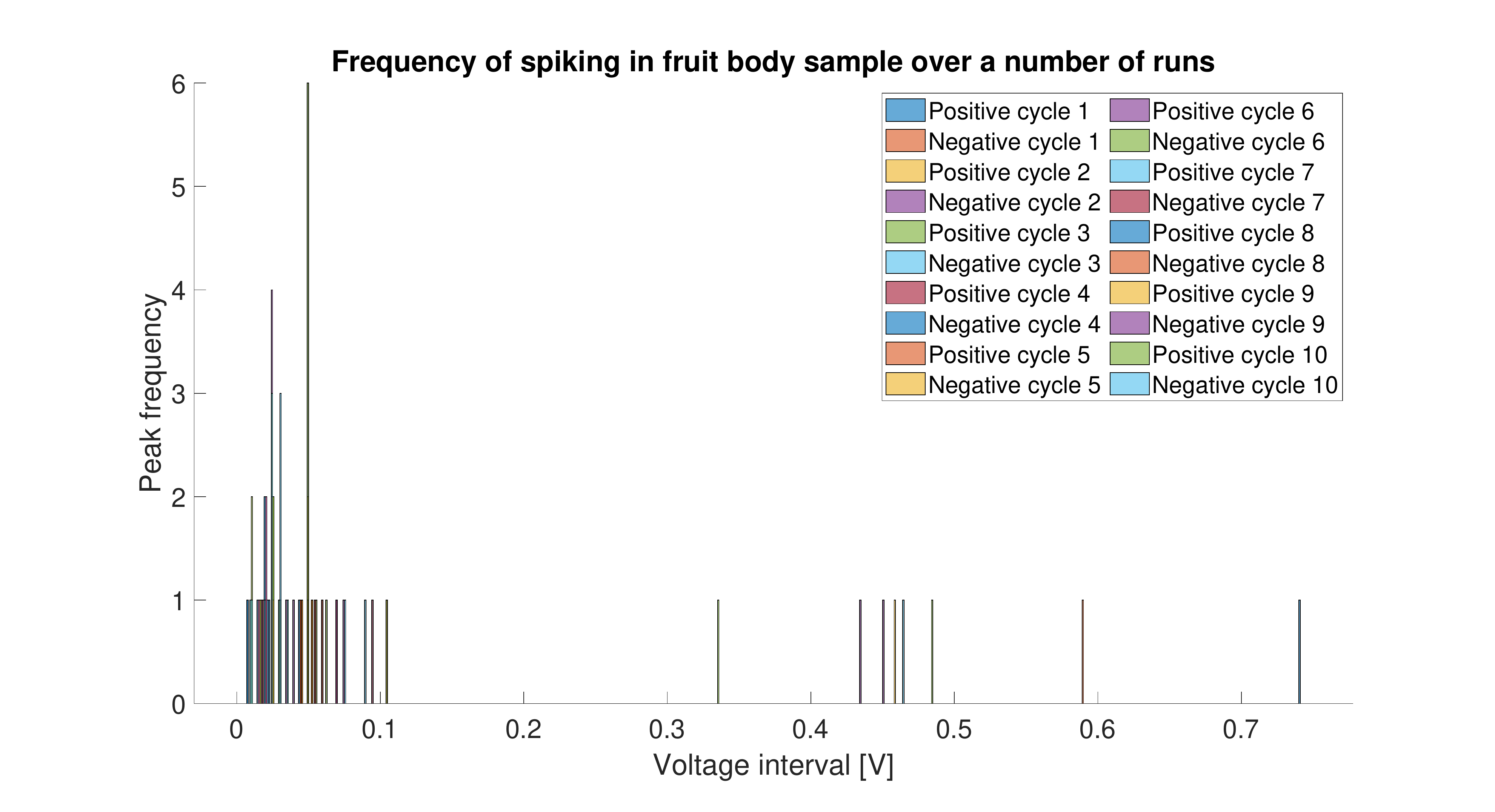}}
    \subfigure[]{\includegraphics[width=0.45\textwidth]{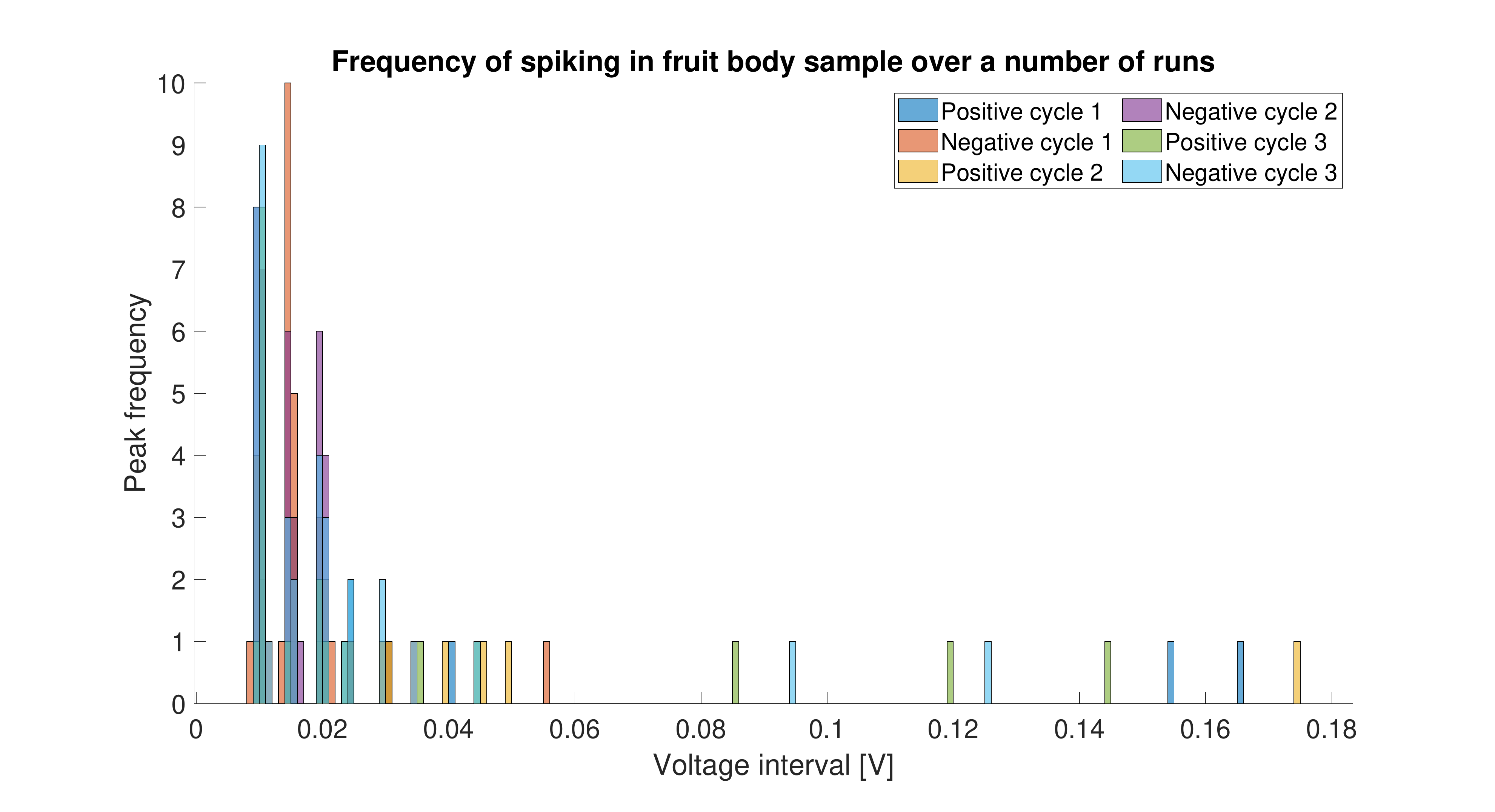}}
    \subfigure[]{\includegraphics[width=0.45\textwidth]{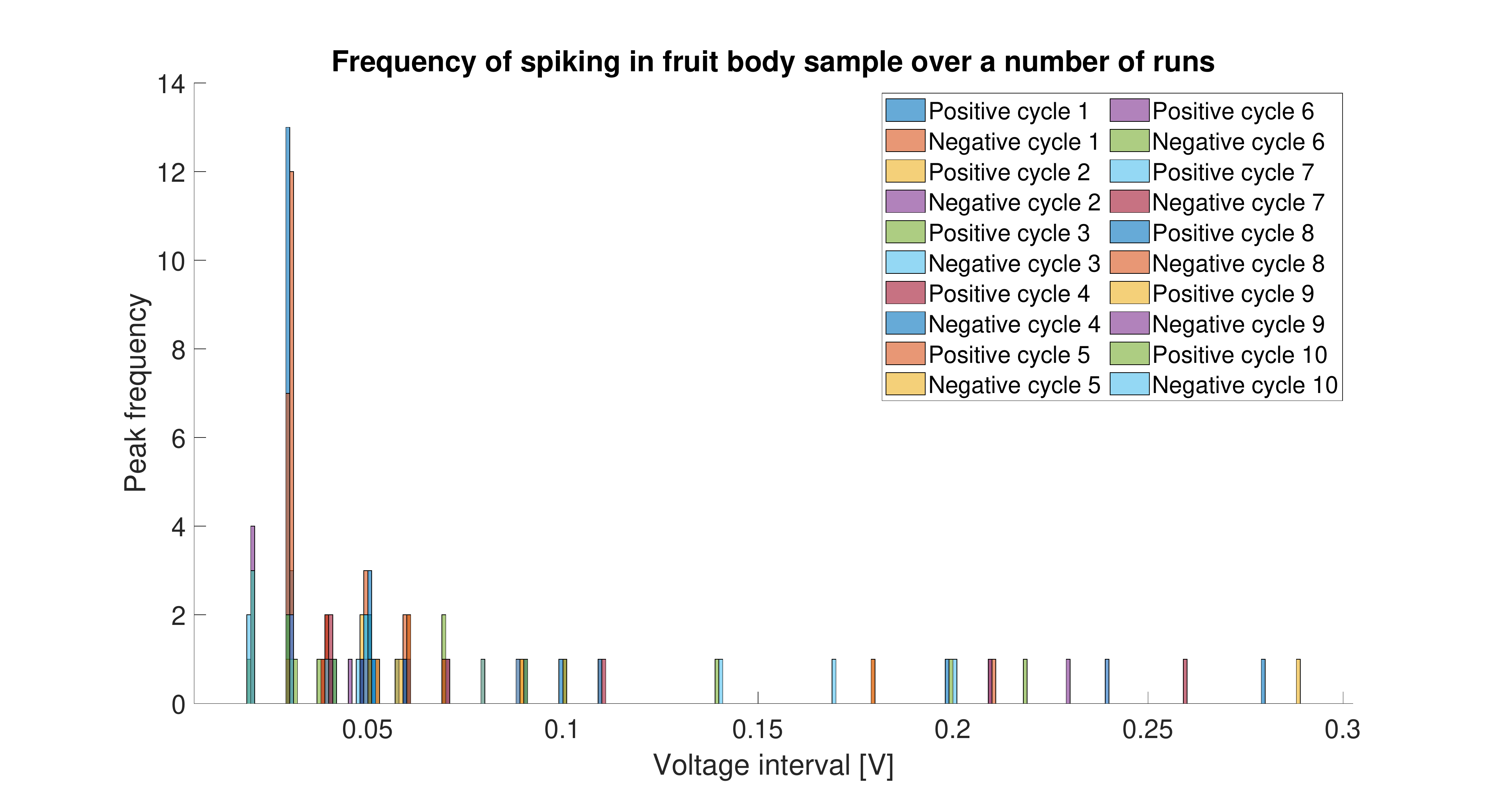}}
    \subfigure[]{\includegraphics[width=0.45\textwidth]{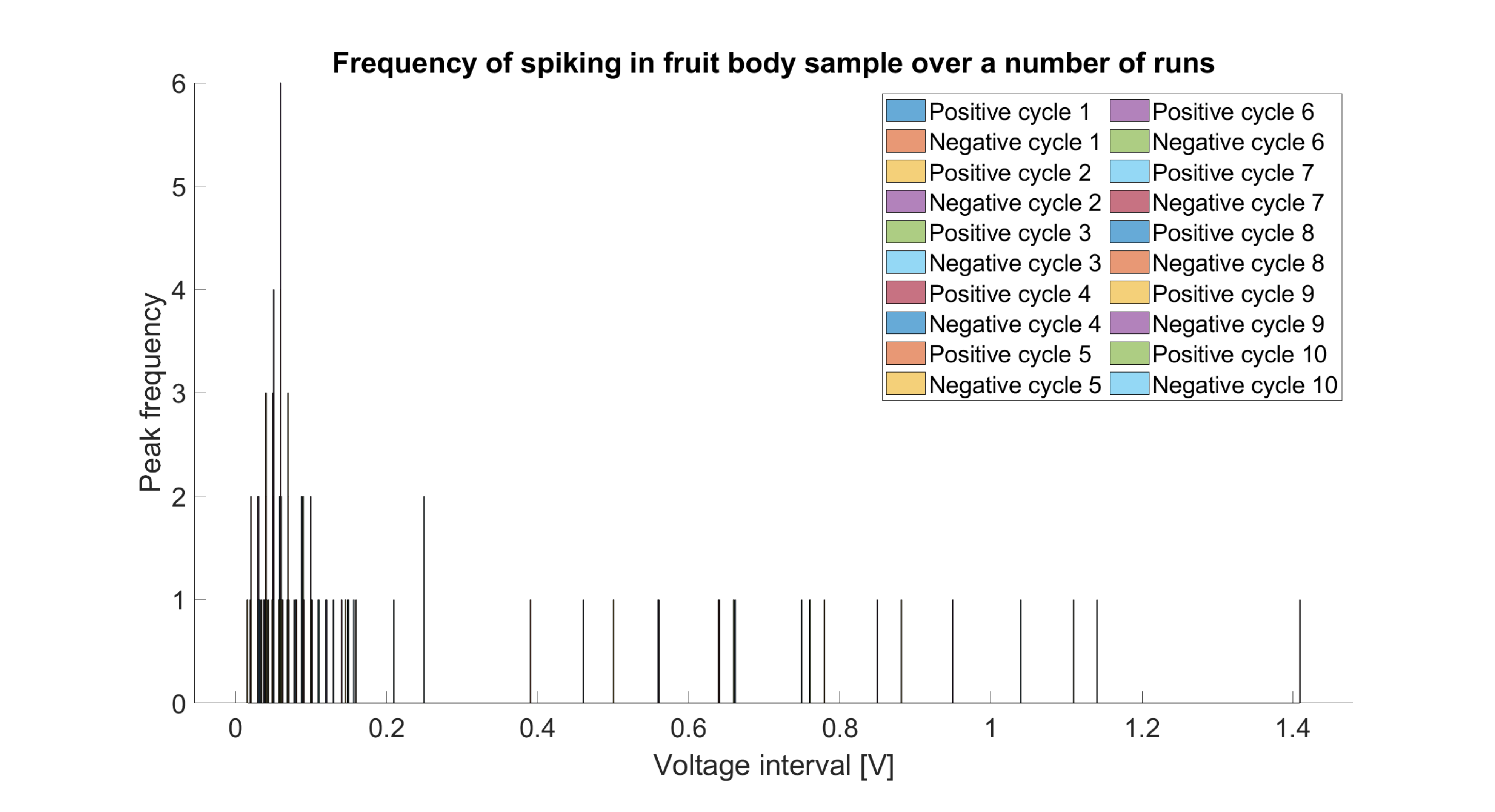}}
    \subfigure[]{\includegraphics[width=0.45\textwidth]{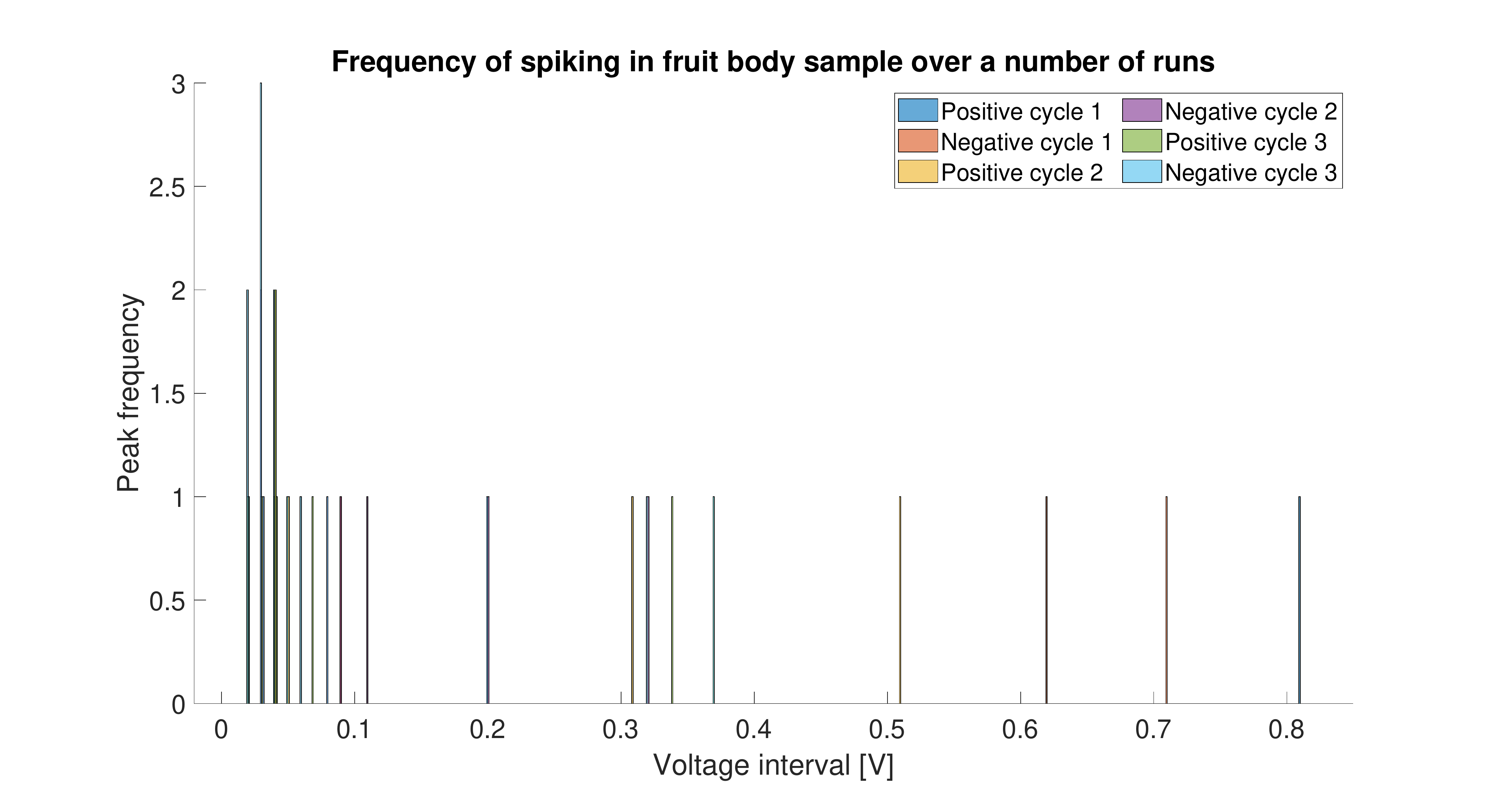}}
    \caption{Frequency of voltage interval between spikes for cyclic voltammetry of \emph{Echeveria pulidonis} under the following conditions: (a) -0.5V to 0.5V, sample delay 10ms, (b) -0.5V to 0.5V, sample delay 100ms, (c) -0.5V to 0.5V, sample delay 1000ms, (d) -1V to 1V, sample delay 10ms, (e) -1V to 1V, sample delay 100ms, (f) -1V to 1V, sample delay 1000ms }
    \label{fig:spikes_echeveria}
\end{figure}

\newpage

\section{Modelling Onion Memfractance}
\label{sec:model}

Here we report the I-V characteristics of onion (-0V5 to 0V5) with a delay of 10ms between steps (Fig.~\ref{fig:RMS18}). It is evident from the results that onion displays memristive behaviour. Although this vegetable typically does not demonstrate the `pinching' property of an ideal memristor~\cite{chua2014if}, it can be clearly seen that the biological matter exhibits memory properties when the electrical potential across the substrate is swept. A positive sweep yields a higher magnitude current when the applied voltage is positive; and a smaller magnitude current when the applied voltage is negative.\par
Fractional Order Memory Elements (FOME) are proposed as a combination of Fractional Order Mem-Capacitors (FOMC) and Fractional Order Mem-Inductors (FOMI)~\cite{Abdelouahab2014memfractance}. The FOME (1) is based on the generalised Ohm's law and parameterised as follows: $\alpha_1$, $\alpha_2$ are arbitrary real numbers which satisfy $ 0 \leq \alpha_1,\alpha_2 \leq 2$ and models the solution space by~\cite{Khalil2019Fractional}, $F_M^{\alpha_1,\alpha_2}$ is the memfractance, $q(t)$ is the time dependent charge, $\varphi(t)$ is the time dependent flux. Therefore, the memfractance $(F_M^{\alpha_1,\alpha_2}$ is an interpolation between four points: \emph{MC} --- memcapacitance, $R_M$ - memristor, \emph{MI} --- meminductance, and $R_{2M}$ - the second order memristor. Full derivations for the generalised FOME model are given by~\cite{Abdelouahab2014memfractance,Khalil2019Fractional}. The definition of memfractance can be straightforward generalised to any value of $\alpha_1$, $\alpha_2$ (see [1, Fig. 27]).\par

\begin{equation}
    D_t^{\alpha_1}\varphi(t) = F_M^{\alpha_1, \alpha_2}(t) D_t^{\alpha_2} q(t)
\end{equation}

The appearance of characteristics from various memory elements in the onion I-V curves supports the assertion that the onion is a memfractor where $\alpha_1$ and $\alpha_2$ are both greater than 0 and less than 2.\par
There is no biological reason for memfractance of onion, be a usual closed formula. Therefore, one can get only a mathematical approximation of this function. In this section, we propose two alternatives to obtain the best approximation for memfractance in the case of onion I-V characteristics for averaged cyclic voltammetry of Fig.~\ref{fig:RMS18}.\par

\subsection{Single Polynomial Approximation on the whole interval}

Raw data include the time, voltage and intensity of each reading characteristics of onion with a delay of 10\,ms between steps. There are 401 readings for each run. The process of these data, in order to obtain a mathematical approximation of memfractance, in the first alternative, takes 4 steps as follows. First step: approximate $v(t)$ by a thirty-degree polynomial.\par
\begin{figure}
    \centering
    \includegraphics[width=\textwidth]{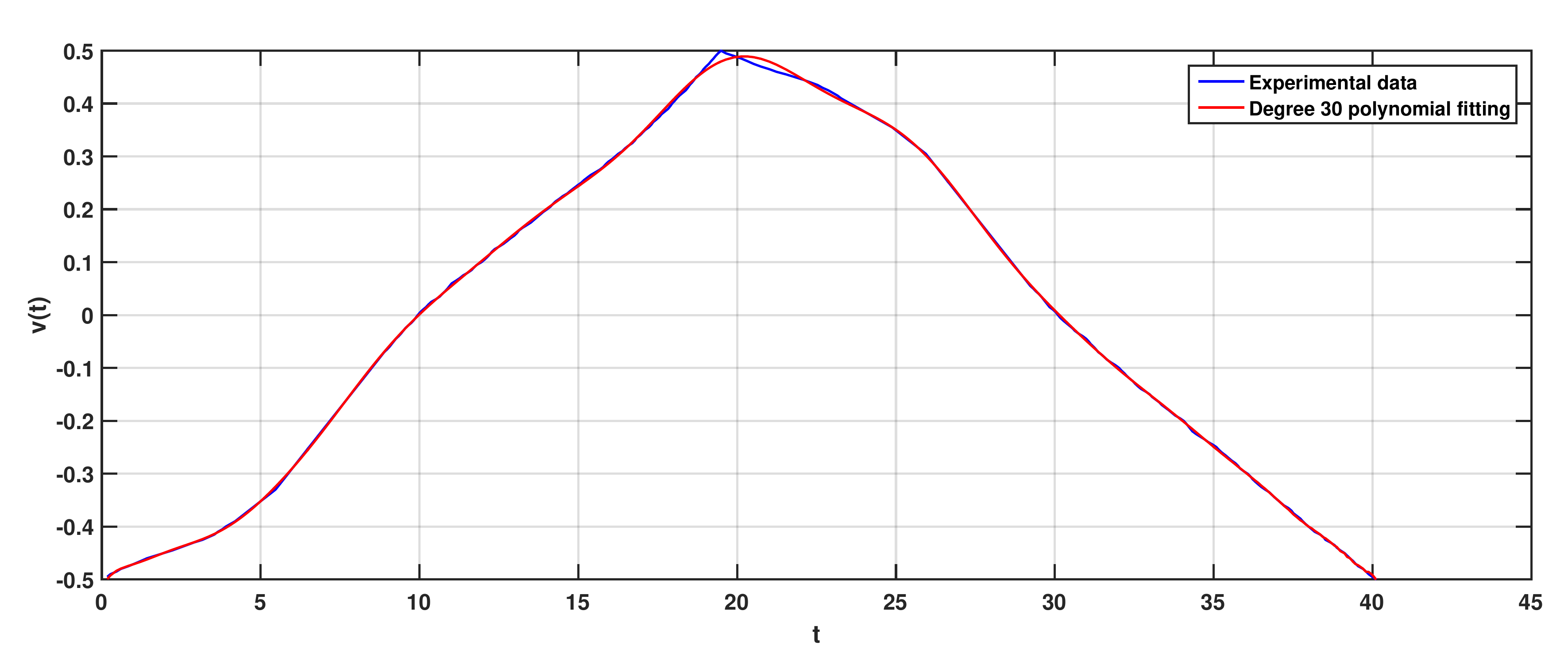}
    \caption{Voltage versus time and its approximation by a 30-degree polynomial.}
    \label{fig:RMS8}
\end{figure}

First step: approximate v(t) by a thirty-degree polynomial (Fig.~\ref{fig:RMS8}) whose coefficients are given in table~\ref{tab:1}.\par

\begin{equation}
    v(t) \approx P(t) = \sum_{j=0}^{j=30}a_jt^j
\end{equation}

\begin{table}[!tbp]
    \centering
    \caption{Coefficient of P(t)}
    \begin{tabular}{|c|c|c|c|}
        \hline 
        $a_0$ & -1.94717918941007e-37 & $a_{16}$ & 2.12544609190968e-10 \\ \hline
        $a_1$ & 7.05370777962793e-35 & $a_{17}$ & -7.02417186413251e-09 \\ \hline
        $a_2$ & -1.09608654138843e-32 & $a_{18}$ & 1.49798679659499e-07 \\ \hline
        $a_3$ & 9.17166664784253e-31 & $a_{19}$ & -2.25915349408741e-06 \\ \hline
        $a_4$ & -3.96814860043400e-29 & $a_{20}$ & 2.48111180885677e-05 \\ \hline
        $a_5$ & 2.72250233614171e-28 & $a_{21}$ & -0.000198753598325262 \\ \hline
        $a_6$ & 6.27868812867892e-26 & $a_{22}$ & 0.00113701417453570 \\ \hline
        $a_7$ & -3.22386131288943e-24 & $a_{23}$ & -0.00438858300276266 \\ \hline
        $a_8$ & 5.35491902163618e-23 & $a_{24}$ & 0.00970067769202458 \\ \hline
        $a_9$ & 5.67150904115691e-22 & $a_{25}$ & -0.00277217499616736 \\ \hline
        $a_{10}$ & -2.10303480491708e-20 & $a_{26}$ & -0.0499326832665322 \\ \hline
        $a_{11}$ & -1.15343437980144e-19 & $a_{27}$ & 0.145011436255266 \\ \hline
        $a_{12}$ & -3.79360135264957e-17 & $a_{28}$ & -0.184313854027240 \\ \hline
        $a_{13}$ & 3.22192992558703e-15 & $a_{29}$ & 0.131713253328233 \\ \hline
        $a_{14}$ & -6.04224119899083e-14 & $a_{30}$ & -0.517083007531048\\ \hline
        $a_{15}$ & -2.79837373009902e-12 & & \\ \hline
    \end{tabular}
    
    \label{tab:1}
\end{table}

\begin{table}[!tpb]
    \centering
    \caption{Quality of fitness.}
    \begin{tabular}{|c|c|c|}
        \hline
         Sum of squared estimate of errors & $SSE = \sum_{j=1}^{j=n}(v_j - \hat{v_j})^2$ & 0.00433091975307024  \\ \hline
         Sum of squared residuals & $SSR = \sum_{j=1}^{j=n}(\hat{v_j} - \bar{v})^2$ & 33.3245185894048 \\ \hline
         Sum of square total & $SST = SSE + SSR$ & 33.3288495091578 \\ \hline
         Coefficient of determination & $R-square = \frac{SSR}{SST}$ &  0.999870054927882 \\ \hline
    \end{tabular}
    \label{tab:2}
\end{table}

The polynomial fits very well the experimental voltage curve, as the statistical indexes show in table~\ref{tab:2}.\par
\begin{figure}[!tpb]
    \centering
    \includegraphics[width=\textwidth]{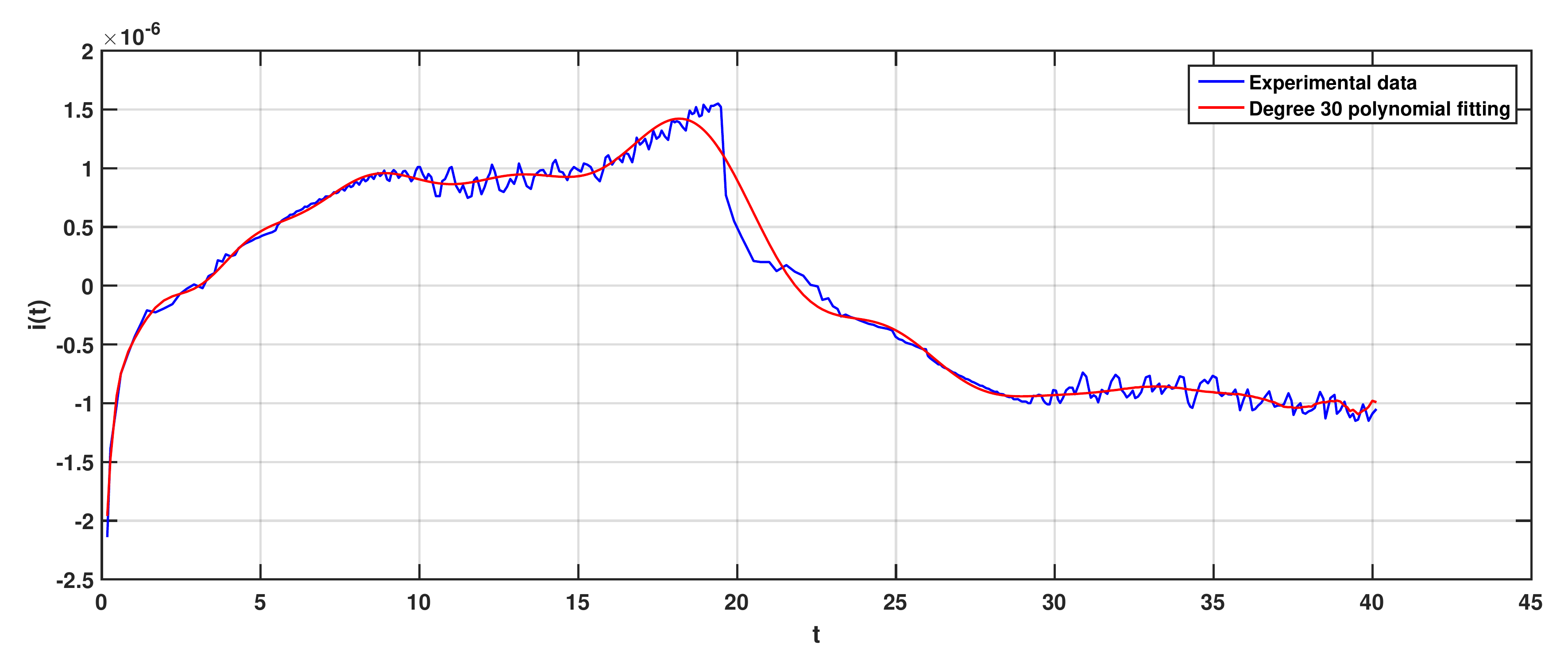}
    \caption{Current versus time and its approximation by a 30-degree polynomial.}
    \label{fig:RMS9}
\end{figure}

Step 2: in the same way approximate the current $i(t)$ using a thirty-degree polynomial (Fig.~\ref{fig:RMS9}) whose coefficients are given in Tab.~\ref{tab:3}.\par

\begin{equation}
    i(t) \approx Q(t) = \sum_{j=0}^{j=30}b_jt^j
\end{equation}

\begin{table}[!tpb]
    \centering
    \caption{Coefficient of Q(t)}
    \begin{tabular}{|c|c|c|c|}
        \hline
        $b_0$ & 6.68985942609987e-43 & $b_{16}$ & -6.07388014701807e-15 \\ \hline
        $b_1$ & -2.71853051938565e-40 & $b_{17}$ & 2.25277616546114e-13 \\ \hline
        $b_2$ & 4.74482538144499e-38 & $b_{18}$ & -5.52975396257029e-12 \\ \hline
        $b_3$ & -4.51931699470006e-36 & $b_{19}$ & 9.83125923574317e-11 \\ \hline
        $b_4$ & 2.37196020005930e-34 & $b_{20}$ & -1.31005323328773e-09 \\ \hline
        $b_5$ & -4.77095462655549e-33 & $b_{21}$ & 1.32446639560140e-08 \\ \hline
        $b_6$ & -1.72460618050188e-31 & $b_{22}$ & -1.01592865082809e-07 \\ \hline
        $b_7$ & 1.31453124268333e-29 & $b_{23}$ & 5.86034919766866e-07 \\ \hline
        $b_8$ & -3.41852207771790e-28 & $b_{24}$ & -2.50008266407625e-06 \\ \hline
        $b_9$ & 1.33534616806178e-26 & $b_{25}$ & 7.69320497766088e-06 \\ \hline
        $b_{10}$ & -1.05754298813303e-24 & $b_{26}$ & -1.65013726134833e-05 \\ \hline
        $b_{11}$ & 4.06179631752752e-23 & $b_{27}$ & 2.35982843687423e-05 \\ \hline
        $b_{12}$ & 5.95988497909884e-23 & $b_{28}$ & -2.12898753184482e-05  \\ \hline
        $b_{13}$ & -5.64419454512205e-20 & $b_{29}$ & 1.14379593758435e-05 \\ \hline
        $b_{14}$ & 1.13468538605641e-18 & $b_{30}$ & -3.40155939059945e-06 \\ \hline
        $b_{15}$ & 7.60350529110369e-17 & &  \\ \hline
    \end{tabular}
    
    \label{tab:3}
\end{table}

\begin{table}[!tpb]
    \centering
    \caption{Quality of fitness.}
    \begin{tabular}{|c|c|}
        \hline 
        Sum of squared estimate of errors SSE & 2.97429625769270e-12 \\ \hline
        Sum of squared residuals SSR & 3.15876438461950e-10 \\ \hline 
        Sum of square total  SST & 3.18850734719642e-10 \\ \hline
        Coefficient of determination  R-square & 0.990671822474212 \\ \hline
    \end{tabular}

    \label{tab:4}
\end{table}

Again, the polynomial fits well the experimental intensity curve, as displayed in Tab.~\ref{tab:4}.\par
Step 3: From (1) used under the following form when $D_t^{\alpha_2}q(t) \neq 0$.\par

\begin{equation}
    F_M^{\alpha_1,\alpha_2}(t) = \frac{D_t^{\alpha_1}\varphi(t)}{D_t^{\alpha_2}q(t)}
\end{equation}
and the Rieman-Liouville fractional derivative defined by~\cite{FDEPodlubny1999}.\par

\begin{equation}
    {^{RL}_{0}}D_{t}^{\alpha}f(t) = \frac{1}{\Gamma(m-\alpha)}\frac{d^m}{dt^m}\int_{0}^{t}(t-s)^{m - \alpha - 1}f(s)ds\text{, m - 1} < \alpha < m  
\end{equation}

together with the formula for the power function\par

\begin{equation}
    {^{RL}_{0}}D_t^{\alpha}\left(a t^{\beta}\right) = \frac{a \Gamma(\beta + 1)}{\Gamma (\beta - \alpha + 1)}t^{\beta - \alpha}\text{, }\beta > -1, \alpha > 0,
\end{equation}

we obtain the closed formula of $F_M^{\alpha_1,\alpha_2}(t)$, approximation of the true biological memfractance of onions.\par

\begin{equation}
   F_M^{\alpha_1, \alpha_2}(t) = \frac{D_t^{\alpha_1}\varphi(t)}{D_t^{\alpha_2}\varphi(t)} = \frac{{^{RL}_{0}}D_t^{\alpha_1}\sum_{j=0}^{j=30}\frac{a_j}{j+1}t^{j+1}}{{^{RL}_{0}}D_{t}^{\alpha_2}\sum_{j=0}^{j=30}\frac{b_j}{j+1}t^{j+1}} = \frac{\sum_{j=0}^{j=30}\frac{a_j\Gamma(j+1)}{\Gamma(j+2-\alpha_1)}t^{j+1-\alpha_1}}{\sum_{j=0}^{j=30}\frac{b_j\Gamma(j+1)}{\Gamma(j+2-\alpha_2)}t^{j+1-\alpha_2}}
\end{equation}

Step 4 choice of parameter $\alpha_1$ and $\alpha_2$: We are looking for the best value of these parameters in the range $(\alpha_1,\alpha_2)\in[0,2]^2$. In this goal, we are considering first the singularities of $F_M^{\alpha_1,\alpha_2}(t)$ in order to avoid their existence, using suitable values of the parameters. Secondly, we will choose the most regular approximation.\par
\begin{figure}
    \centering
    \includegraphics[width=\textwidth]{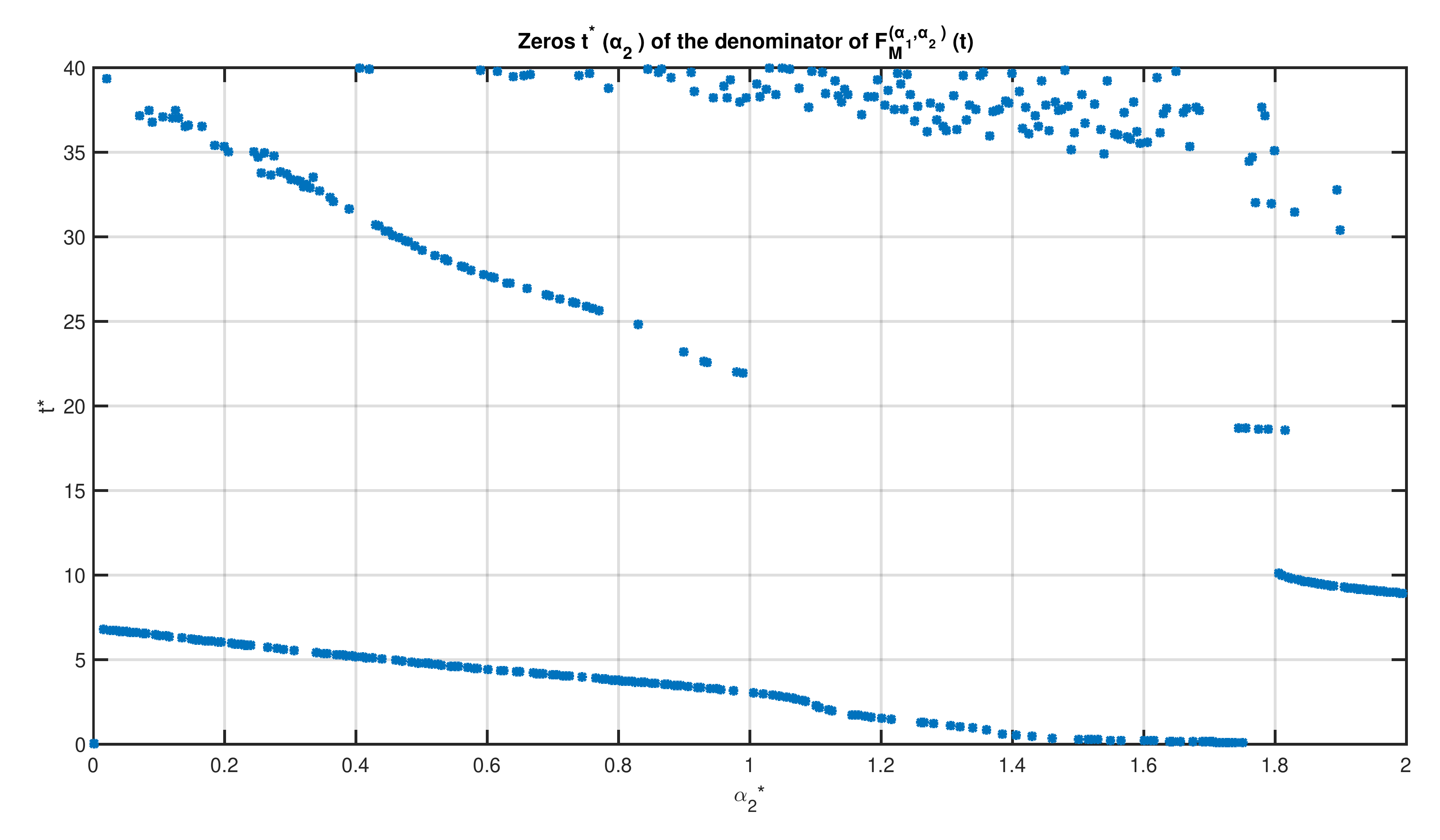}
    \caption{Zeros $t^*(\alpha_2)$ of the denominator of $F_M^{\alpha_1,\alpha_2}(t)$.}
    \label{fig:RMS10}
\end{figure}

We compute numerically, the values $t^*(\alpha_2)$  which vanish the denominator of $F_M^{\alpha_1,\alpha_2}(t)$ (Fig.~\ref{fig:RMS10}).\par

\begin{figure}[!tpb]
    \centering
    \includegraphics[width=\textwidth]{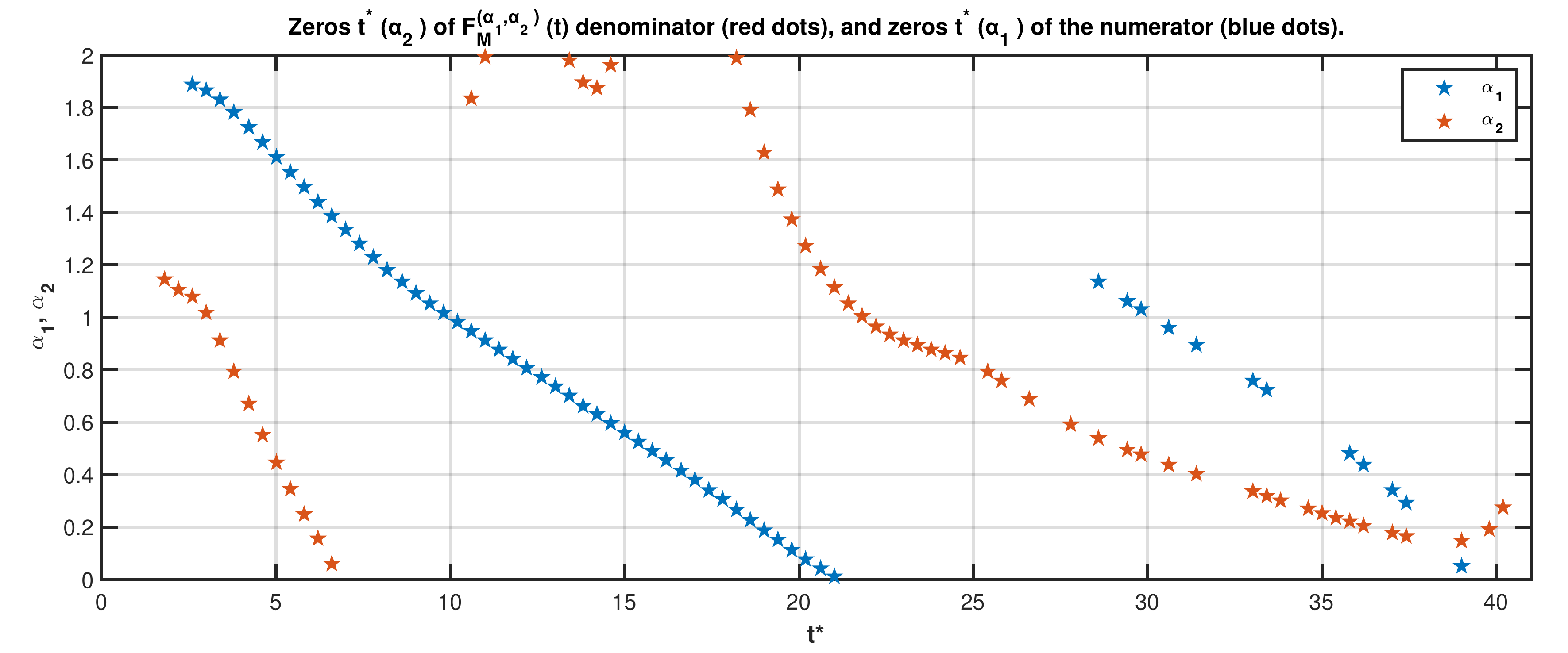}
    \caption{Zeros $t^*(\alpha_2)$ of $F_M^{\alpha_1,\alpha_2}(t)$ denominator (red dots), and zeros $t^*(\alpha_1)$ of the numerator (blue dots).}
    \label{fig:RMS11}
\end{figure}

\begin{figure}[!tpb]
    \centering
    \includegraphics[width=\textwidth]{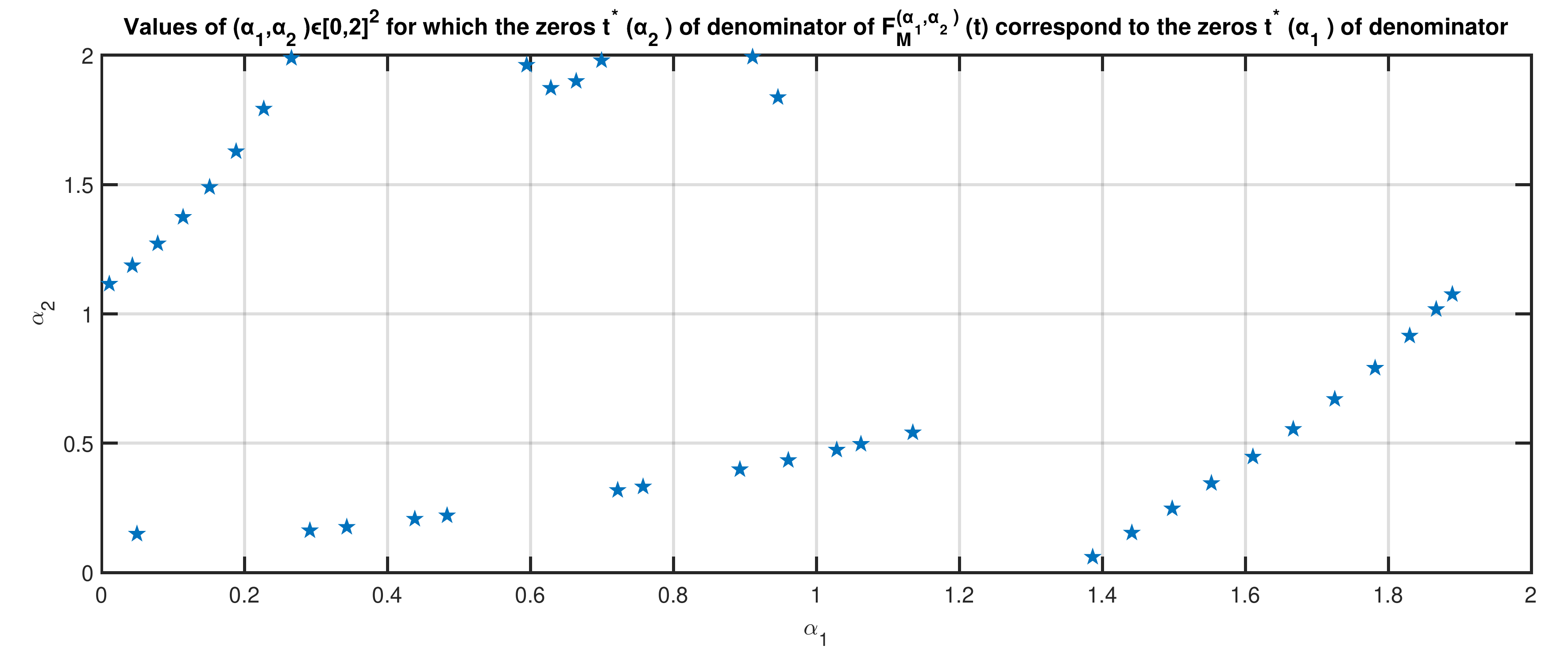}
    \caption{Values of $(\alpha_1,\alpha_2)\in[0,2]^2$ for which the zeros $t^*(\alpha_2)$ of denominator of $F_M^{\alpha_1,\alpha_2}(t)$ correspond to the zeros $t^*(\alpha_1)$ of denominator.}
    \label{fig:RMS12}
\end{figure}

We observe one, two or three coexisting solutions depending on the value of $\alpha_2$. Moreover, there is no value of $\alpha_2$ without zero of the denominator. Therefore, in order to eliminate the singularities, we need to determine the couples $(\alpha_1,\alpha_2)\in[0,2]^2$, vanishing simultaneously denominator and numerator of $F_M^{\alpha_1,\alpha_2}(t)$ (Figs.~\ref{fig:RMS11} and \ref{fig:RMS12}).\par

\begin{figure}[!tpb]
    \centering
    \includegraphics[width=\textwidth]{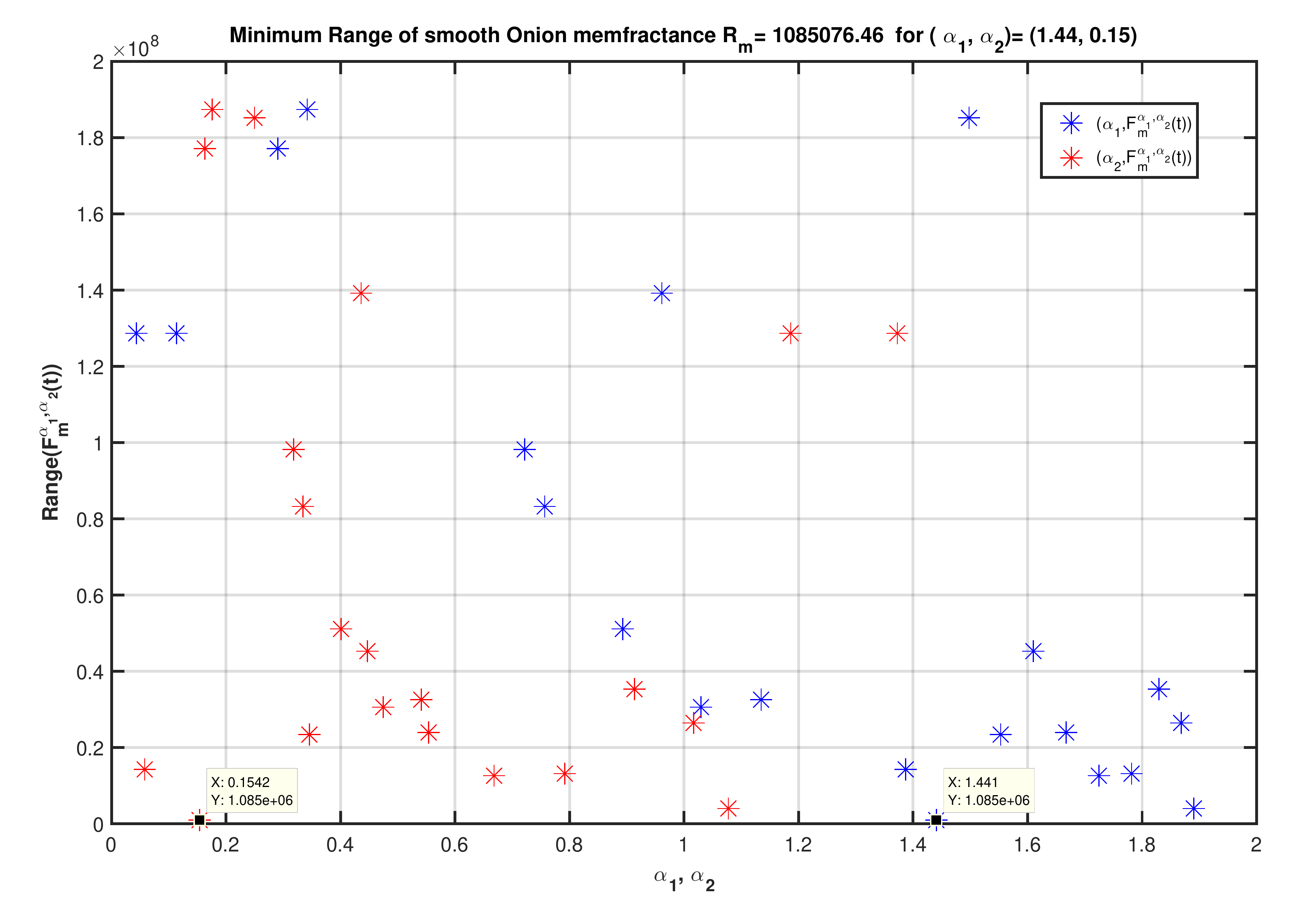}
    \caption{Values of range $(F_M^{\alpha_1,\alpha_2}(t))$  for $(\alpha_1,\alpha_2 )\in[0,2]^2.$}
    \label{fig:RMS13}
\end{figure}

In the second part of step 4, we choose the most regular approximation. We consider that the most regular approximation is the one for which the function range $(F_M^{\alpha_1,\alpha_2}(t))$ is minimal (Fig.~\ref{fig:RMS13}).\par

\begin{equation}
    \text{range}\left(F_M^{\alpha_1,\alpha_2}(t)\right) = \max_{t \in [0,41]}\left(F_M^{\alpha_1,\alpha_2}(t)\right) - \min_{t \in [0,41]}\left(F_M^{\alpha_1,\alpha_2}(t)\right)
\end{equation}

\begin{figure}
    \centering
    \includegraphics[width=\textwidth]{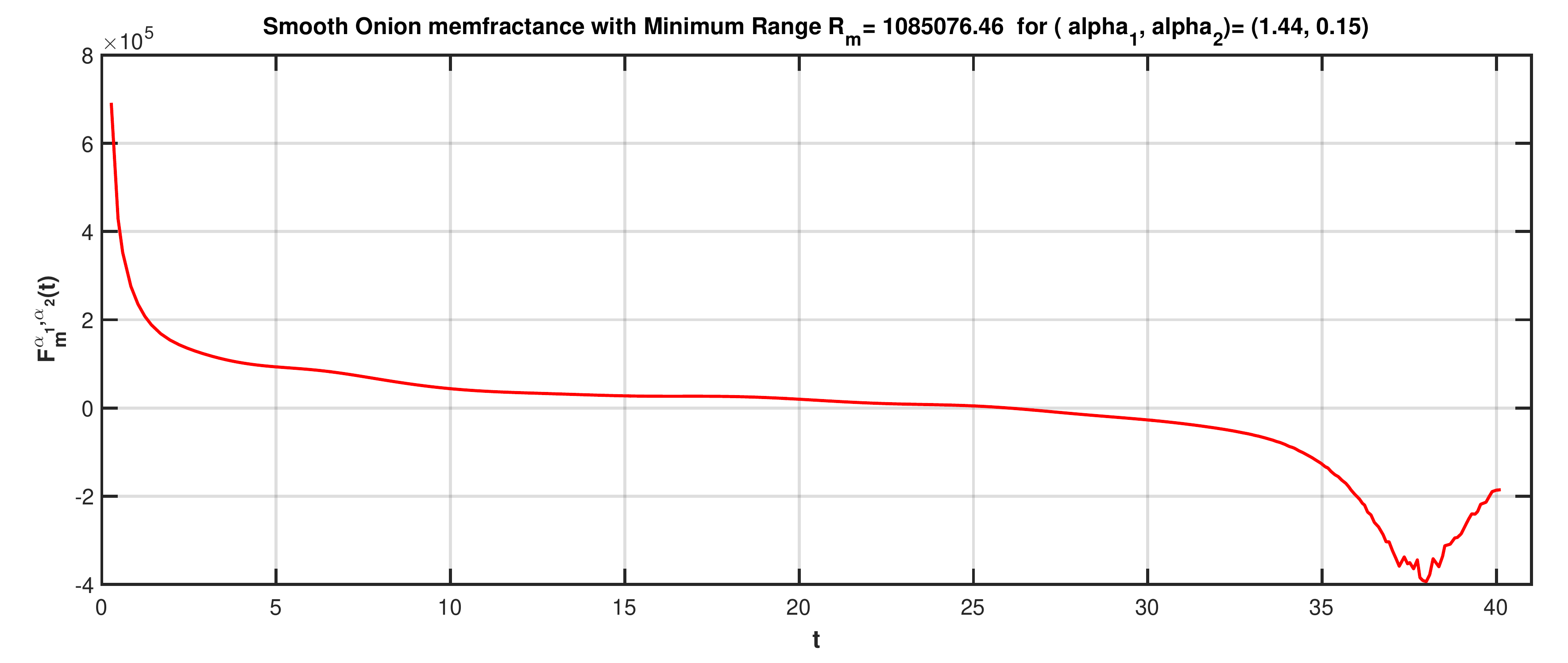}
    \caption{Memfractance for $(\alpha_1,\alpha_2)$ given in Tab.~\ref{tab:5}.}
    \label{fig:RMS14}
\end{figure}

From the numerical results, the best couple $(\alpha_1,\alpha_2)$ and the minimum range of  $F_M^{\alpha_1,\alpha_2}(t)$ are given in table~\ref{tab:5}, and the corresponding Memfractance is displayed in Fig.~\ref{fig:RMS14}.\par

\begin{table}[!tpb]
    \centering
    \caption{Minimum values of $\alpha$}
    \begin{tabular}{|c|c|c|}
        \hline
        $\alpha_1$ & $\alpha_2$ & Minimum range of $F_M^{\alpha_1, \alpha_2}(t)$ \\ \hline
        1.441224116 & 0.154232123 & 1085076.46348631 \\ \hline
    \end{tabular}
    
    \label{tab:5}
\end{table}

\begin{figure}[!tpb]
    \centering
    \includegraphics[width=\textwidth, trim=4 4 4 4,clip]{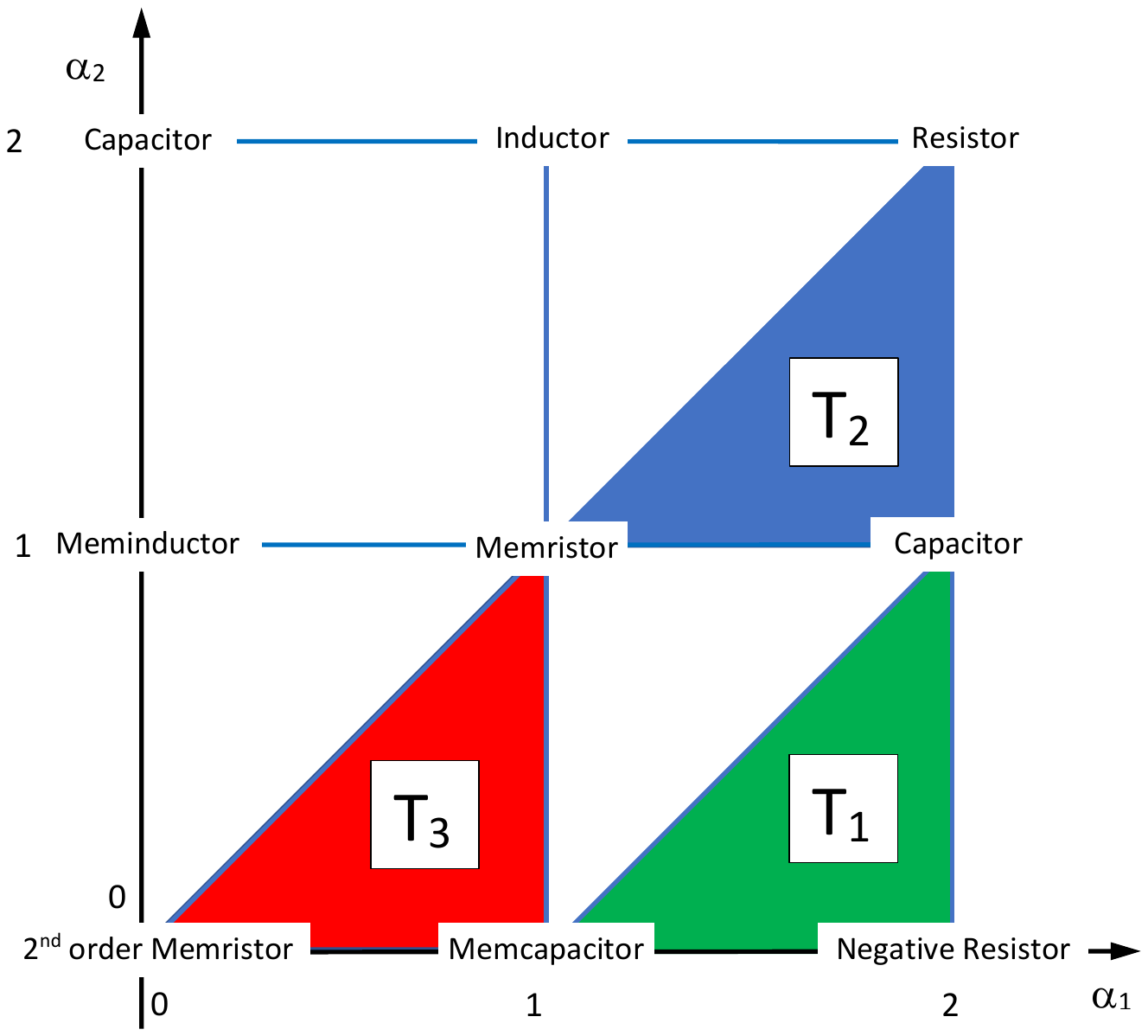}
    \caption{Non-binary solution space showing mem-fractive properties of a memory element}
    \label{fig:RMS15}
\end{figure}

The value of $(\alpha_1,\alpha_2)$ given in Table~\ref{tab:5} belongs to the triangle $T_1$ of Fig.~\ref{fig:RMS15}, whose vertices are memcapacitor, capacitor and negative-resistor, which means that Onion is like a mix of such basic electronic devices.\par

\begin{figure}
    \centering
    \includegraphics[width=\textwidth]{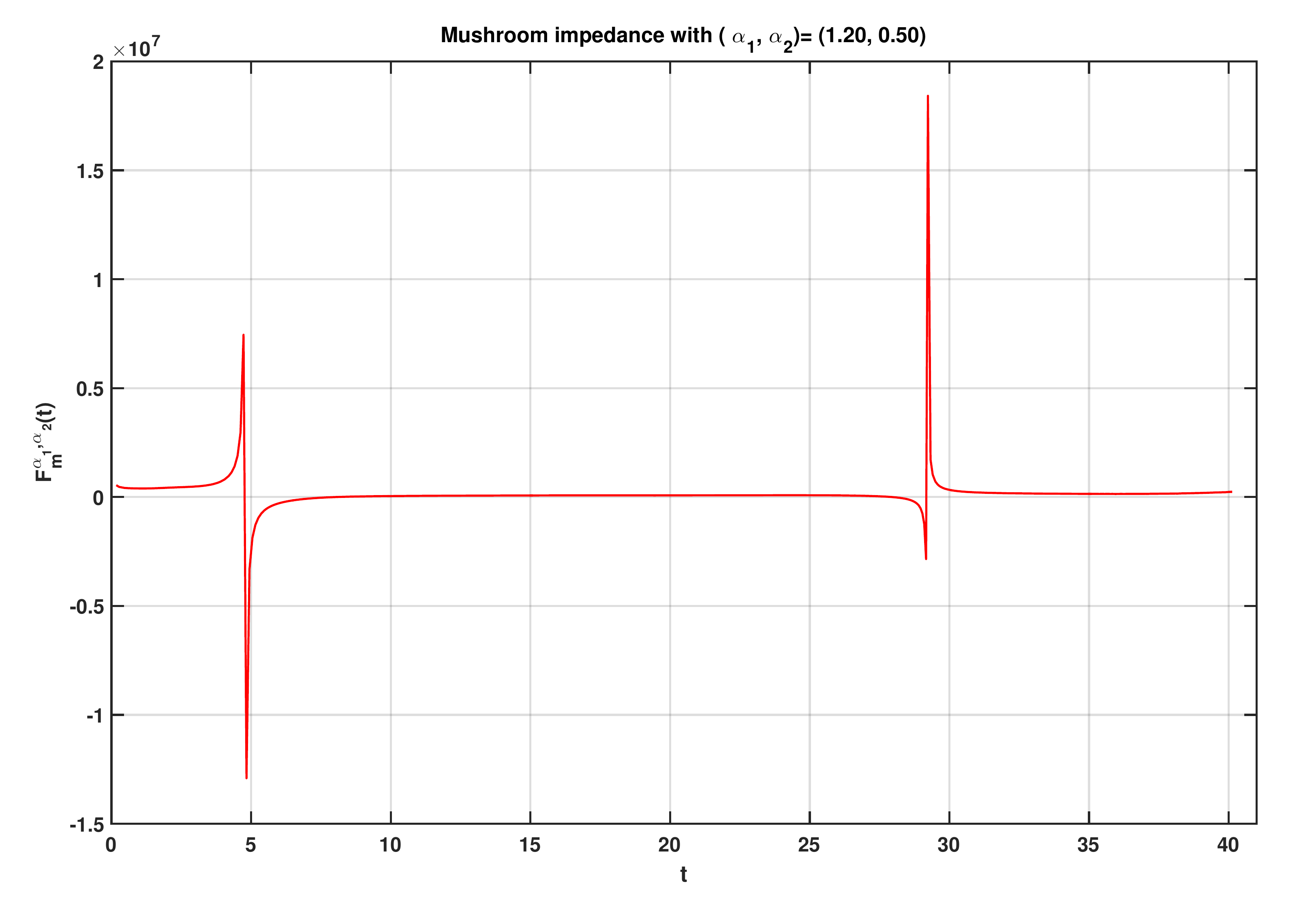}
    \caption{Memfractance with two singularities for $(\alpha_1,\alpha_2)=(1.2, 0.5)$.}
    \label{fig:RMS16}
\end{figure}

As a counter-example of our method for choosing the best possible memfractance, Fig.~\ref{fig:RMS16} displays, the memfractance for a non-optimal couple $(\alpha_1,\alpha_2)=(1.2, 0.5)$ which presents two singularities.\par 

\begin{figure}[!tpb]
    \centering
    \includegraphics[width=\textwidth]{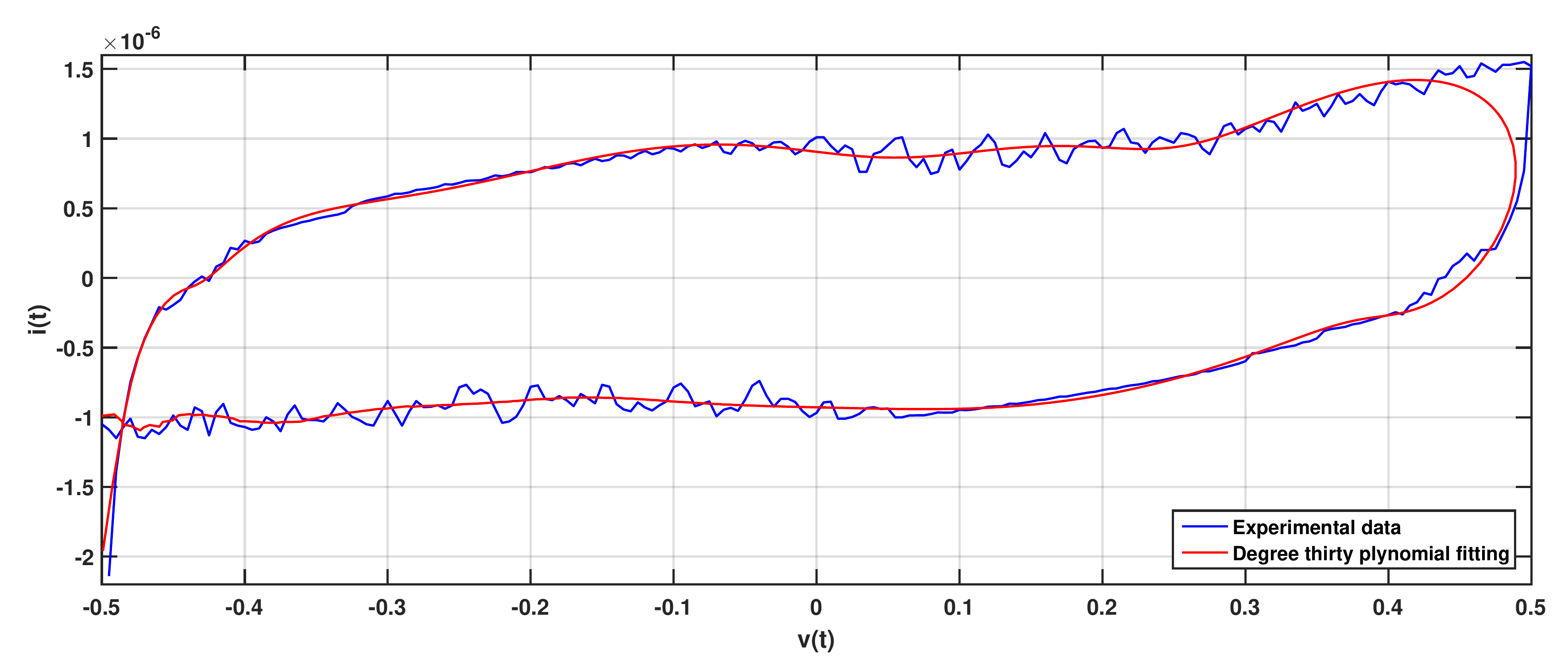}
    \caption{Comparison between average experimental data of cyclic voltammetry performed over -0.5\,V to 0.5\,V, placement, and approximate values of $v(t)$ and $i(t)$.}
    \label{fig:RMS17}
\end{figure}

The comparison of average experimental data of cyclic voltammetry performed over -0.5\,V to 0.5\,V, and closed approximate formula is displayed in Fig.~\ref{fig:RMS17}, showing a very good agreement between both curves except near the maximum value of $v(t)$ and $i(t)$.\par

\begin{figure}[!tpb]
    \centering
    \includegraphics[width=\textwidth]{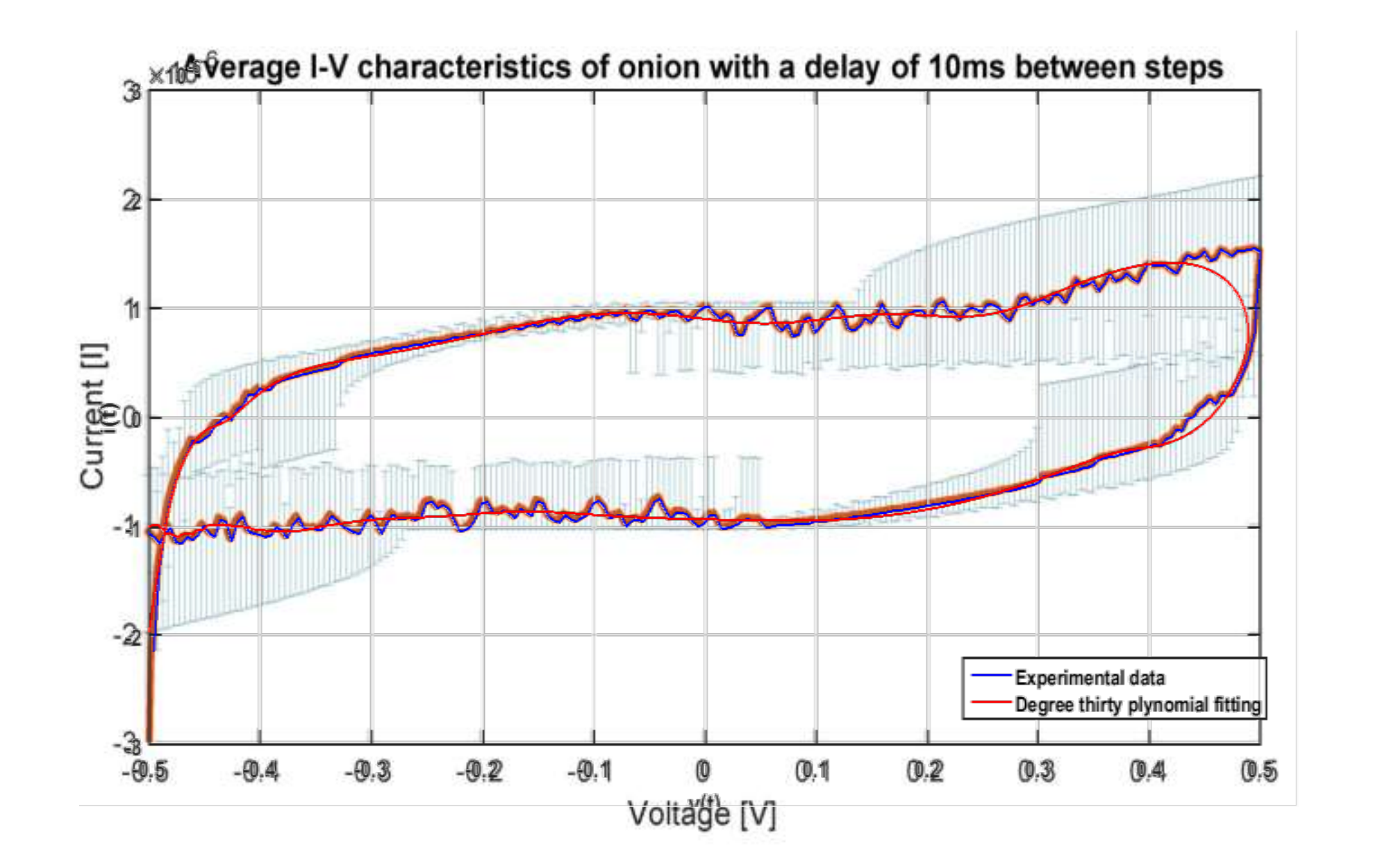}
    \caption{Both average experimental data curve and the curve computed from closed approximate formula are nested into the histogram of data of all runs.}
    \label{fig:RMS18}
\end{figure}

Figure~\ref{fig:RMS18} shows that the curve computed from closed approximate formula belongs to the histogram of data of all runs.\par

\subsection{Alternative approximation of the cycling voltammetry}

Due to the way of conducting the experiments, the voltage curve presents a vertex, that means that the function $v(t)$ is non-differentiable for $T=19.4897504$ (Fig.~\ref{fig:RMS8}). In fact, the value of $T$ is the average value of the non-differentiable points for the 10 runs. The value $T_{\text{max}} = 40.12276457$ is the maximum time of the experimentation, $0\leq t \leq T_{\text{max}}$.\par 
In this alternative approximation, we follow the same 4 steps as in 4.1, changing the approximation by a thirty-degree polynomial to an approximation by a 2-piecewise D-degree-polynomial, for both $v(t)$ and $i(t)$. Here $D = 10$.\par
\begin{figure}[!tpb]
    \centering
    \includegraphics[width=\textwidth]{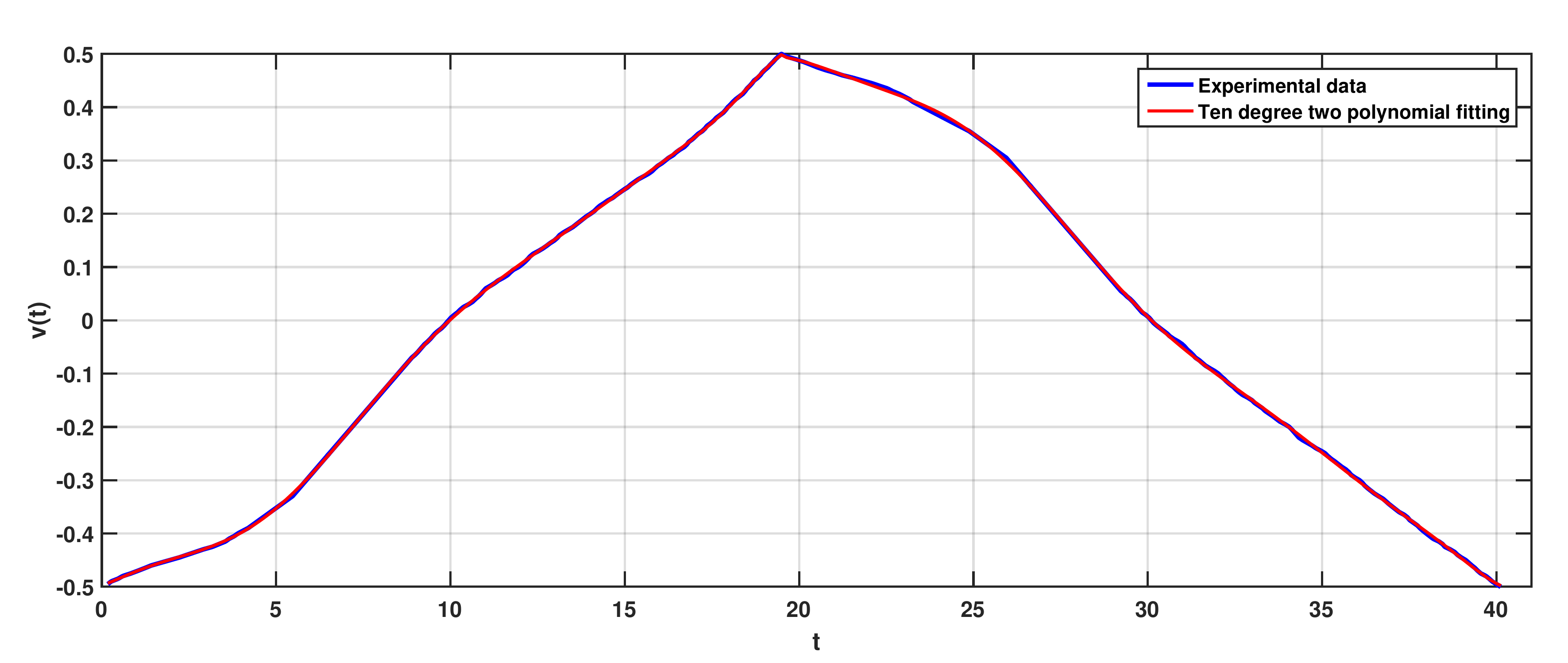}
    \caption{Voltage versus time and its approximation by 2-piecewise tenth degree polynomial.}
    \label{fig:RMS19}
\end{figure}

First step: approximation of $v(t)$ by a 2-piecewise tenth-degree-polynomial (Fig.~\ref{fig:RMS19}) whose coefficients are given in table~\ref{tab:6}.\par

\begin{equation}
    v(t) = \begin{cases}
    P_1(t) = \sum_{j=0}^{j=D}a_jt^j\text{, for }0 \leq t \leq T \\
    P_2(t) = \sum_{j=0}^{j=D}a_j^{\prime}t^j\text{, for }t \leq t < T_{\text{max}}
    \end{cases}
\end{equation}

\begin{table}[!tpb]
    \centering
    \caption{Coefficients of $a$ and $a^{\prime}$}
    \begin{tabular}{|c|c|c|c|}
        \hline
        Coefficient & Value for $0 \leq t \leq T$ & Coefficient &  Value for $T < t < T_{\text{max}}$ \\ \hline
        $a_0$ & -5.51E-11 & $a^{\prime}_0 $ & 6.87E-11 \\ \hline
        $a_1$ & 4.46E-09 & $a^{\prime}_1 $ & -2.02E-08 \\ \hline
        $a_2$ & -1.39E-07 & $a^{\prime}_2 $ & 2.64E-06  \\ \hline
        $a_3$ & 1.90E-06 & $a^{\prime}_3 $ & -0.000202629 \\ \hline
        $a_4$ & -5.30E-06 & $a^{\prime}_4 $ & 0.010116533 \\ \hline
        $a_5$ & -0.000154541 & $a^{\prime}_5 $ & -0.342890106 \\ \hline
        $a_6$ & 0.001735895 & $a^{\prime}_6 $ & 7.988613675 \\ \hline
        $a_7$ & -0.006015086 & $a^{\prime}_7 $ & -126.3046069 \\ \hline
        $a_8$ & 0.00583923 & $a^{\prime}_8 $ & 1296.836674 \\ \hline
        $a_9$ & 0.025851171 & $a^{\prime}_9 $ & -7808.105752 \\ \hline
        $a_{10}$ & -0.497985462 & $a^{\prime}_{10} $ & 20935.49797 \\ \hline
    \end{tabular}
    \label{tab:6}
\end{table}

The flux is obtained integrating $v(t)$ versus time. \par

\begin{equation}
    \varphi(t) = \begin{cases}
        IP_1(t) = \sum_{j=0}^{j=D}\frac{a_j}{j+1}t^{j+1}\text{, for }0 \leq t \leq T \\
        IP_2(t) = \sum_{j=0}^{j=D}\frac{a^{\prime}_j}{j+1}t^{j+1}\text{, for }T \leq t < T_{\text{max}} 
    \end{cases}
\end{equation}

The polynomial fits very well the experimental voltage curve, as the statistical indexes show in Tab.~\ref{tab:7}.\par

\begin{table}[!tpb]
    \centering
    \caption{Goodness of fit}
    \begin{tabular}{|c|c|c|}
        \hline
        Approximation interval & $t < T$ & $t > T$ \\ \hline
        Sum of squared estimate of errors SSE & 0.000390862 & 0.001120834 \\ \hline
        Sum of squared residuals SSR & 16.66147438 & 16.66067277\\ \hline
        Sum of square total SST & 16.66186524 & 16.66179361\\ \hline
        Coefficient of determination  R-square & 0.999976541518186 & 0.999932730288331\\ \hline
    \end{tabular}
    \label{tab:7}
\end{table}

\begin{figure}[!tpb]
    \centering
    \includegraphics[width=\textwidth]{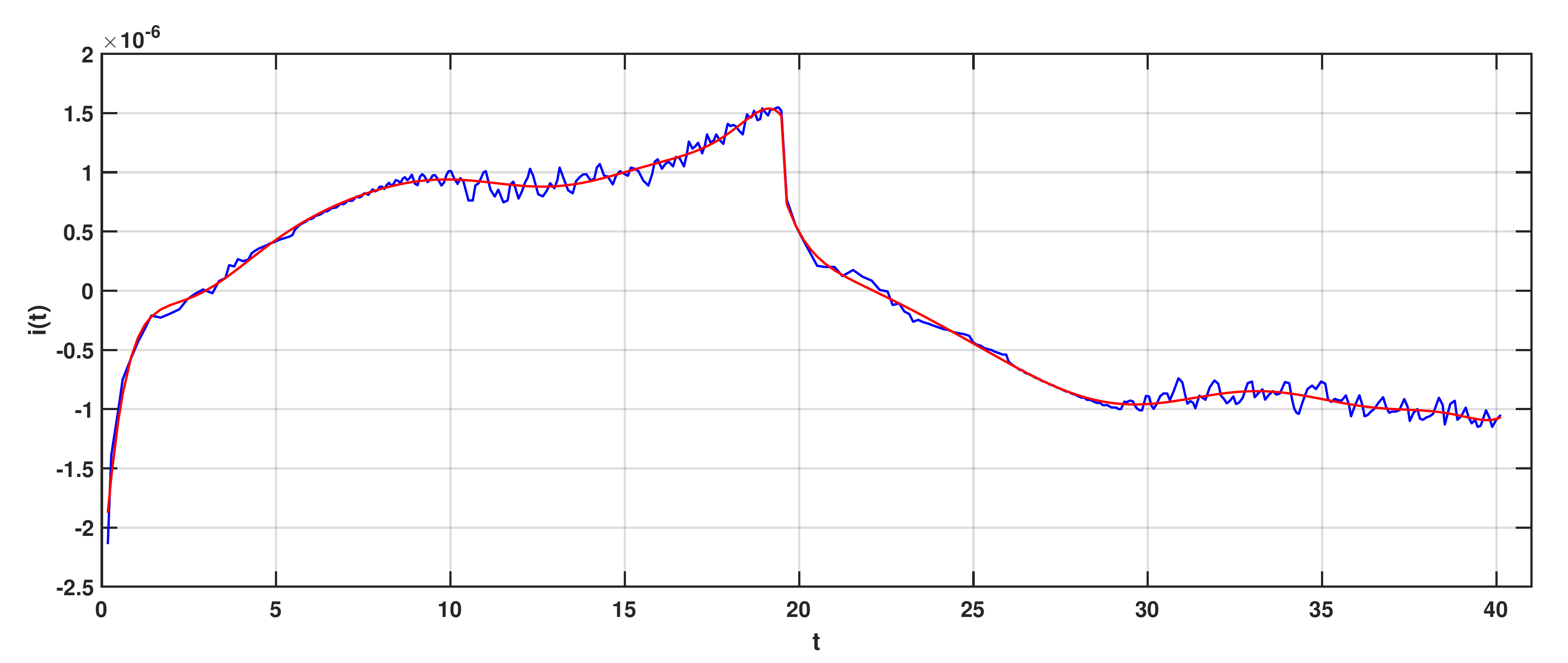}
    \caption{Current versus time and its approximation by 2-piecewise tenth degree polynomial.}
    \label{fig:RMS20}
\end{figure}

Step 2: in the same way, one approximates the current $i(t)$ using a 2-piecewise tenth degree polynomial (Fig.~\ref{fig:RMS20}) whose coefficients are given in Tab.~\ref{tab:8}.\par

\begin{equation}
    i(t) = \begin{cases}
    P_3(t) = \sum_{j=0}^{j=D} b_jt^j\text{, for }0 \leq t \leq T \\
    P_4(t) = \sum_{j=0}^{j=D} b^{\prime}_jt^j\text{, for }T \leq t < T_{\text{max}}
    \end{cases}
\end{equation}

\begin{table}[!tpb]
    \centering
    \caption{Coefficients of $b$ and $b^{\prime}$.}
    \begin{tabular}{|c|c|c|c|}
        \hline
        Coefficient & Value for $0 \leq t \leq T$ & Coefficient & Value for $T < t < T_{\text{max}}$  \\ \hline
        $b_0$ & 3.13E-10 & $b^{\prime}_0$ & -2.13E-09 \\ \hline
        $b_1$ & -4.86E-09 & $b^{\prime}_1$ & 1.09E-07 \\ \hline
        $b_2$ & 4.88E-08 & $b^{\prime}_2$ & -3.78E-06 \\ \hline
        $b_3$ & -3.18E-07 & $b^{\prime}_3$ & 9.07E-05 \\ \hline
        $b_4$ & 1.31E-06 & $b^{\prime}_4$ & -0.001481936 \\ \hline
        $b_5$ & -3.20E-06 & $b^{\prime}_5$ & 0.015797711 \\ \hline
        $b_6$ & 4.26E-06 & $b^{\prime}_6$ & -0.099223872 \\ \hline
        $b_7$ & -2.52E-06 & $b^{\prime}_7$ & 0.278895403 \\ \hline
        $b_8$ & 3.13E-10 & $b^{\prime}_8$ & -2.13E-09 \\ \hline
        $b_9$ & -4.86E-09 & $b^{\prime}_9$ & 1.09E-07 \\ \hline
        $b_{10}$ & 4.88E-08 & $b^{\prime}_{10}$ & -3.78E-06 \\ \hline
    \end{tabular}
    \label{tab:8}
\end{table}

Again, the polynomial fits very well the experimental voltage curve, as the statistical indexes show in Tab.~\ref{tab:9}.\par

\begin{table}[!tpb]
    \centering
    \caption{Quality of fitness.}
    \begin{tabular}{|c|c|c|}
        \hline
        Approximation interval & $t < T$ & $t > T$ \\ \hline
        Sum of squared estimate of errors SSE & 6.63E-13 & 5.92E-13 \\ \hline
        Sum of squared residuals SSR & 4.97E-11 & 2.36E-11\\ \hline
        Sum of square total SST & 5.04E-11 & 2.41E-11 \\ \hline
        Coefficient of determination R-square & 0.986832207 & 0.97550029 \\ \hline
    \end{tabular}
    \label{tab:9}
\end{table}

Therefore, the charge is given by\par

\begin{equation}
    q(t) = \begin{cases}
        IP_3(t) = \sum_{j=0}^{j=D}\frac{b_j}{j+1}t^{j+1}\text{, for }0 \leq t \leq T\\
        IP_4(t) = \sum_{j=0}^{j=D}\frac{b^{\prime}_j}{j+1}t^{j+1}\text{, for }T \leq t < T_{\text{max}}
    \end{cases}
\end{equation}

Step 3: Following the same calculus as before with (4), one obtains\par

\begin{equation}
    \text{for } 0 \leq t \leq T\text{, }F_M^{\alpha_1,\alpha_2}(t) = \frac{{^{RL}_{0}}D_t^{\alpha_1}\varphi(t)}{{^{RL}_{0}}D_t^{\alpha_2}q(t)} = \frac{{^{RL}_{0}}D_t^{\alpha_1}[IP_1(t)]}{{^{RL}_{0}D_t^{\alpha_2}[IP_3(t)]}} = \frac{\sum_{j = 1}^{j=D}\frac{a_j\Gamma(j+1)}{\Gamma(j+2-\alpha_1)}t^{j+1-\alpha_1}}{\sum_{j=0}^{j=D}\frac{b_j\Gamma(j+1)}{\Gamma(j+2-\alpha_2)}t^{j+1-\alpha_2}}
\end{equation}

However, because fractional derivative has memory effect, for $T < t < T_{\text{max}}$, the formula is slightly more complicated\par

\begin{equation}
    \begin{split}
    F_M^{\alpha_1,\alpha_2}(t) & = \frac{{^{RL}_{0}}D_t^{\alpha_1}\varphi(t)}{{^{RL}_{0}}D_t^{\alpha_2}q(t)} = \frac{\frac{1}{\Gamma(m_1-\alpha_1)}\frac{d^{m_1}}{dt^{m_1}}\int_{0}^{t}(t-s)^{m_1 - \alpha_1 - 1}\varphi(s)ds}{\frac{1}{\Gamma(m_2-\alpha_2)}\frac{d^{m_2}}{dt^{m_2}}\int_{0}^{t}(t-s)^{m_2-\alpha_2-1}q(s)ds}\text{, }m_1 - 1 < \alpha_1 < m_1\text{ and }m_2 - 1 < \alpha_2 < m_2 \\
    & = \frac{\frac{1}{\Gamma(m_1 - \alpha_1)}\frac{d^{m_1}}{dt^{m_1}}\left[ \int_{0}^{T}(t-s)^{m_1 - \alpha_1 - 1} IP_1(s)ds + \int_T^t(t-s)^{m_1 - \alpha_1 - 1} IP_2(s)ds \right]}{\frac{1}{\Gamma(m_2-\alpha_2)}\frac{d^{m_2}}{dt^{m_2}}\left[ \int_0^T(t-s)^{m_2 - \alpha_2 - 1}IP_3(s)ds + \int_{T}^{t}(t-s)^{m_2 - \alpha_2 - 1}IP_4(s)ds \right]}\\
    & = \frac{\frac{1}{\Gamma(m_1 - \alpha_1)}\frac{d^{m_1}}{dt^{m_1}} \sum_{j=0}^{j=D} \left[ \frac{a_j}{j+1}\int_{0}^{T}(t-s)^{m_1 - \alpha_1 - 1}s^{j+1}ds + \frac{a^{\prime}_j}{j+1}\int_{T}^{t}(t-s)^{m_1 - \alpha_1 - 1}s^{j+1}ds \right]}{ \frac{1}{\Gamma(m_2 - \alpha_2)} \frac{d^{m_2}}{dt^{m_2}} \sum_{j=0}^{j=D}\left[ \frac{b_j}{j+1} \int_{0}^{T}(t-s)^{m_2 - \alpha_2 - 1}s^{j+1}ds + \frac{b^{\prime}_j}{j+1} \int_{T}^{t}(t-s)^{m_2 - \alpha_2 - 1} s^{j+1}ds \right] }
    \end{split}
\end{equation}

Using integration by part repeatedly $D+1$ times we obtain

\begin{landscape}
\begin{equation}
    \label{eq:big_boi}
    \begin{split}
        & F_M^{\alpha_1,\alpha_2}(t) \\
        & = \frac{ \frac{1}{\Gamma(m_1 - \alpha_1)} \frac{d^{m_1}}{dt^{m_1}} \sum_{j=0}^{j=D}\left[ \frac{a_j}{j+1} \left[ \sum_{k=0}^{k=j+1} \left[ \frac{-(j+1)!\Gamma(m_1-\alpha_1)(t_T)^{m_1 + k - \alpha_1} T^{j+1-k} }{(j+1-k)!\Gamma(m_1 + k + 1 - \alpha_1} \right] + \frac{(j+1)!\Gamma(m_1 - \alpha_1)t^{m_1+k-\alpha_1}}{\Gamma(m_1 + j + 1 - \alpha_1)} \right] + \frac{a^{\prime}_j}{j+1}\left[ \sum_{k=0}^{k=j+1} \frac{(j+1)!\Gamma(m_1 - \alpha_1)(t-T)^{m_1+k-\alpha_1}T^{j+1-k}}{(j+1-k)!\Gamma(m_1 + k + 1 -\alpha_1)} \right] \right] }{ \frac{1}{\Gamma(m_2-\alpha_2)} \frac{d^{m_2}}{dt^{m_2}} \sum_{j=0}^{j=D} \left[ \frac{b_j}{j+1} \left[ \sum_{k=0}^{k=j+1} \left[ \frac{-(j+1)!\Gamma(m_2 - \alpha_2)(t-T)^{m_2 + k - \alpha_2} T^{j+1-k}}{(j+1-k)!\Gamma(m_2 + k + 1 - \alpha_2)} \right] + \frac{(j+1)!\Gamma(m_2 - \alpha_2)t^{m_2 + k - \alpha_2}}{\Gamma(m_2 + j + 1 - \alpha_2)} \right] + \frac{b^{\prime}_j}{j+1}\left[ \sum_{k=0}^{k=j+1} \frac{(j+1)!\Gamma(m_2 - \alpha_2)(t-T)^{m_2 + k - \alpha_2}T^{j+1-k}}{(j+1-k)!\Gamma(m_2 + k + 1 - \alpha_2)} \right] \right]}\\
        & = \frac{ \frac{1}{\Gamma(m_1 - \alpha_1)} \frac{d^{m_1}}{dt^{m_1}} \sum_{j=0}^{j=D} \left[ (a^{\prime}_j - a_j) \sum_{k=0}^{k = j + 1} \left[ \frac{j!\Gamma(m_1 - \alpha_1)(t - T)^{m_1 + k - \alpha_1} T^{j + 1 - k}}{(j+1-k)!\Gamma(m_1 + k + 1 - \alpha_1)} \right] + a_j \frac{j! \Gamma(m_1 - \alpha_1)t^{m_1 + j + 1 - \alpha_1}}{\Gamma(m_1 + j + 2 - \alpha_1} \right]}{ \frac{1}{\Gamma(m_2 - \alpha_2)} \frac{d^{m_2}}{dt^{m_2}} \sum_{j=0}^{j=D} \left[ (b^{\prime}_j - b_j) \sum_{k=0}^{k=j+1} \left[ \frac{j!\Gamma(m_2 - \alpha_2)(t-T)^{m_2 + k - \alpha_2}T^{j+1-k}}{(j+1-k)!\Gamma(m_2 + k + 1 - \alpha_2)} \right] + b_j \frac{j!\Gamma(m_2 - \alpha_2) t^{m_2 + j + 1 - \alpha_2}}{\Gamma(m_2 + j + 2 - \alpha_2)} \right] }\\ 
        & = \frac{\sum_{j=0}^{j=D} \left[ (a^{\prime}_j - a_j) \sum_{k=0}^{k=j+1} \left[ \frac{j!(t-T)^{k-\alpha_1}T^{j+1-k}}{(j+1-k)!\Gamma(k+1-\alpha_1)} \right] + a_j \frac{j!t^{j+1-\alpha_1}}{\Gamma(j+2-\alpha_1)}\right] }{\sum_{j=0}^{j=D} \left[ (b^{\prime}_j - b_j) \sum_{k=0}^{k=j+1} \left[ \frac{j!(t-T)^{k - \alpha_2}T^{j+1-k}}{(j+1-k)!\Gamma(k+1-\alpha_2)} \right] + b_j \frac{j!t^{j+1-\alpha_2}}{\Gamma(j+2-\alpha_2)} \right]}
    \end{split}
\end{equation}
\end{landscape}

In this 2-piece wise approximation, the vertex is non-differentiable, this implies that (\ref{eq:big_boi}) expression has a singularity at $T$ (because $(t-T)^{-\alpha_{1,2} }\rightarrow \infty)$.

It could be possible to avoid this singularity, using a 3-piece wise approximation, smoothing the vertex. However, the calculus are very tedious. We will explain, below, what our simpler choice implies.

\begin{equation}
    \begin{split}
    \text{Then }F_M^{\alpha_1,\alpha_2}(t)& = \frac{(t-T)^{-\alpha_1}\left[ \sum_{j=0}^{j=D}\left[ (a^{\prime}_j - a_j) \sum_{k=0}^{k=j+1} \left[ \frac{j!(t-T)^kT^{j+1-k}}{(j+1-k)!\Gamma(k+1-\alpha_1}) \right]  + a_j \frac{j!t^{j+1 - \alpha_1}(t-T)^{\alpha_1}}{\Gamma(j+2-\alpha_1)} \right] \right] }{ (t-T)^{-\alpha_2} \left[ \sum_{j=0}^{j=D} \left[ (b^{\prime}_j - b_j) \sum_{k=0}^{k=j+1} \left[ \frac{j!(t-T)^k T^{j+1-k}}{(j+1-k)!\Gamma(k+1-\alpha_2)} \right] + b_j \frac{j!t^{j+1-\alpha_2}(t-T)^{\alpha_2}}{\Gamma(j+2-\alpha_2)} \right]  \right] }\\
    & = \frac{ \sum_{j=0}^{j=D} \left[ (a^{\prime}_j - a_j) \sum_{k=0}^{k=j+1} \left[ \frac{j!(t-T)^kT^{j+1-k}}{(j+1-k)!\Gamma(k+1-\alpha_1)} \right] + a_j\frac{j!t^{j+1-\alpha_1}(t-T)^{\alpha_1}}{\Gamma(j+2-\alpha_1)} \right] }{ (t-T)^{\alpha_1 - \alpha_2} \sum_{j=0}^{j=D}\left[ (b^{\prime}_j - b_j) \sum_{k=0}^{k=j+1} \left[ \frac{j!(t-T)^kT^{j+1-k}}{(j+1-k)!\Gamma(k+1-\alpha_2)} \right] + b_j\frac{j!t^{j+1-\alpha_2}(t-T)^{\alpha_2}}{\Gamma(j+2 - \alpha_2)} \right] }
    \end{split}
\end{equation}

Finally

\begin{equation}
    F_M^{\alpha_1,\alpha_2}(t) = \begin{cases}
    \frac{\sum_{j=0}^{j=D} \frac{a_j\Gamma(j+1)}{\Gamma(j+2-\alpha_1)}t^{j+1-\alpha_1} }{\sum_{j=0}^{j=D}\frac{b_j\Gamma(j+1)}{\Gamma(j+2-\alpha_2)}t^{j+1-\alpha_2}}\text{,} & \text{for } 0 \leq t \leq T \\
    \frac{\sum_{j=0}^{j=D} \left[ (a^{\prime}_j - a_j) \sum_{k=0}^{k=j+1} \left[ \frac{j!(t-T)^kT^{j+1-k}}{(j+1-k)!\Gamma(k+1-\alpha_1)} \right] + a_j \frac{j!t^{j+1-\alpha_1}(t-T)^{\alpha_1}}{\Gamma(j+2-\alpha_1)}\right] }{(t-T)^{\alpha_1-\alpha_2} \sum_{j=0}^{j=D} \left[ (b^{\prime}_j - b_j) \sum_{k=0}^{k=j+1} \left[ \frac{j!(t-T)^kT^{j+1-k}}{(j+1-k)!\Gamma(k+1-\alpha_2)} \right] + b_j\frac{j!t^{j+1-\alpha_2(t-T)^{\alpha_2}}}{\Gamma(j+2-\alpha_2)} \right] }\text{,} & \text{for } T < t < T_{\text{max}}
    \end{cases}
\end{equation}

\begin{figure}[!tpb]
    \centering
    \includegraphics[width=\textwidth]{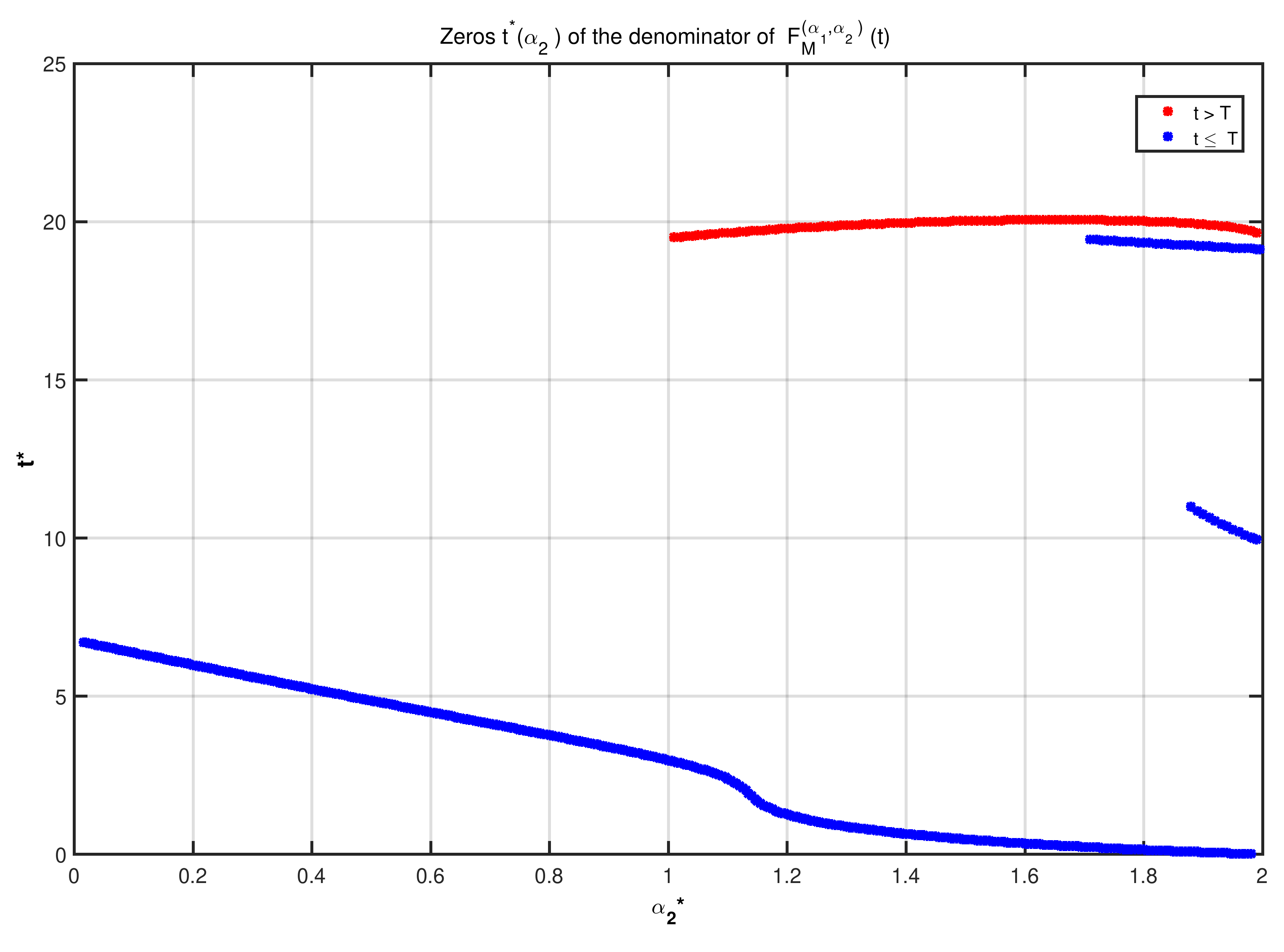}
    \caption{The zeros $t^*(\alpha_2)$, of the denominator of $F_M^{\alpha_1,\alpha_2}(t)$, as function of $\alpha_2$.}
    \label{fig:RMS21}
\end{figure}

Step 4 choice of parameter $\alpha_1$ and $\alpha_2$: Following the same idea as for the first alternative, we try to avoid singularity for $F_M^{\alpha_1,\alpha_2}(t)$, except of course the singularity near $T$, which is of mathematical nature (non-differentiability of voltage and intensity at $t = T$). Figure~\ref{fig:RMS21} display zeros $t^*(\alpha_2)$, of the denominator of $F_M^{\alpha_1,\alpha_2}(t)$.\par

\begin{figure}[!tpb]
    \centering
    \includegraphics[width=\textwidth]{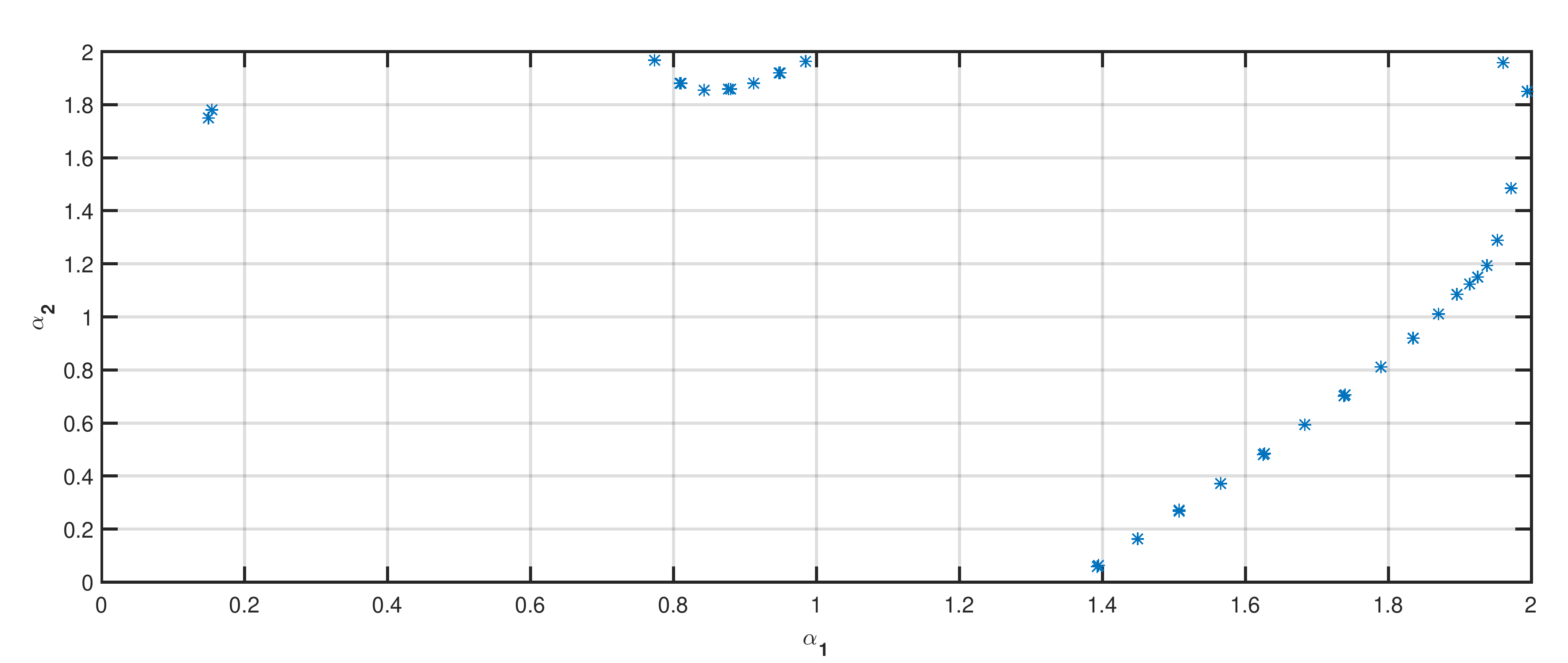}
    \caption{Couples $(\alpha_1,\alpha_2)$ for which the denominator and numerator of $F_M^{\alpha_1,\alpha_2}(t)$ are null simultaneously.}
    \label{fig:RMS22}
\end{figure}

Figure~\ref{fig:RMS22} displays the curves of couples $(\alpha_1,\alpha_2)$ for which the denominator and numerator of $F_M^{\alpha_1,\alpha_2}(t)$ are null simultaneously  for $t < T$ and $t > T$. On this figure, the value of $\alpha_1$ that corresponds to $\alpha_2=1.483238482$ is $\alpha_1 \approx 1.971795208$. The corresponding memfractance is displayed in Fig.~\ref{fig:RMS24}.\par

\begin{figure}[!tpb]
    \centering
    \includegraphics[width=\textwidth]{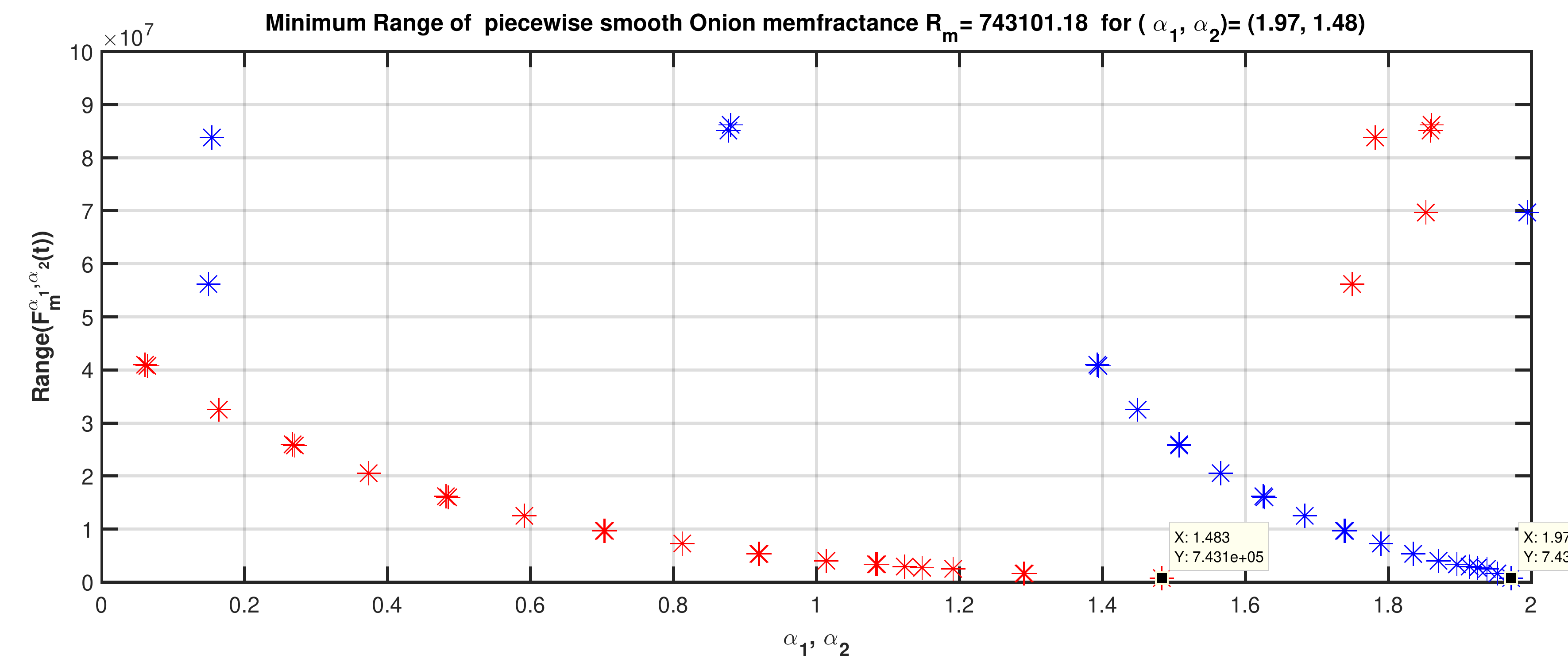}
    \caption{Values of range $(F_M^{\alpha_1,\alpha_2}(t))$  for  $(\alpha_1,\alpha_2)\in[0,2]^2$.}
    \label{fig:RMS23}
\end{figure}

\begin{figure}[!tpb]
    \centering
    \includegraphics[width=\textwidth]{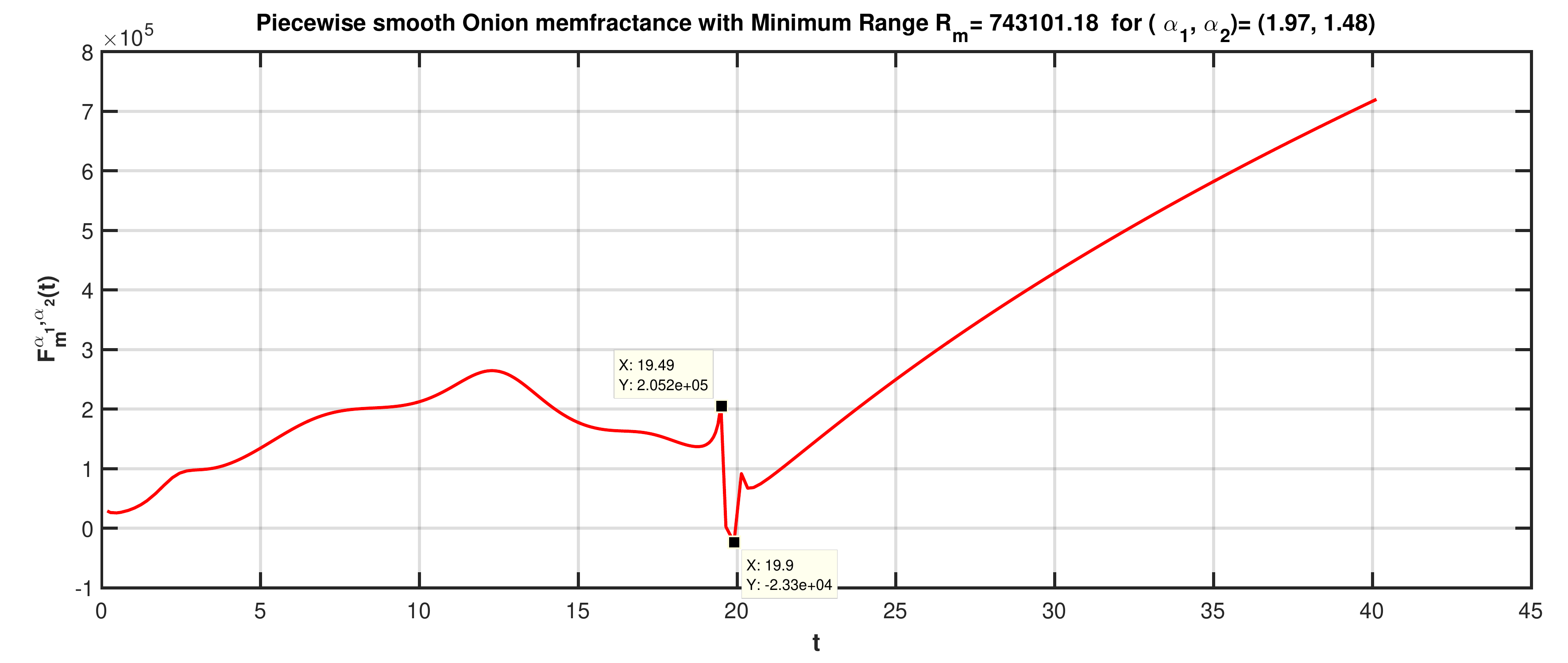}
    \caption{Memfractance for $(\alpha_1=1.971795208,\alpha_2=1.483238482)$ given in Tab.~\ref{tab:10}.}
    \label{fig:RMS24}
\end{figure}

\begin{table}[!tpb]
    \centering
     \caption{Minimum values of $\alpha$.}
    \begin{tabular}{|c|c|c|}
        \hline
        $\alpha_1$ & $\alpha_2$ & Minimum range of $F_M^{\alpha_1,\alpha_2}(t)$  \\ \hline
        1.971795208 & 1.483238482 & 743101.176733524 \\ \hline 
    \end{tabular}
    \label{tab:10}
\end{table}

The singularity observed in Figs.~\ref{fig:RMS24} is due to the non-differentiability of both voltage and intensity functions at point $T$.\par  
The value of $(\alpha_1=1.971795208,\alpha_2=1.483238482)$ belongs to the triangle $T_2$ of Fig.~\ref{fig:RMS15}, whose edges are resistor, memristor, and capacitor. Therefore the mem-fractance property of onion is a combination of those of these electric components. \par
\begin{figure}[!tpb]
    \centering
    \includegraphics[width=\textwidth]{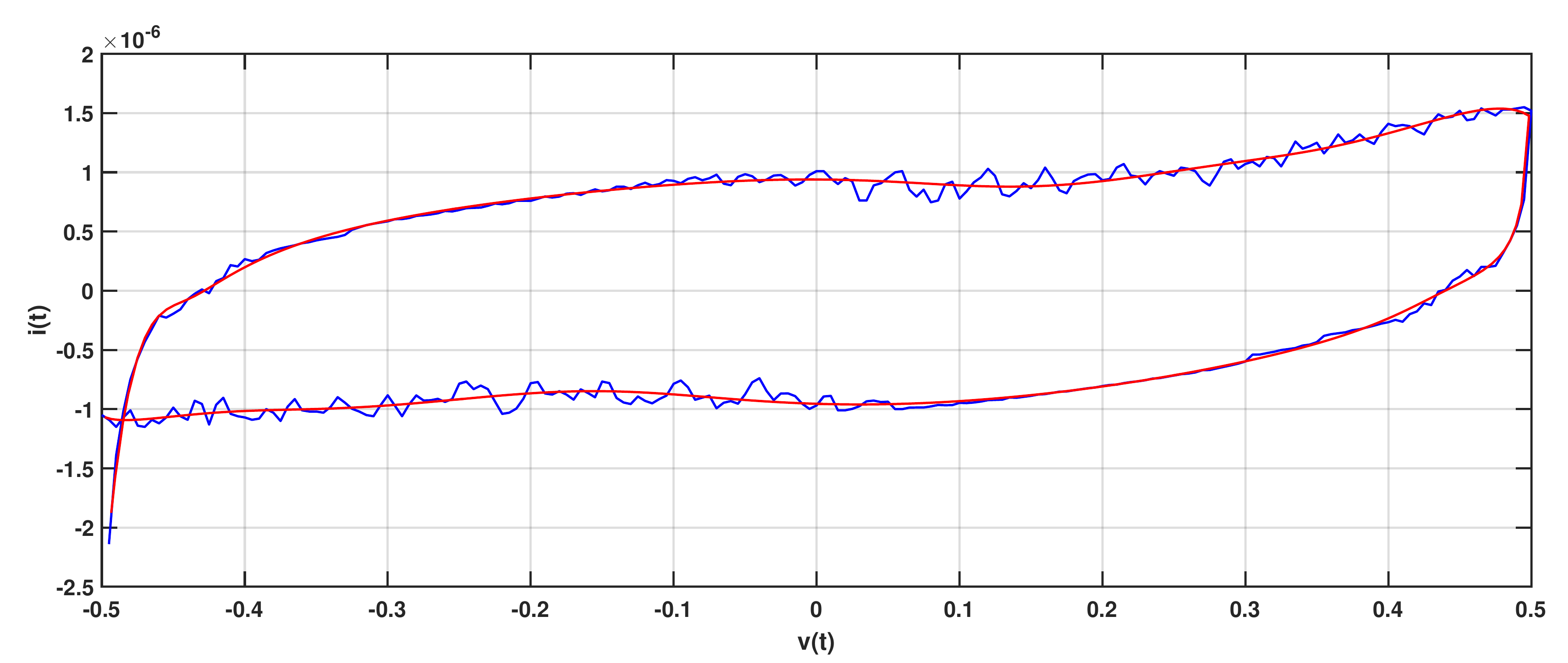}
    \caption{Comparison between average experimental data of cyclic voltammetry performed over -0.5\,V to 0.5\,V, Stem-to-cap electrode placement, and closed approximative formula.}
    \label{fig:RMS25}
\end{figure}

The comparison of average experimental data of cyclic voltammetry performed over -0.5\,V to 0.5\,V and closed approximative formula is displayed in Fig.~\ref{fig:RMS25}, showing a very good agreement between both curves.\par

\subsection{Alternative approximation of the cycling voltammetry of onion with a delay of 1s between steps}

Due to the way of conducting the experiments, the voltage curve presents a vertex, that means that the function $v(t)$ is non-differentiable for $T=218.835146340667$. In fact, the value of $T$ is the average value of the non-differentiable points for the 3 runs. On has $T_{\text{max}} = 446.142389256667$, and $D = 15$.\par
We perform an approximation by a 2-piecewise D-degree-polynomial, for both $v(t)$ and $i(t)$. \par

\begin{figure}[!tpb]
    \centering
    \includegraphics[width=\textwidth]{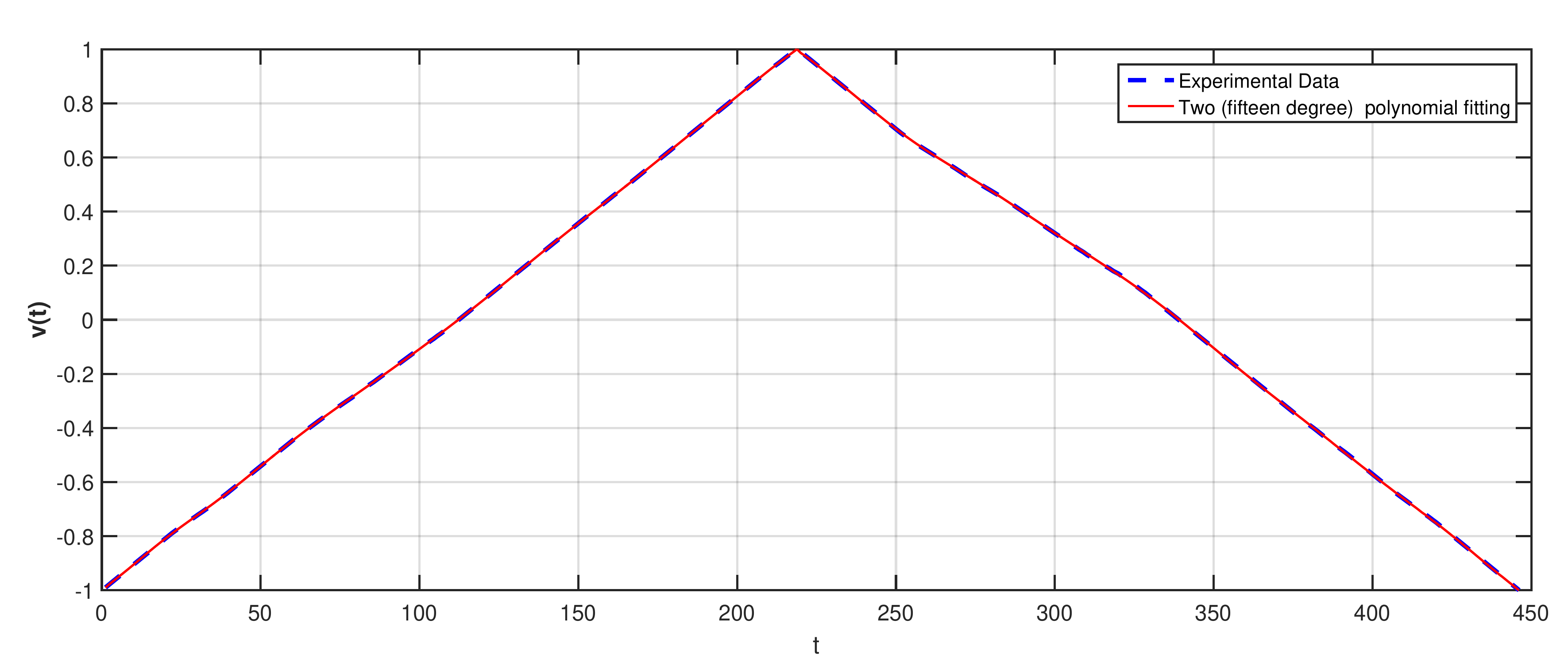}
    \caption{Voltage versus time and its approximation by 2-piecewise fifteenth degree polynomial. }
    \label{fig:RMS26}
\end{figure}

First step: approximation of $v(t)$ by a 2-piecewise fifteen-degree-polynomial defined by (9), (Fig.~\ref{fig:RMS26}) whose coefficients are given in table~\ref{tab:11}.\par

\begin{table}[!tpb]
    \centering
    \caption{Coefficients of $a$ and $a^{\prime}$}
    \begin{tabular}{|c|c|c|c|}
        \hline
        Coefficient & Value for $0 \leq t \leq T$  & Coefficient &  Value for $T < t < 446.142389256667$\\ \hline
        $a_0$ & 3.54420781829105e-30 & $a^{\prime}_0$ & 1.00427487680972e-30 \\ \hline
        $a_1$ & -5.96398936924787e-27 & $a^{\prime}_1$ & -5.20695580801111e-27 \\ \hline
        $a_2$ & 4.52630929682328e-24 & $a^{\prime}_2$ & 1.25279988942641e-23 \\ \hline
        $a_3$ & -2.04721171726301e-21 & $a^{\prime}_3$ & -1.85546677570461e-20 \\ \hline
        $a_4$ & 6.14198657142099e-19 & $a^{\prime}_4$ & 1.89174603354674e-17 \\ \hline
        $a_5$ & -1.28714657924748e-16 & $a^{\prime}_5$ & -1.40637895477368e-14 \\ \hline
        $a_6$ & 1.93189458719508e-14 & $a^{\prime}_6$ & 7.87564822151931e-12 \\ \hline
        $a_7$ & -2.09391204258264e-12 & $a^{\prime}_7$ & -3.38279075621789e-09\\ \hline
        $a_8$ & 1.63042090512474e-10 & $a^{\prime}_8$ & 1.12362332993865e-06\\ \hline
        $a_9$ & -8.95113952596272e-09 & $a^{\prime}_9$ & -0.000288613689878837 \\ \hline
        $a_{10}$ & 3.34546189613670e-07 & $a^{\prime}_{10}$ & 0.0568590592565226 \\ \hline
        $a_{11}$ & -8.03669009474361e-06 & $a^{\prime}_{11}$ & -8.43718319262010 \\ \hline
        $a_{12}$ & 0.000113493644607879 & $a^{\prime}_{12}$ & 912.822831960333 \\ \hline
        $a_{13}$ & -0.000835676981290587 & $a^{\prime}_{13}$ & -67977.9563028956\\ \hline
        $a_{14}$ & 0.0122632177149911 & $a^{\prime}_{14}$ & 3115831.30623377 \\ \hline
        $a_{15}$ & -1.00360065451426 & $a^{\prime}_{15}$ & -66266900.3096982\\ \hline
    \end{tabular}
    \label{tab:11}
\end{table}

The flux is again obtained integrating $v(t)$ versus time (10).\par

The polynomial fits very well the experimental voltage curve, as the statistical indexes show in Tab.~\ref{tab:12}.\par

\begin{table}[!tpb]
    \centering
    \caption{Quality of fitness.}
    \begin{tabular}{|c|c|c|}
        \hline
        Approximation interval & $t < T$ & $t > T$ \\ \hline
        Sum of squared estimate of errors SSE & 8.05206626404267e-05 & 0.000379119559037593 \\ \hline
        Sum of squared residuals SSR & 66.6597470338760 & 66.7305863691811\\ \hline
        Sum of square total SST & 66.6598275545387 & 66.7309654887402\\ \hline
        Coefficient of determination R-square & 0.999998792066143 & 0.999994318686141 \\ \hline
    \end{tabular}
    \label{tab:12}
\end{table}

\begin{figure}[!tpb]
    \centering
    \includegraphics[width=\textwidth]{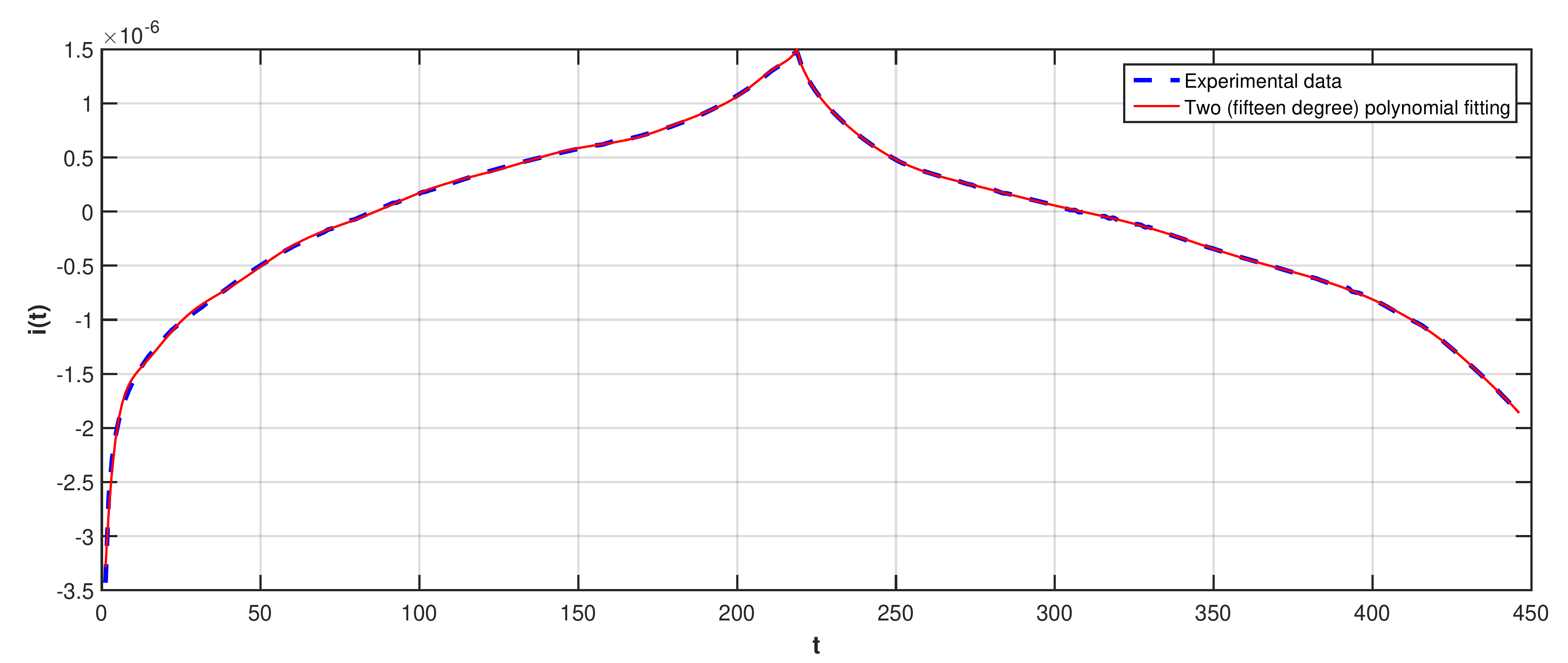}
    \caption{Current versus time and its approximation by 2-piecewise fifteenth degree polynomial.}
    \label{fig:RMS27}
\end{figure}

Step 2: in the same way, one approximates the current $i(t)$ using a 2-piecewise fifteenth degree polynomial defined by (11), (Fig.~\ref{fig:RMS27}) whose coefficients are given in table~\ref{tab:13}.\par

\begin{table}[!tpb]
    \centering
    \caption{Coefficients of $b$ and $b^{\prime}$}
    \begin{tabular}{|c|c|c|c|}
        \hline
        Coefficient & Value for $0 \leq t \leq T$  & Coefficient &  Value for $T < t < 446.142389256667$\\ \hline
        $b_0$ & 9.54254808462533e-35 & $b^{\prime}_0$ & 1.53300080630080e-38\\ \hline
        $b_1$ & -1.63161644309873e-31 & $b^{\prime}_1$ & -2.89601182252080e-34 \\ \hline
        $b_2$ & 1.26206638752557e-28 & $b^{\prime}_2$ & 1.13996788975235e-30 \\ \hline
        $b_3$ & -5.83981425666847e-26 & $b^{\prime}_3$ & -2.25810438491826e-27 \\ \hline
        $b_4$ & 1.80113459097984e-23 & $b^{\prime}_4$ & 2.79988171542492e-24 \\ \hline
        $b_5$ & -3.90531862314380e-21 & $b^{\prime}_5$ &-2.39282614008484e-21 \\ \hline
        $b_6$ & 6.11858793928889e-19 & $b^{\prime}_6$ & 1.48305233409647e-18 \\ \hline
        $b_7$ & -7.01109241829061e-17 & $b^{\prime}_7$ &-6.85633253958774e-16\\ \hline
        $b_8$ & 5.88169906789969e-15 & $b^{\prime}_8$ & 2.39797611614612e-13\\ \hline
        $b_9$ & -3.58165613133728e-13 & $b^{\prime}_9$ & -6.36783612027411e-11\\ \hline
        $b_{10}$ & 1.55442674717007e-11 & $b^{\prime}_{10}$ & 1.27624831718578e-08 \\ \hline
        $b_{11}$ & -4.66705880467474e-10 & $b^{\prime}_{11}$ & -1.89820925444517e-06\\ \hline
        $b_{12}$ & 9.27058527759065e-09 & $b^{\prime}_{12}$ &0.000202910655233761 \\ \hline
        $b_{13}$ & -1.14142215596473e-07 & $b^{\prime}_{13}$ & -0.0147140989613784\\ \hline
        $b_{14}$ & 8.12388017501071e-07 & $^{\prime}_{14}$ & 0.646651137483905\\ \hline
        $b_{15}$ & -4.08952638018235e-06 & $b^{\prime}_{15}$ & -12.9611877781724\\ \hline
    \end{tabular}
    \label{tab:13}
\end{table}

Again, the polynomial fits very well the experimental voltage curve, as the statistical indexes show in Tab.~\ref{tab:14}.\par

\begin{table}[!tpb]
    \centering
    \caption{Goodness of fit}
    \begin{tabular}{|c|c|c|}
        \hline
        Approximation interval & $t < T$ & $t > T$ \\ \hline
        Sum of squared estimate of errors SSE & 7.99361696557309e-14 & 2.86502655937580e-15\\ \hline
        Sum of squared residuals SSR & 1.45875395884222e-10 & 1.02218430381697e-10\\ \hline
        Sum of square total SST & 1.45955332053877e-10 & 1.02221295408256e-10\\ \hline
        Coefficient of determination R-square & 0.999452324430147 & 0.999971972312150\\ \hline
    \end{tabular}
    \label{tab:14}
\end{table}

Therefore, the charge is given by (12).\par

Step 3: Following the same calculus as before with (4), for $0 \leq t \leq T\text{,  }F_M^{\alpha_1,\alpha_2}(t)$ is defined by (13).\par
However, because fractional derivative has memory effect, for $T < t < T_{\text{max}}$, the formula is slightly more complicated. It is defined by (14), (15), (16) and (17).\par

In this 2-piecewise approximation, the vertex is non-differentiable, this implies that (15) expression has a singularity at $T$ (because $(t-T)^{-\alpha_{1,}} \rightarrow \inf )$. 

\begin{figure}[!tpb]
    \centering
    \includegraphics[width=\textwidth]{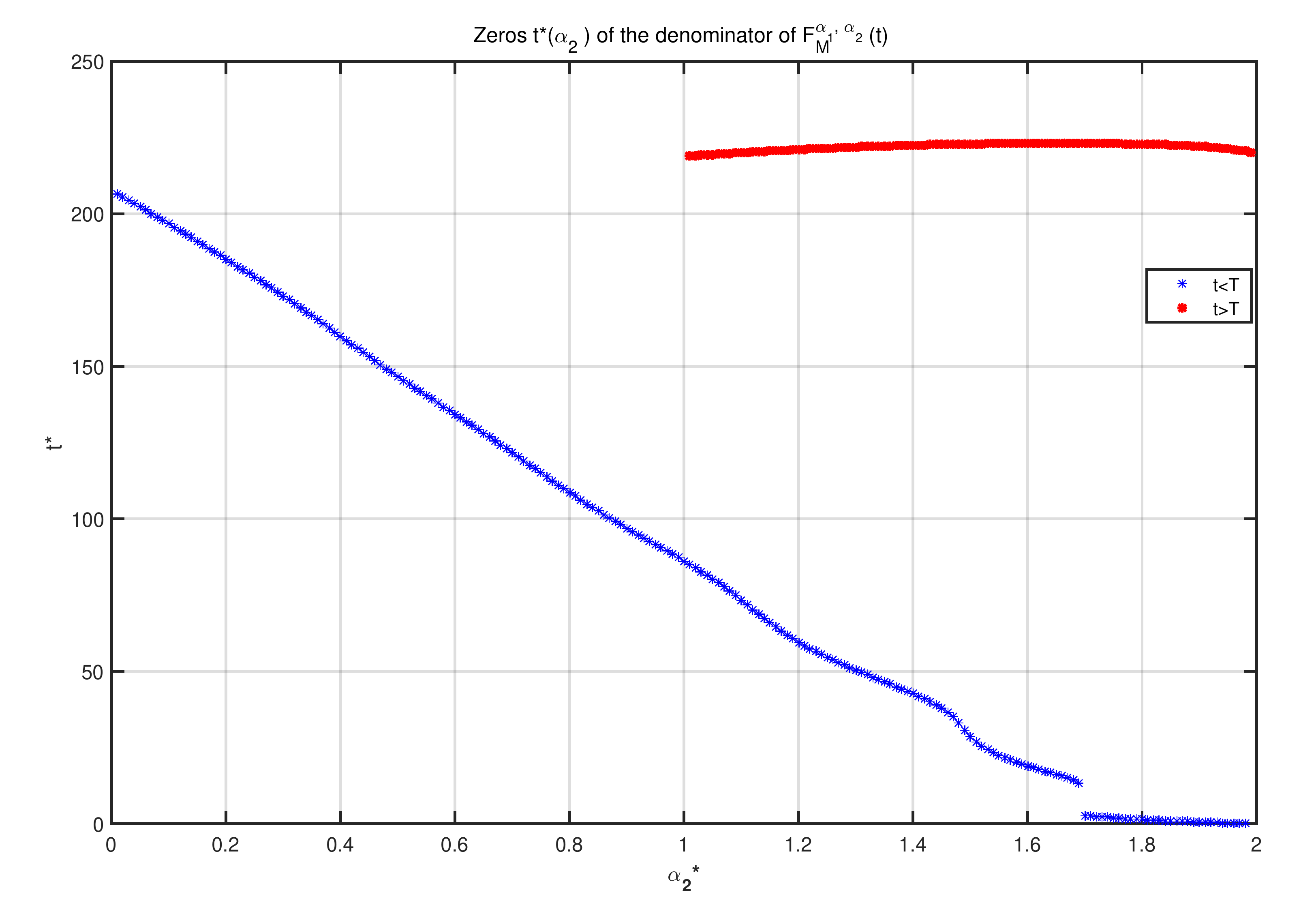}
    \caption{The zeros $t^*(\alpha_2)$, of the denominator of $F_M^{\alpha_1,\alpha_2}(t)$, as function of $\alpha_2$.}
    \label{fig:RMS28}
\end{figure}

Step 4 choice of parameter $\alpha_1$ and $\alpha_2$: Following the same idea as for the first alternative, we try to avoid singularity for $F_M^{\alpha_1,\alpha_2}(t)$, except of course the singularity near $T$, which is of mathematical nature (non-differentiability of voltage and intensity at $t = T$). Figure~\ref{fig:RMS28} display zeros$t^*(\alpha_2)$, of the denominator of $F_M^{\alpha_1,\alpha_2}(t)$.\par

\begin{figure}[!tpb]
    \centering
    \includegraphics[width=\textwidth]{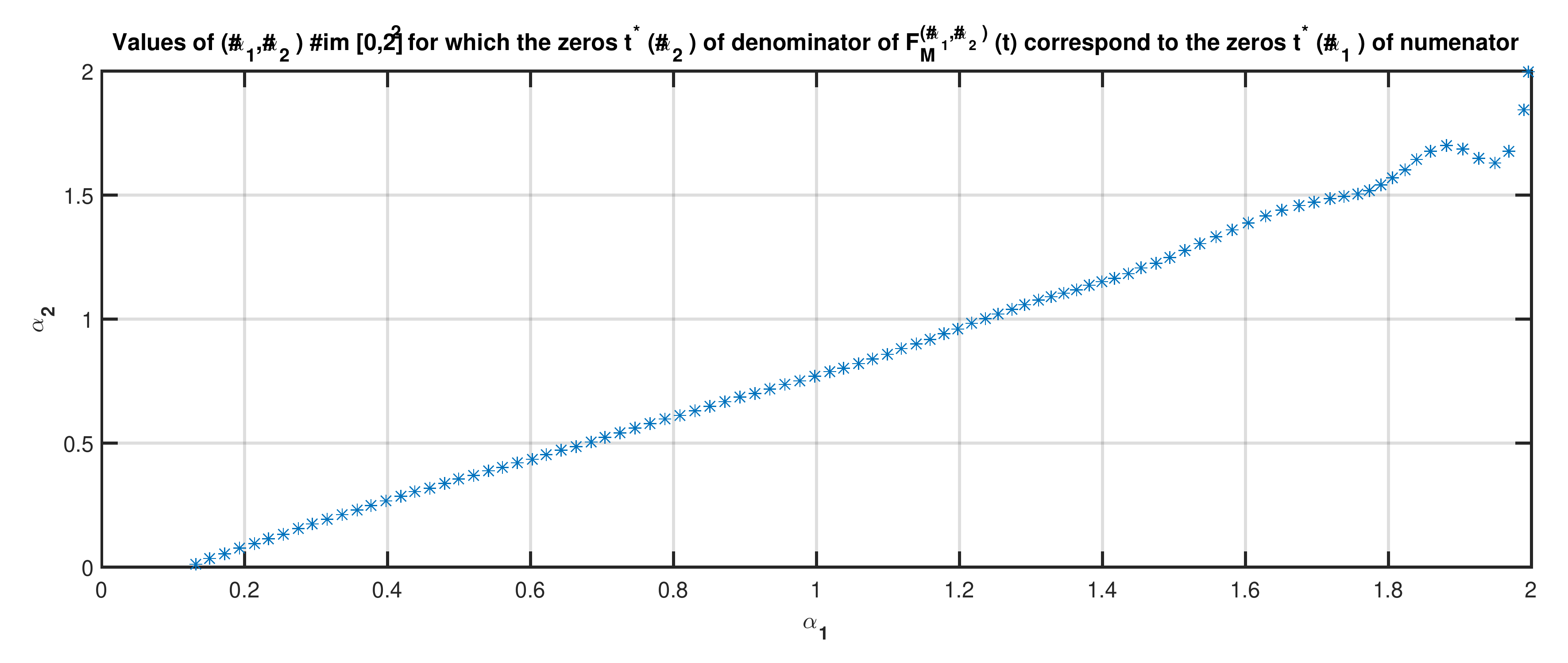}
    \caption{Couples $(\alpha_1,\alpha_2)$ for which the denominator and numerator of $F_M^{\alpha_1,\alpha_2}(t)$ are null simultaneously.}
    \label{fig:RMS29}
\end{figure}

Figure~\ref{fig:RMS29} displays the curves of couples $(\alpha_1,\alpha_2)$ for which the denominator and numerator of $F_M^{\alpha_1,\alpha_2}(t)$ are null simultaneously for $t < T$ and $t > T$.\par 

\begin{figure}[!tpb]
    \centering
    \includegraphics[width=\textwidth]{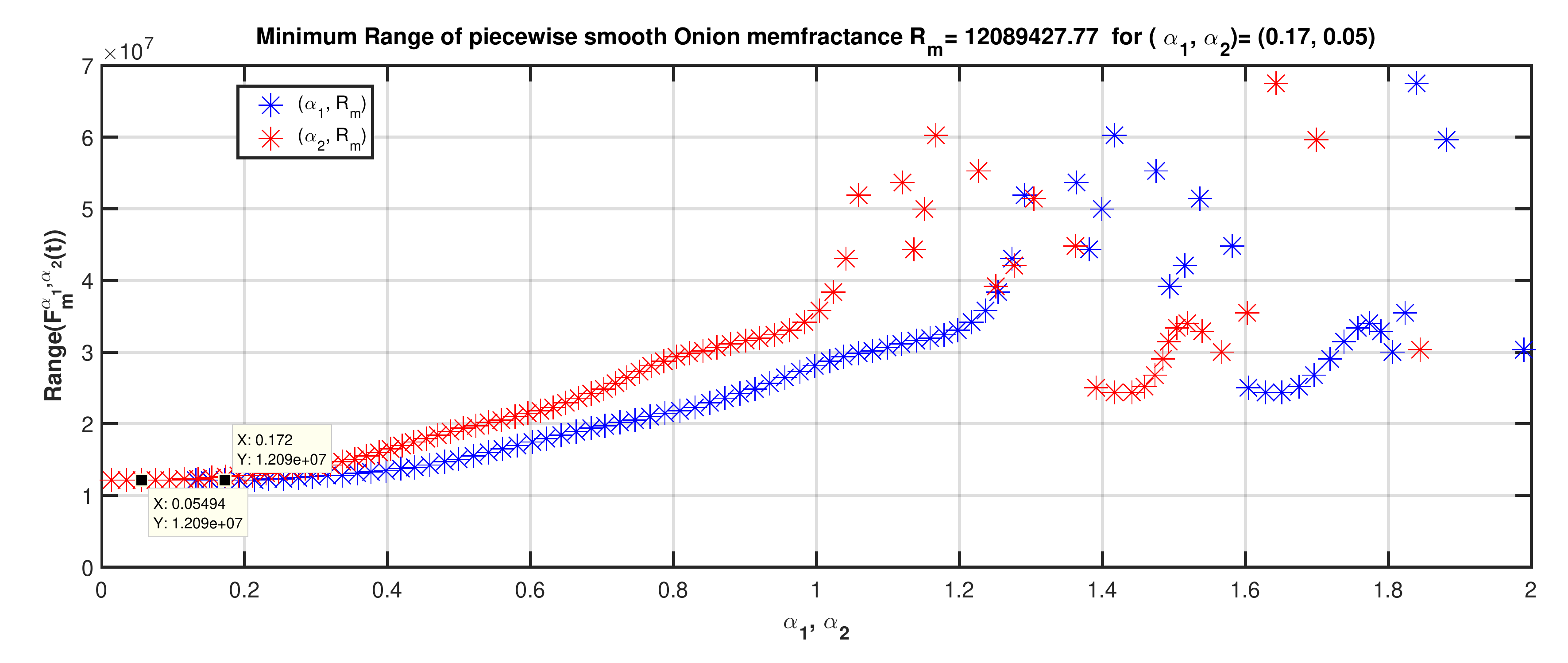}
    \caption{Values of range $(F_M^{\alpha_1,\alpha_2}(t))$ for $(\alpha_1,\alpha_2 )\in[0,2]^2$.}
    \label{fig:RMS30}
\end{figure}

\begin{figure}[!tpb]
    \centering
    \includegraphics[width=\textwidth]{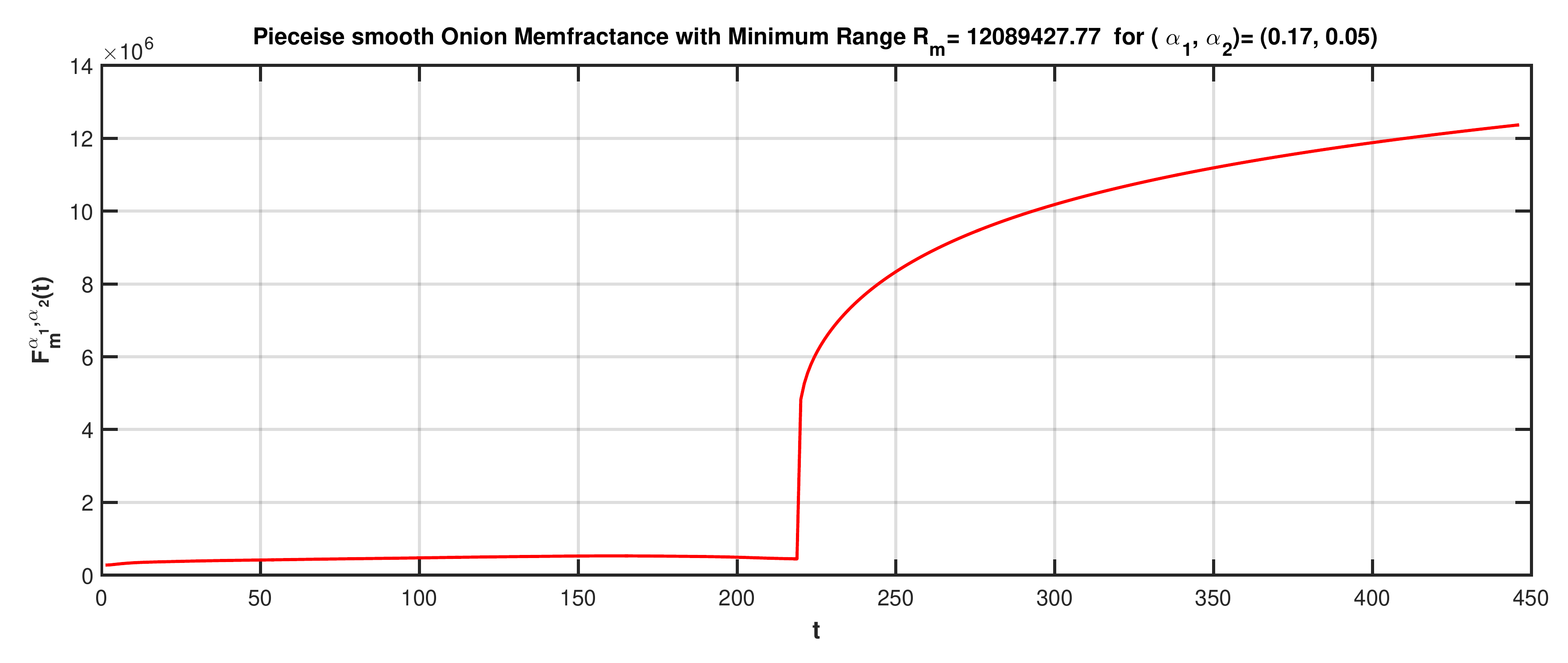}
    \caption{Memfractance for $(\alpha_1=0.171972381,\alpha_2=0.054935584)$ given in Tab.~\ref{tab:15}.}
    \label{fig:RMS31}
\end{figure}

\begin{table}[!tpb]
    \centering
    \caption{Values of $\alpha$}
    \begin{tabular}{|c|c|c|}
        \hline
        $\alpha_1$ & $\alpha_2$ & Minimumg range of $F_M^{\alpha_1,\alpha_2}(t)$\\ \hline
        0.171972381 & 0.054935584 & 12089427.7744264 \\ \hline
    \end{tabular}
    \label{tab:15}
\end{table}

The singularity observed in Figs.~\ref{fig:RMS31} is due to the non-differentiability of both voltage and intensity functions at point $T$.\par  
The value of $(\alpha_1=0.171972381,\alpha_2=0.054935584)$ belongs to the triangle $T_3$ of Fig.~\ref{fig:RMS15}, whose extremities are 2nd memristor, memristor, and memcapacitor. In this experiment, onion has property related to these basic electric devices.\par
\begin{figure}[!tpb]
    \centering
    \includegraphics[width=\textwidth]{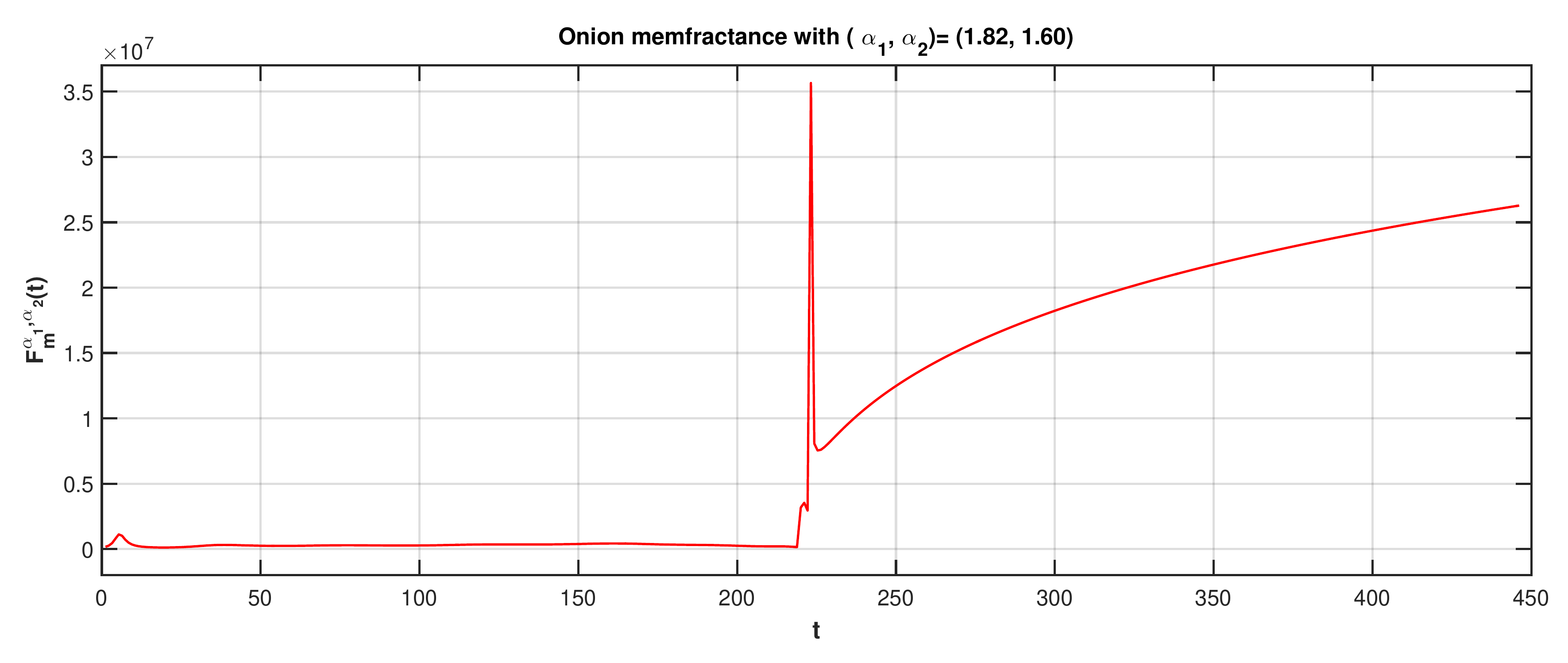}
    \caption{Memfractance with singularity for $(\alpha_1,\alpha_2)=(1.82, 1.6)$.}
    \label{fig:RMS32}
\end{figure}

As a counter-example of our method for choosing the best possible memfractance, Fig.
~\ref{fig:RMS32} displays, the memfractance for a non-optimal couple $(\alpha_1,\alpha_2)=(1.82, 1.6)$ which presents two singularities.\par

\begin{figure}[!tpb]
    \centering
    \includegraphics[width=\textwidth]{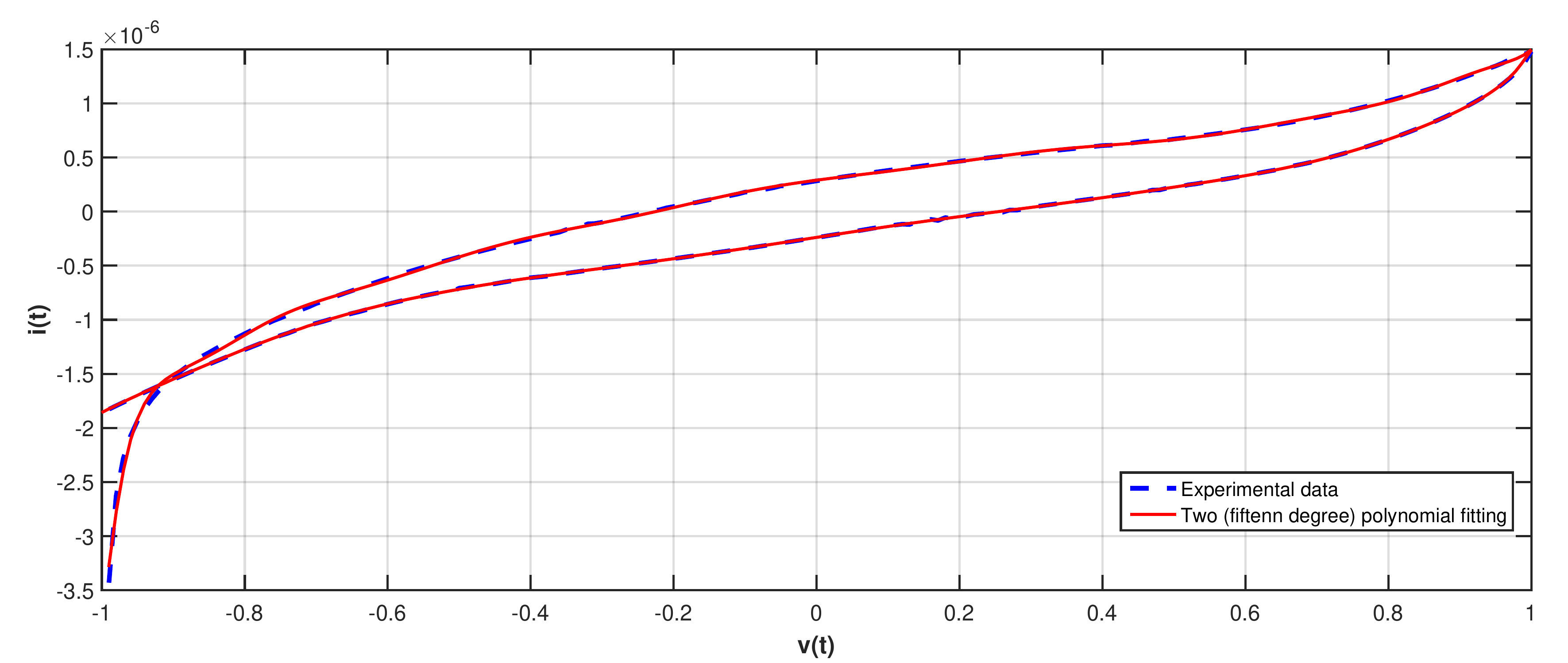}
    \caption{Comparison between average experimental data of cyclic voltammetry performed over -1\,V to 1\,V and closed approximative formula.}
    \label{fig:RMS33}
\end{figure}

The comparison of average experimental data of cyclic voltammetry performed over -1\,V to 1\,V, Stem-to-cap electrode placement, and closed approxmating formula is displayed in Fig.~\ref{fig:RMS33}, showing a very good agreement between both curves.\par

\newpage
\begin{figure}[!tbp]
    \centering
    \includegraphics[width=0.7\textwidth]{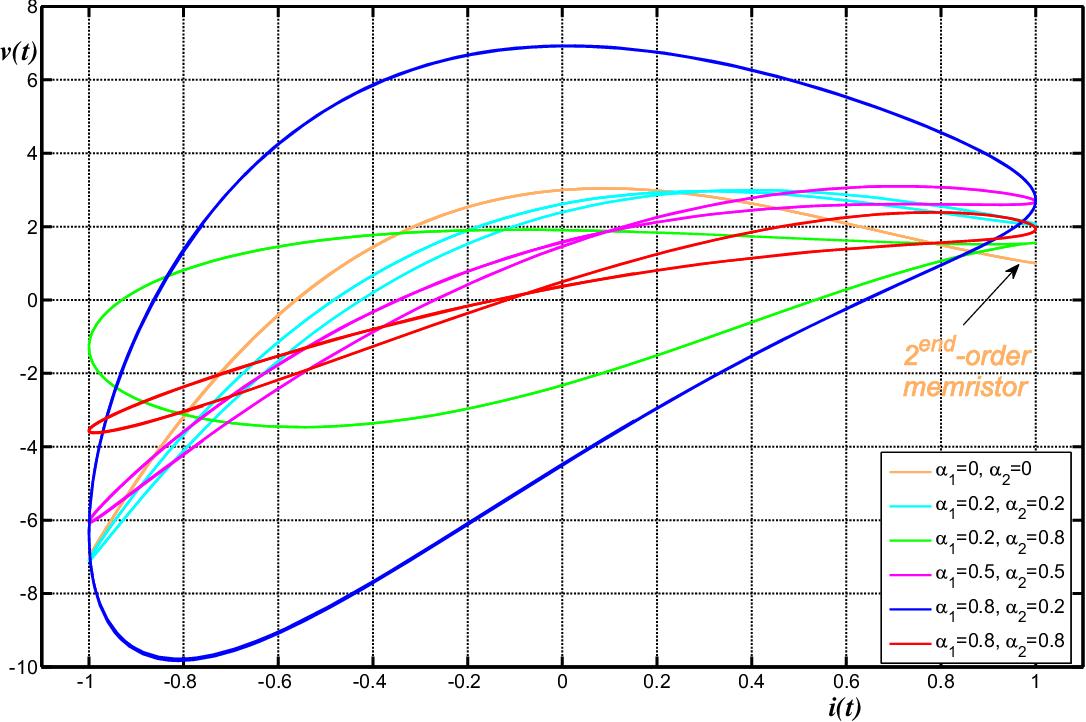}
    \caption{Ideal plots of mem-fractance}
    \label{fig:memfractancePots}
\end{figure}

\newpage

\section{Discussions}
\label{sec:discussions}

It has been shown in this paper that items taken from nature all exhibit memristive properties wherein the conducted current from the positive half of a cyclic voltammetry sweep does not match the conducted current from the negative cycle. This is in line with previously published results on I-V characterisation of organic and biological substrates, which indicate memristive properties of organic polymers~\cite{erokhin2005hybrid}, skin~\cite{martinsen2010memristance}, blood~\cite{kosta2011human}, slime mould \emph{Physarum polycephalum}~\cite{gale2015slime},
plants~\cite{volkov2014memristors}, fruits~\cite{paper:apples_memristor}, and  
tubulin microtubules~\cite{del2019two,chiolerio2020resistance}.

In addition, the level as to the divergence between the positive and negative cycles is also demonstrated to be a function of the sweep frequency (time). These memristive properties have also been observed in medium such as damp wood shavings and water. It is therefore proposed that any living system,  a large proportion of which is water acting as a charge carrier, is able to exhibit memristive properties, the degree of which can be expressed on a continuous scale.

It is also observed that the crossing behaviour expected from an ideal, passive, memristor is not often seen from the naturally occurring specimens. A naturally occurring specimen is capable of generating its own potential, which has an associated conducted current, both of which have been measured as part of the cyclic voltammetry. Increasing the delay time between consecutive readings, will produce an I-V characteristic where the negative and positive phases `pinch' closer together.\par

In all instances, the fingerprints of memristive devices are observed in specimens taken from nature. It can be expressed that memristance is indeed not a binary feature, however it exists more of a continuum [0,1] - 0 representing pure resistance, 1 representing the ideal memristor --- each device can then be assigned a number on the scale [0,1] to characterise its `degree of memristance'.\par

More generally mem-fractance which is a general paradigm linking memristive, mem-capacitive and mem-inductive property of electric elements, should be the adequate frame for the mathematical modelling all plants and fungi. The use of fractional derivatives to analyse the mem-fractance, is obvious if one considers that fractional derivatives have memory, which allow a perfect modelling of memristive elements. Their handling is however delicate if one wants to avoid any flaw.
In Section 4, the case of onion is analysed. Increasing the frequency of sampling of the current intensity and changing the range of voltage leads to slightly different mem-fractance. With low frequency and higher voltage, the onion has properties which is a combination of memristor, mem-capacitor and second order memristor. In the case of high frequency, and low voltage, the onion is merely more a mix of resistor, capacitor and memristor, showing less memory effect!\par

Additionally, current oscillations during the cyclic voltammetry are produced by all sample specimens. Typically, the oscillatory effect can be observed only on one phase of the voltammetry for a given voltage range which is, again, a behaviour that can be associated to a device whose resistance is a function of its previous resistance. Although, it is worth stating that some samples do produce overlapping oscillatory effects (both phases of voltammetry) for certain conditions --- however this can be controlled through careful selection of voltammetry conditions. This spiking activity is typical of a device that exhibits memristive behaviours as having been previously observed in experiments with the electrochemical devices with a graphite reference electrodes~\cite{erokhin2008electrochemically}, in experiments with electrode metal on solution-processed flexible titanium dioxide memristors~\cite{gale2015effect}, see also analysis in~\cite{gale2014emergent}.  The spiking properties of the meristive devices can be utilized in the field of neurmorphic systems~\cite{serrano2013proposal,indiveri2013integration,prezioso2016spiking,pickett2013scalable,linares2011spike,indiveri2015memory}. \par

\section*{Acknowledgement}

This project has received funding from the European Union's Horizon 2020 research and innovation programme FET OPEN ``Challenging current thinking'' under grant agreement No 858132. \par
With special thanks to Neil Phillips and Andrew Geary (UWE, Bristol, UK) for providing the data on tap water.

\bibliographystyle{unsrt}
\bibliography{bibliography,references_fungal_memristor, references_model}

\newpage
\section{Appendix}
\label{sec:Appendix}

%\subsection{Fruiting Bodies}

\begin{figure}[!hbt]
    \centering
    \subfigure[]{\includegraphics[width=0.68\textwidth]{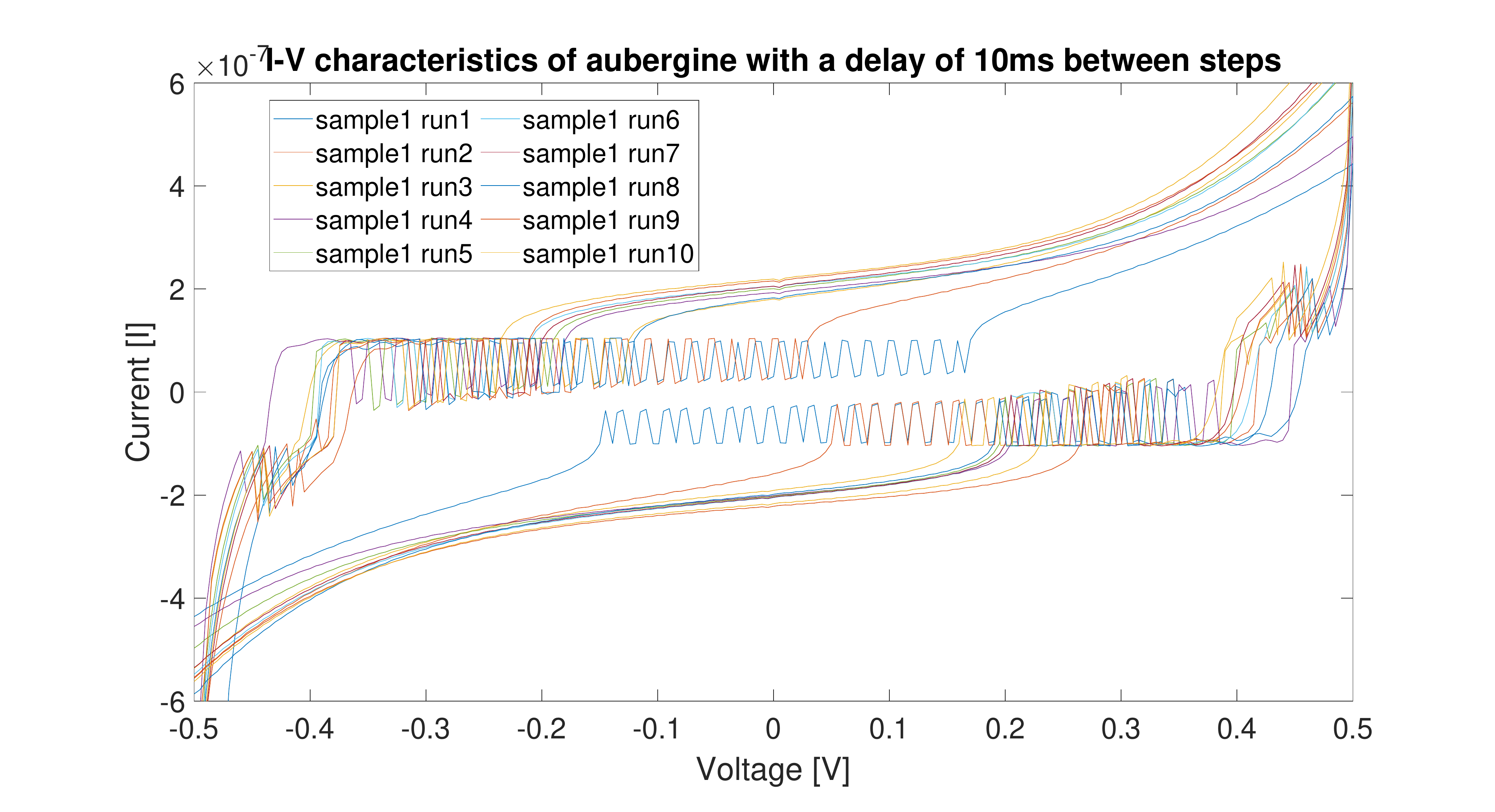}}
    \subfigure[]{\includegraphics[width=0.68\textwidth]{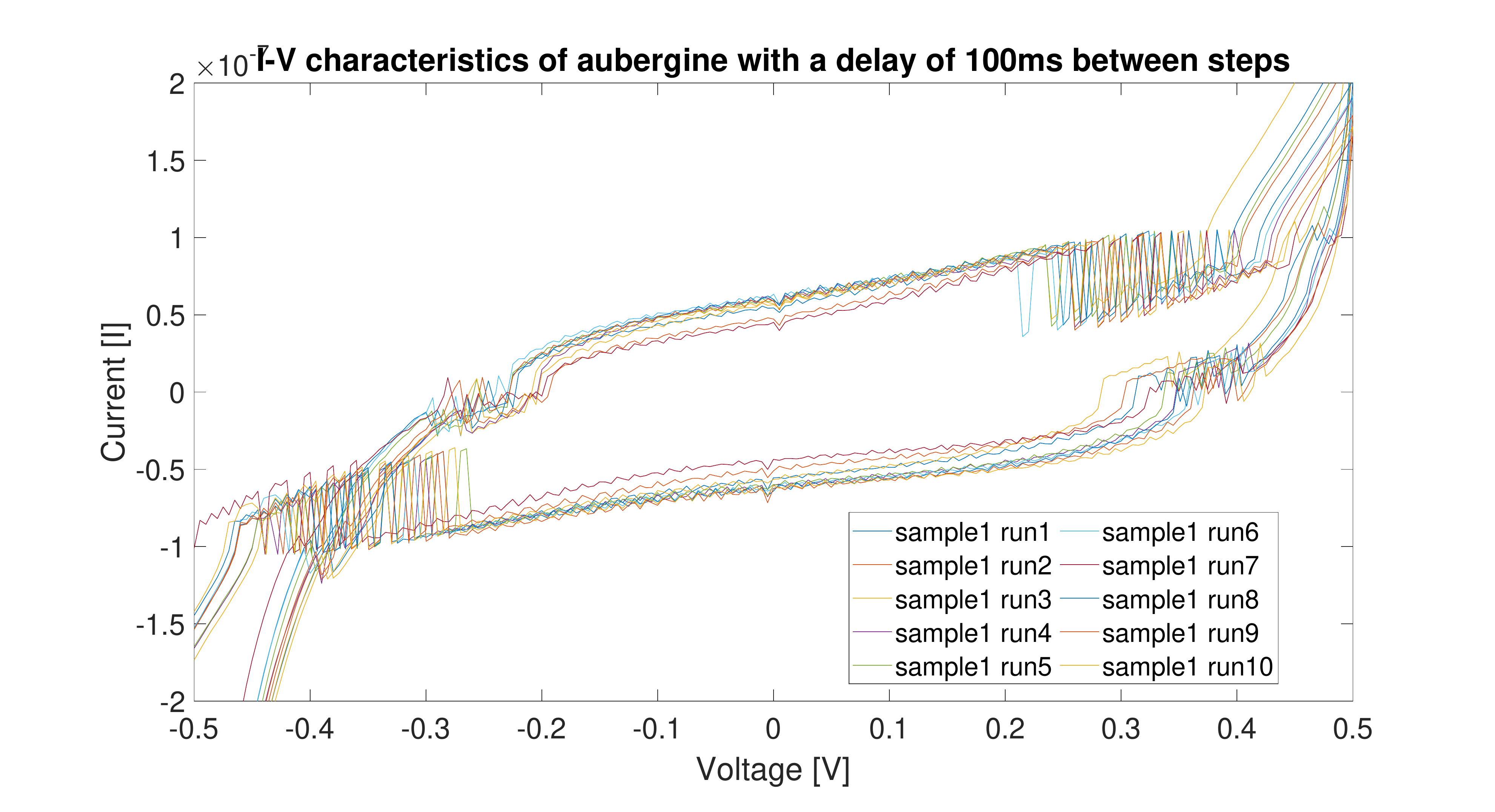}}
    \subfigure[]{\includegraphics[width=0.68\textwidth]{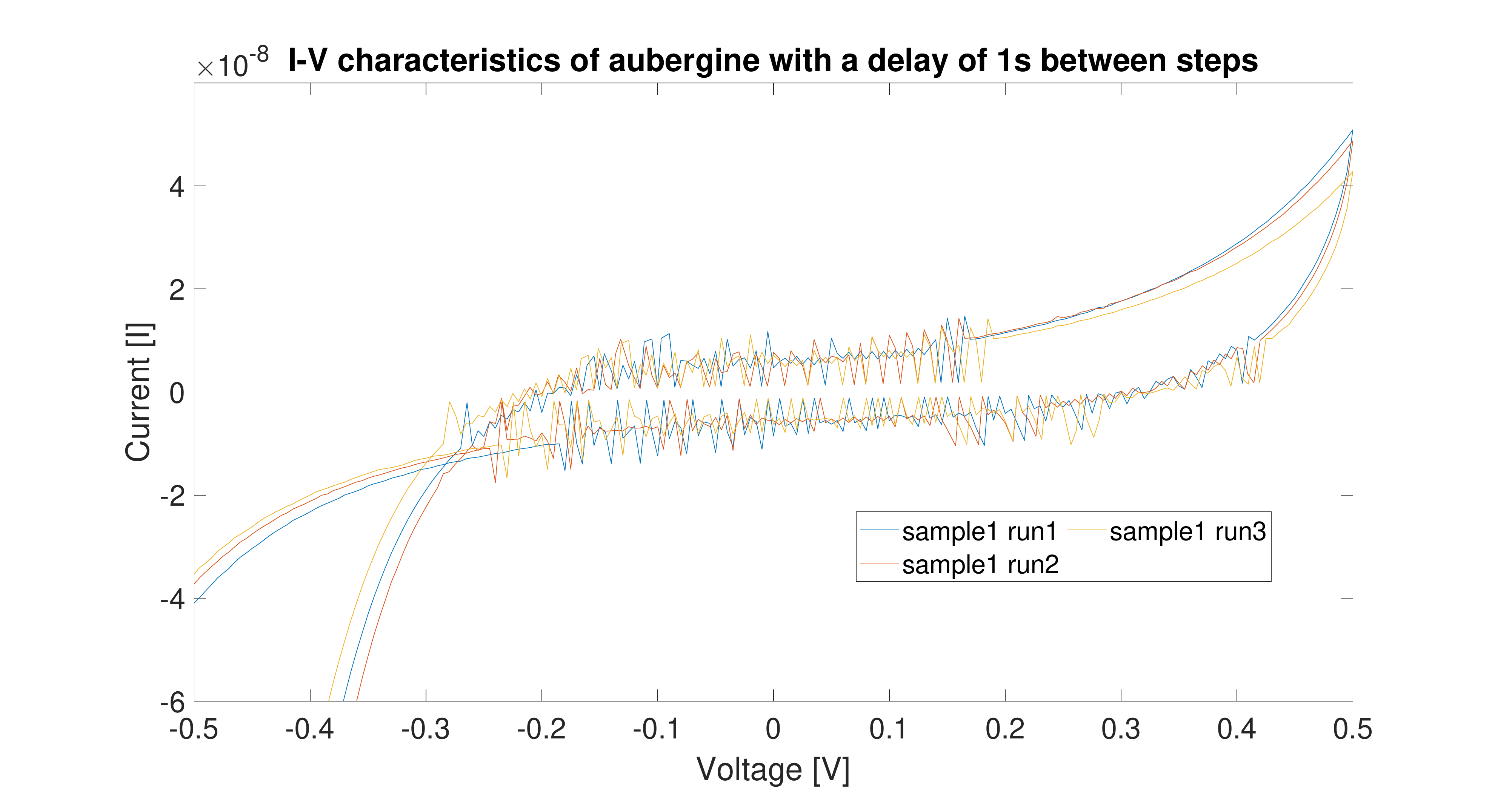}}
    \caption{Cyclic voltammetry (-0V5 to 0V5) of aubergine. (a) delay time between settings is 10ms, (b) delay time between settings is 100ms, (c) delay time between settings is 1000ms }
    \label{fig:aubergine1Vpp}
\end{figure}

\begin{figure}[!hbt]
    \centering
    \subfigure[]{\includegraphics[width=0.7\textwidth]{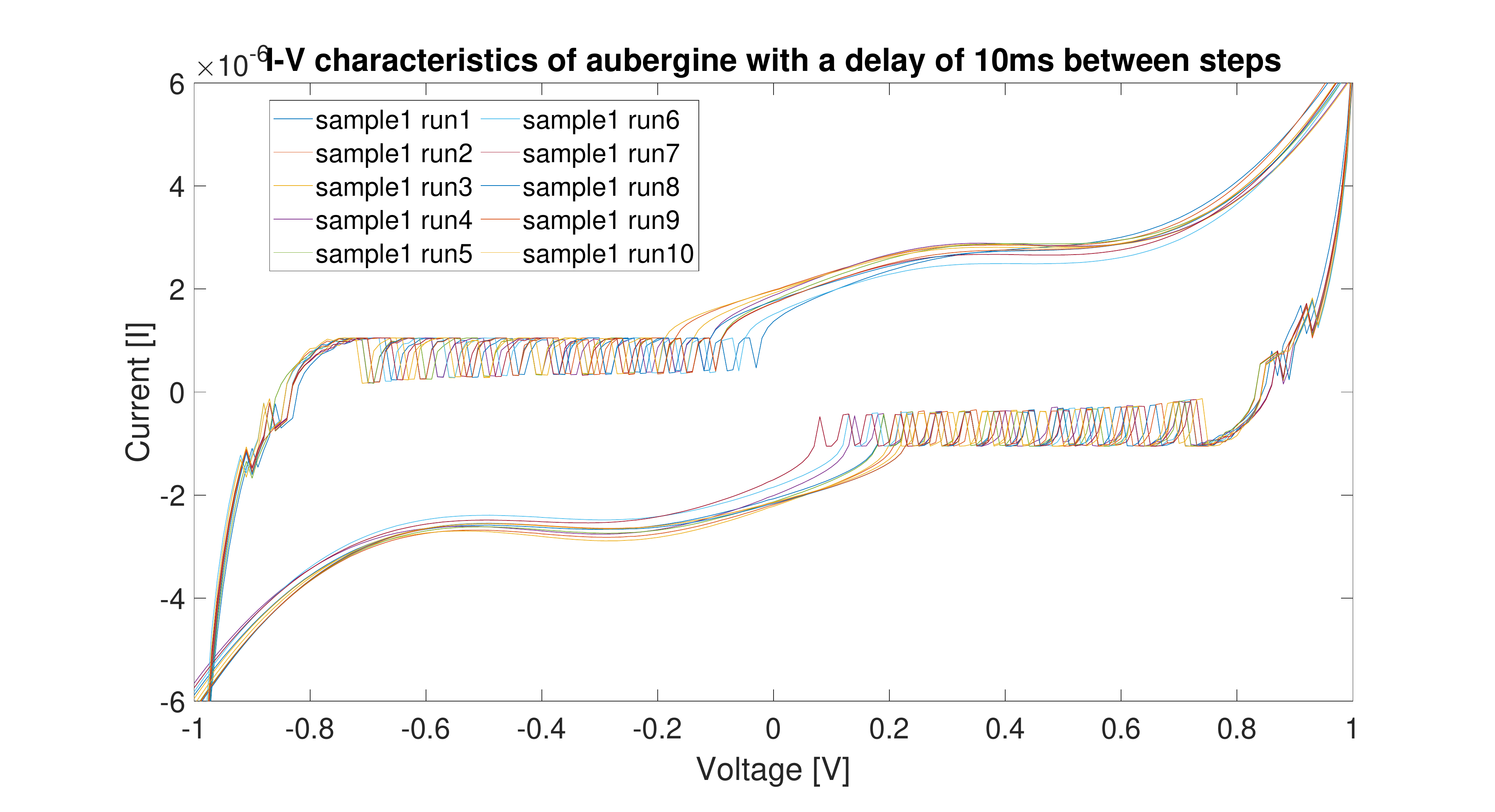}}
    \subfigure[]{\includegraphics[width=0.7\textwidth]{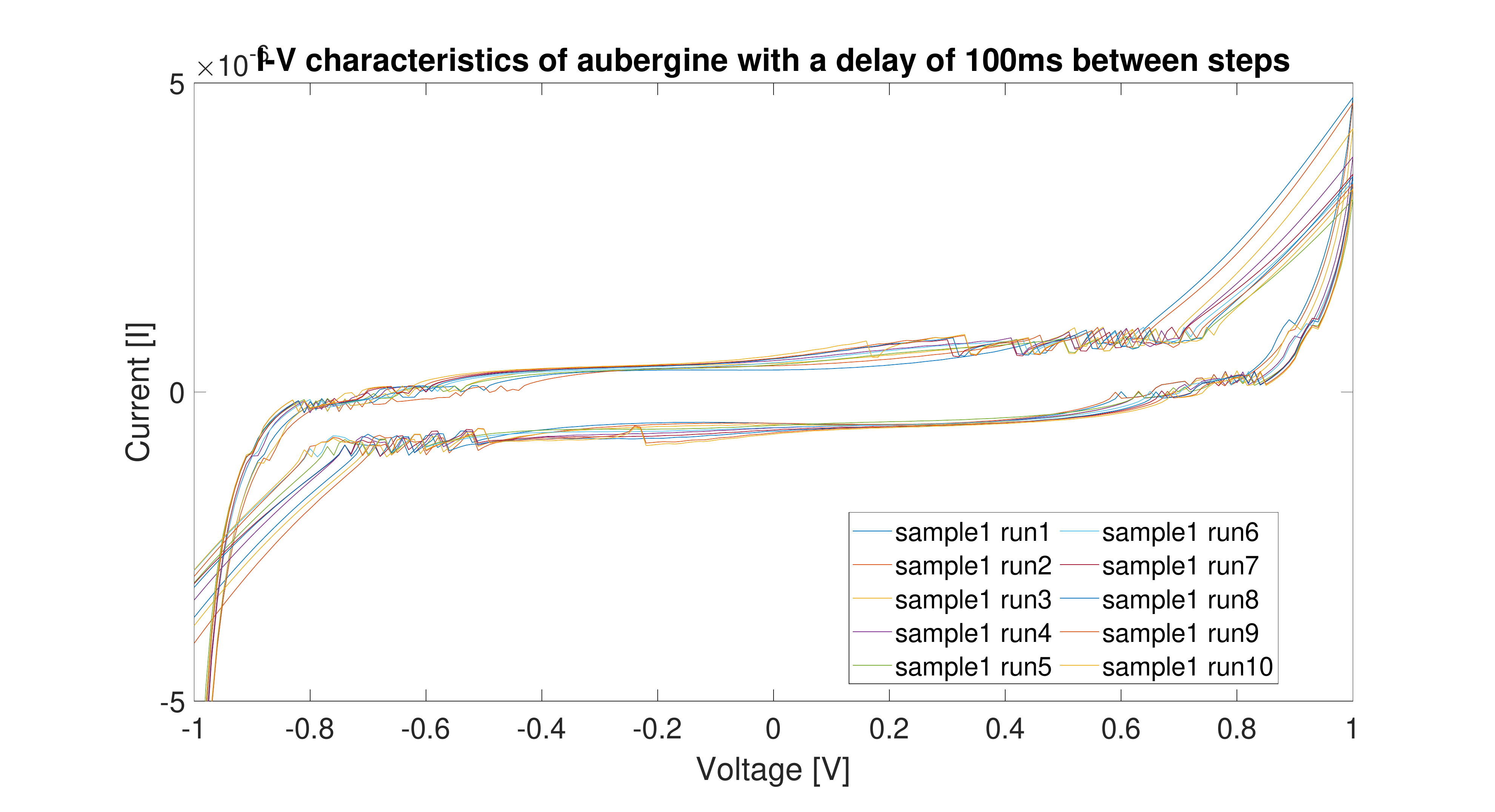}}
    \subfigure[]{\includegraphics[width=0.7\textwidth]{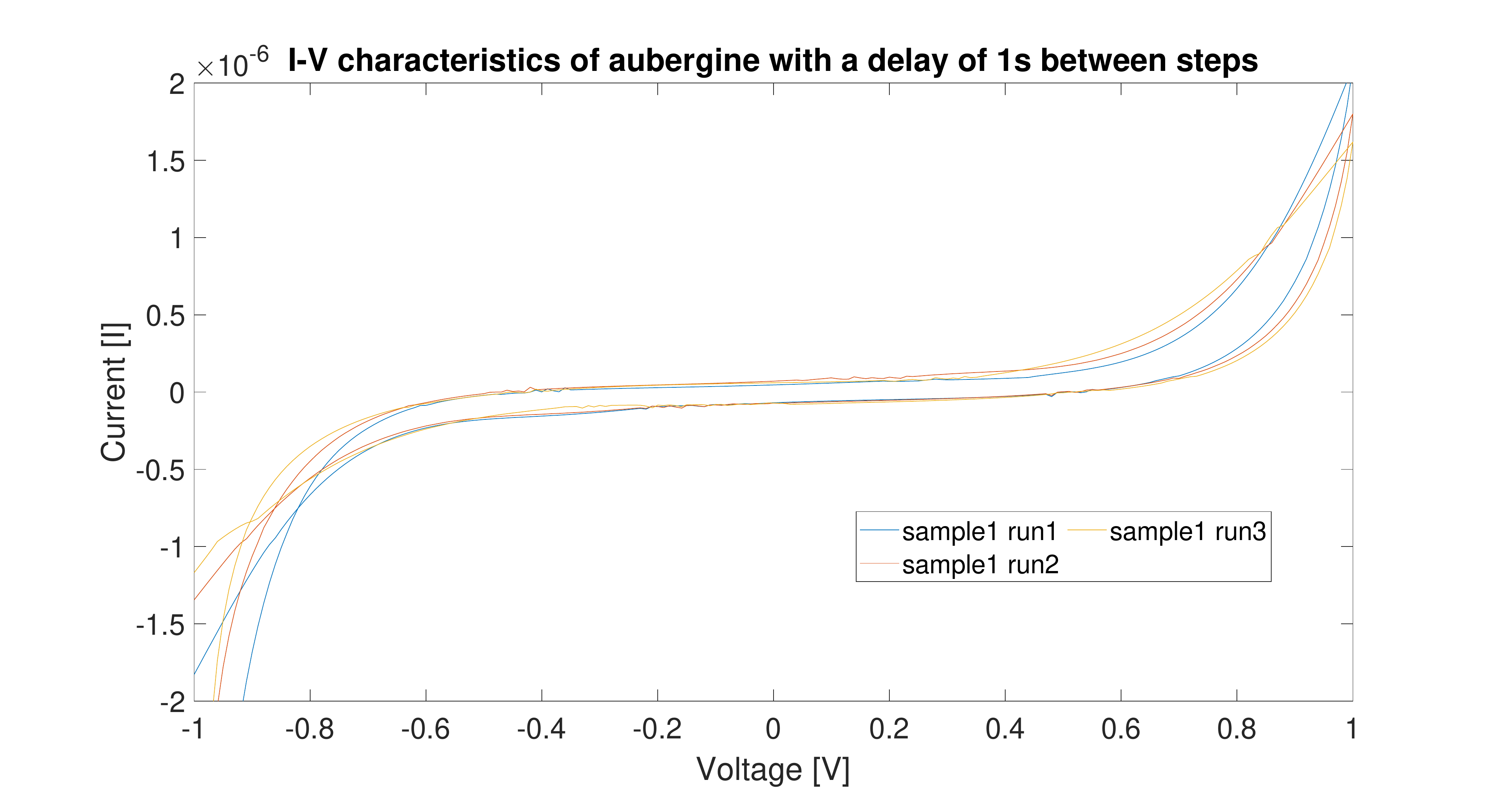}}
    \caption{Cyclic voltammetry (-1V to 1V) of aubergine. (a) delay time between settings is 10ms, (b) delay time between settings is 100ms, (c) delay time between settings is 1000ms }
    \label{fig:aubergine2Vpp}
\end{figure}

\begin{figure}[!hbt]
    \centering
    \subfigure[]{\includegraphics[width=0.7\textwidth]{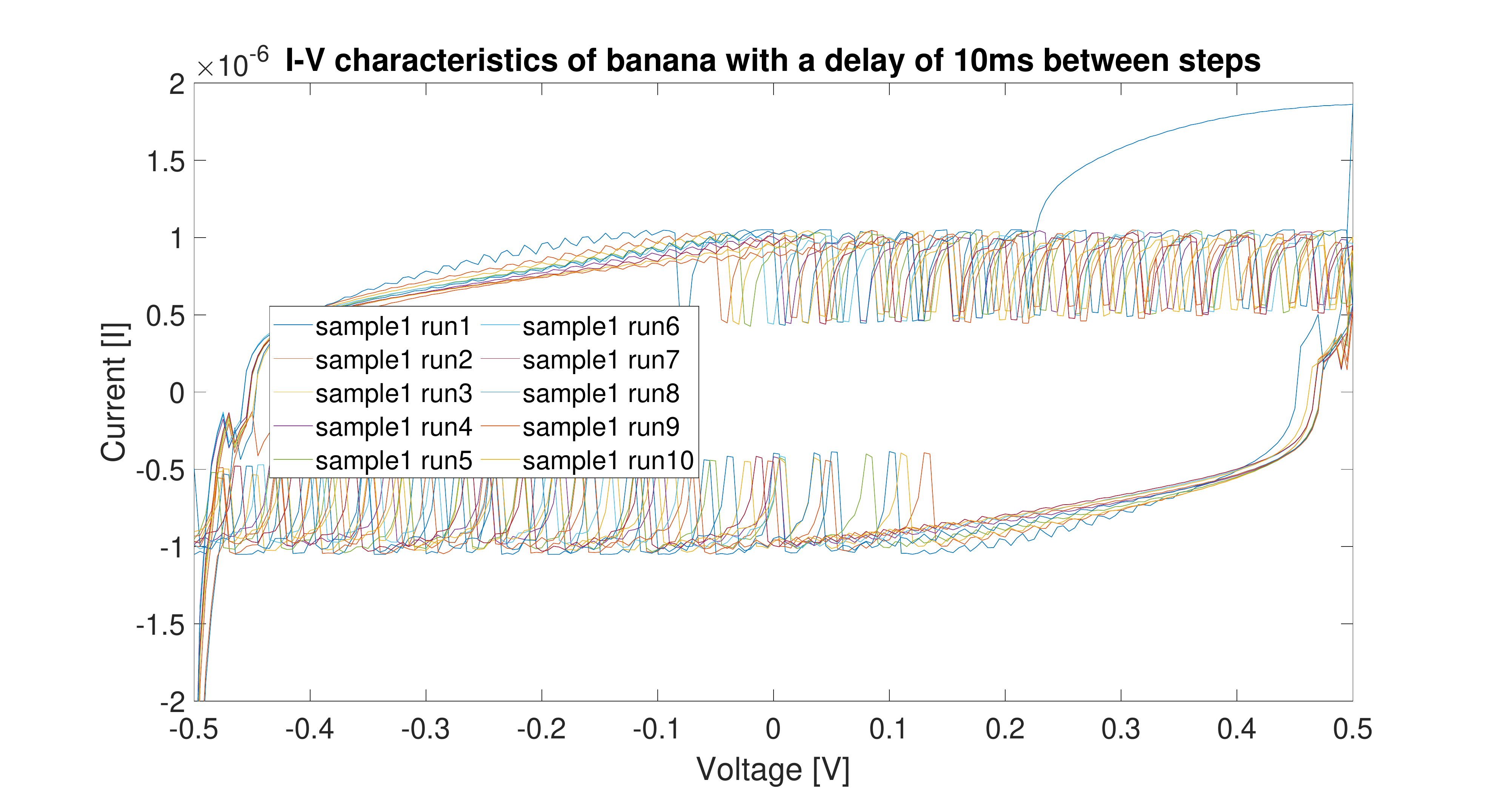}}
    \subfigure[]{\includegraphics[width=0.7\textwidth]{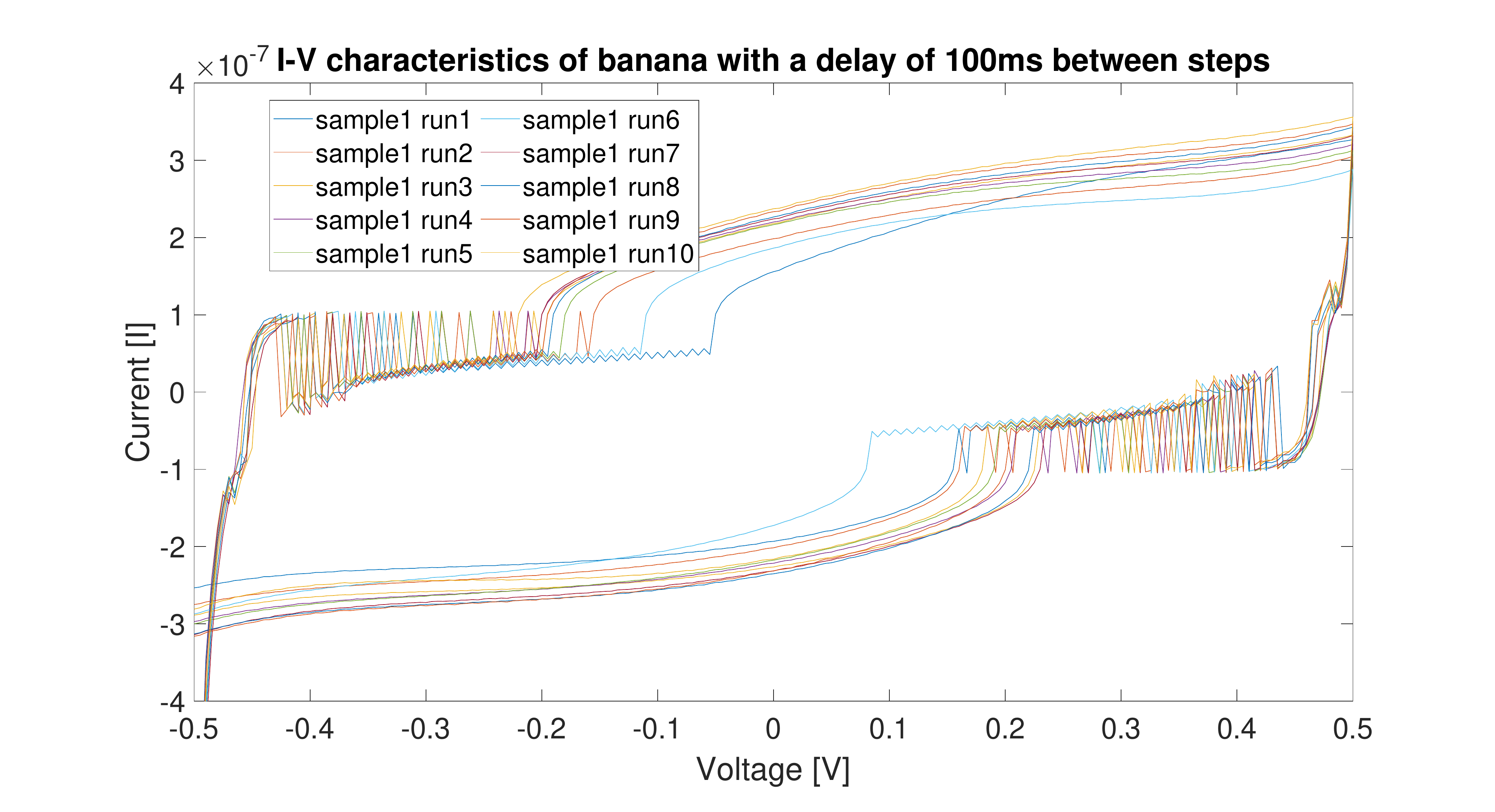}}
    \subfigure[]{\includegraphics[width=0.7\textwidth]{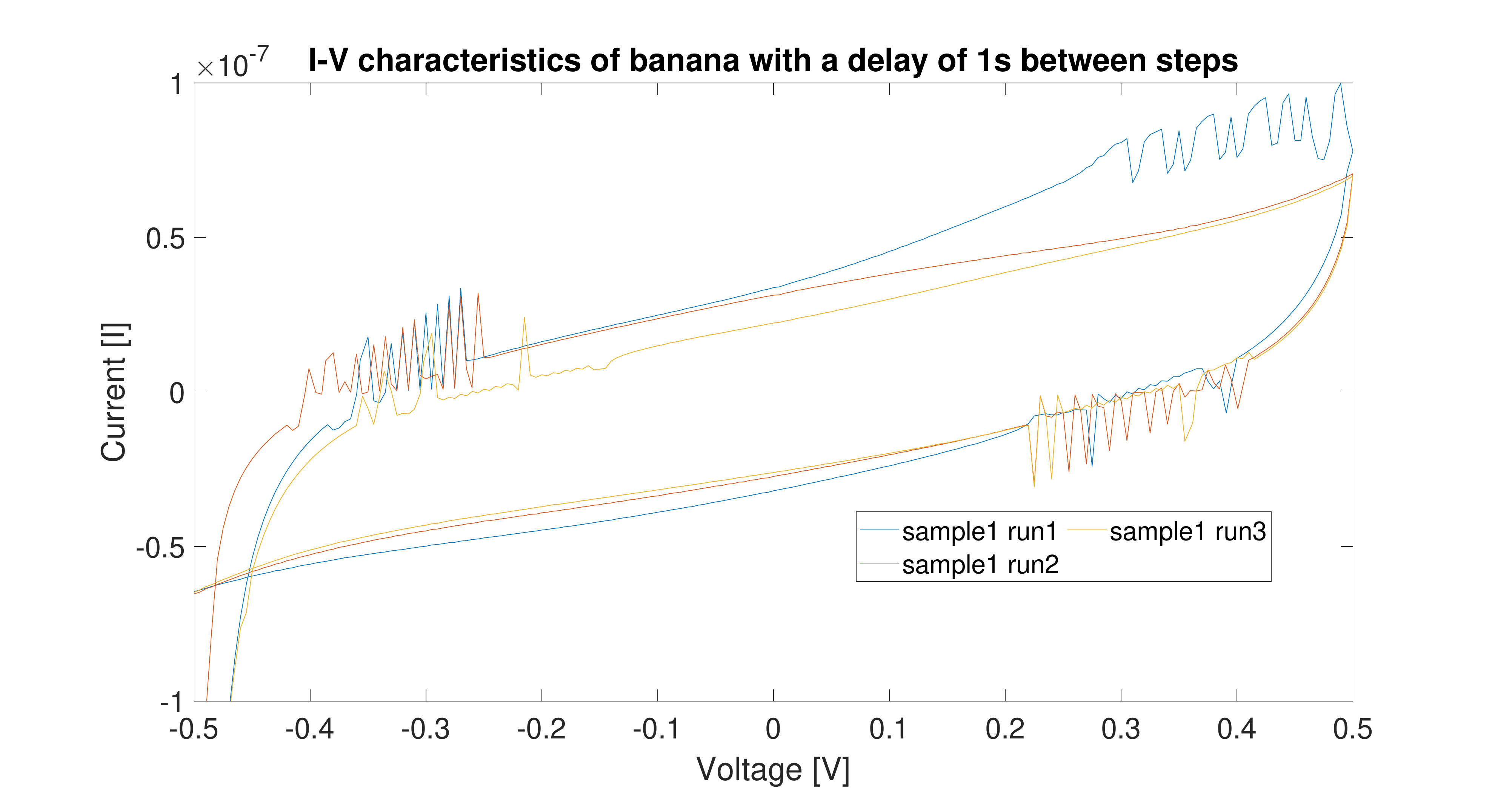}}
    \caption{Cyclic voltammetry (-0V5 to 0V5) of banana. (a) delay time between settings is 10ms, (b) delay time between settings is 100ms, (c) delay time between settings is 1000ms }
    \label{fig:banana1Vpp}
\end{figure}

\begin{figure}[!hbt]
    \centering
    \subfigure[]{\includegraphics[width=0.7\textwidth]{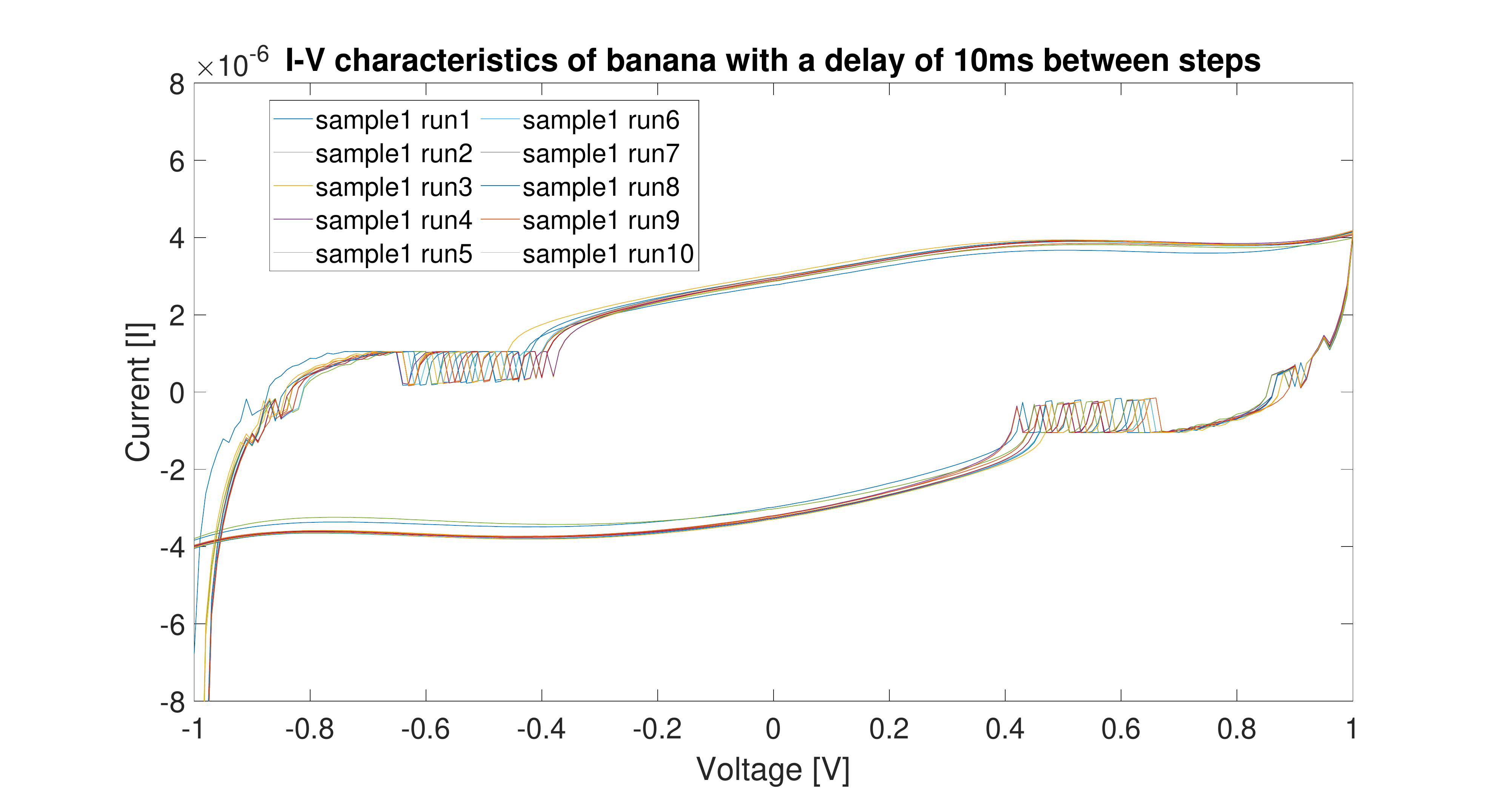}}
    \subfigure[]{\includegraphics[width=0.7\textwidth]{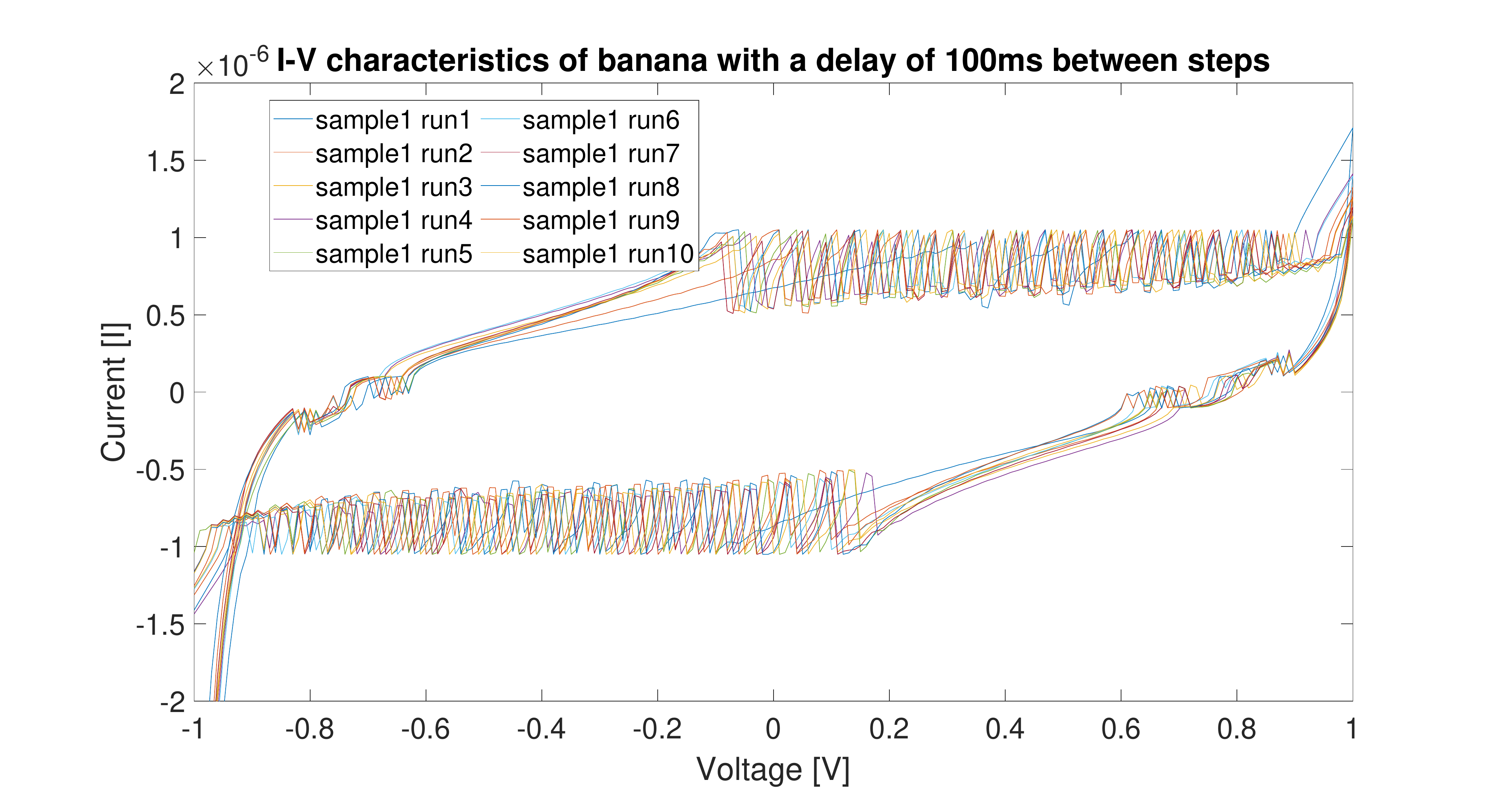}}
    \subfigure[]{\includegraphics[width=0.7\textwidth]{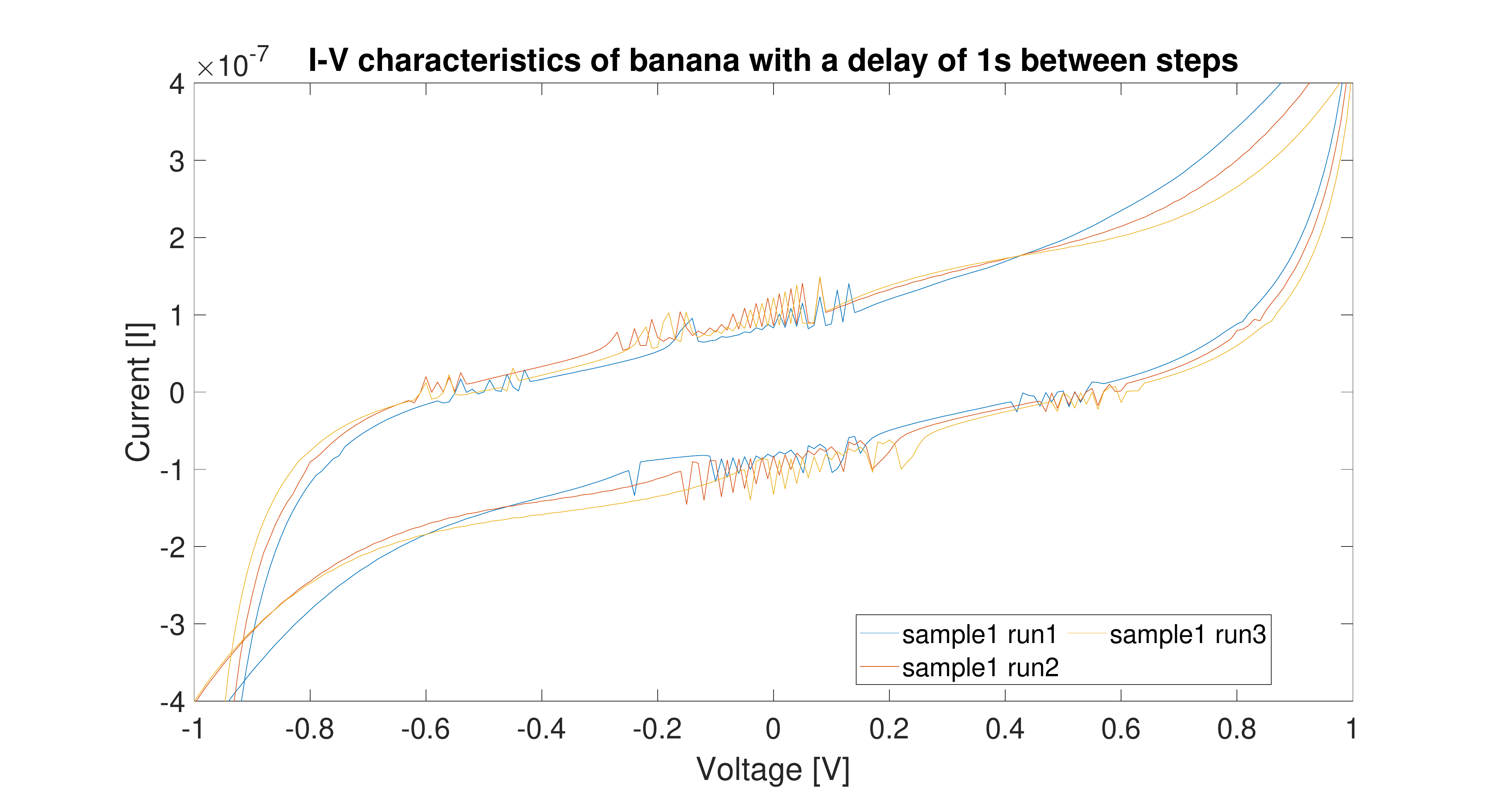}}
    \caption{Cyclic voltammetry (-1V to 1V) of banana. (a) delay time between settings is 10ms, (b) delay time between settings is 100ms, (c) delay time between settings is 1000ms }
    \label{fig:banana2Vpp}
\end{figure}

\begin{figure}[!hbt]
    \centering
    \subfigure[]{\includegraphics[width=0.7\textwidth]{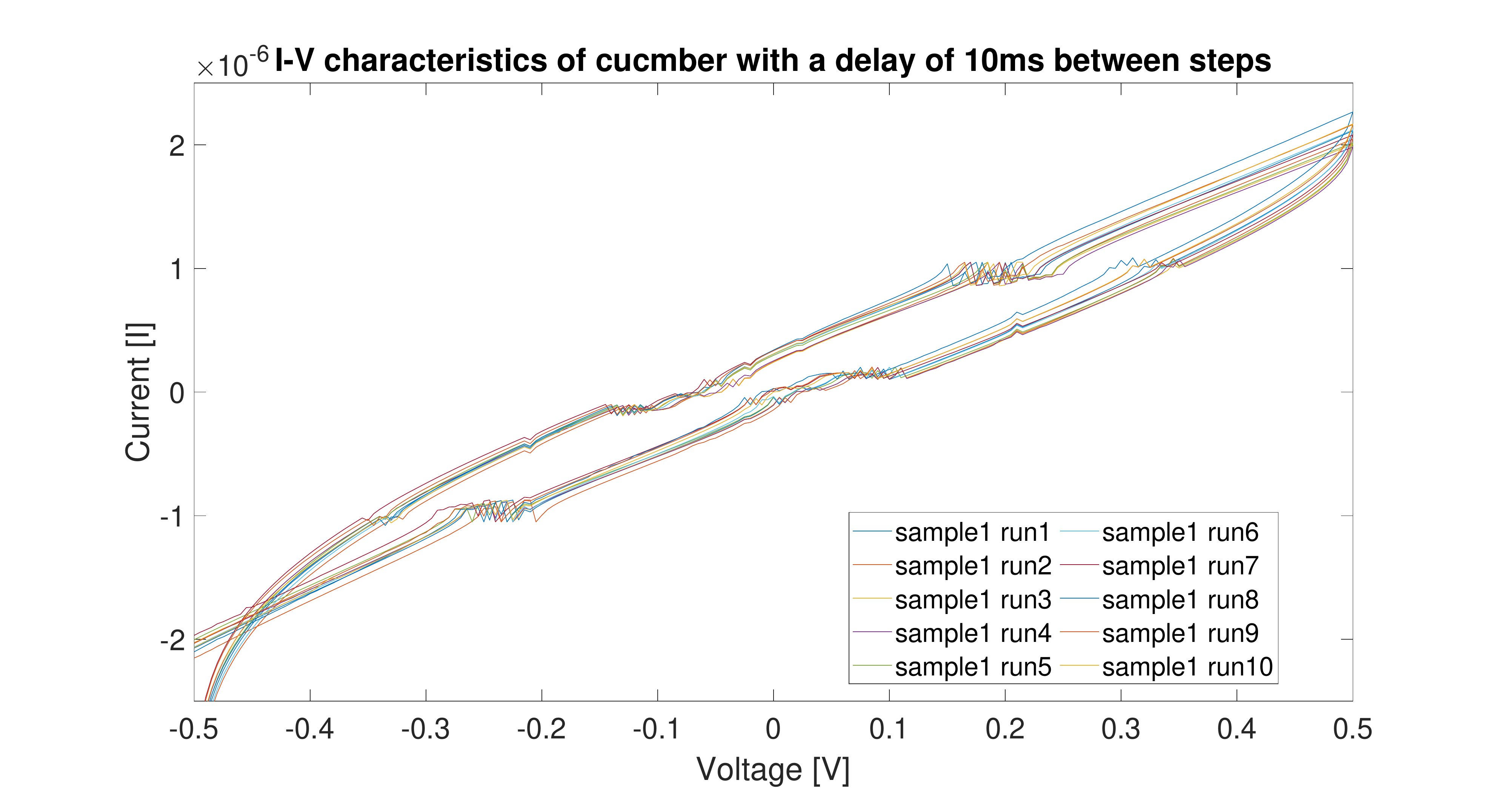}}
    \subfigure[]{\includegraphics[width=0.7\textwidth]{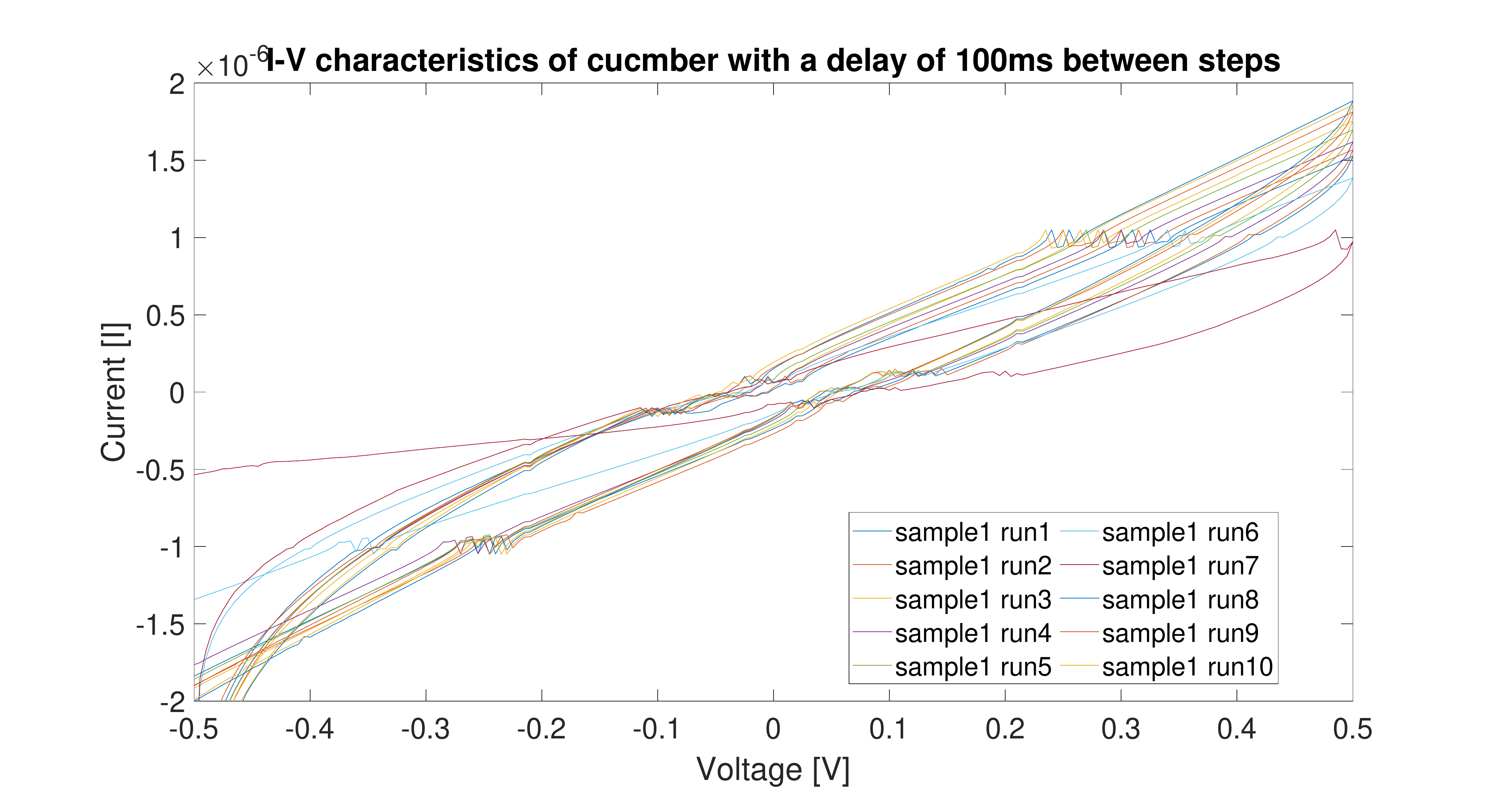}}
    \subfigure[]{\includegraphics[width=0.7\textwidth]{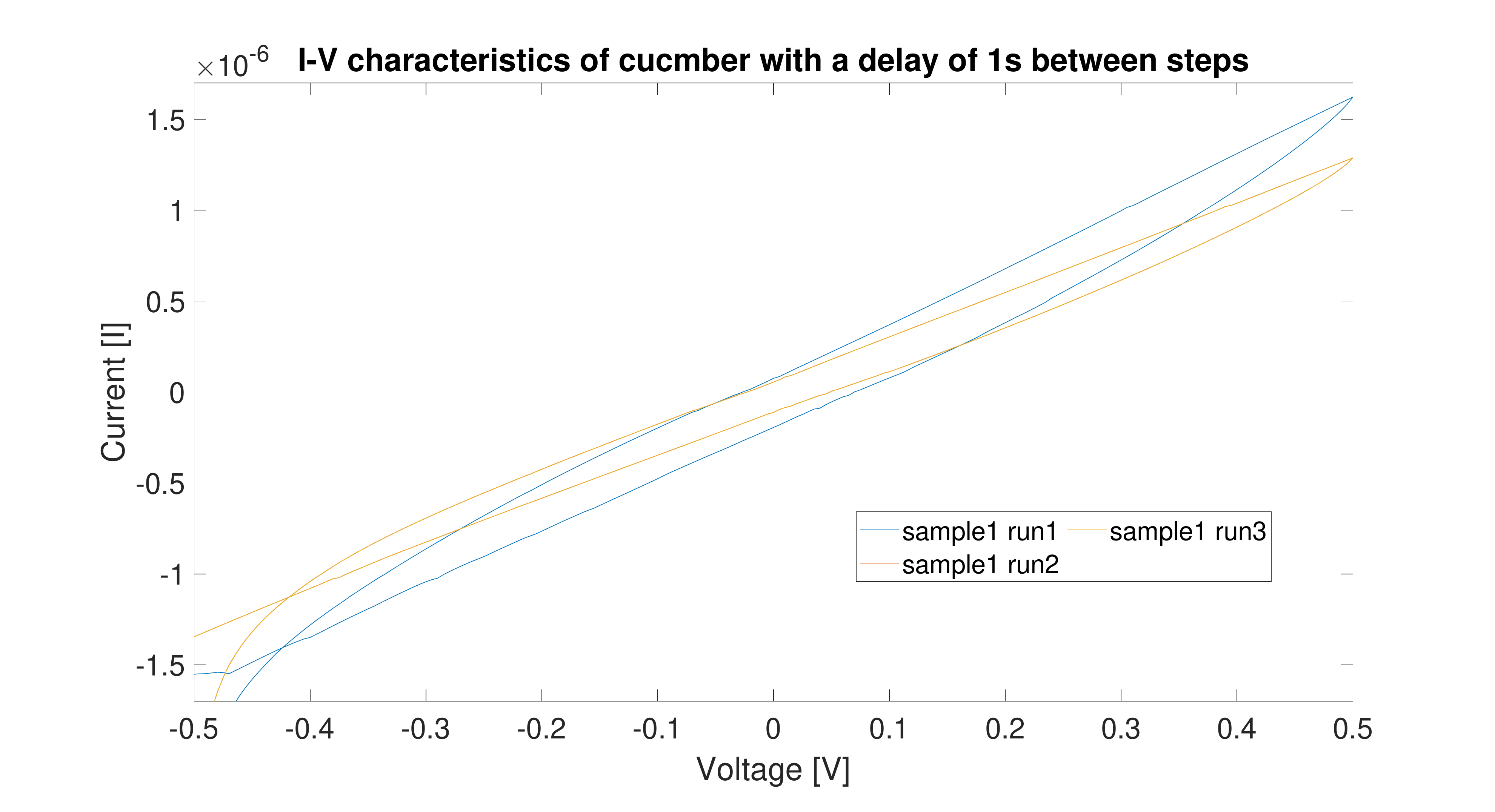}}
    \caption{Cyclic voltammetry (-0V5 to 0V5) of cucumber. (a) delay time between settings is 10ms, (b) delay time between settings is 100ms, (c) delay time between settings is 1000ms }
    \label{fig:cucumber1Vpp}
\end{figure}

\begin{figure}[!hbt]
    \centering
    \subfigure[]{\includegraphics[width=0.7\textwidth]{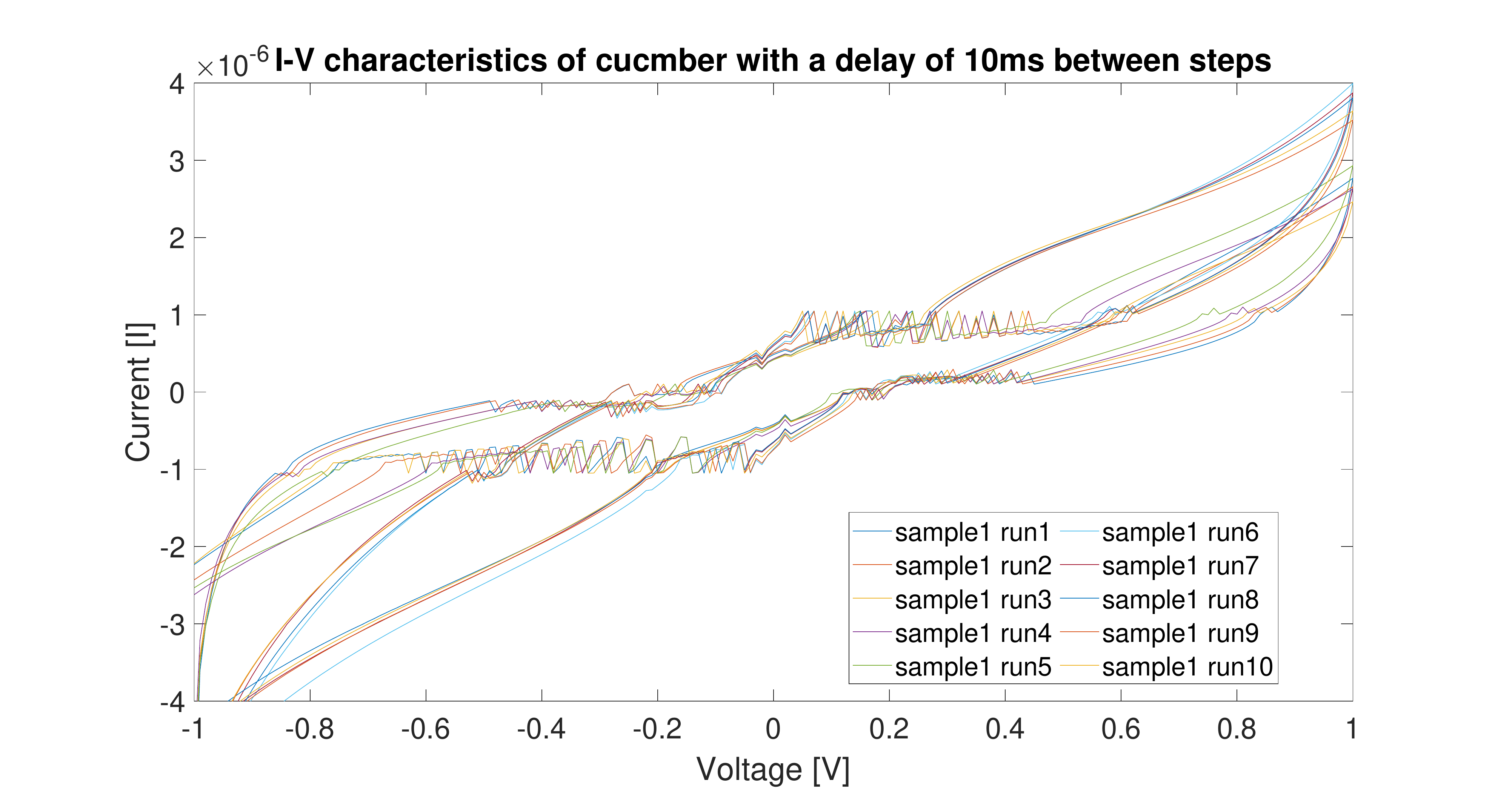}}
    \subfigure[]{\includegraphics[width=0.7\textwidth]{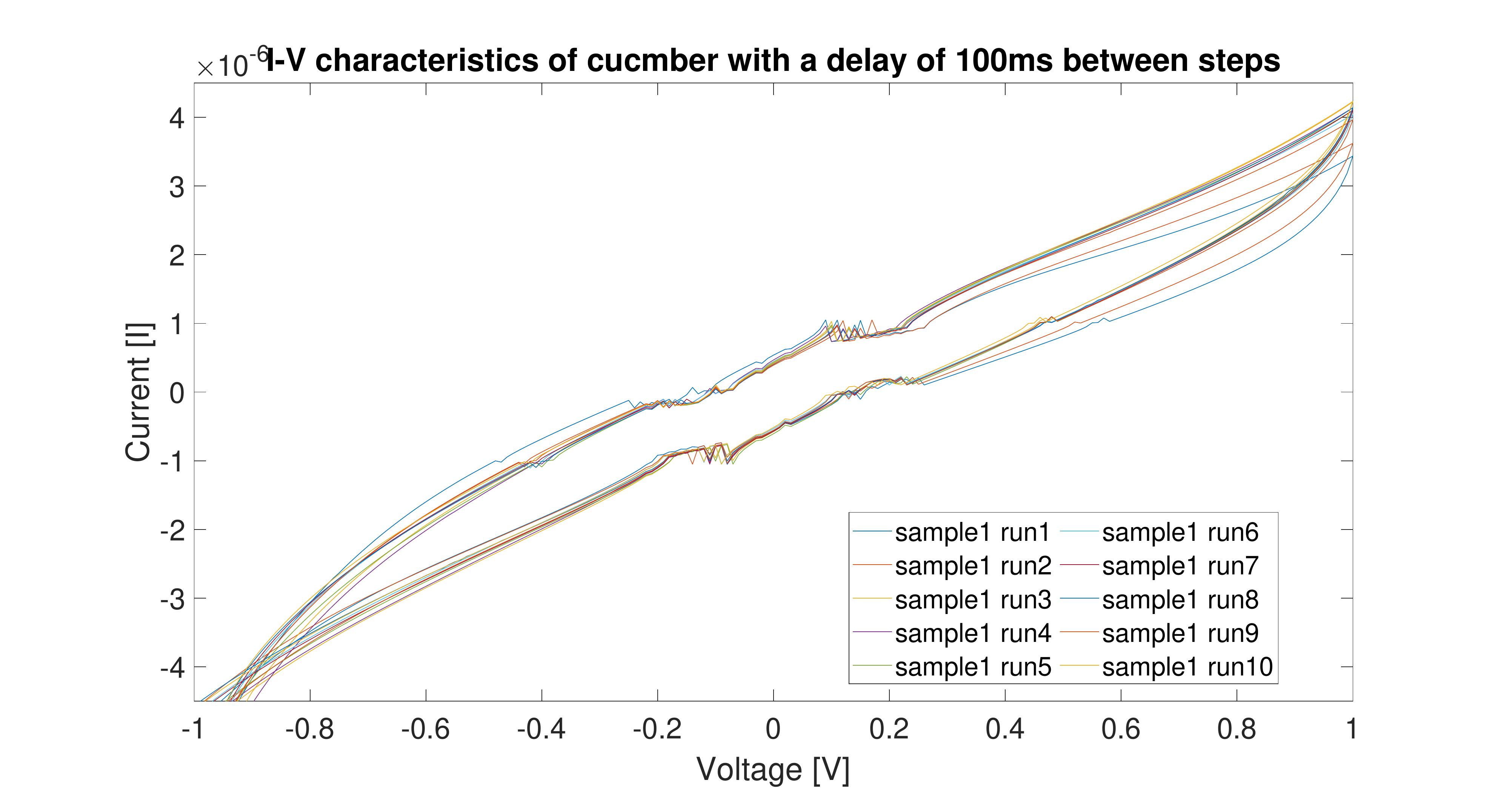}}
    \subfigure[]{\includegraphics[width=0.7\textwidth]{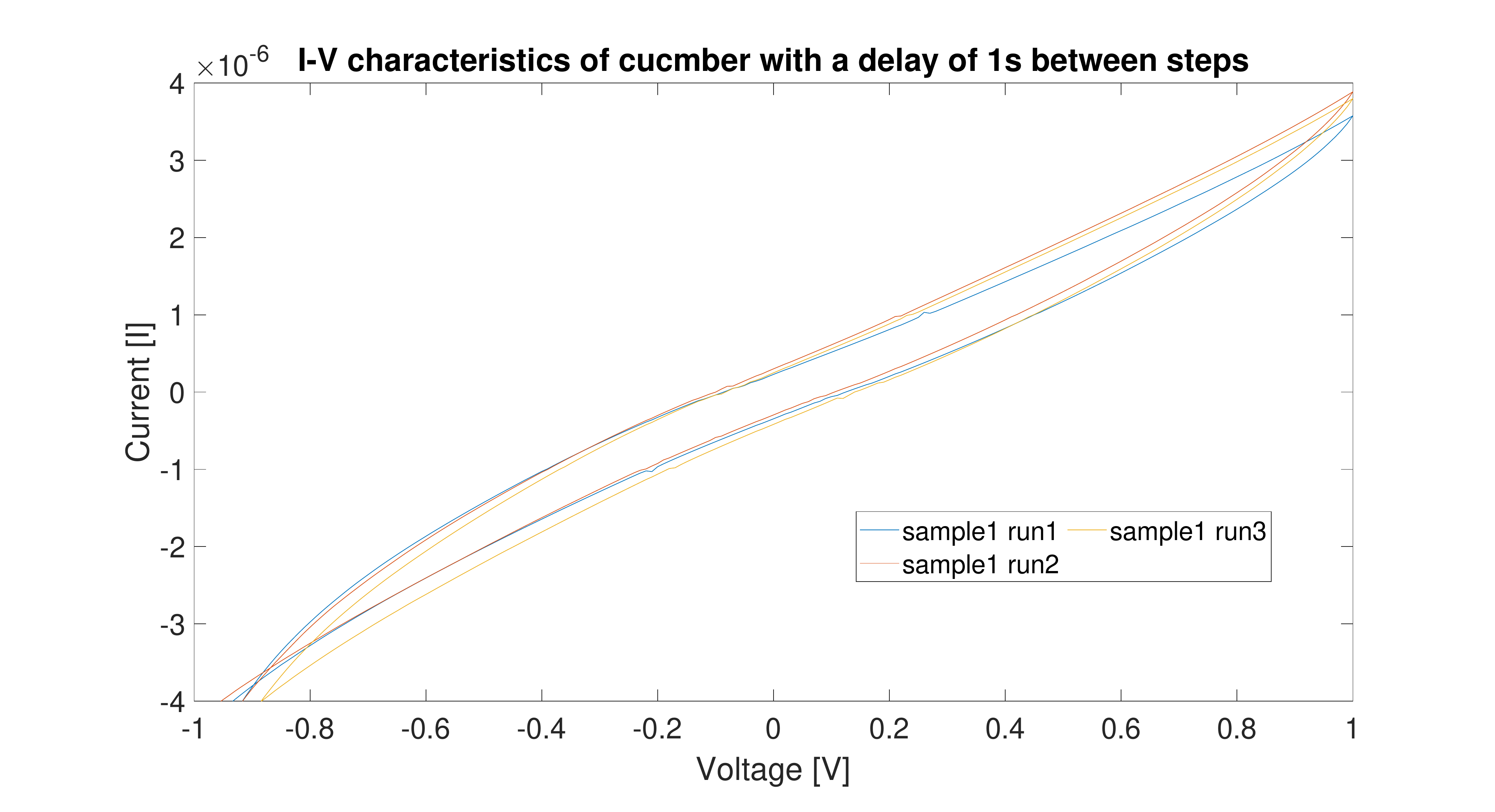}}
    \caption{Cyclic voltammetry (-1V to 1V) of cucumber. (a) delay time between settings is 10ms, (b) delay time between settings is 100ms, (c) delay time between settings is 1000ms }
    \label{fig:cucumber2Vpp}
\end{figure}

\begin{figure}[!hbt]
    \centering
    \subfigure[]{\includegraphics[width=0.7\textwidth]{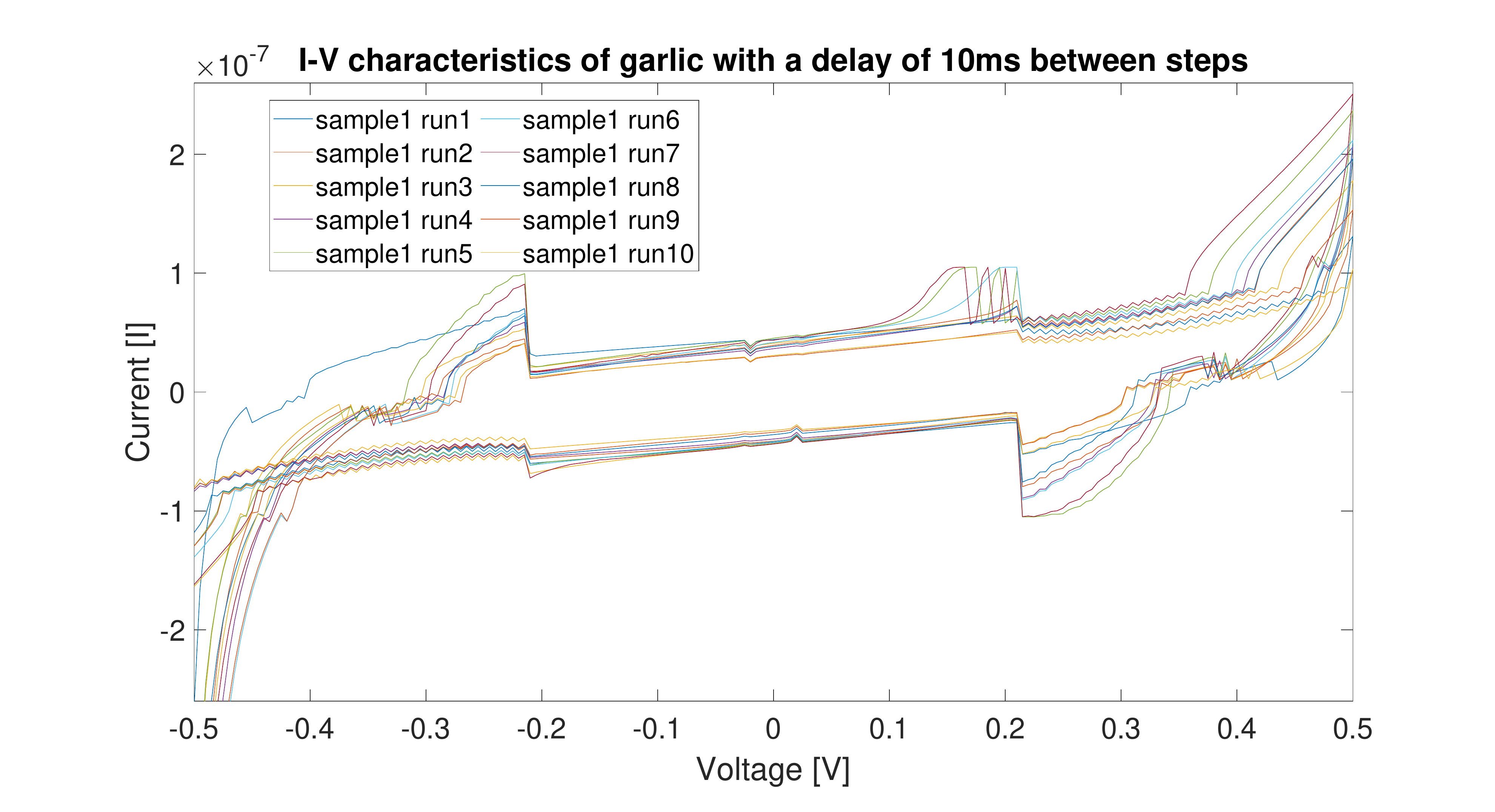}}
    \subfigure[]{\includegraphics[width=0.7\textwidth]{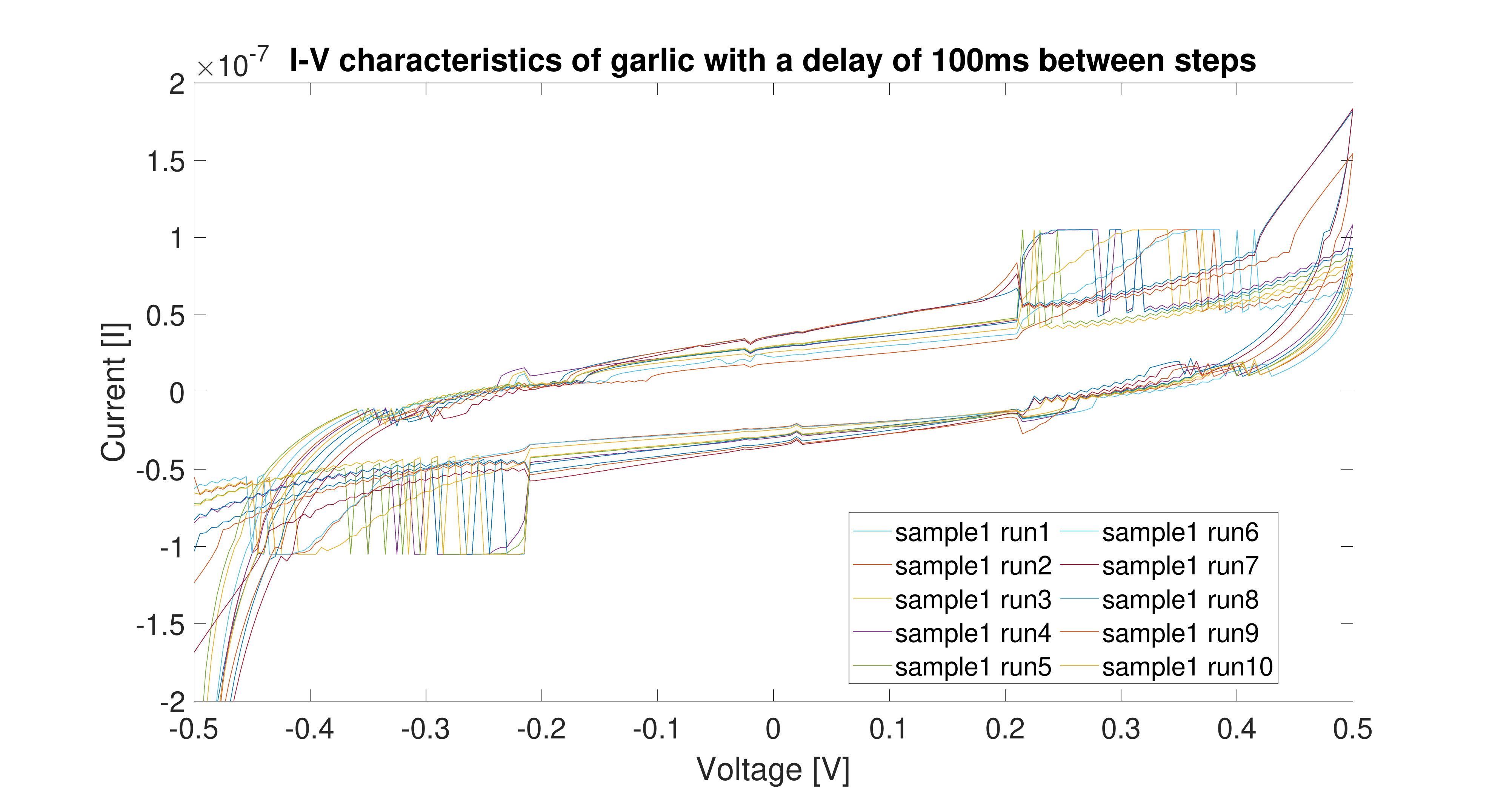}}
    \subfigure[]{\includegraphics[width=0.7\textwidth]{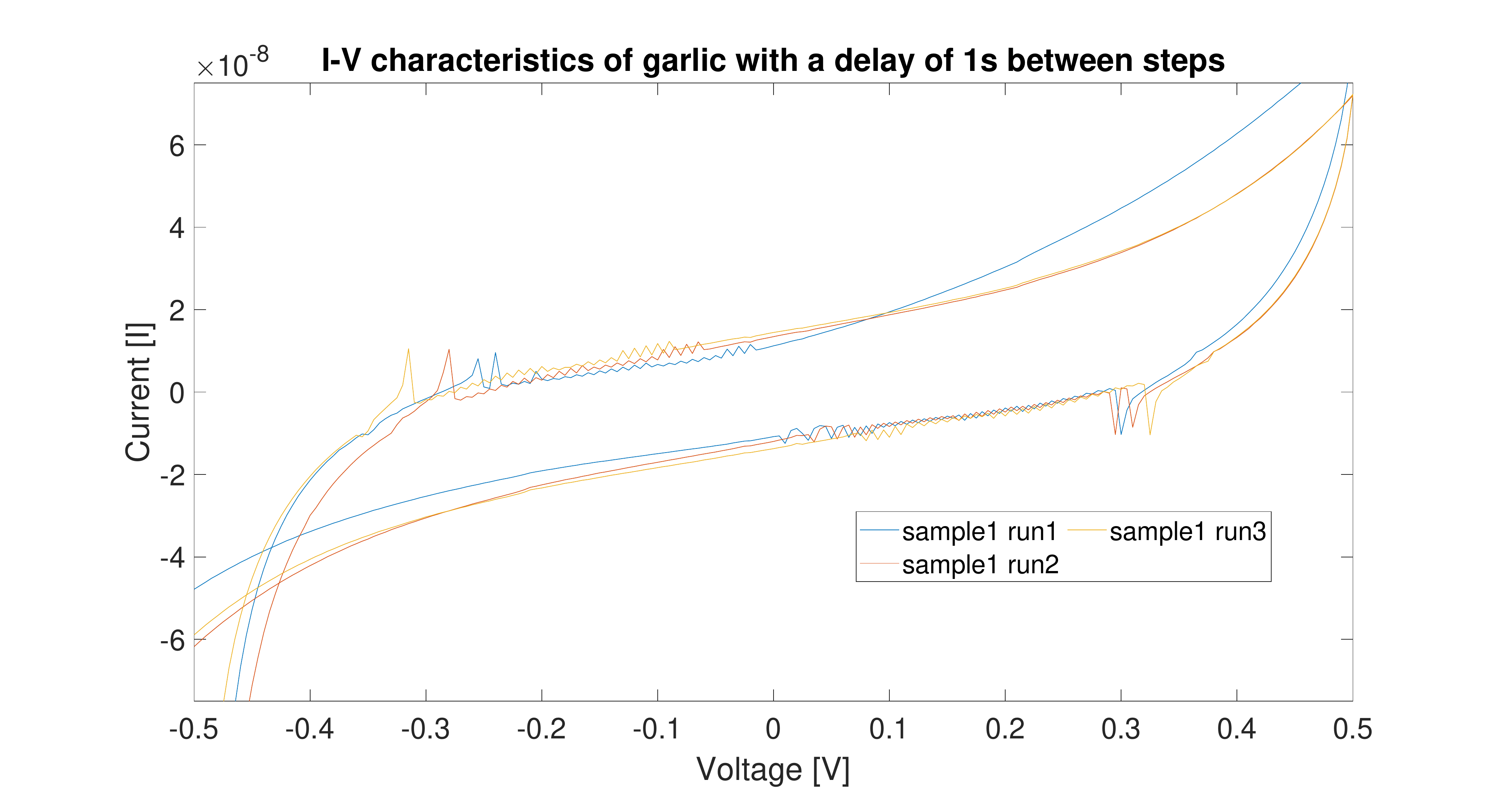}}
    \caption{Cyclic voltammetry (-0V5 to 0V5) of garlic. (a) delay time between settings is 10ms, (b) delay time between settings is 100ms, (c) delay time between settings is 1000ms }
    \label{fig:garlic1Vpp}
\end{figure}

\begin{figure}[!hbt]
    \centering
    \subfigure[]{\includegraphics[width=0.7\textwidth]{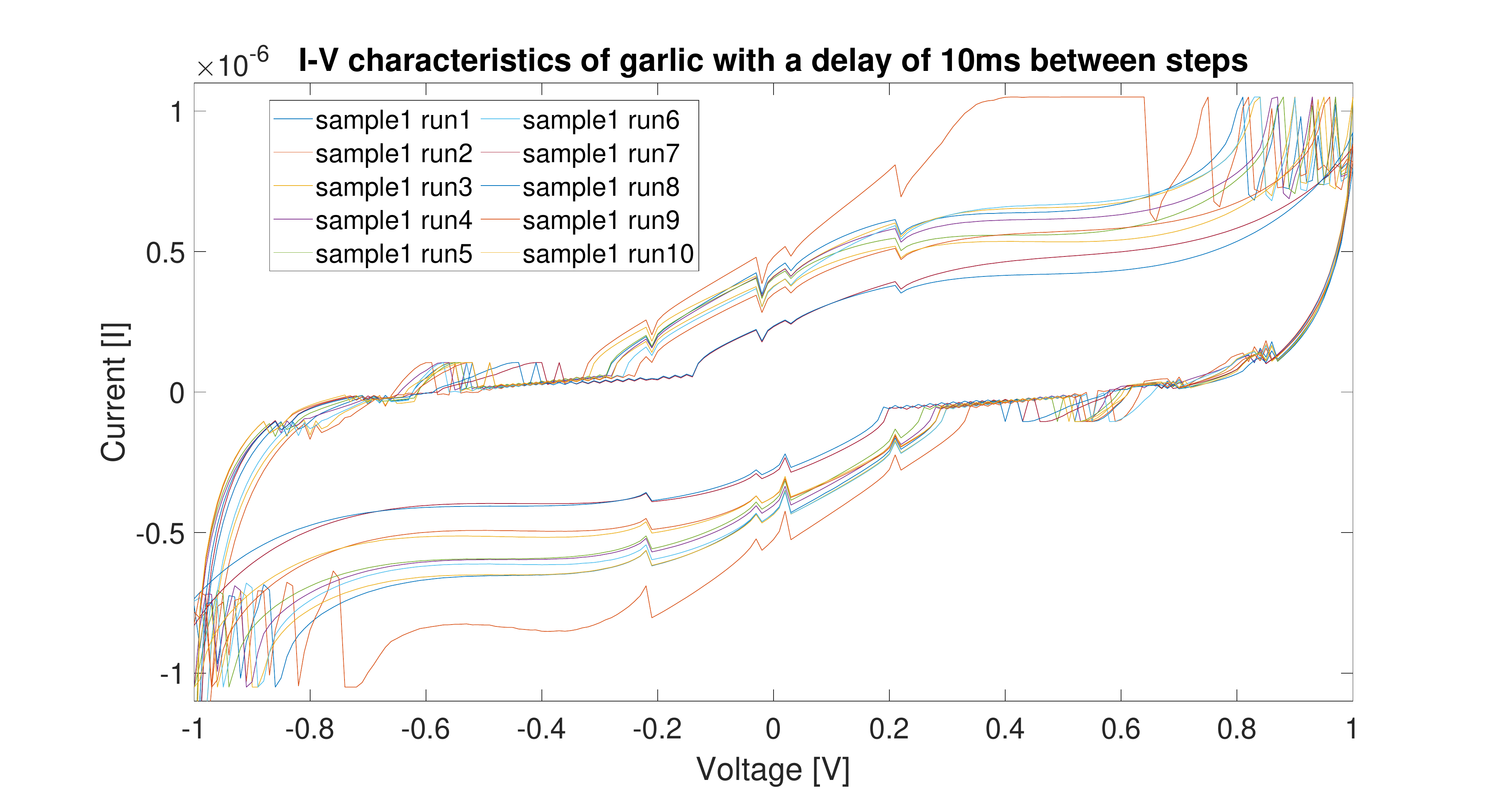}}
    \subfigure[]{\includegraphics[width=0.7\textwidth]{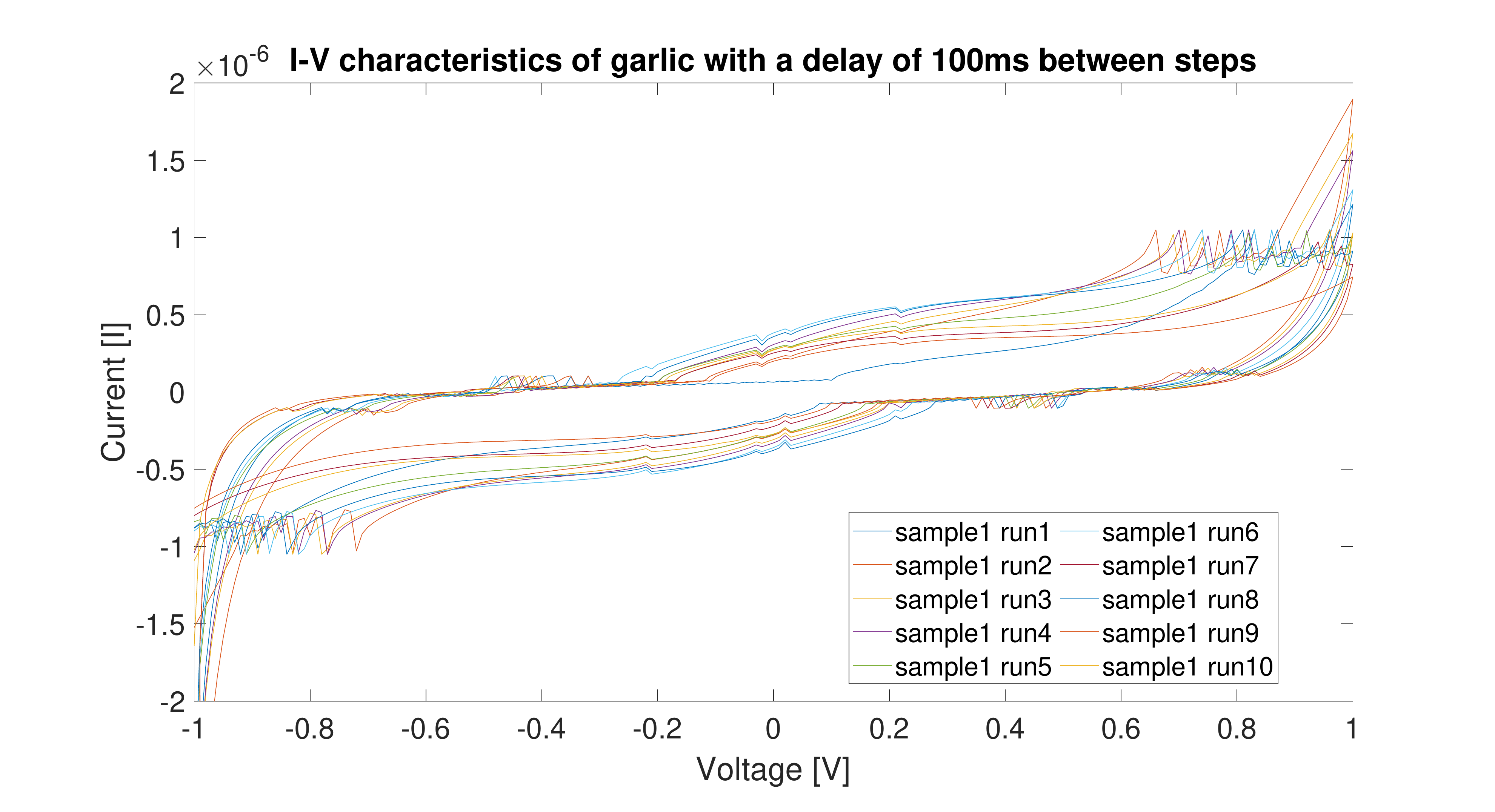}}
    \subfigure[]{\includegraphics[width=0.7\textwidth]{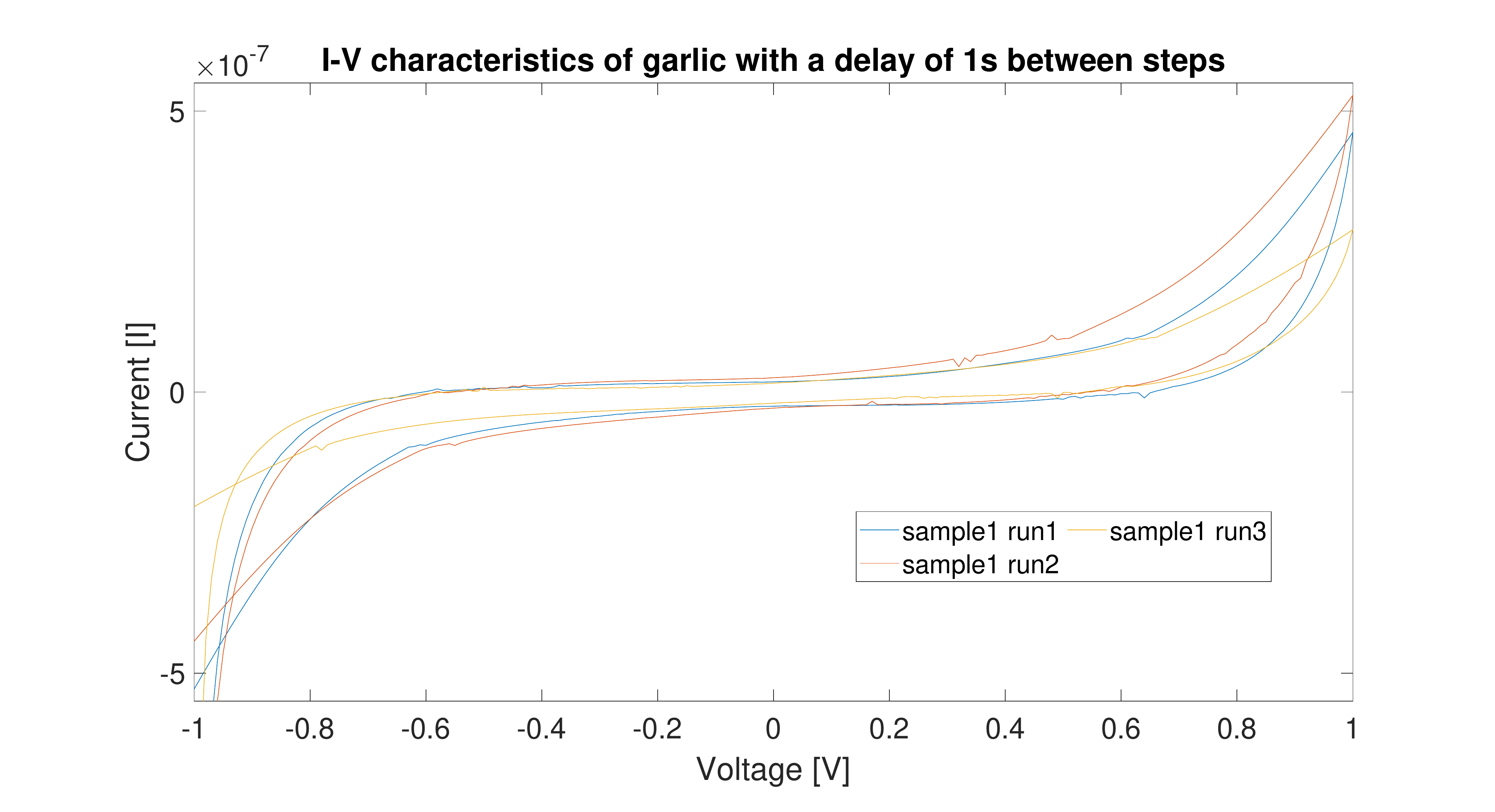}}
    \caption{Cyclic voltammetry (-1V to 1V) of garlic. (a) delay time between settings is 10ms, (b) delay time between settings is 100ms, (c) delay time between settings is 1000ms }
    \label{fig:garlic2Vpp}
\end{figure}

\begin{figure}[!hbt]
    \centering
    \subfigure[]{\includegraphics[width=0.7\textwidth]{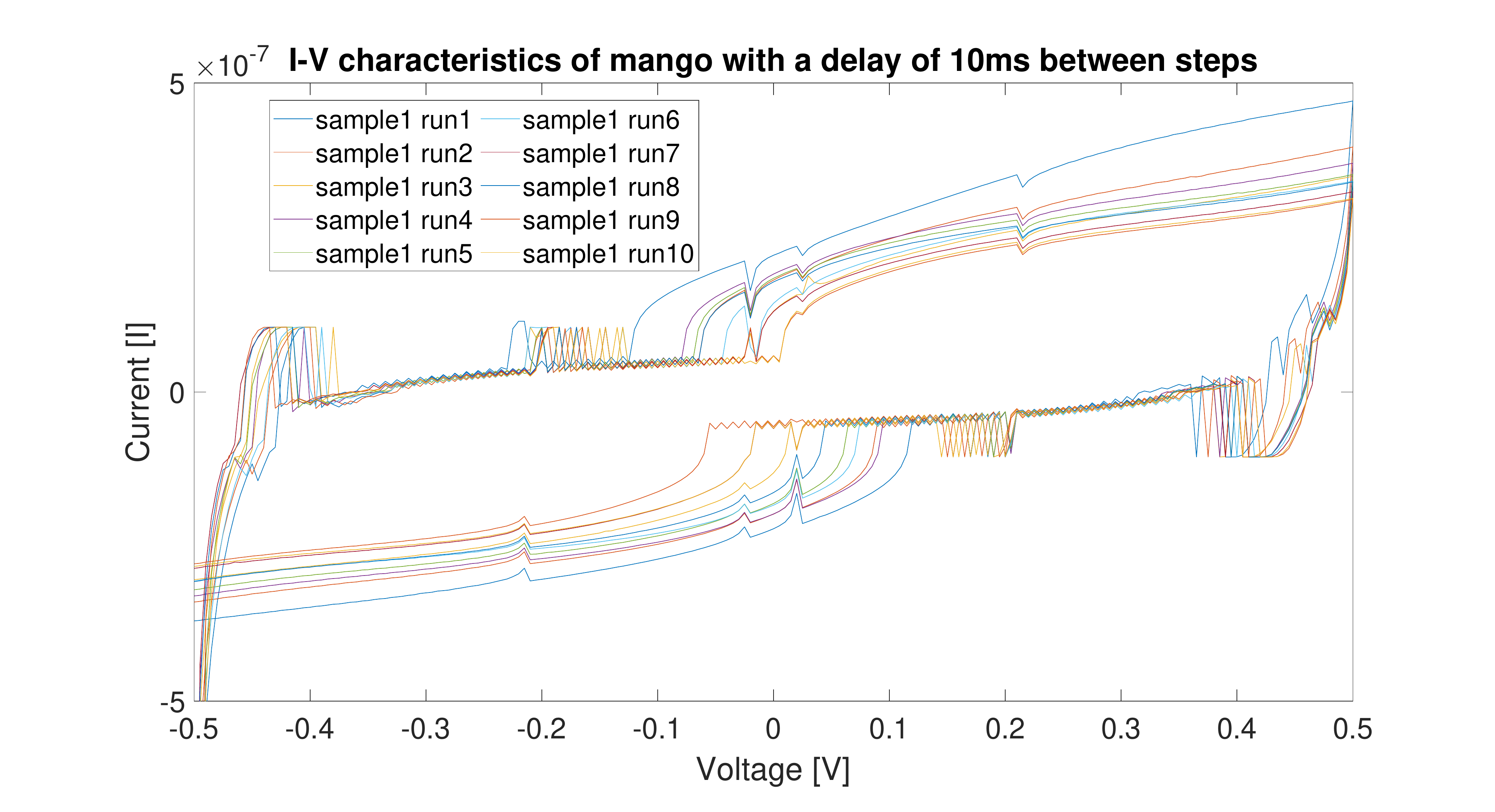}}
    \subfigure[]{\includegraphics[width=0.7\textwidth]{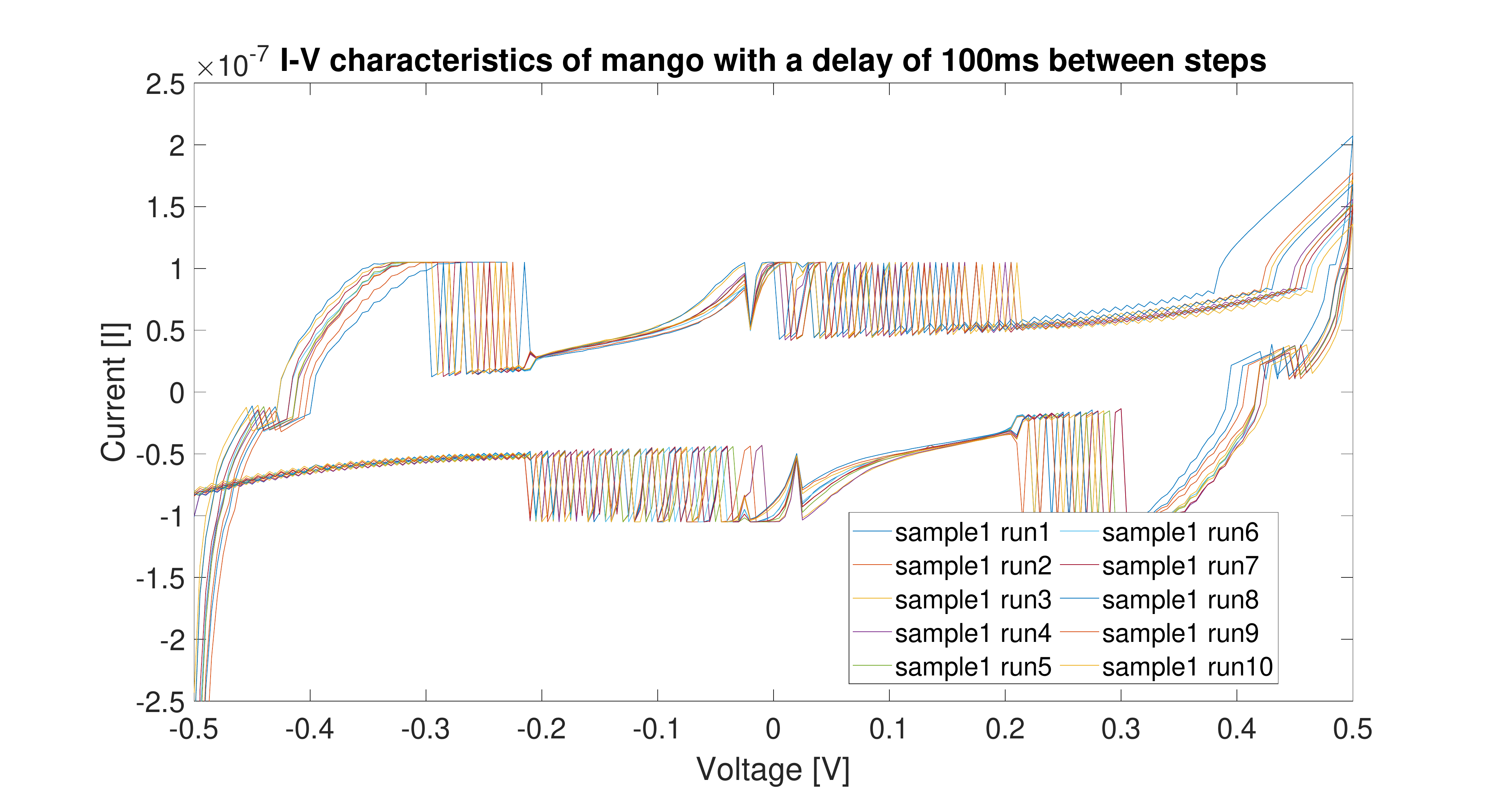}}
    \subfigure[]{\includegraphics[width=0.7\textwidth]{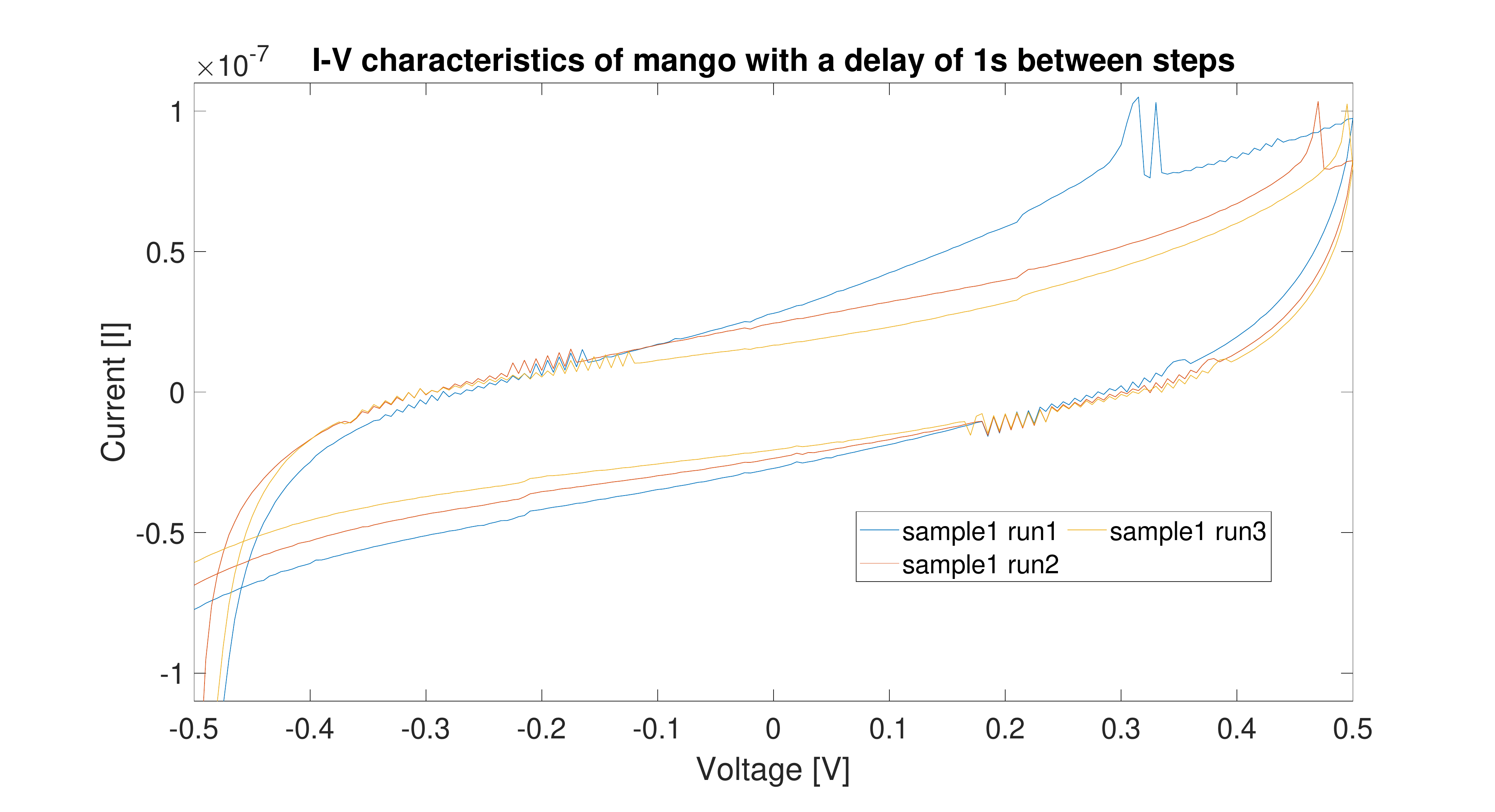}}
    \caption{Cyclic voltammetry (-0V5 to 0V5) of mango. (a) delay time between settings is 10ms, (b) delay time between settings is 100ms, (c) delay time between settings is 1000ms }
    \label{fig:mango1Vpp}
\end{figure}

\begin{figure}[!hbt]
    \centering
    \subfigure[]{\includegraphics[width=0.7\textwidth]{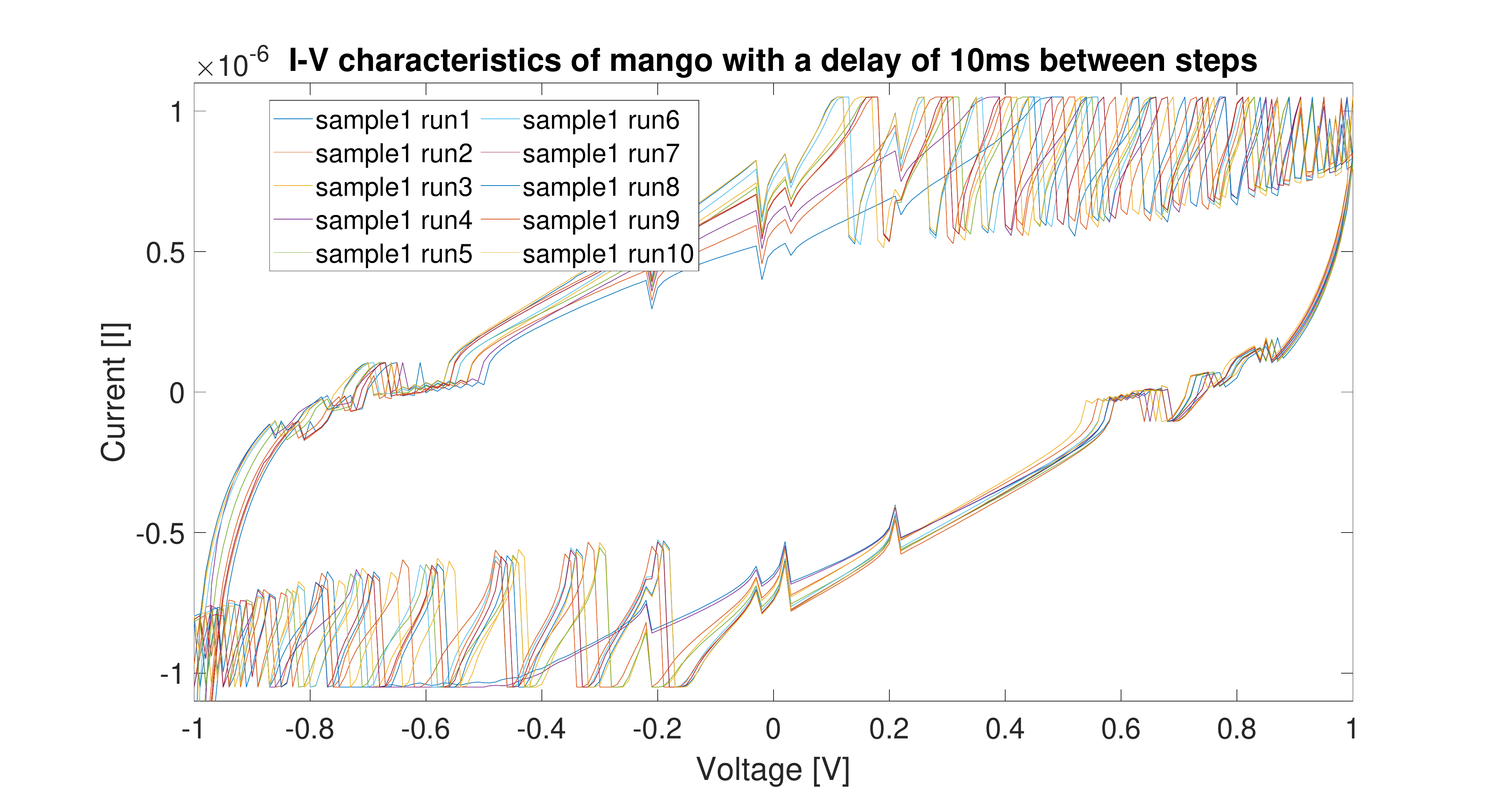}}
    \subfigure[]{\includegraphics[width=0.7\textwidth]{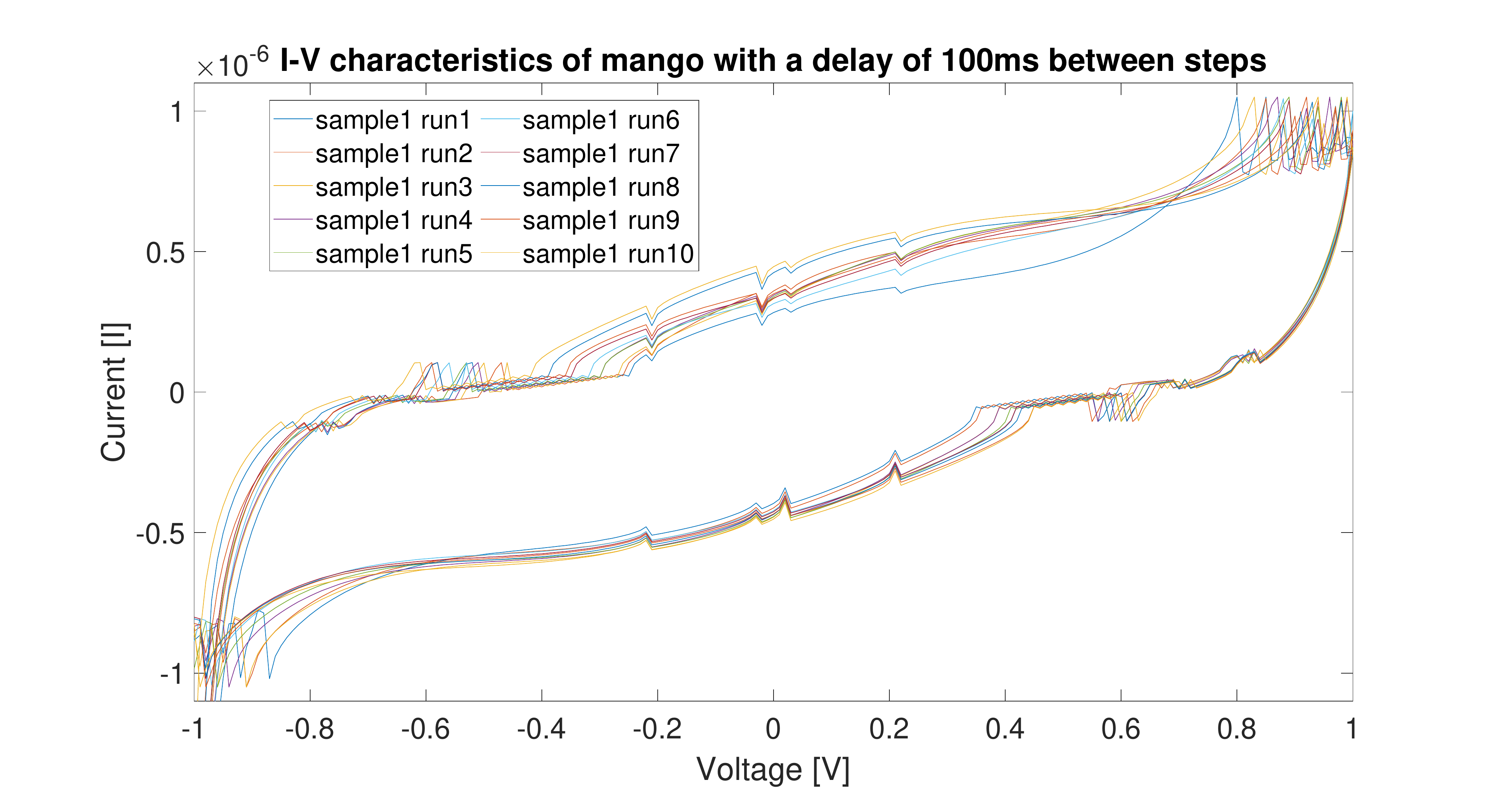}}
    \subfigure[]{\includegraphics[width=0.7\textwidth]{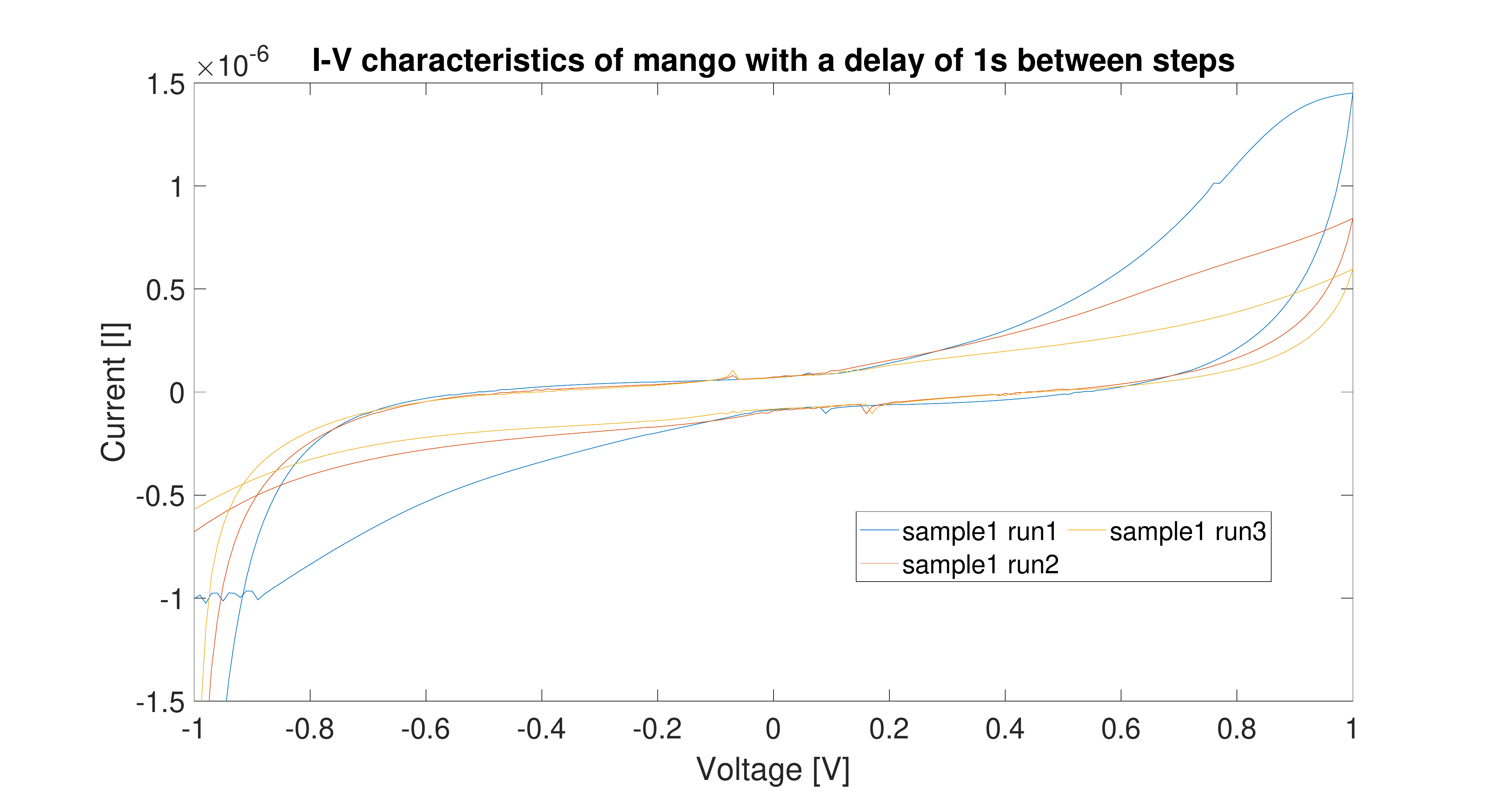}}
    \caption{Cyclic voltammetry (-1V to 1V) of mango. (a) delay time between settings is 10ms, (b) delay time between settings is 100ms, (c) delay time between settings is 1000ms }
    \label{fig:mango2Vpp}
\end{figure}

\begin{figure}[!hbt]
    \centering
    \subfigure[]{\includegraphics[width=0.7\textwidth]{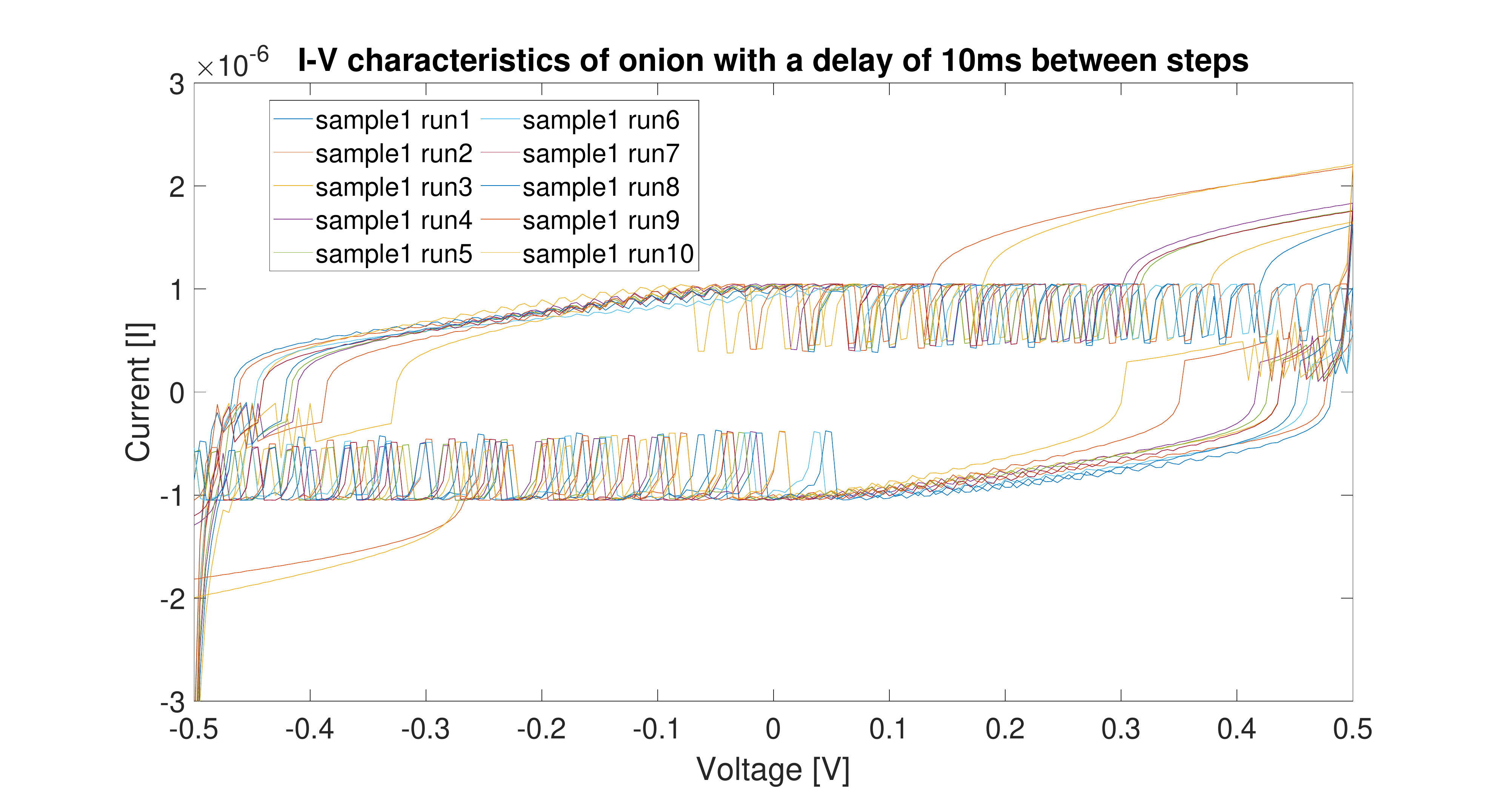}}
    \subfigure[]{\includegraphics[width=0.7\textwidth]{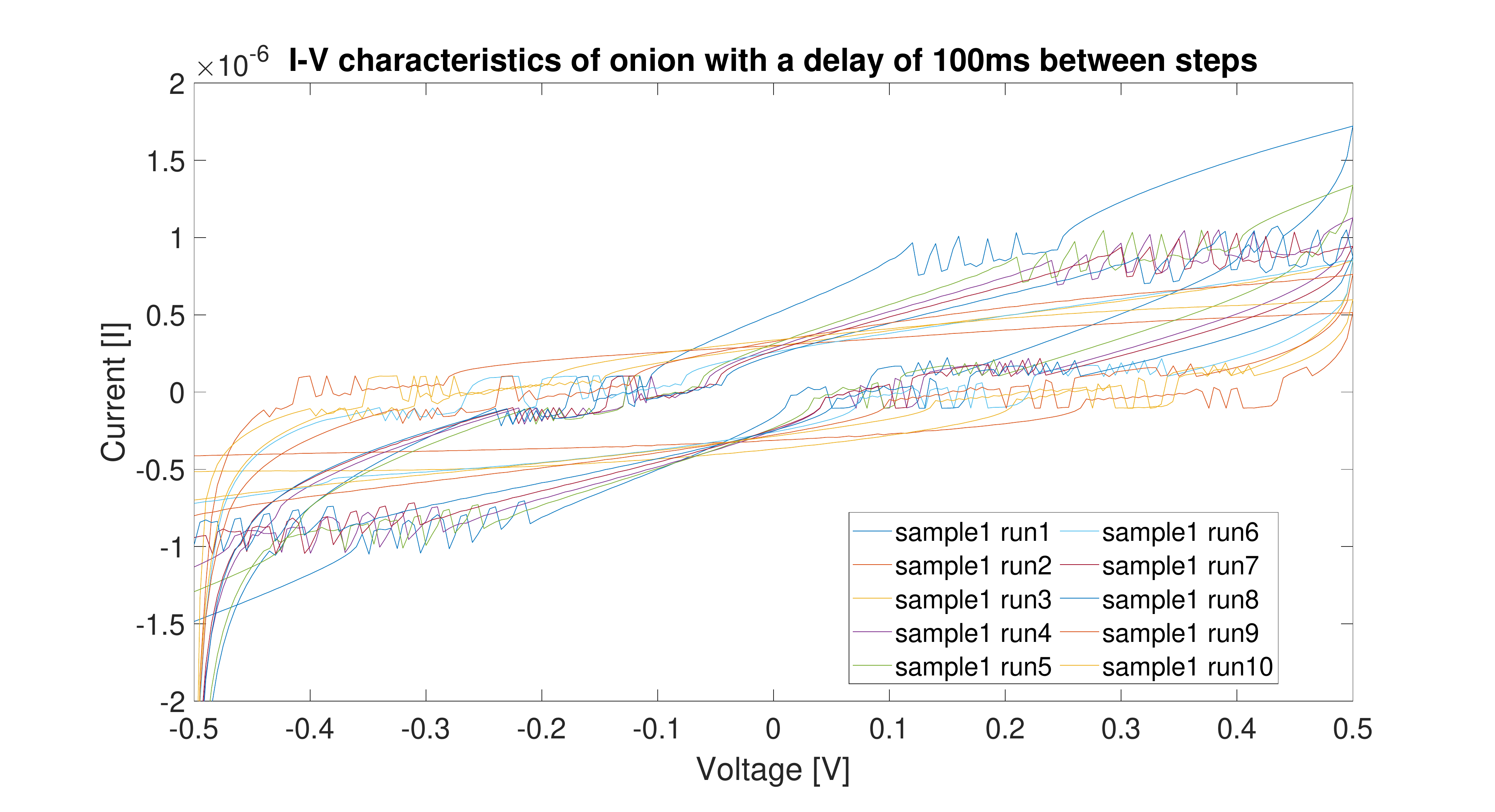}}
    \subfigure[]{\includegraphics[width=0.7\textwidth]{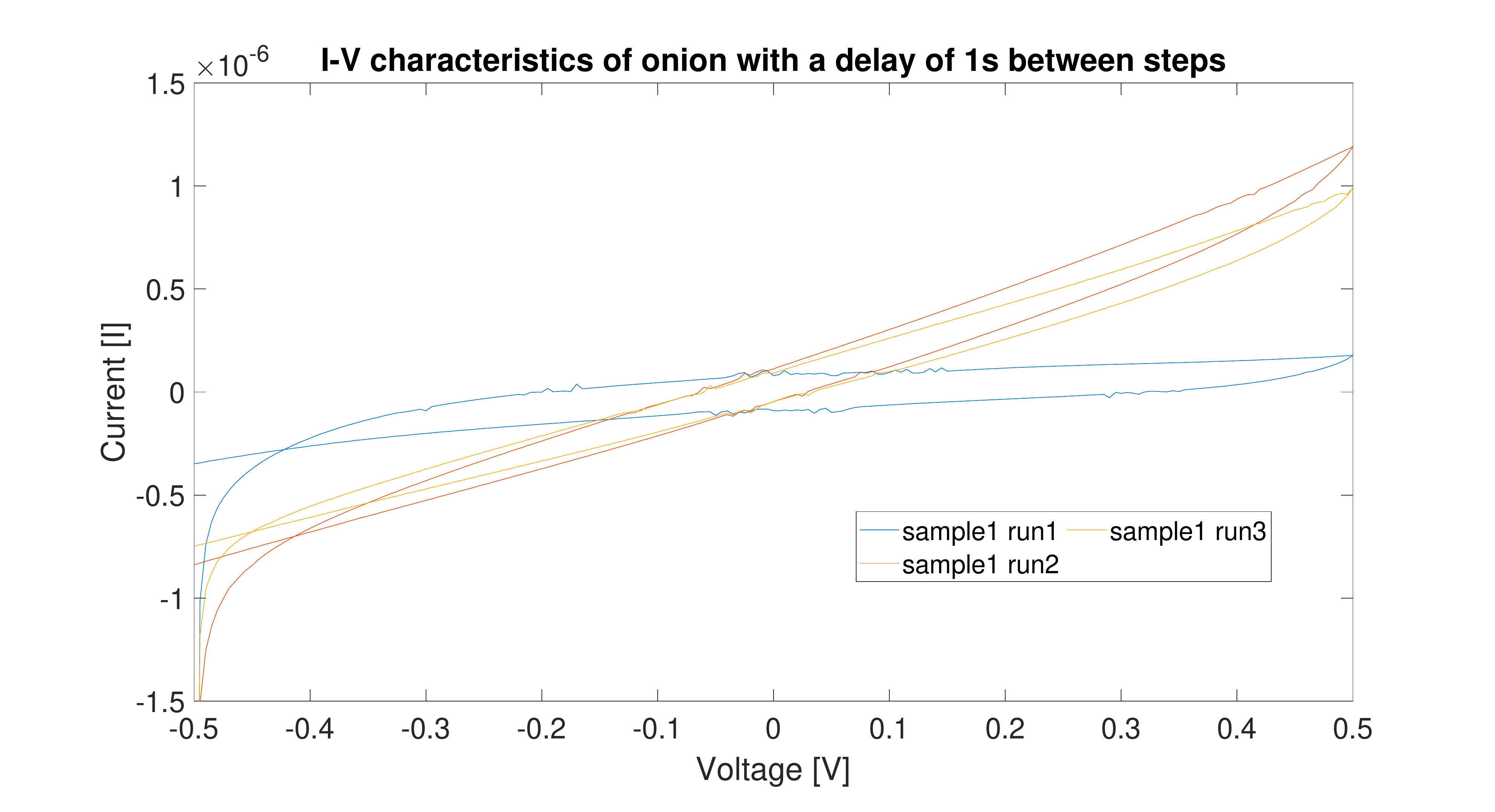}}
    \caption{Cyclic voltammetry (-0V5 to 0V5) of onion. (a) delay time between settings is 10ms, (b) delay time between settings is 100ms, (c) delay time between settings is 1000ms }
    \label{fig:onion1Vpp}
\end{figure}

\begin{figure}[!hbt]
    \centering
    \subfigure[]{\includegraphics[width=0.7\textwidth]{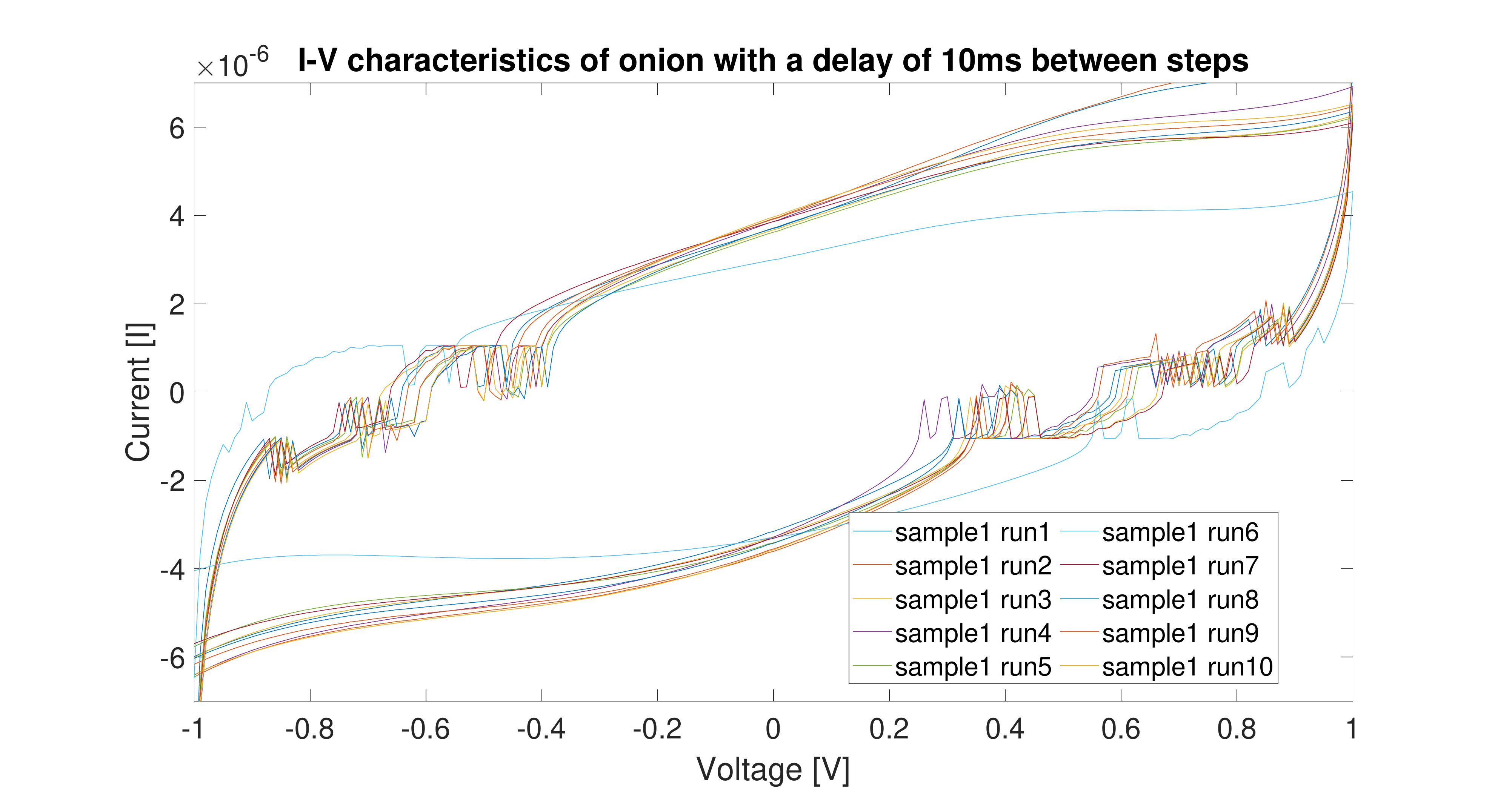}}
    \subfigure[]{\includegraphics[width=0.7\textwidth]{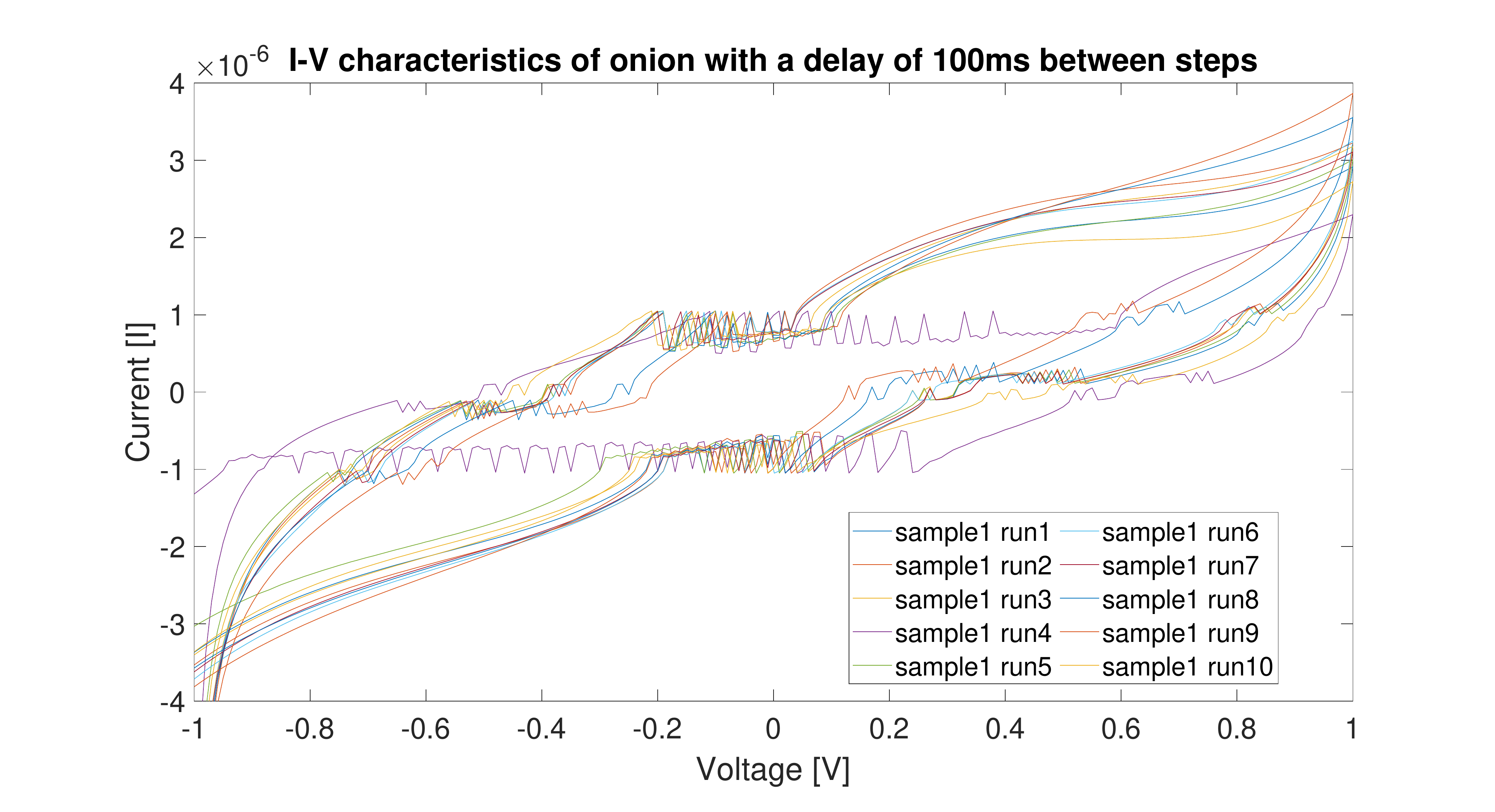}}
    \subfigure[]{\includegraphics[width=0.7\textwidth]{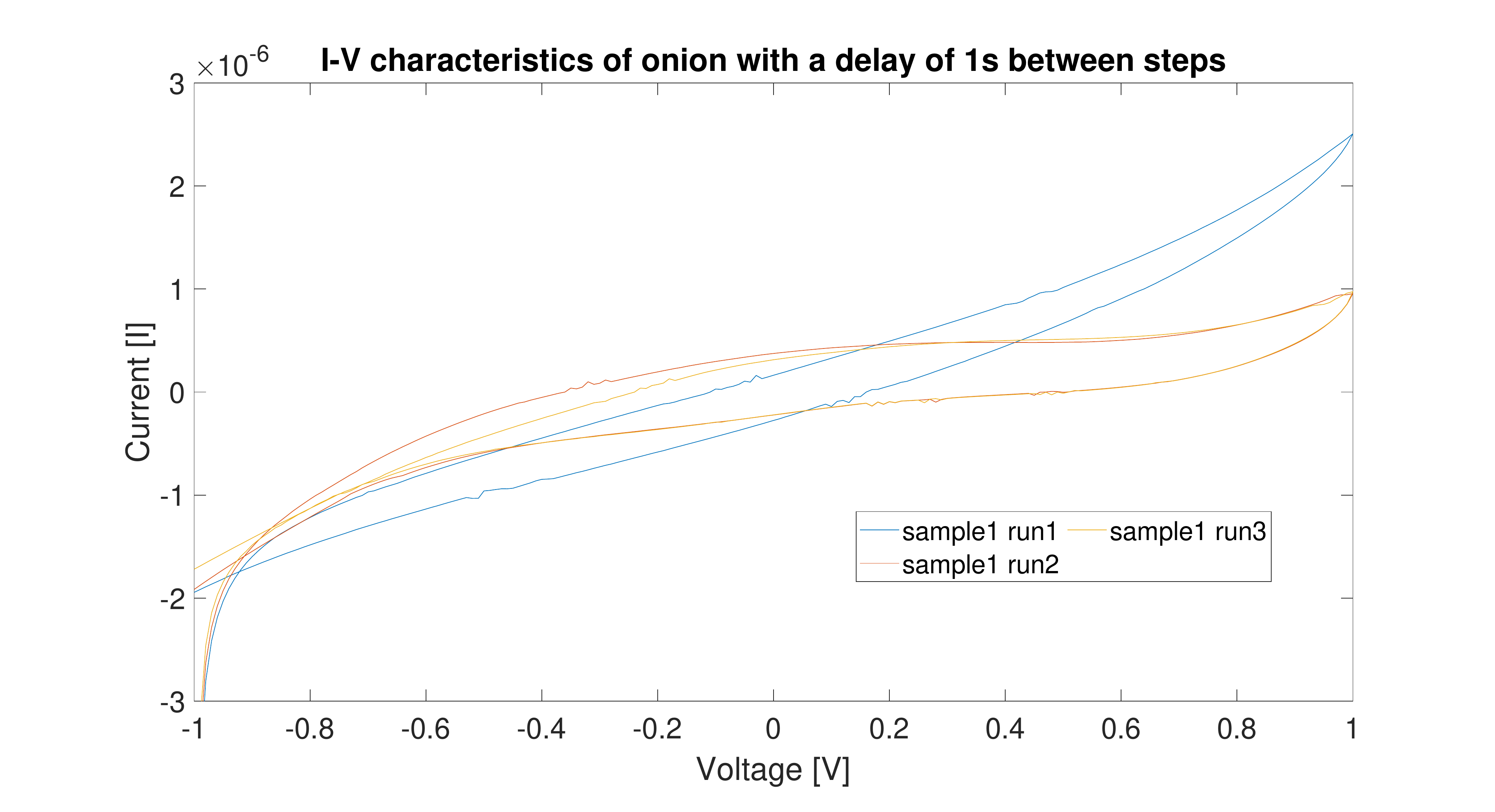}}
    \caption{Cyclic voltammetry (-1V to 1V) of onion. (a) delay time between settings is 10ms, (b) delay time between settings is 100ms, (c) delay time between settings is 1000ms }
    \label{fig:onion2Vpp}
\end{figure}

\begin{figure}[!hbt]
    \centering
    \subfigure[]{\includegraphics[width=0.7\textwidth]{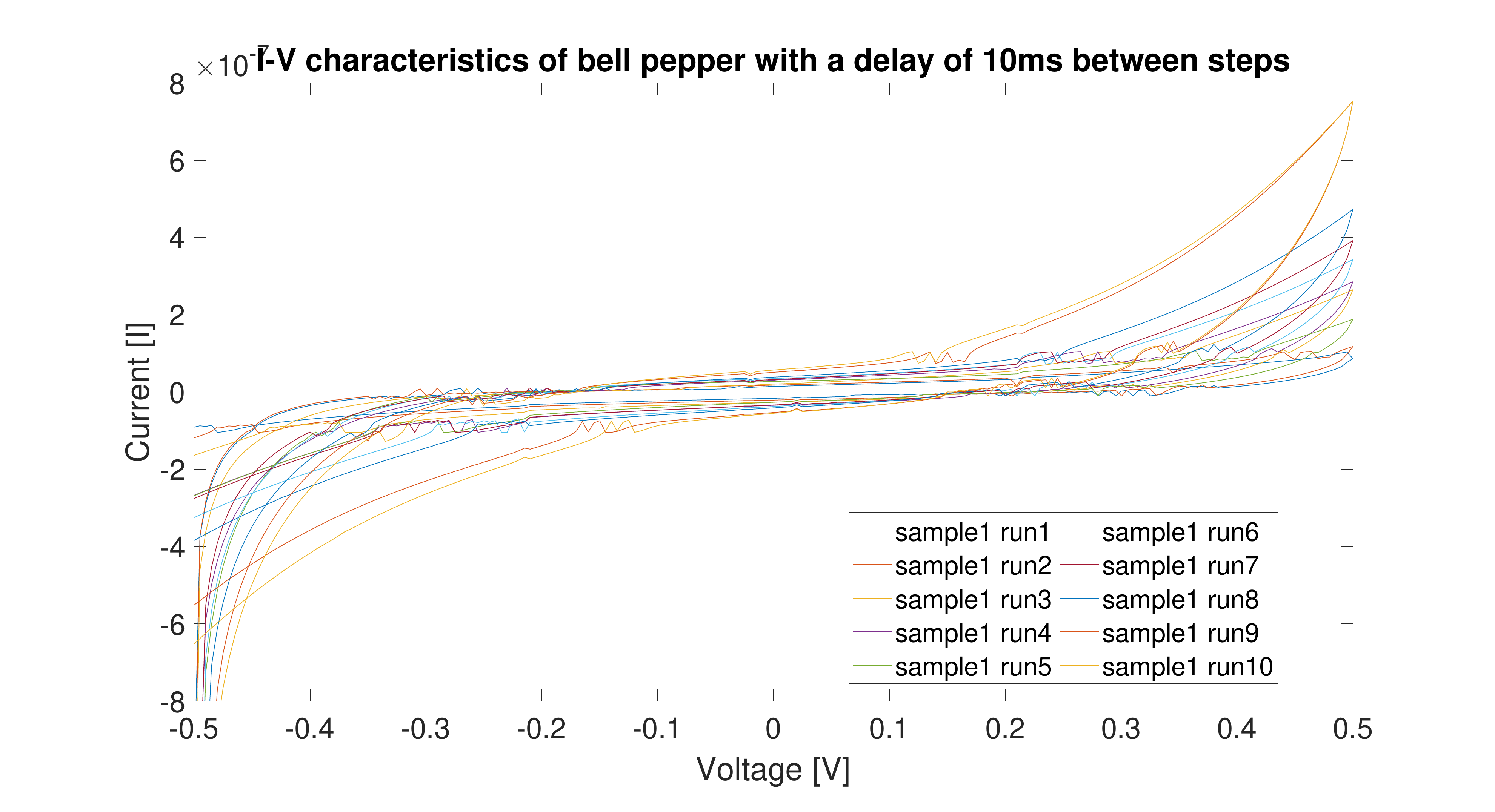}}
    \subfigure[]{\includegraphics[width=0.7\textwidth]{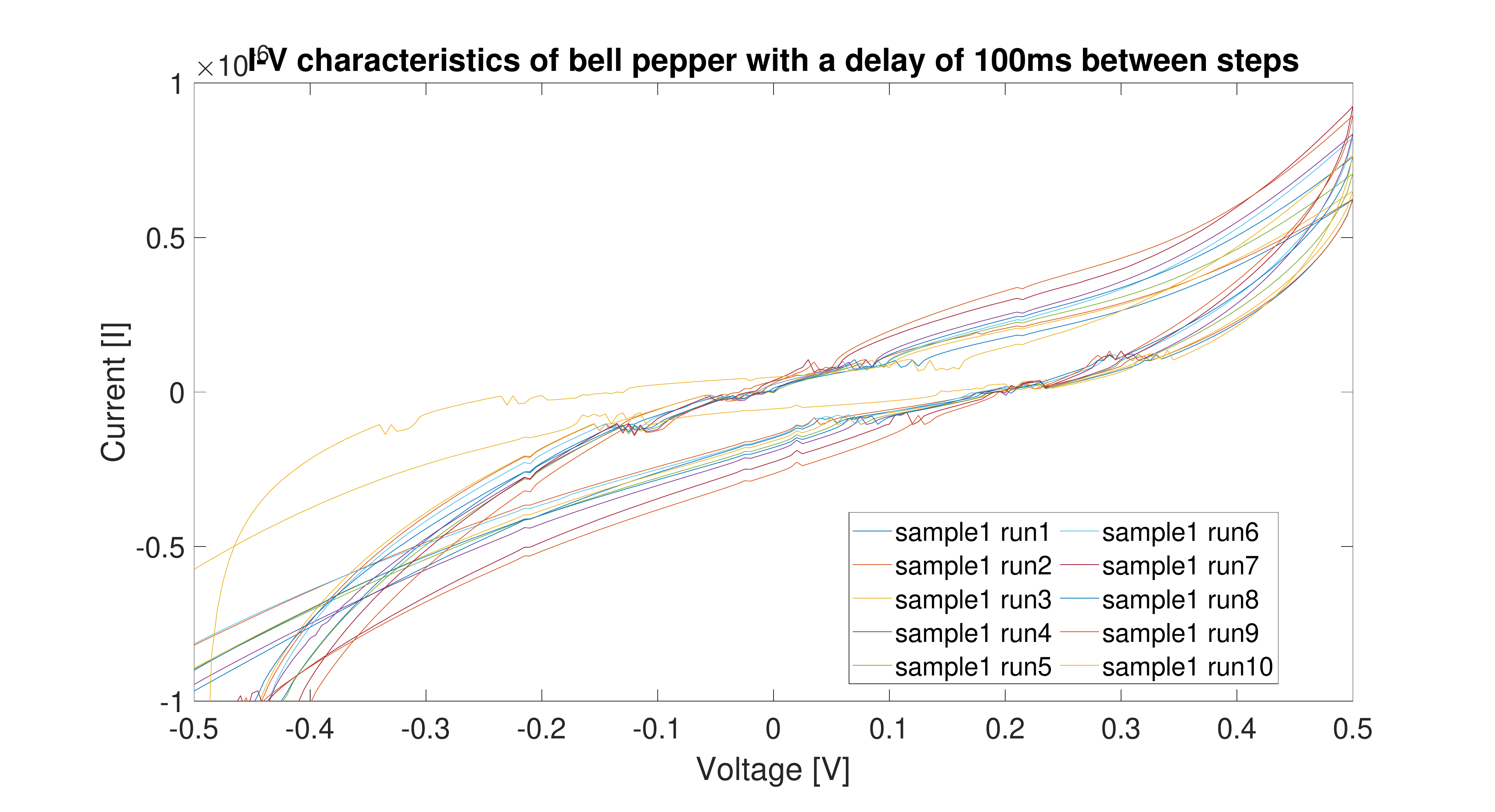}}
    \subfigure[]{\includegraphics[width=0.7\textwidth]{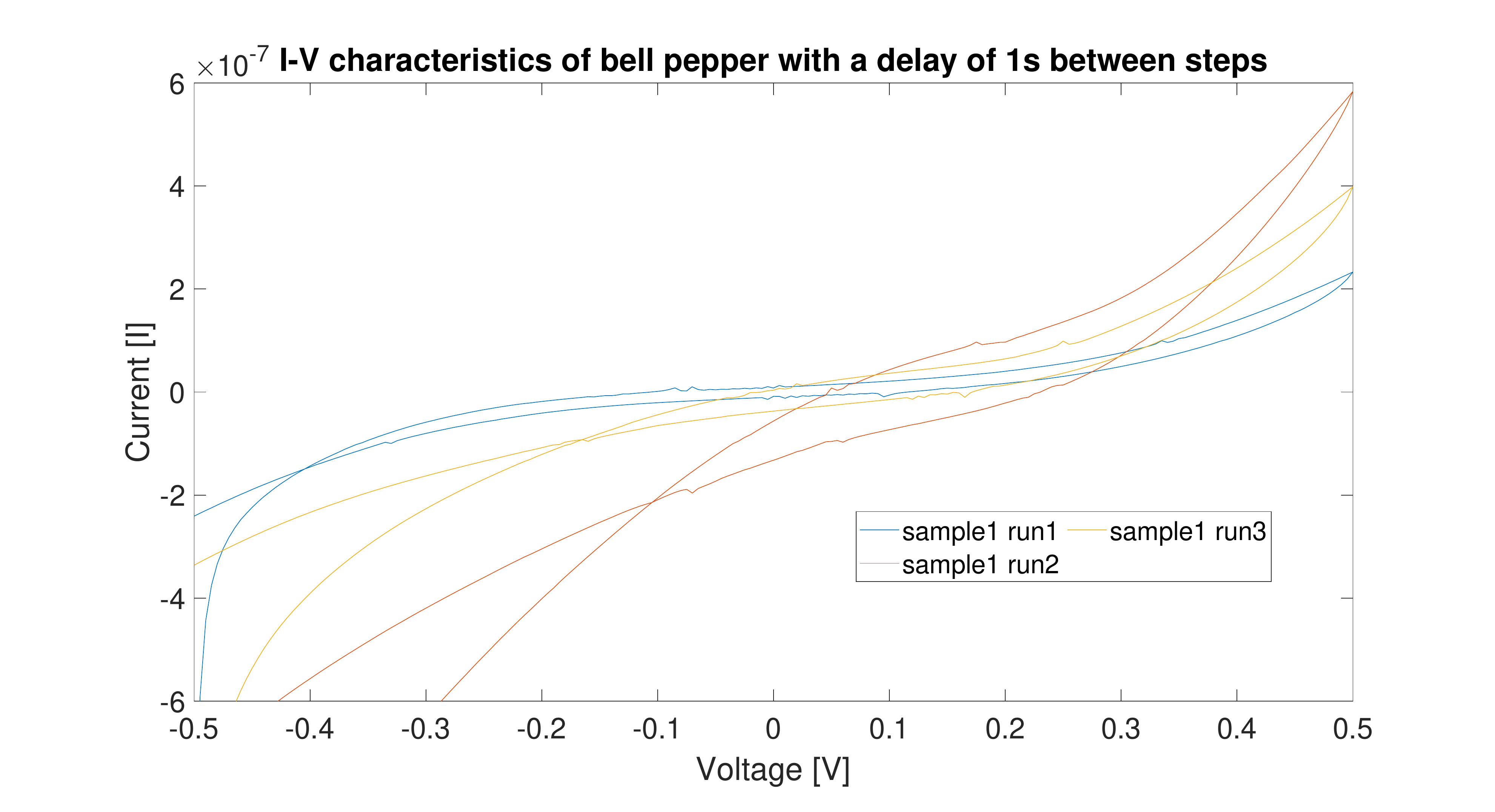}}
    \caption{Cyclic voltammetry (-0V5 to 0V5) of pepper. (a) delay time between settings is 10ms, (b) delay time between settings is 100ms, (c) delay time between settings is 1000ms }
    \label{fig:pepper1Vpp}
\end{figure}

\begin{figure}[!hbt]
    \centering
    \subfigure[]{\includegraphics[width=0.7\textwidth]{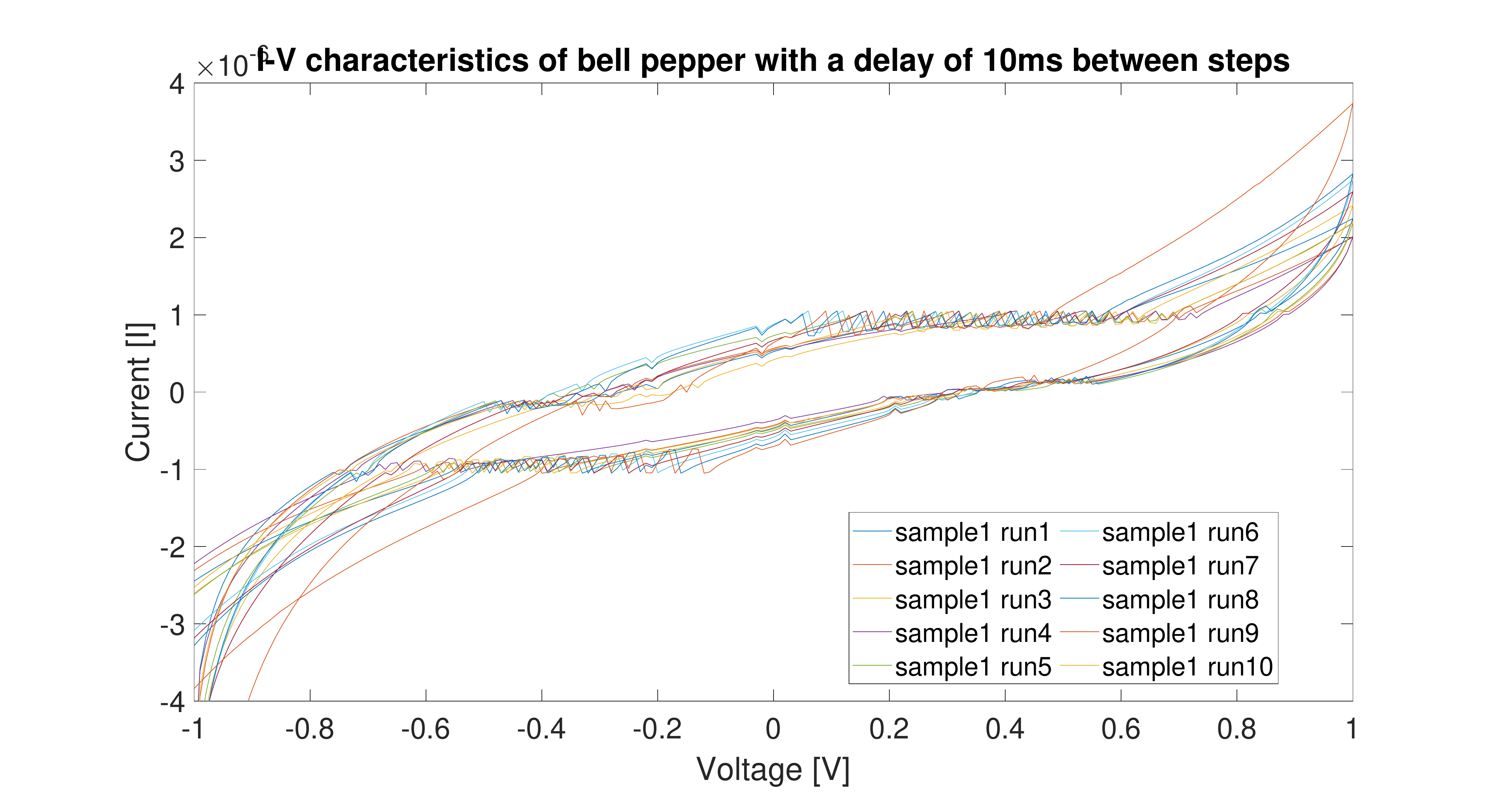}}
    \subfigure[]{\includegraphics[width=0.7\textwidth]{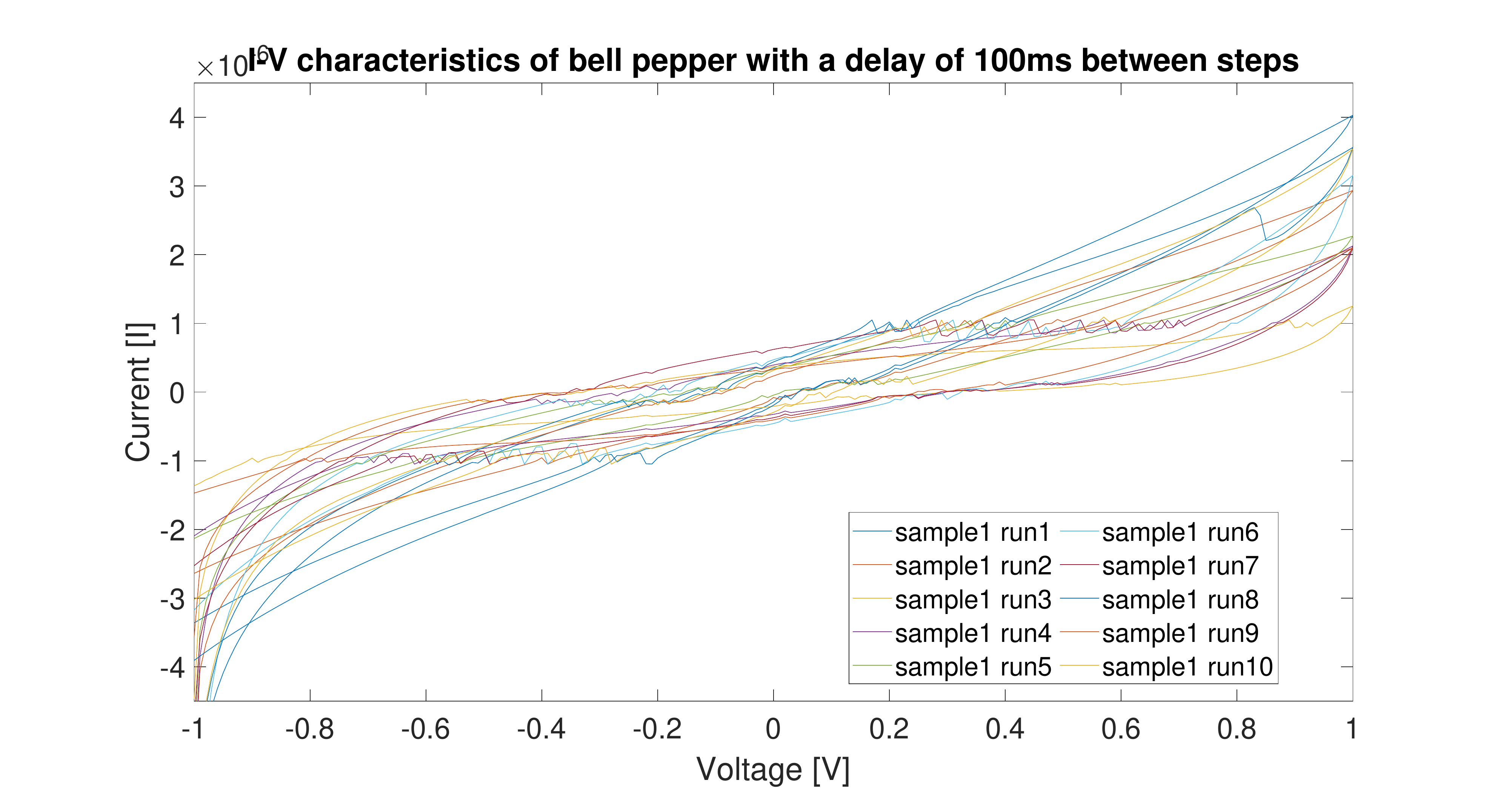}}
    \subfigure[]{\includegraphics[width=0.7\textwidth]{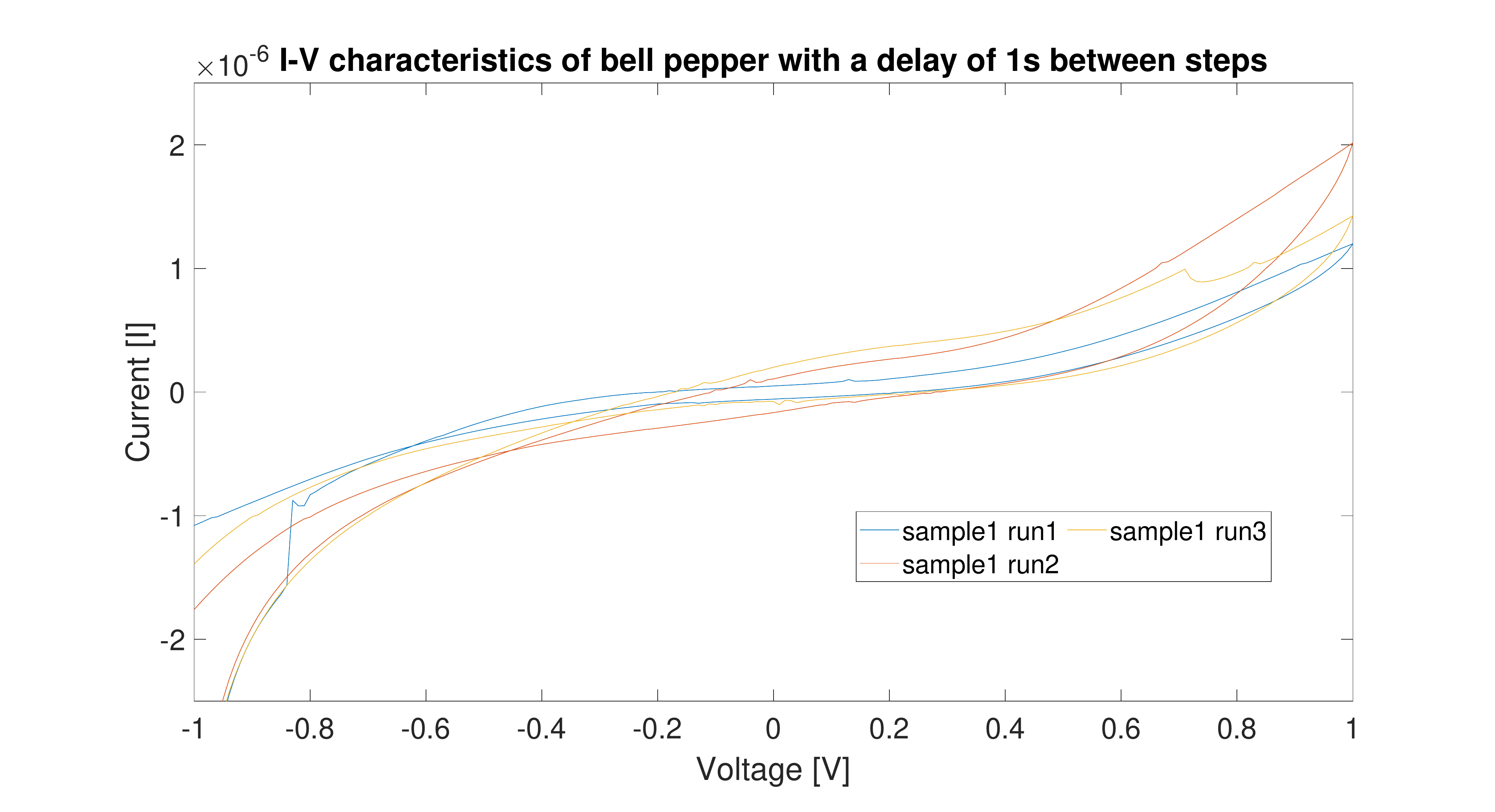}}
    \caption{Cyclic voltammetry (-1V to 1V) of pepper. (a) delay time between settings is 10ms, (b) delay time between settings is 100ms, (c) delay time between settings is 1000ms }
    \label{fig:pepper2Vpp}
\end{figure}

\begin{figure}[!hbt]
    \centering
    \subfigure[]{\includegraphics[width=0.7\textwidth]{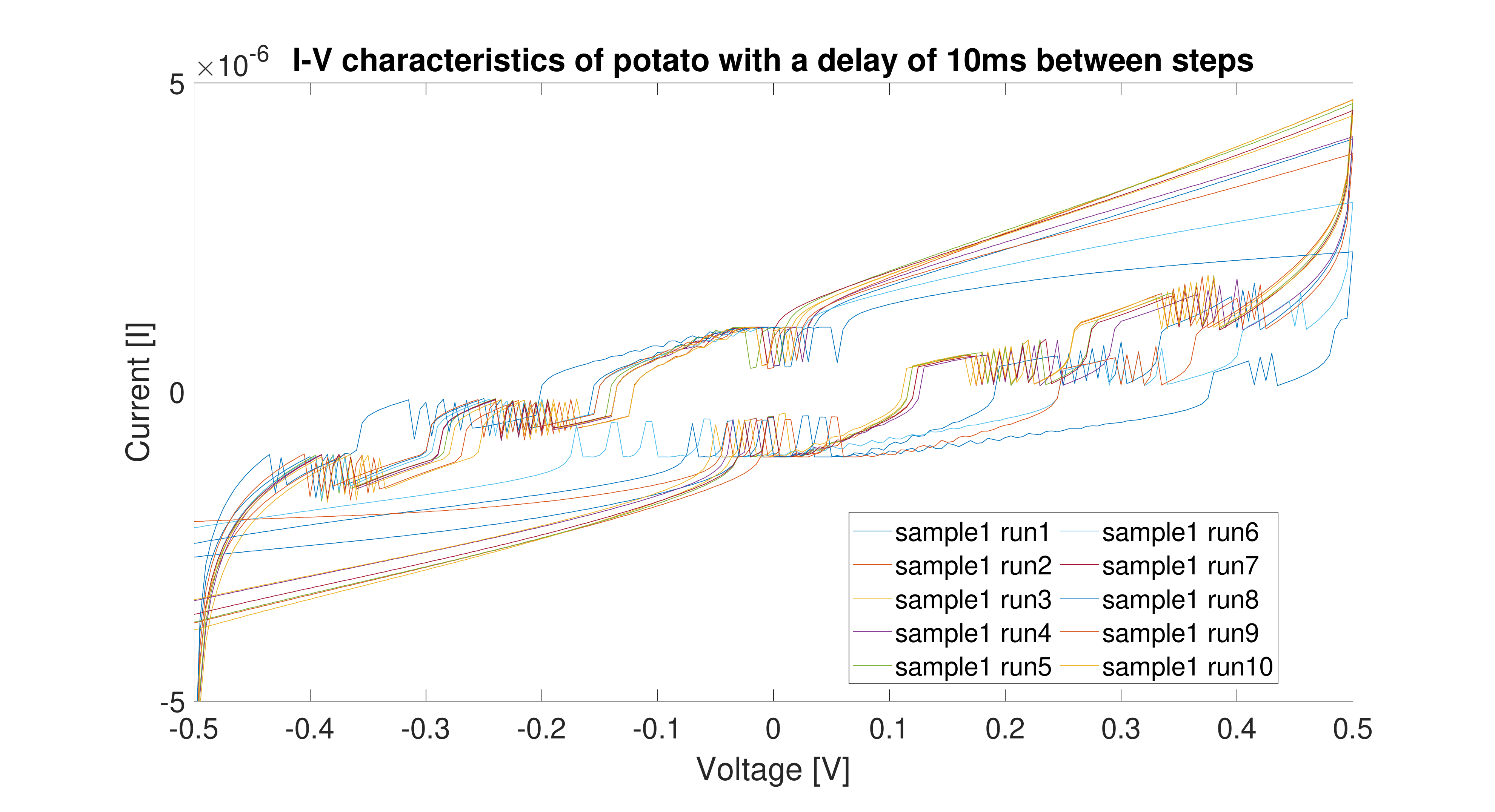}}
    \subfigure[]{\includegraphics[width=0.7\textwidth]{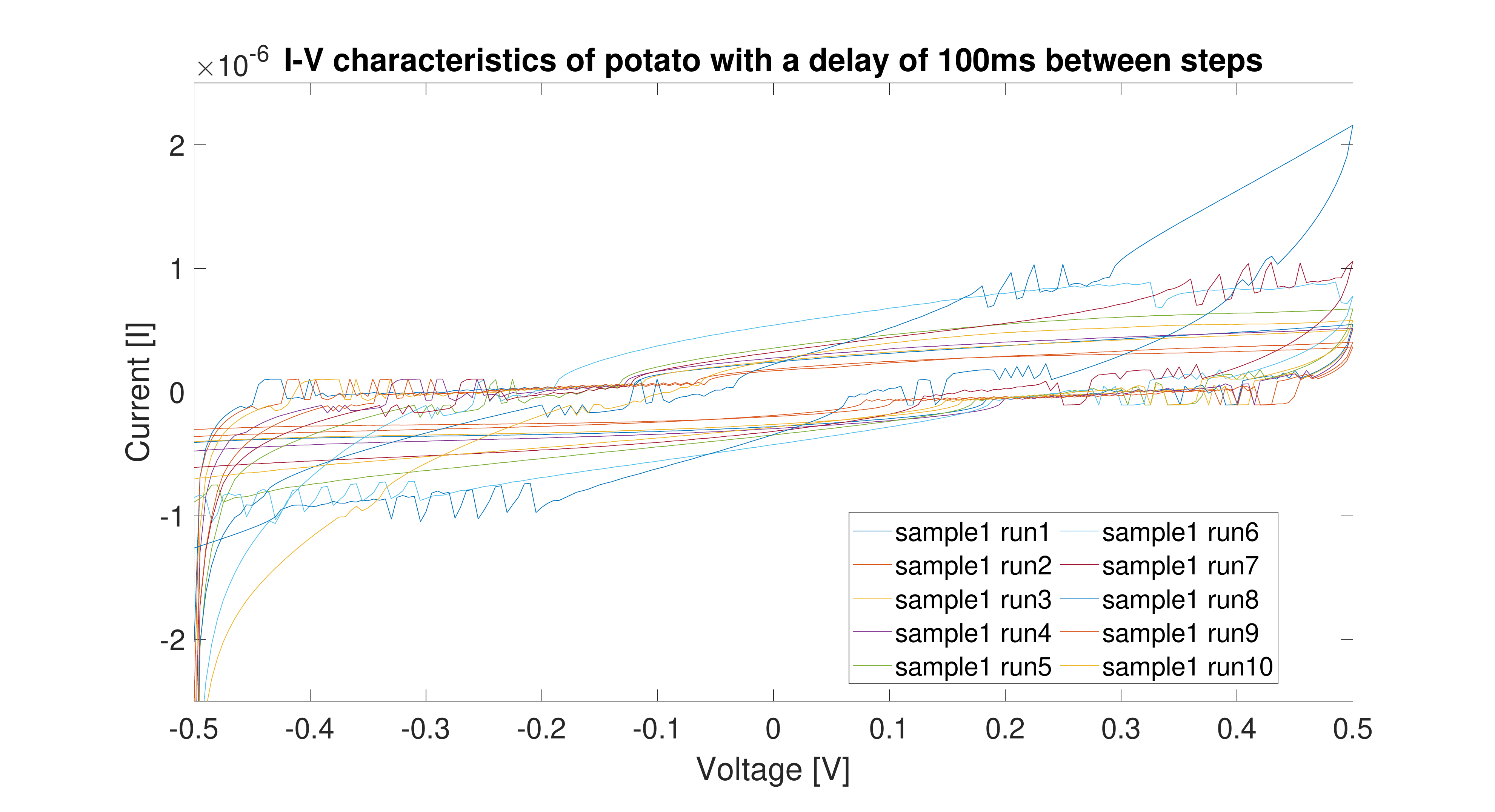}}
    \subfigure[]{\includegraphics[width=0.7\textwidth]{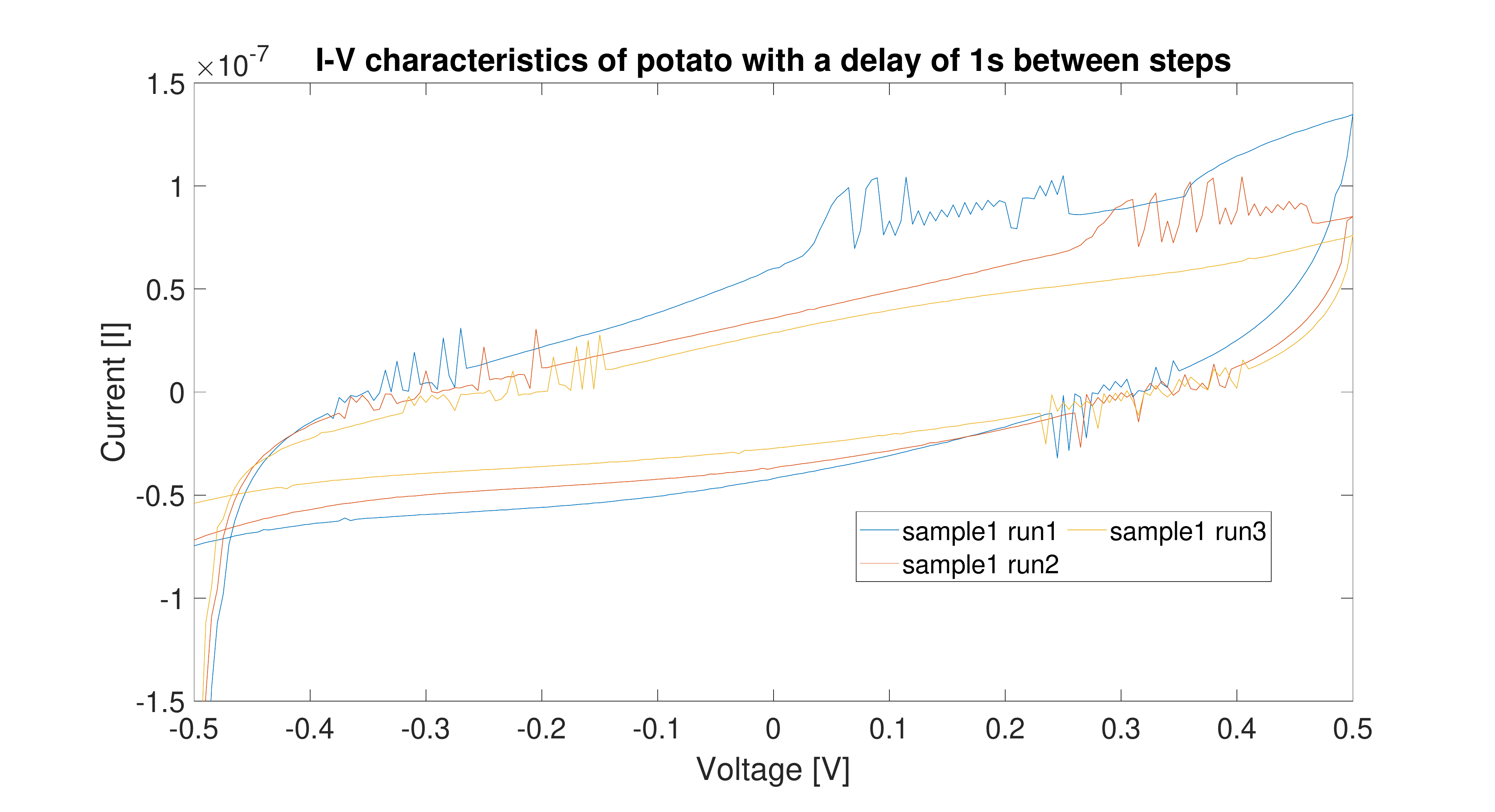}}
    \caption{Cyclic voltammetry (-0V5 to 0V5) of potato. (a) delay time between settings is 10ms, (b) delay time between settings is 100ms, (c) delay time between settings is 1000ms }
    \label{fig:potato1Vpp}
\end{figure}

\begin{figure}[!hbt]
    \centering
    \subfigure[]{\includegraphics[width=0.7\textwidth]{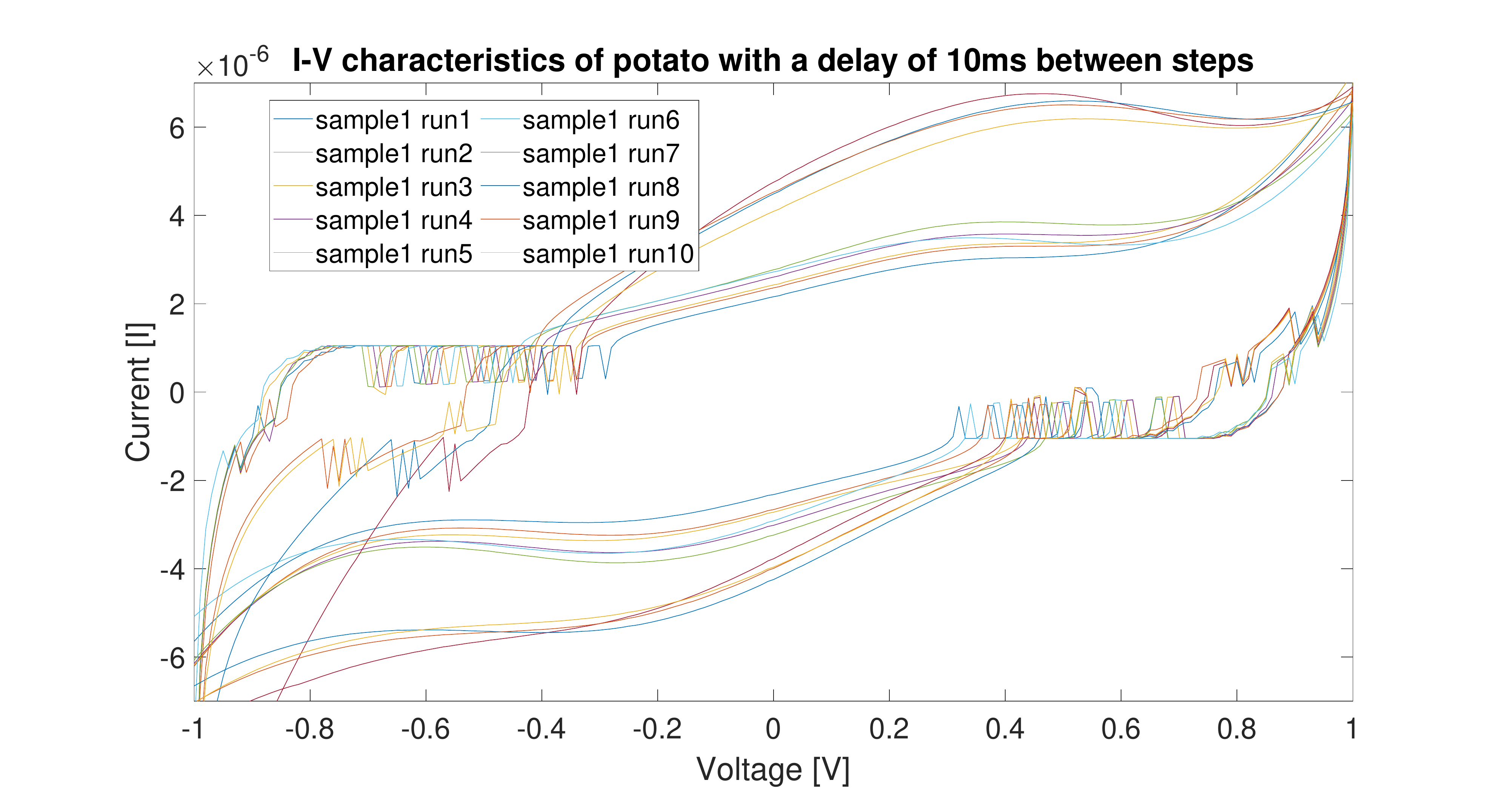}}
    \subfigure[]{\includegraphics[width=0.7\textwidth]{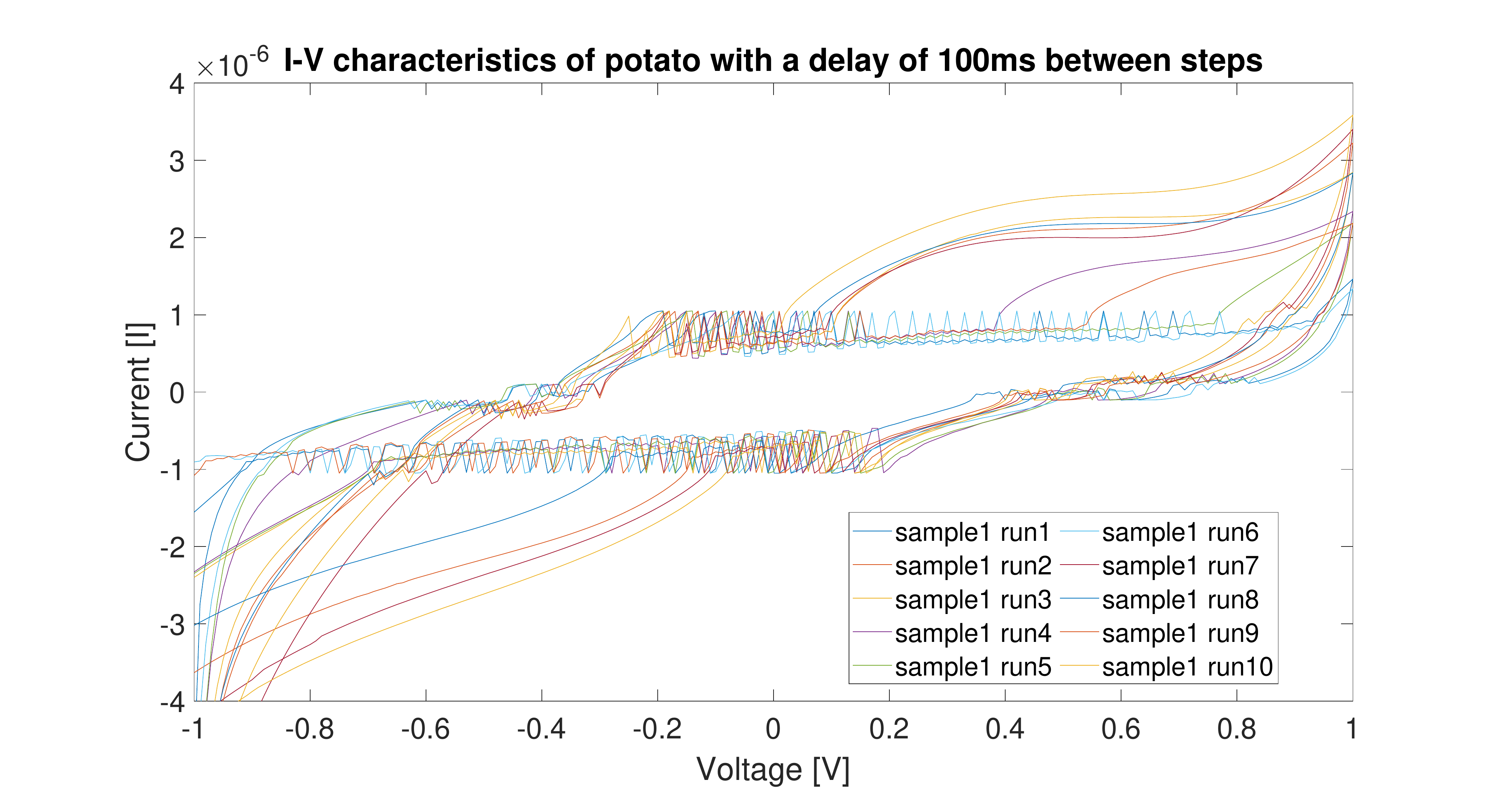}}
    \subfigure[]{\includegraphics[width=0.7\textwidth]{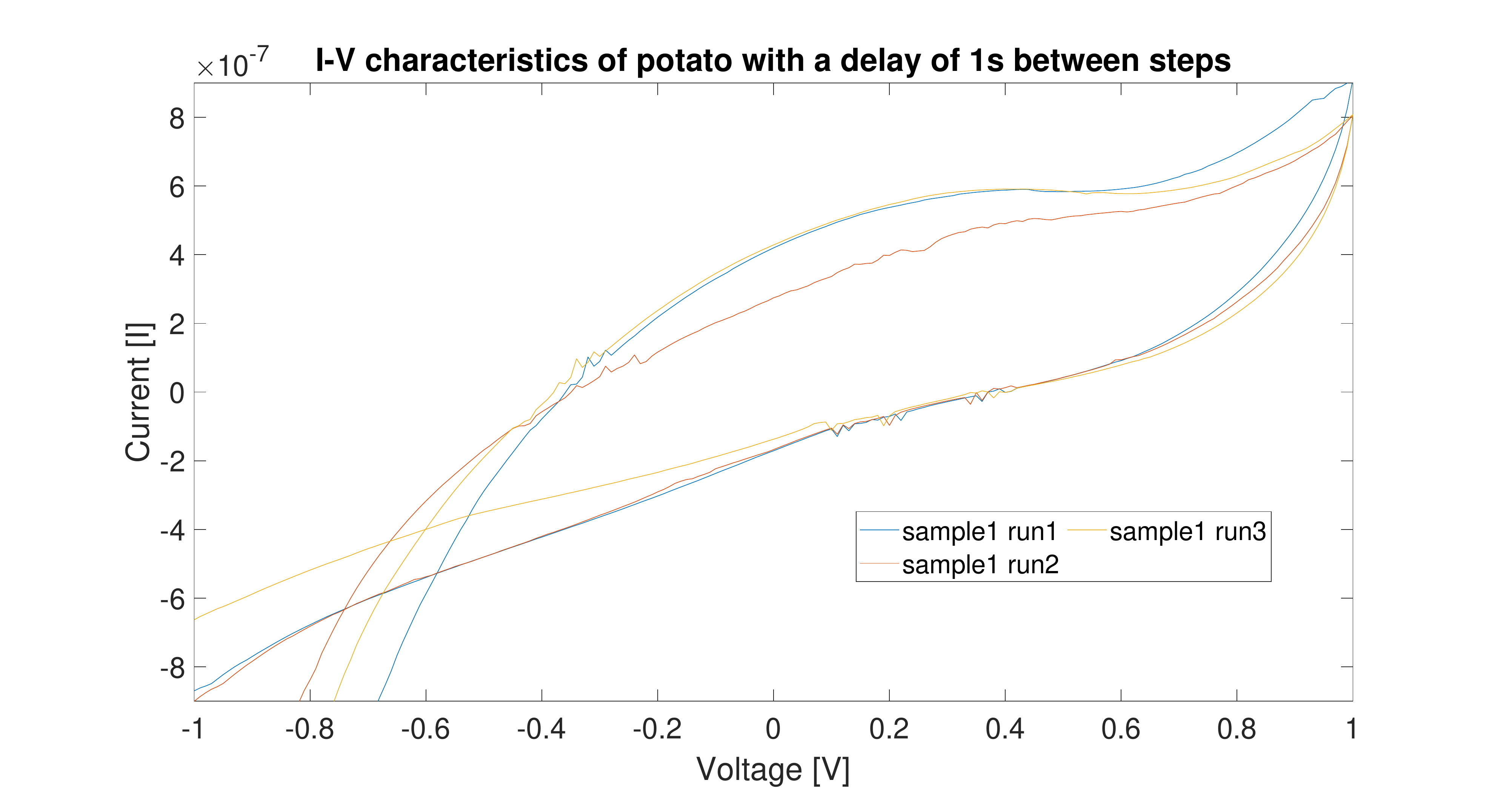}}
    \caption{Cyclic voltammetry (-1V to 1V) of potato. (a) delay time between settings is 10ms, (b) delay time between settings is 100ms, (c) delay time between settings is 1000ms }
    \label{fig:potato2Vpp}
\end{figure}

%\subsection{Mycelium}

\begin{figure}[!hbt]
    \centering
    \subfigure[]{\includegraphics[width=0.7\textwidth]{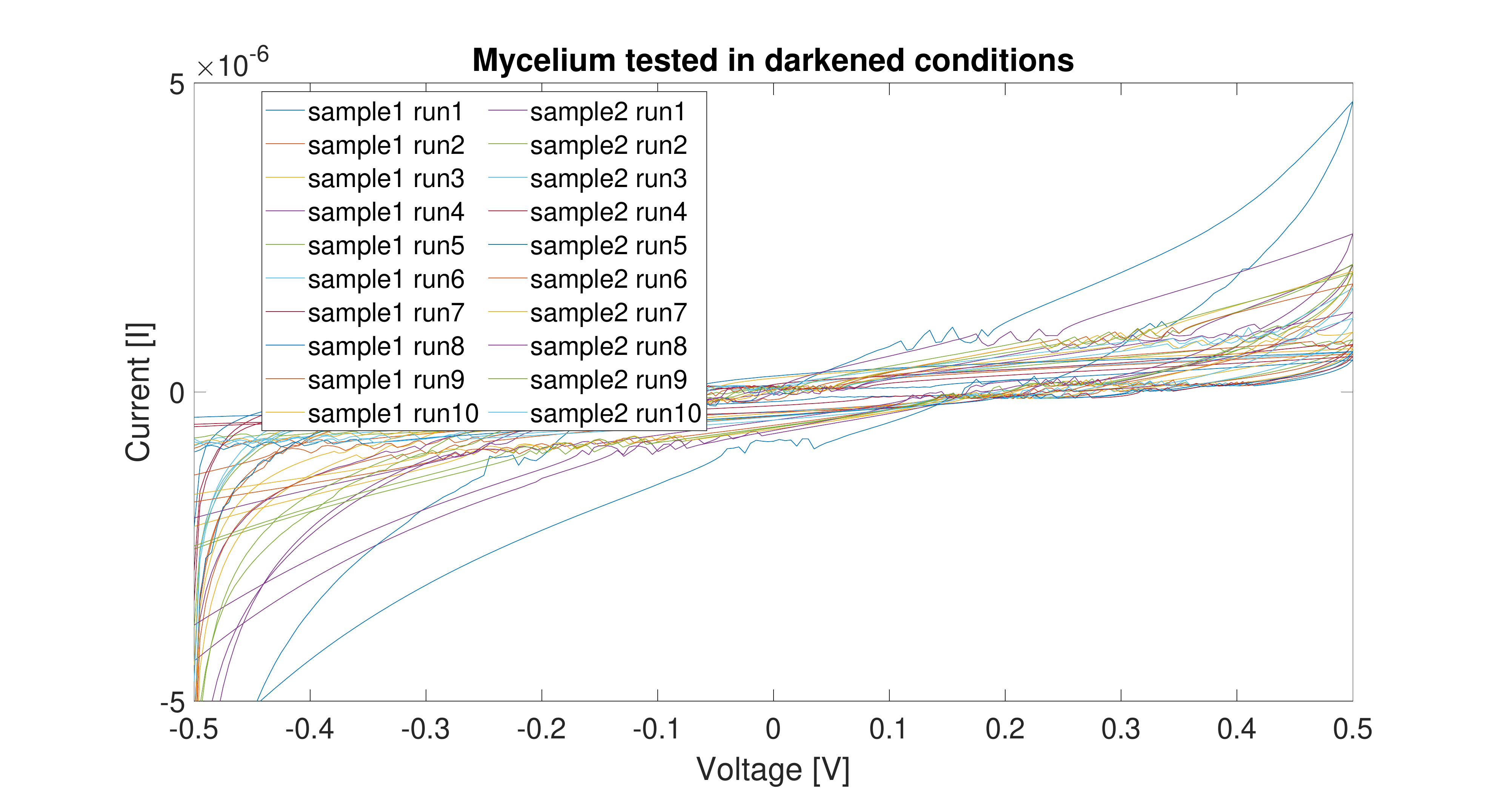}}
    \subfigure[]{\includegraphics[width=0.7\textwidth]{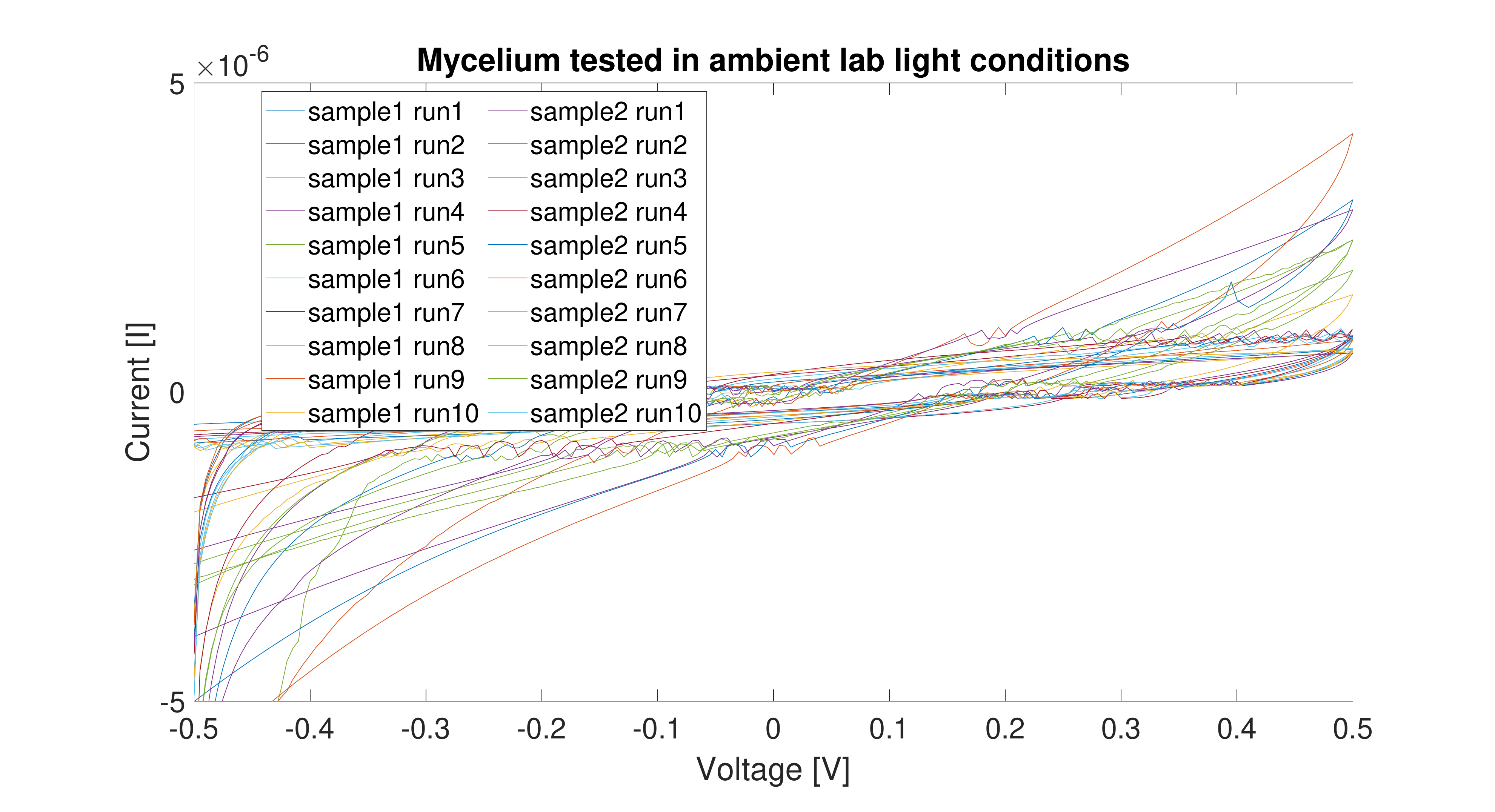}}
    \subfigure[]{\includegraphics[width=0.7\textwidth]{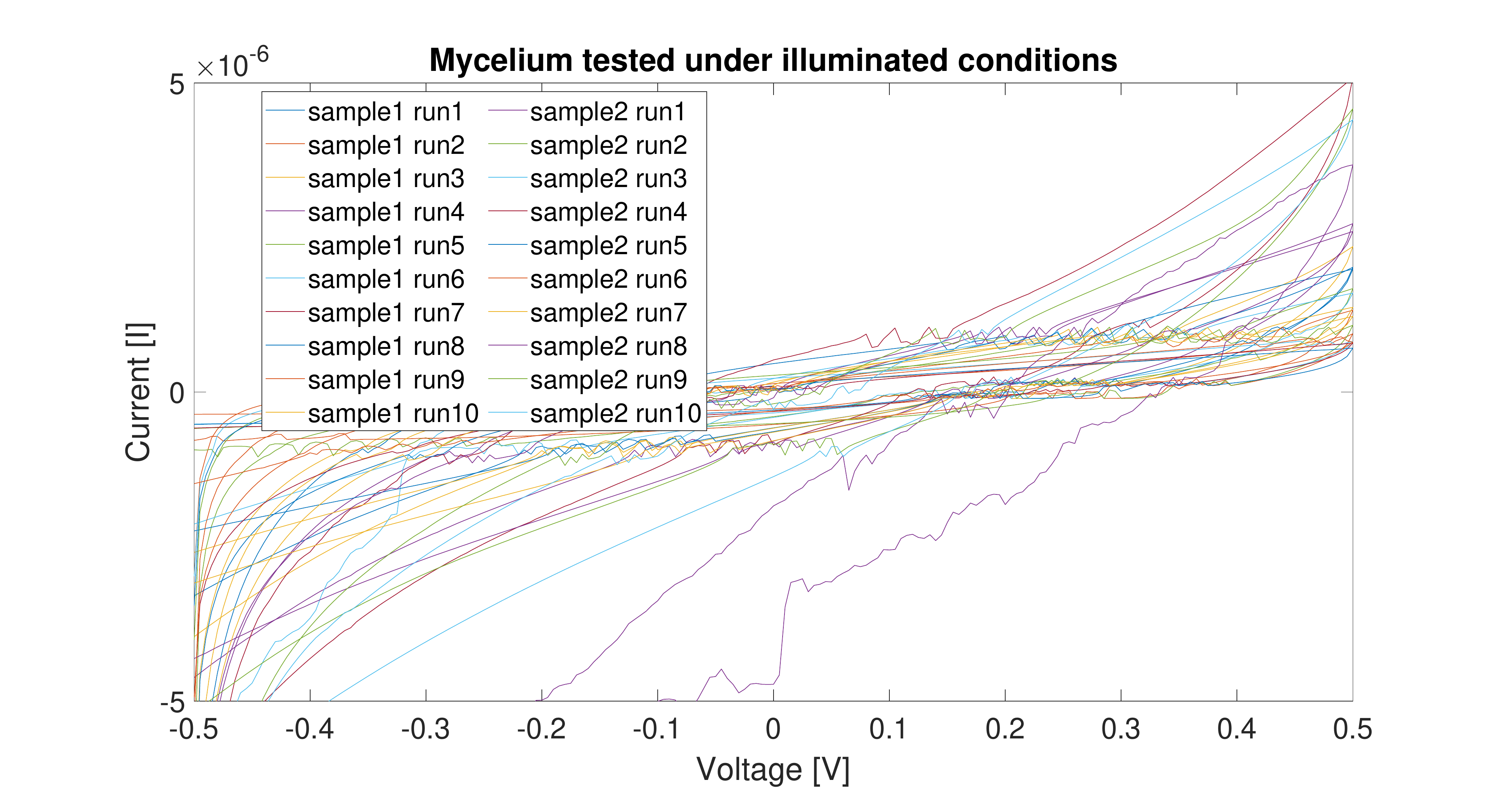}}
    \caption{Cyclic voltammetry (-0V5 to 0V5) of mycelium substrate cultivated on damp wood shavings with three different light levels. (a) covered, (b) ambient lab light (965\,Lux), (c) illuminated (1500\,Lux). }
    \label{fig:mycelium1Vpp}
\end{figure}

\begin{figure}[!hbt]
    \centering
    \subfigure[]{\includegraphics[width=0.7\textwidth]{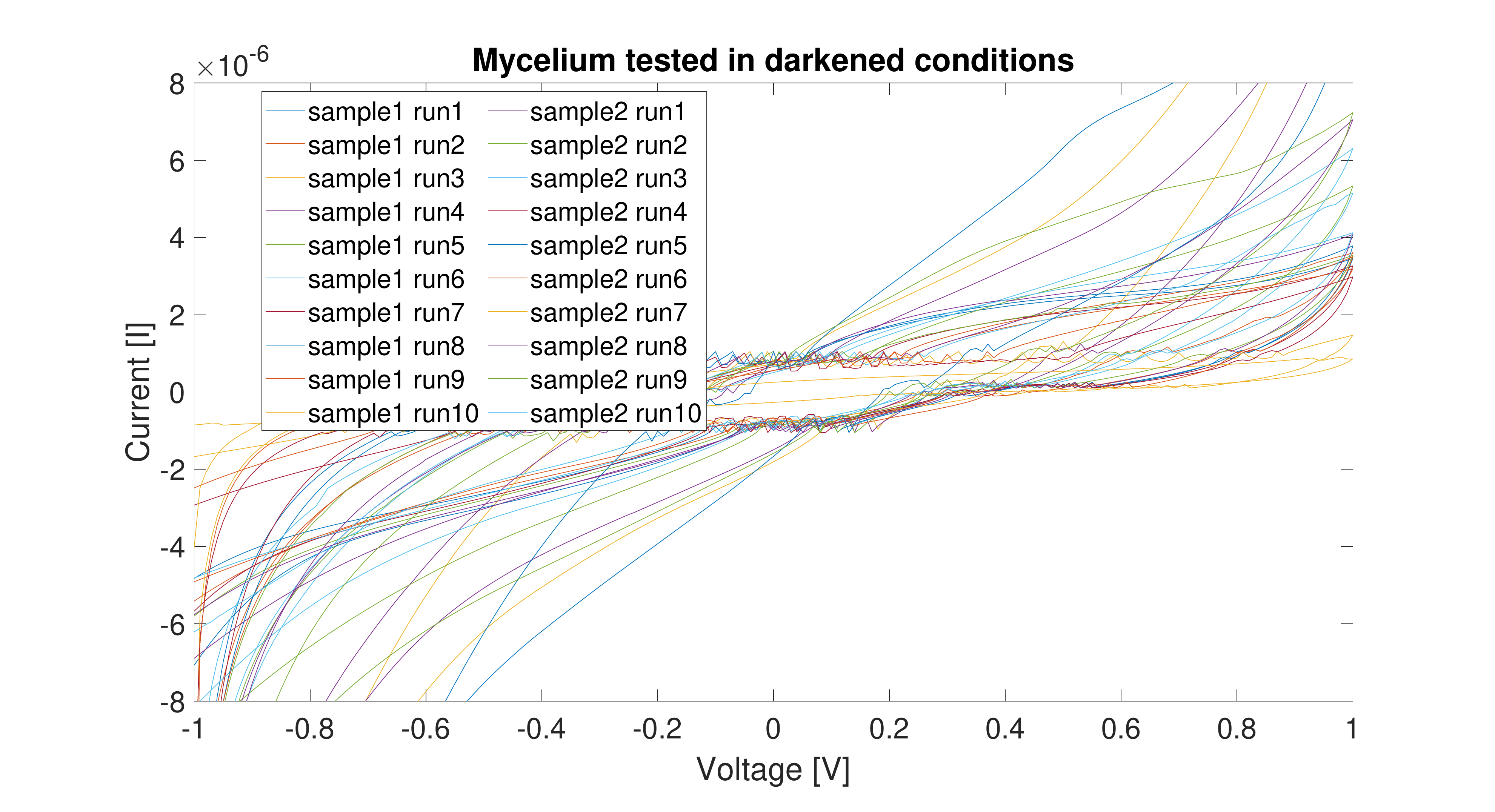}}
    \subfigure[]{\includegraphics[width=0.7\textwidth]{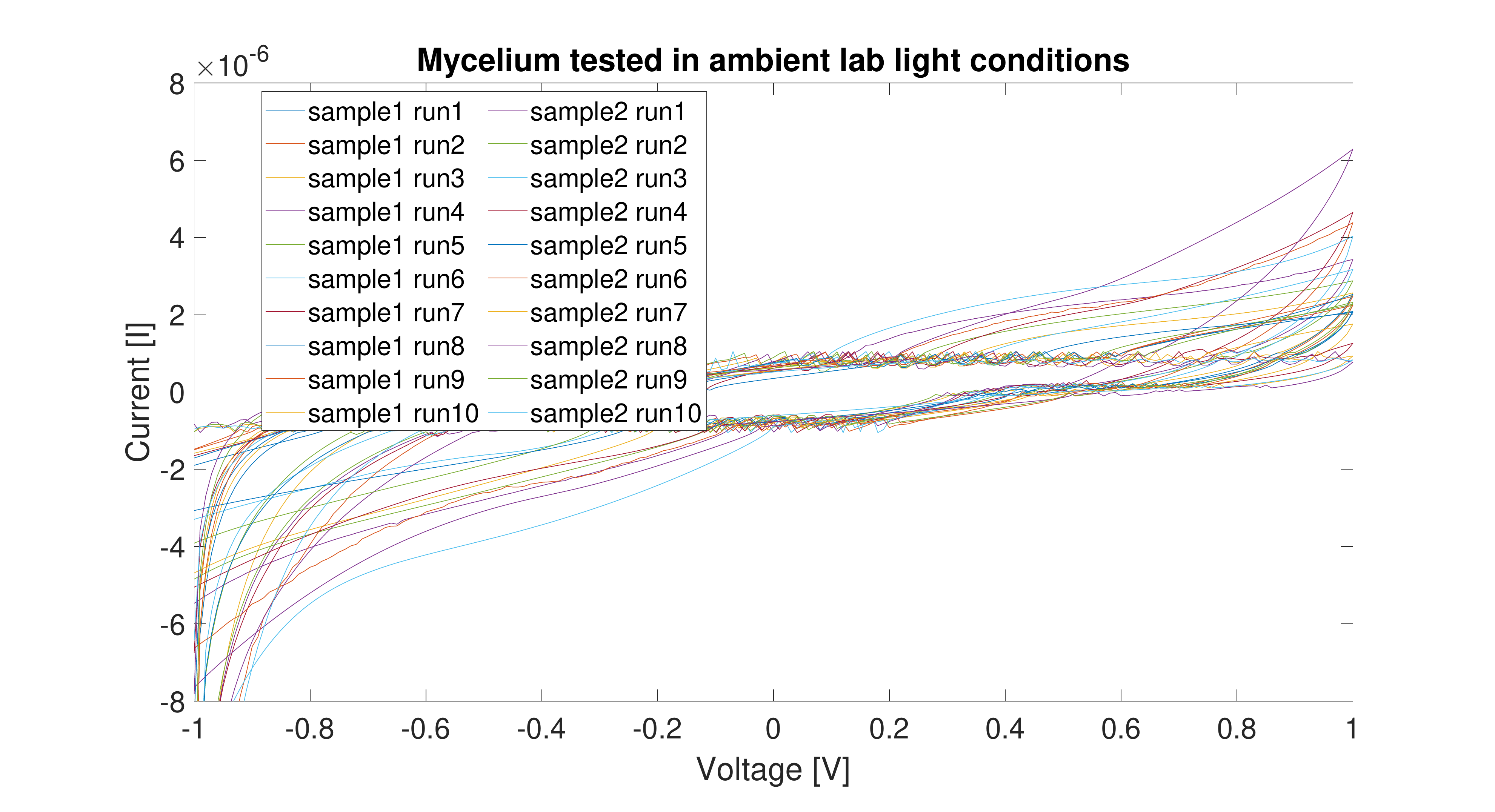}}
    \subfigure[]{\includegraphics[width=0.7\textwidth]{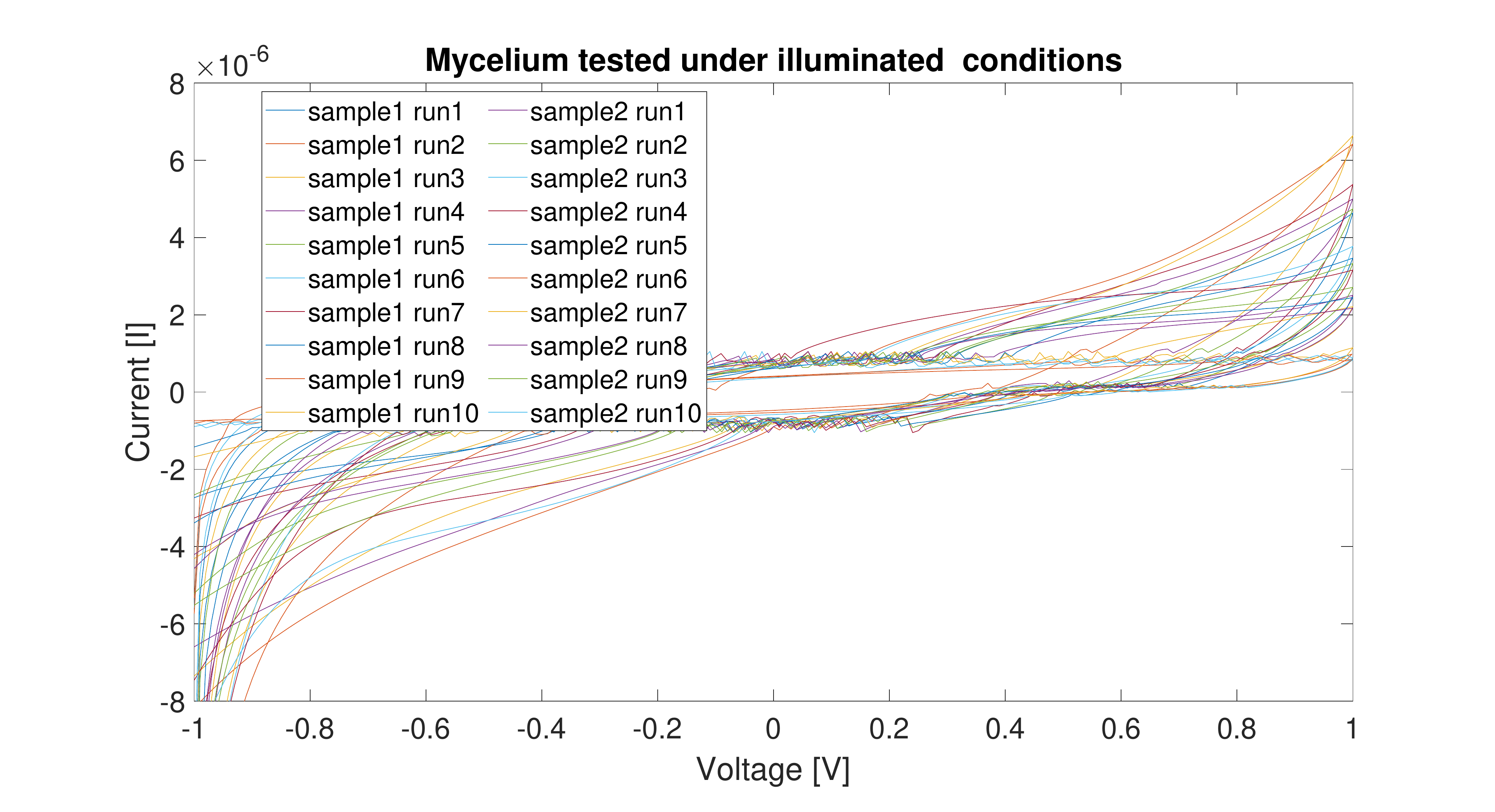}}
    \caption{Cyclic voltammetry (-1v to 1v) of mycelium substrate cultivated on damp wood shavings with three different light levels. (a) covered, (b) ambient lab light (965\,Lux), (c) illuminated (1500\,Lux). }
    \label{fig:mycelium2Vpp}
\end{figure}

\begin{figure}[!hbt]
    \centering
    \subfigure[]{\includegraphics[width=0.7\textwidth]{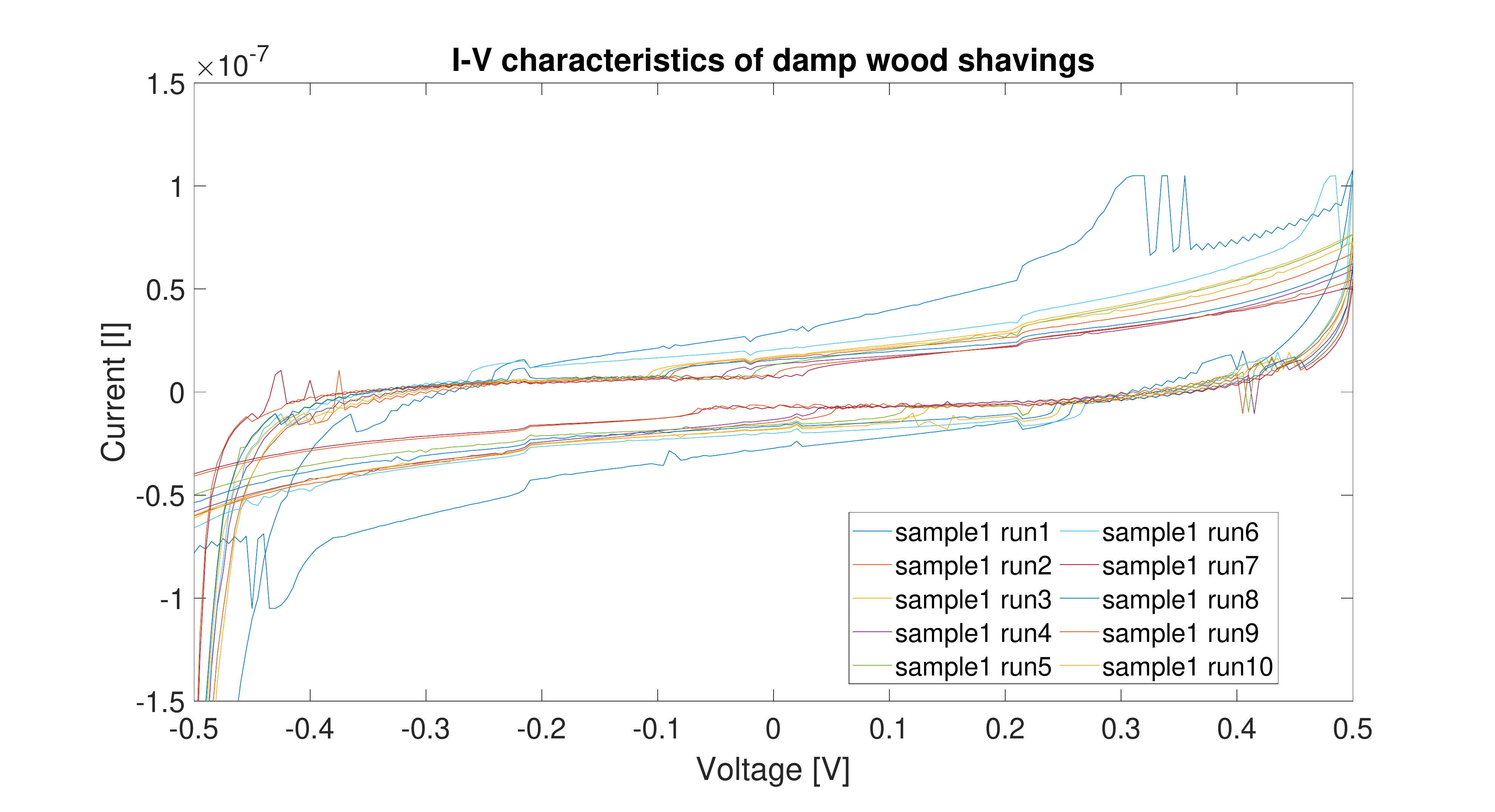}}
    \subfigure[]{\includegraphics[width=0.7\textwidth]{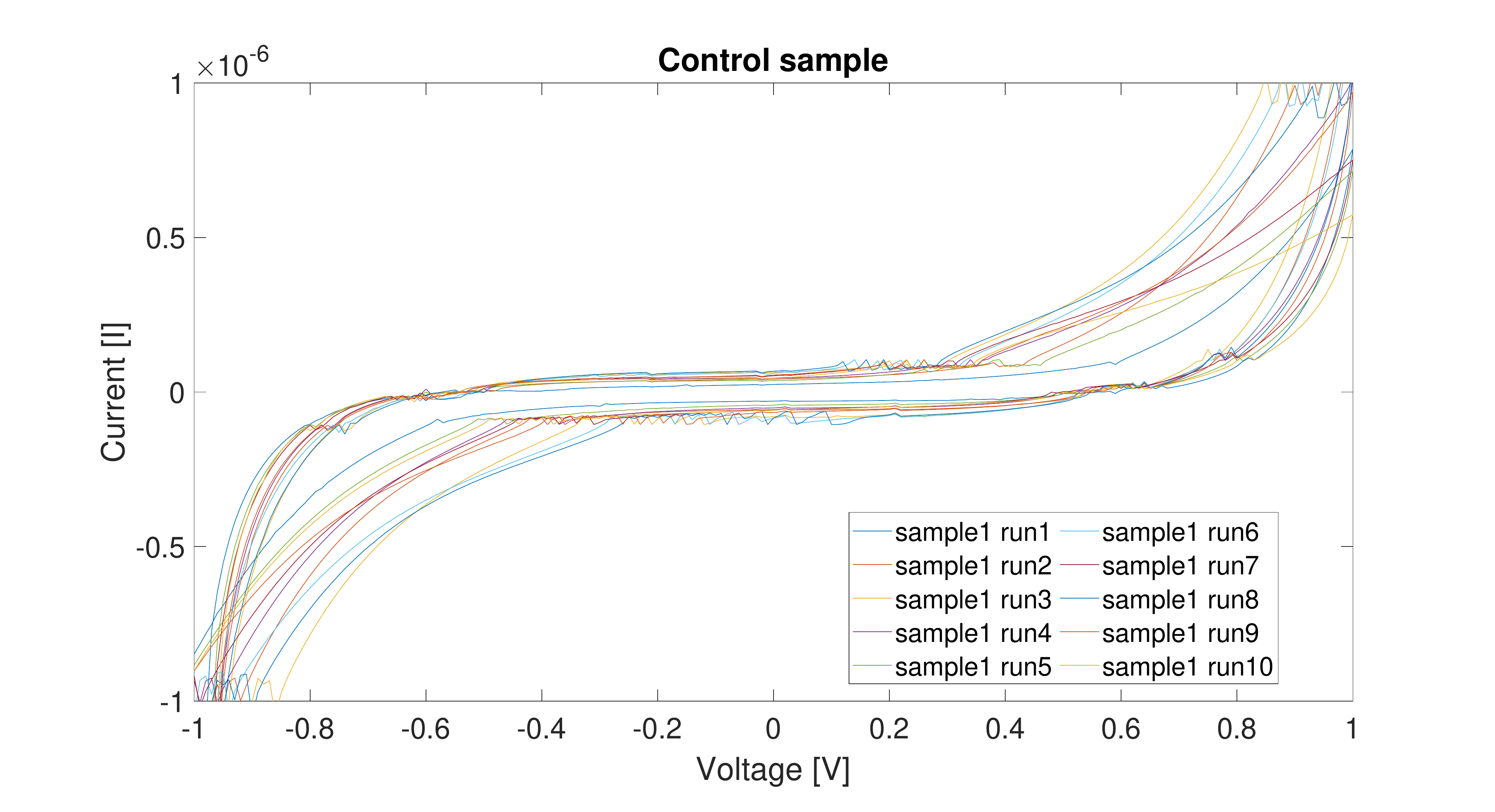}}
    \caption{Cyclic voltammetry of damp wood shavings used as a control for mycelium tests. (a) -0V5 to 0V5, (b) -1V to 1V}
    \label{fig:dampwoodshavings1Vpp}
\end{figure}

\begin{figure}[!hbt]
    \centering
    \subfigure[]{\includegraphics[width=0.7\textwidth]{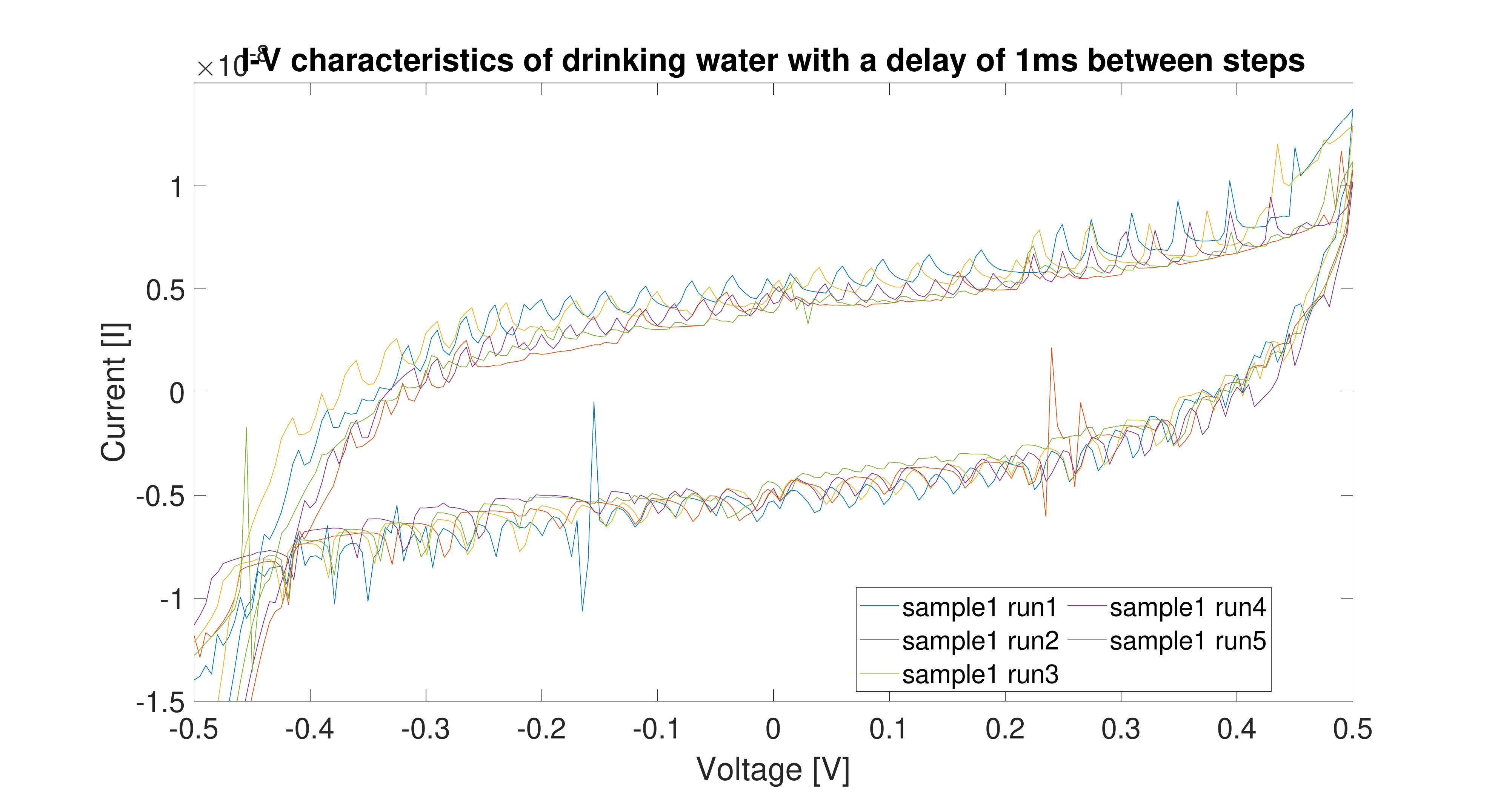}}
    \subfigure[]{\includegraphics[width=0.7\textwidth]{all_data_plots_normal_water_1Vpp_1ms.pdf}}
    \subfigure[]{\includegraphics[width=0.7\textwidth]{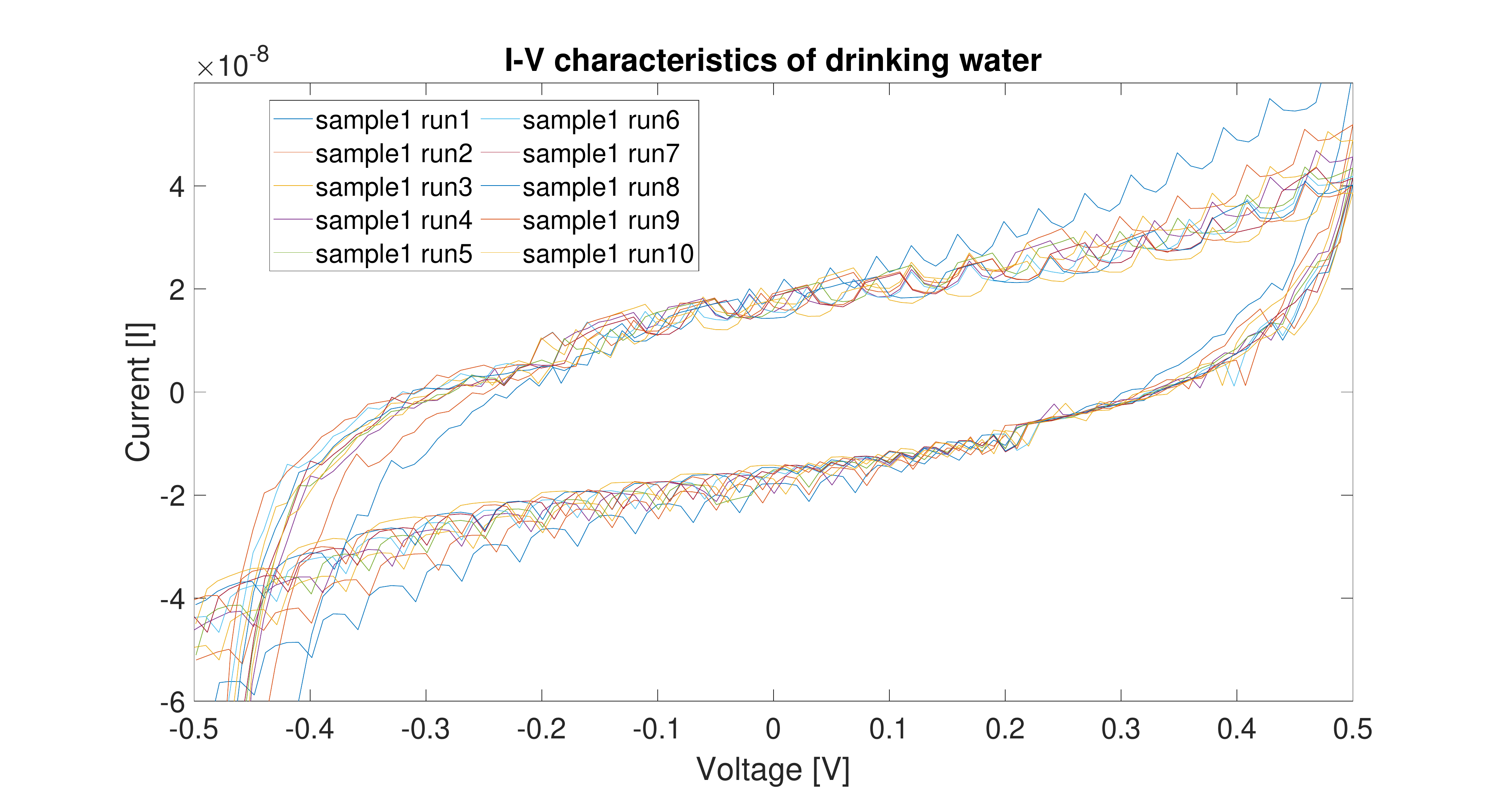}}
    \caption{Cyclic voltammetry (-0V5 to 0V5) of drinking water used as a control for mycelium tests. (a) delay between voltage steps is 1ms, (b) delay between voltage steps is 10ms, (c) delay between voltage steps is 100ms}
    \label{fig:normalwater1Vpp}
\end{figure}

%\subsection{Flora}

\begin{figure}[!hbt]
    \centering
    \subfigure[]{\includegraphics[width=0.7\textwidth]{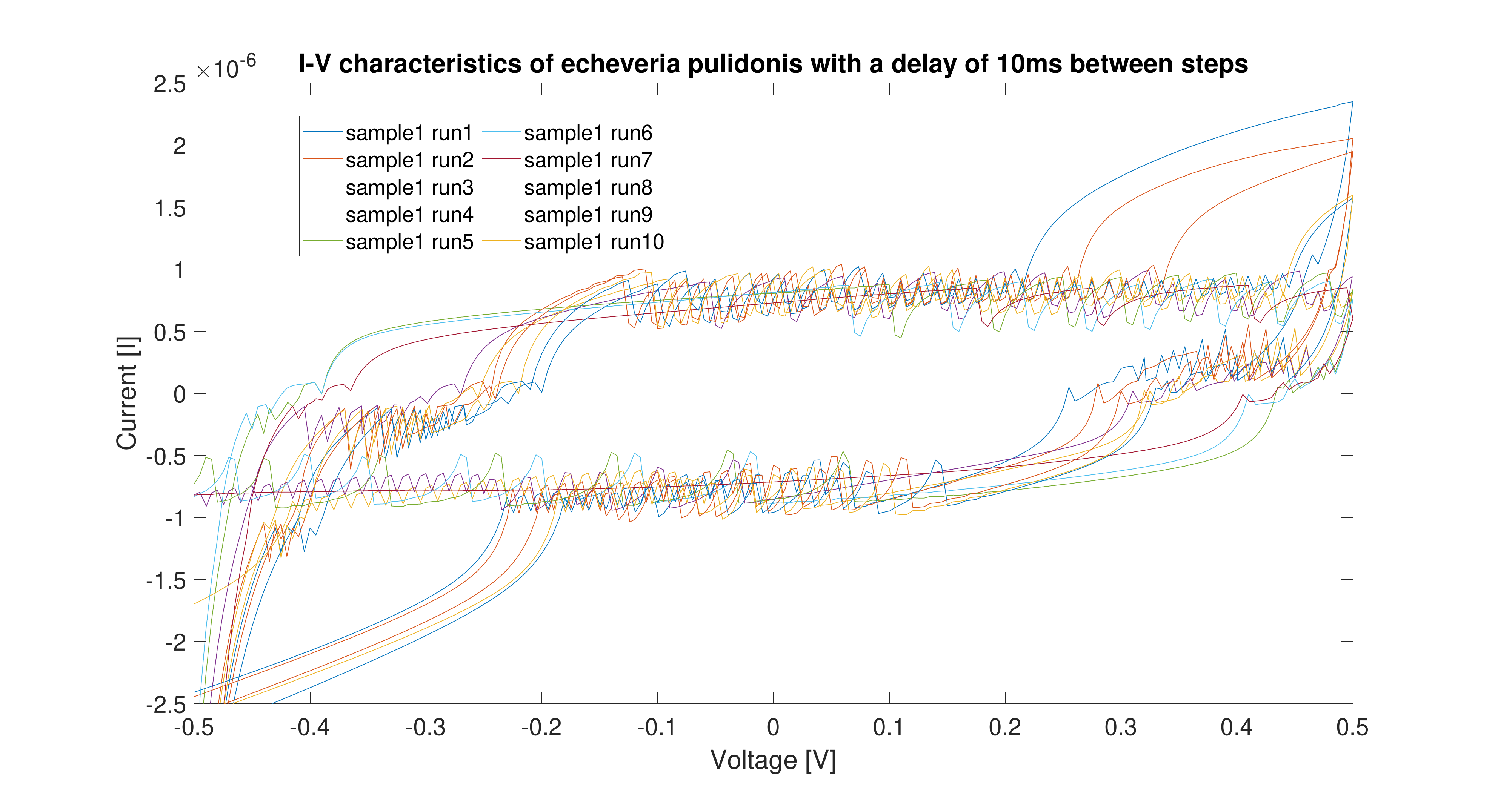}}
    \subfigure[]{\includegraphics[width=0.7\textwidth]{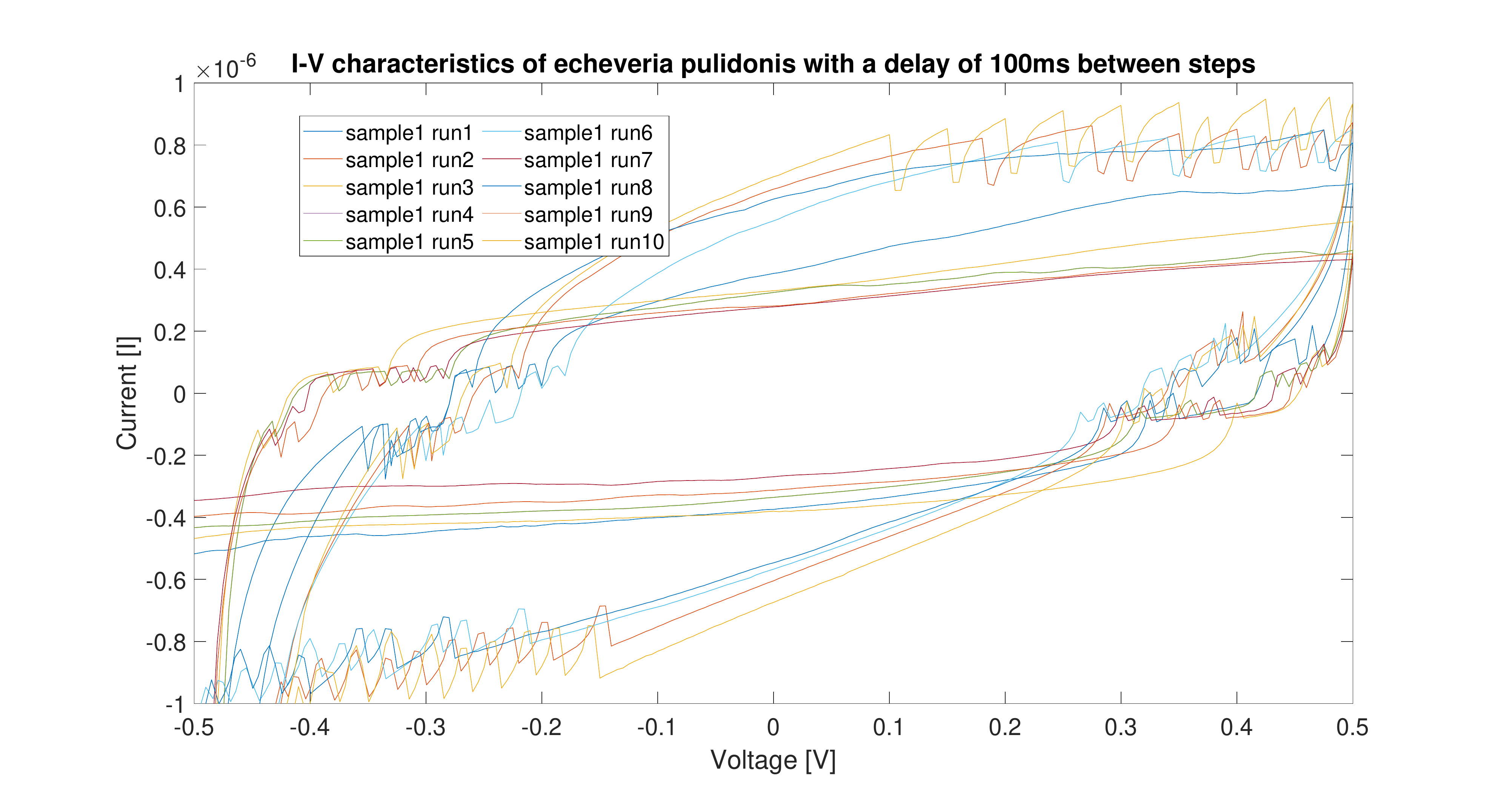}}
    \subfigure[]{\includegraphics[width=0.7\textwidth]{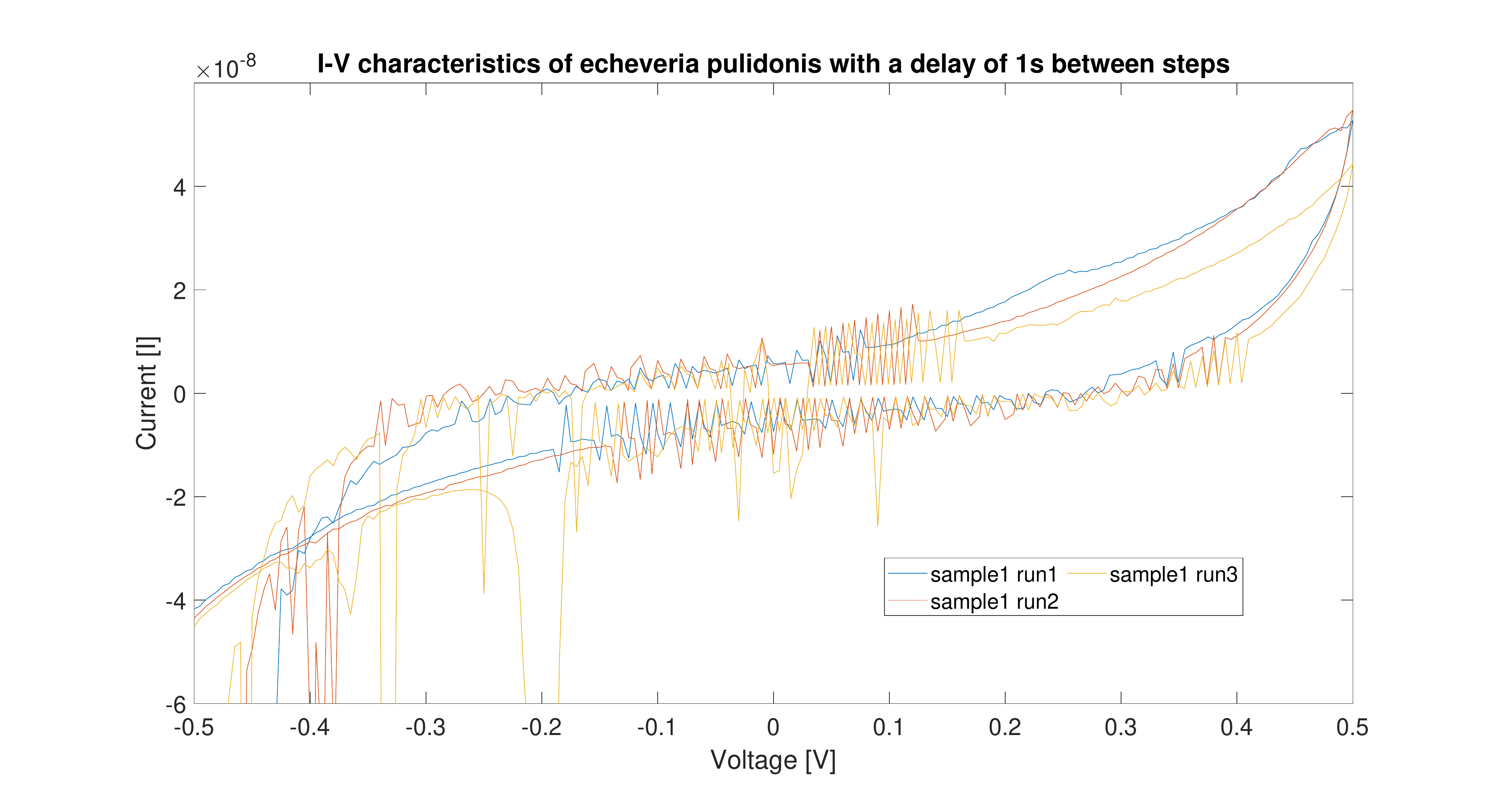}}
    \caption{Cyclic voltammetry (-0V5 to 0V5) of echeveria pulidonis. (a) delay time between settings is 10ms, (b) delay time between settings is 100ms, (c) delay time between settings is 1000ms }
    \label{fig:echeveria1Vpp}
\end{figure}

\begin{figure}[!hbt]
    \centering
    \subfigure[]{\includegraphics[width=0.7\textwidth]{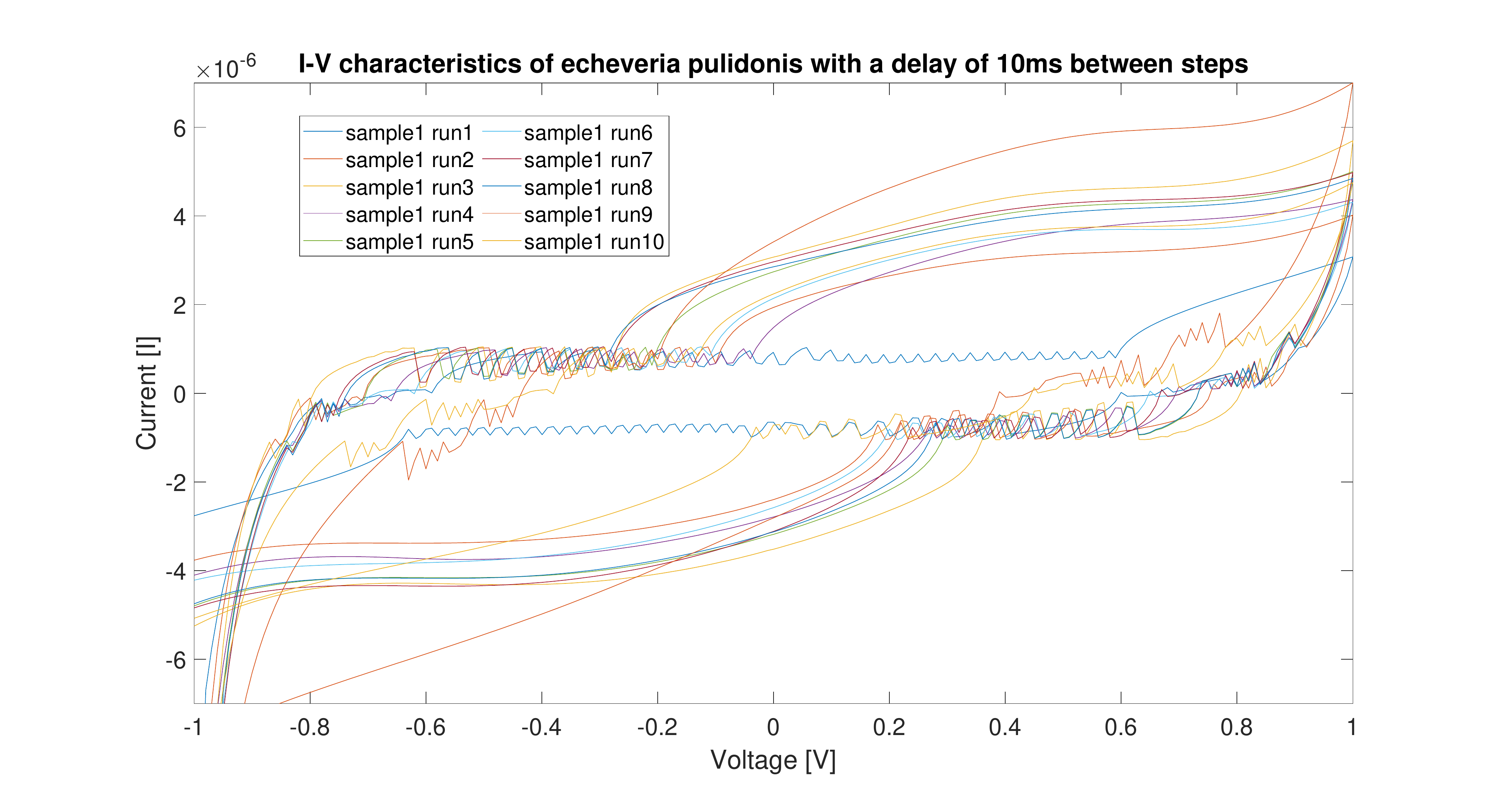}}
    \subfigure[]{\includegraphics[width=0.7\textwidth]{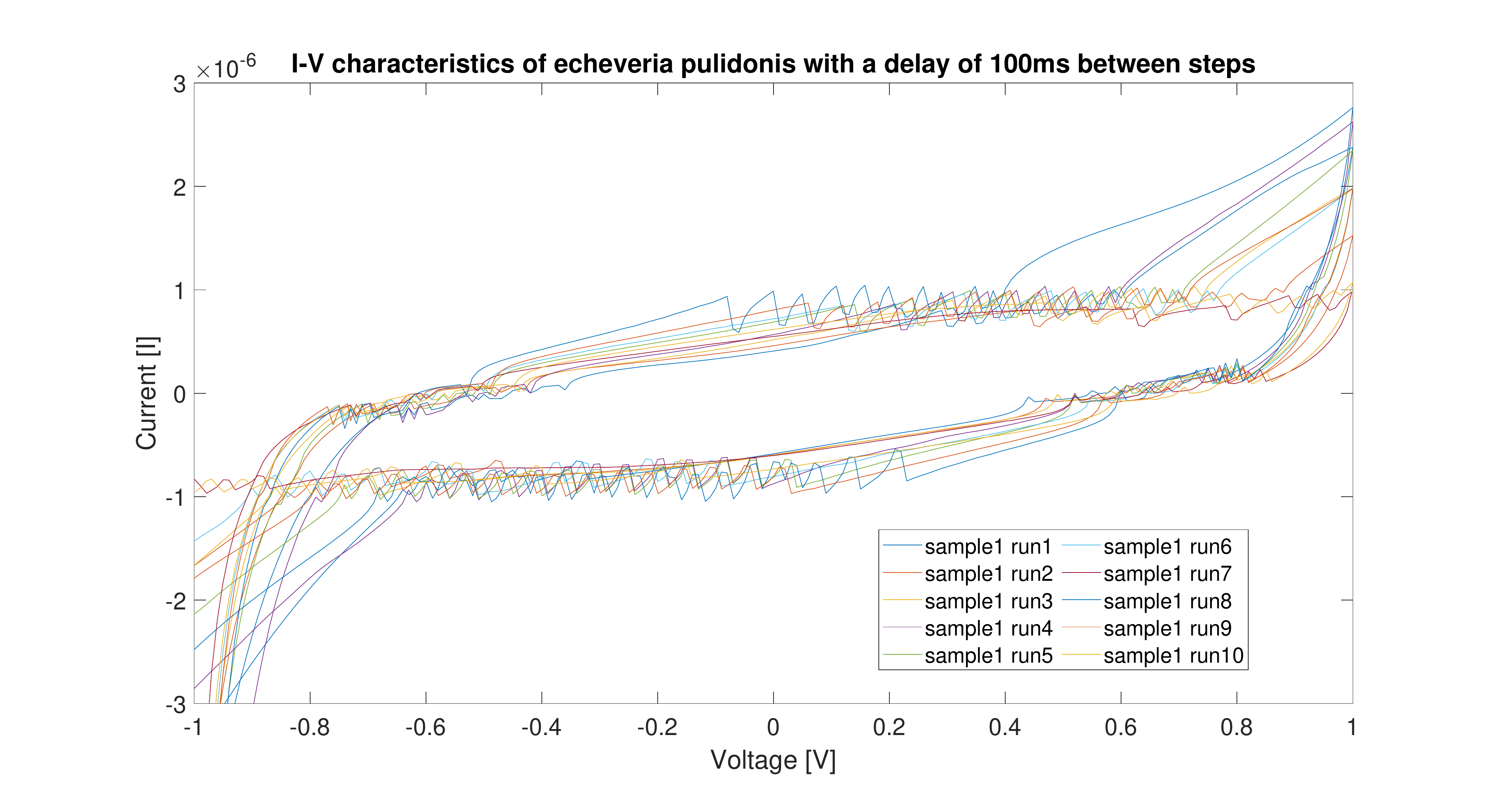}}
    \subfigure[]{\includegraphics[width=0.7\textwidth]{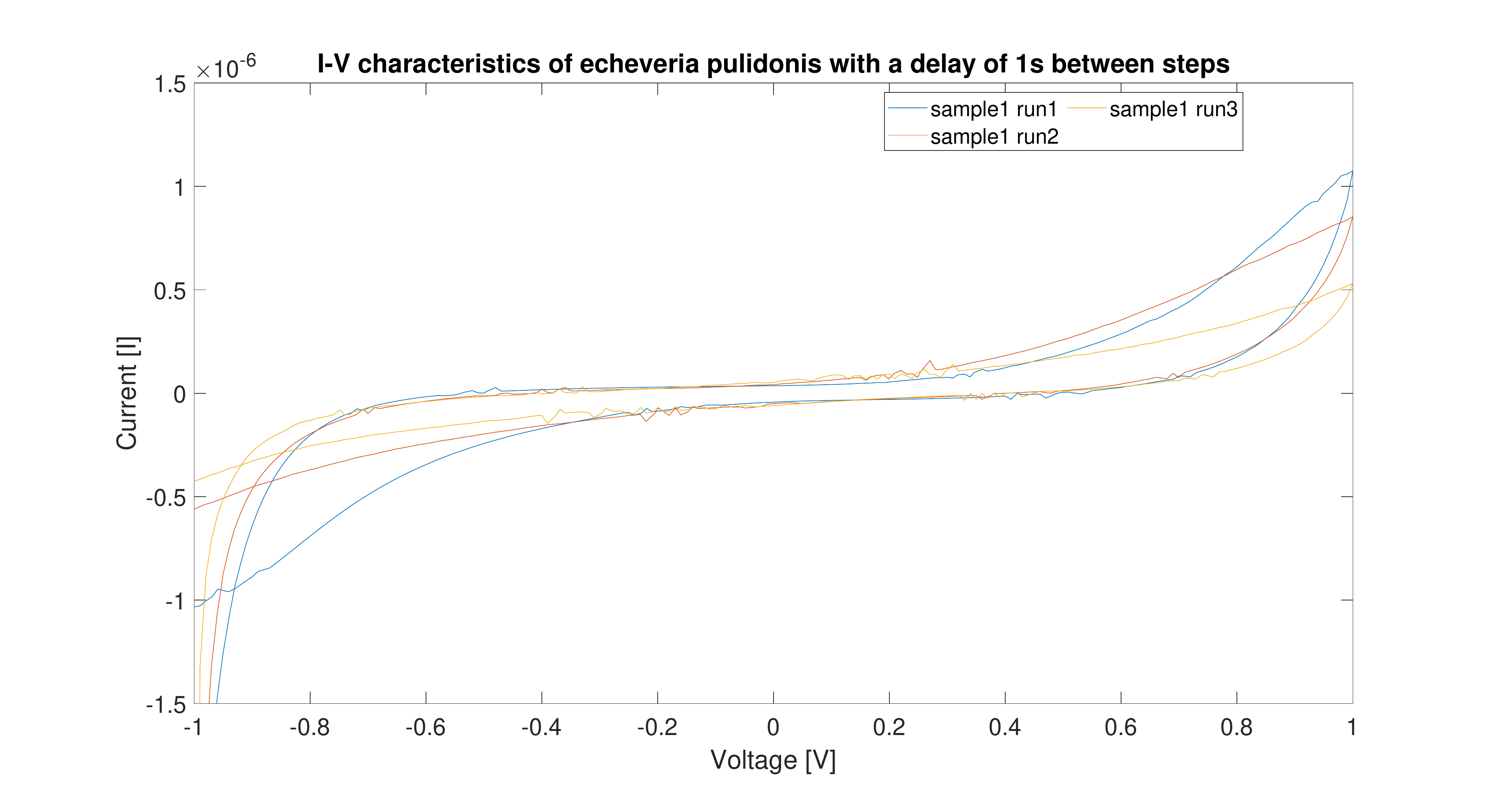}}
    \caption{Cyclic voltammetry (-1V to 1V) of echeveria pulidonis. (a) delay time between settings is 10ms, (b) delay time between settings is 100ms, (c) delay time between settings is 1000ms }
    \label{fig:echeveria2Vpp}
\end{figure}

\begin{figure}[!hbt]
    \centering
    \subfigure[]{\includegraphics[width=0.7\textwidth]{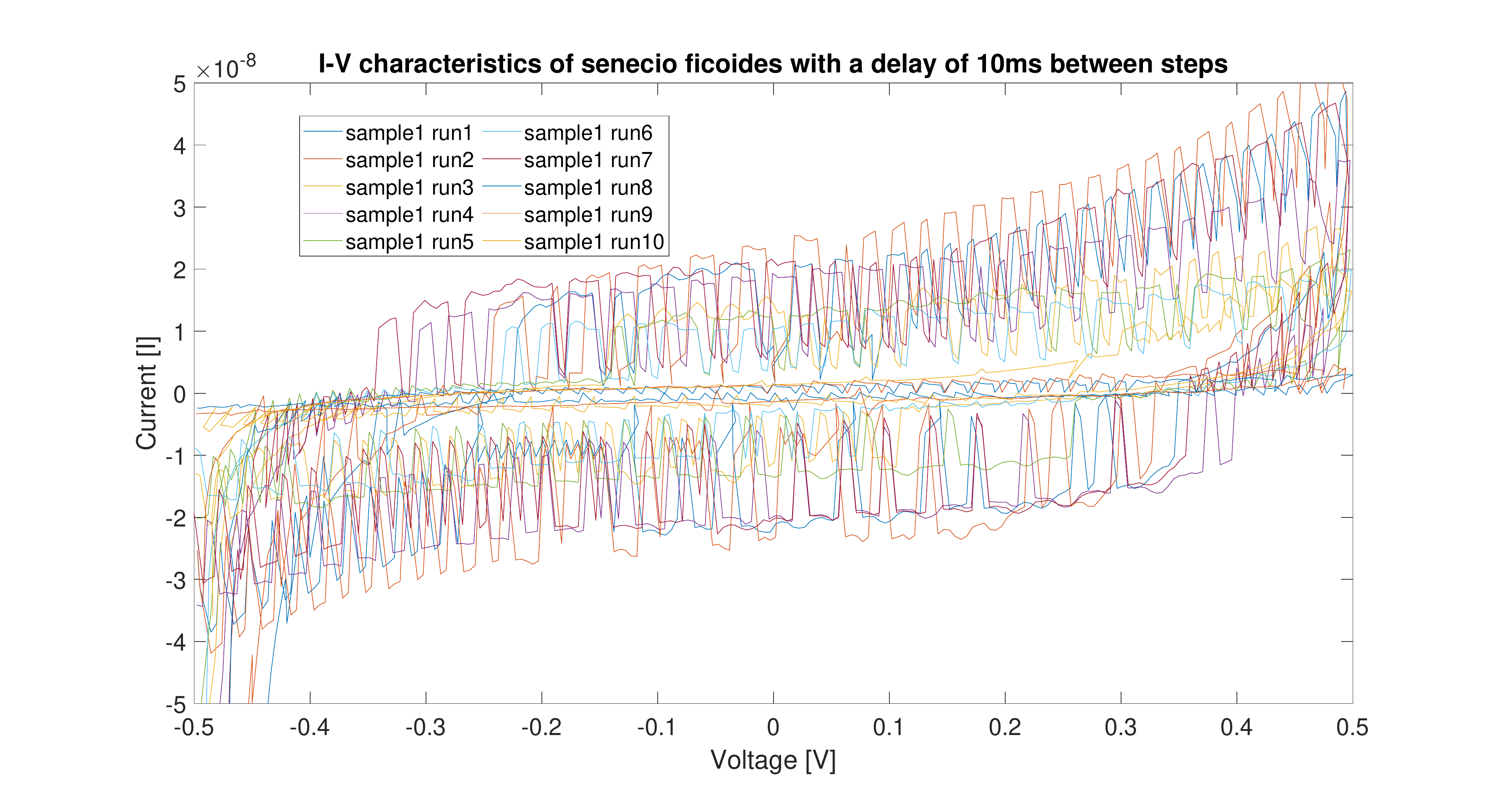}}
    \subfigure[]{\includegraphics[width=0.7\textwidth]{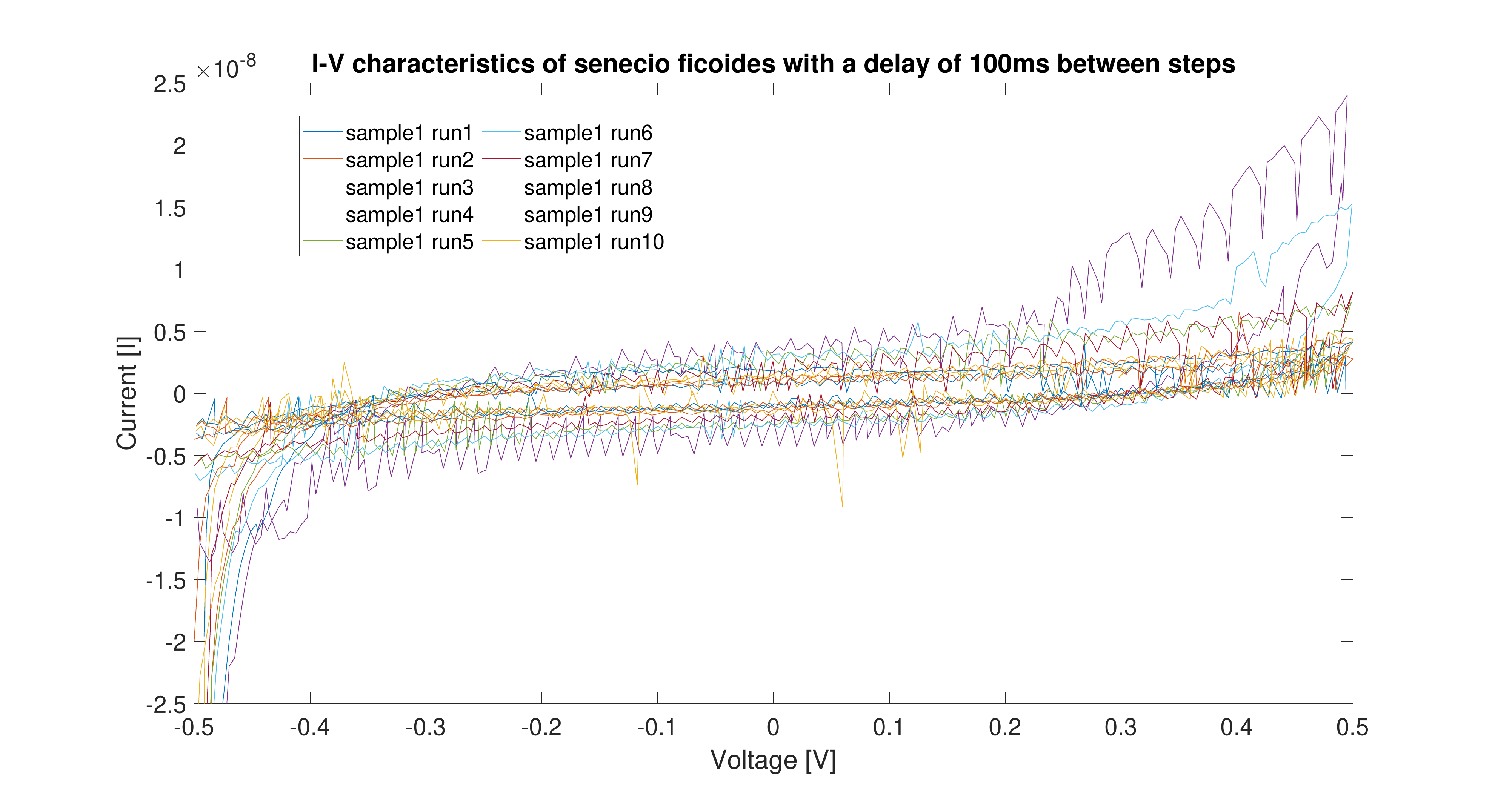}}
    \subfigure[]{\includegraphics[width=0.7\textwidth]{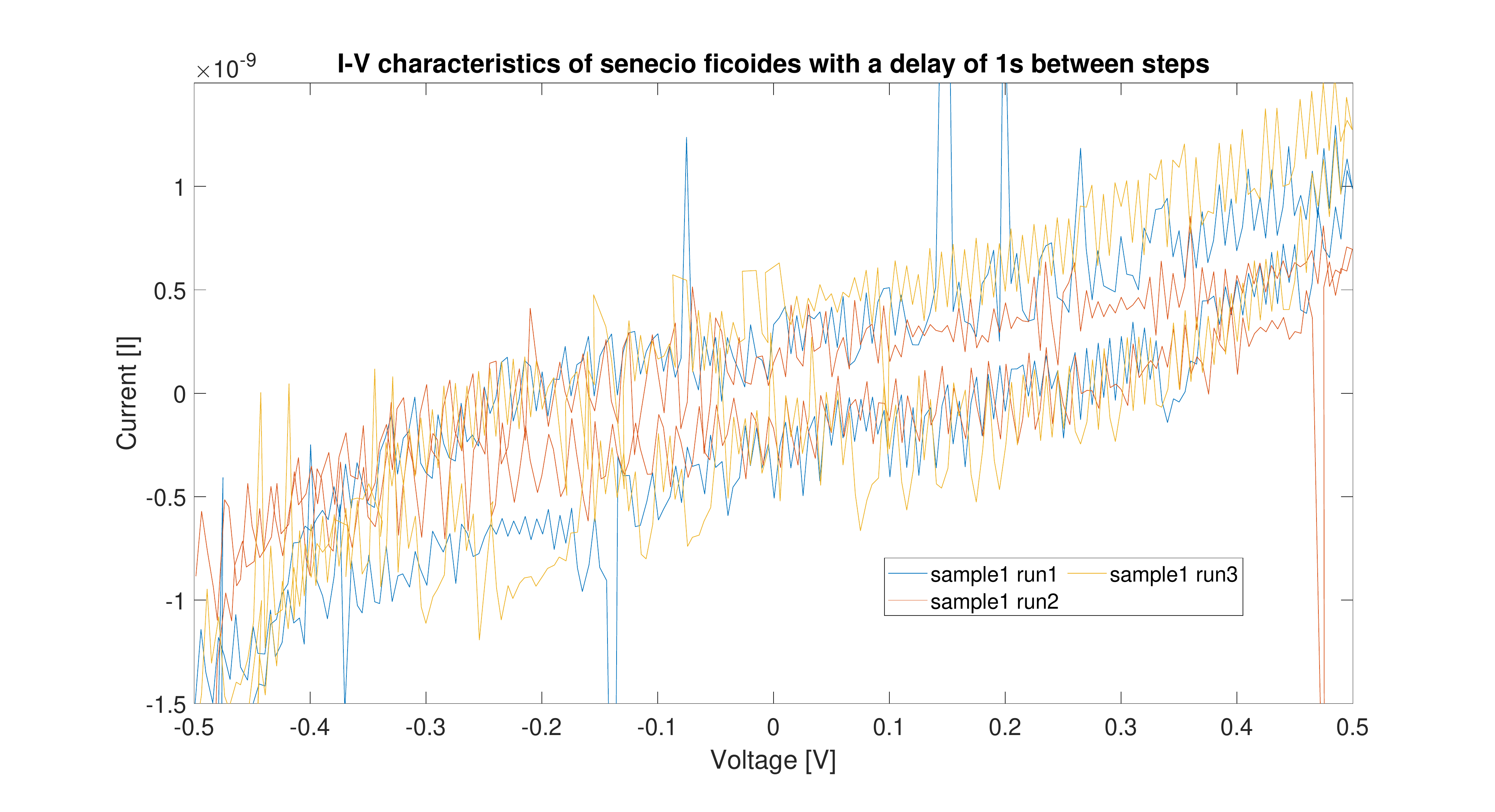}}
    \caption{Cyclic voltammetry (-0V5 to 0V5) of senecio ficoides. (a) delay time between settings is 10ms, (b) delay time between settings is 100ms, (c) delay time between settings is 1000ms }
    \label{fig:senecio1Vpp}
\end{figure}

\begin{figure}[!hbt]
    \centering
    \subfigure[]{\includegraphics[width=0.7\textwidth]{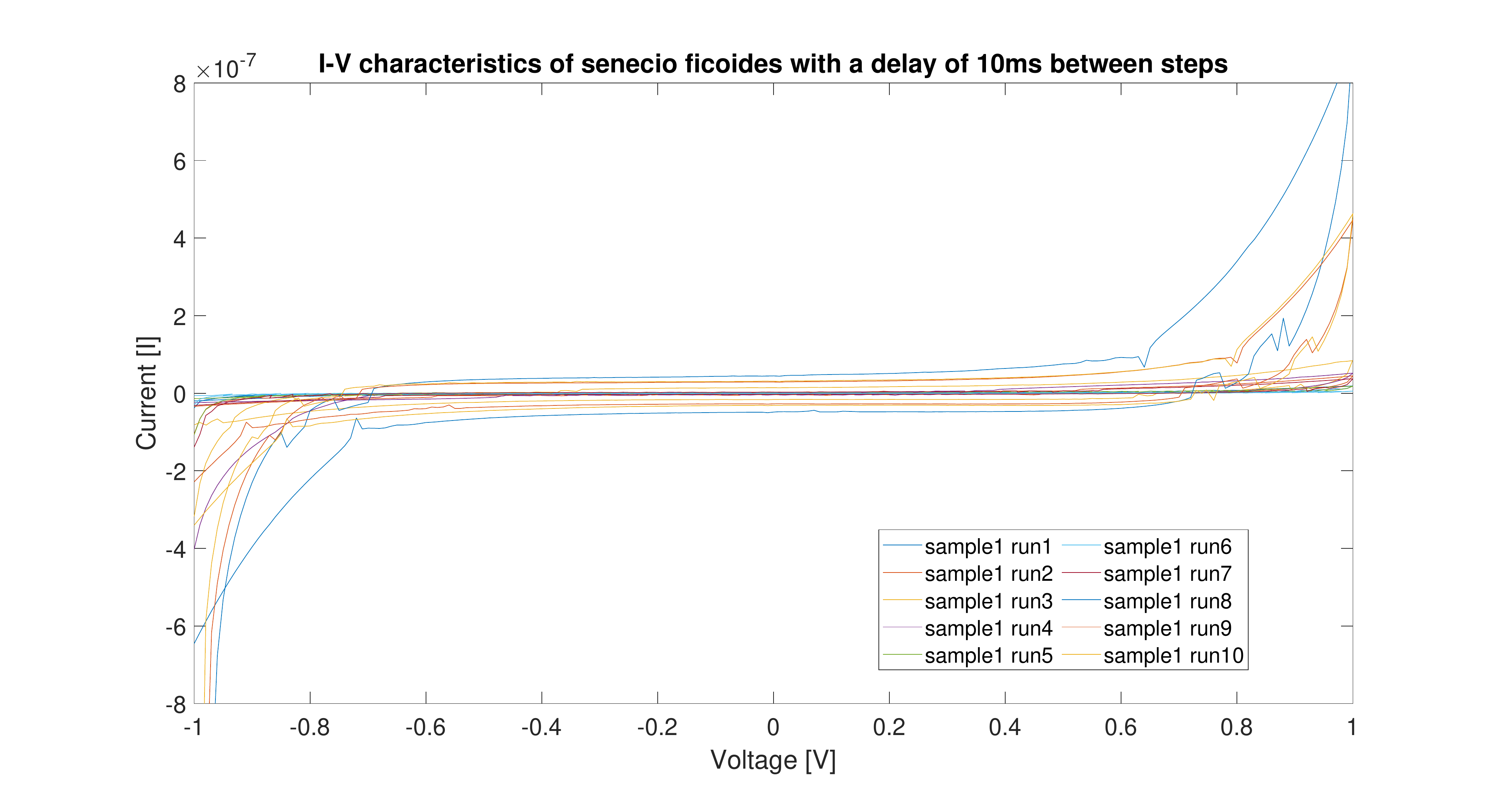}}
    \subfigure[]{\includegraphics[width=0.7\textwidth]{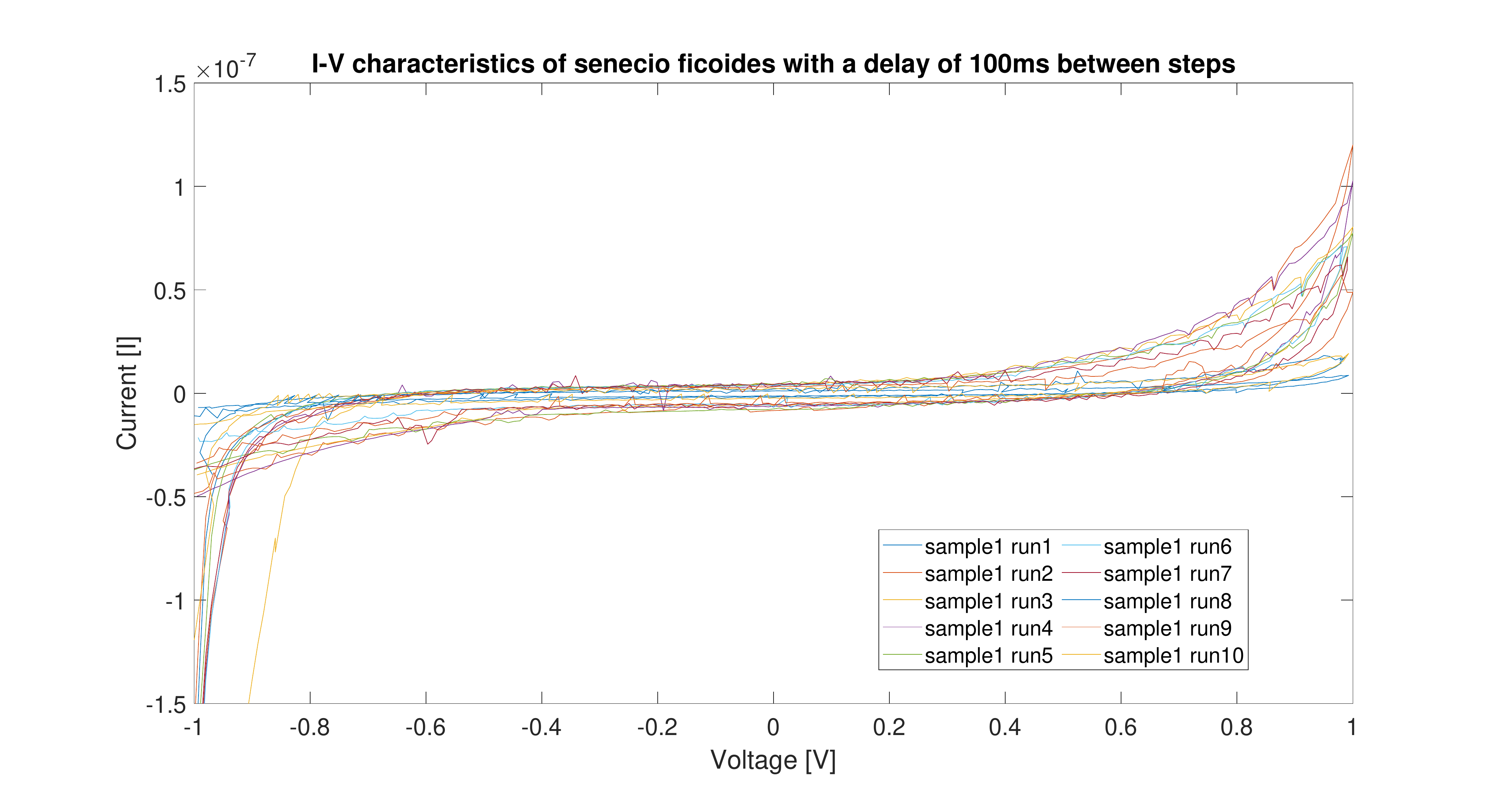}}
    \subfigure[]{\includegraphics[width=0.7\textwidth]{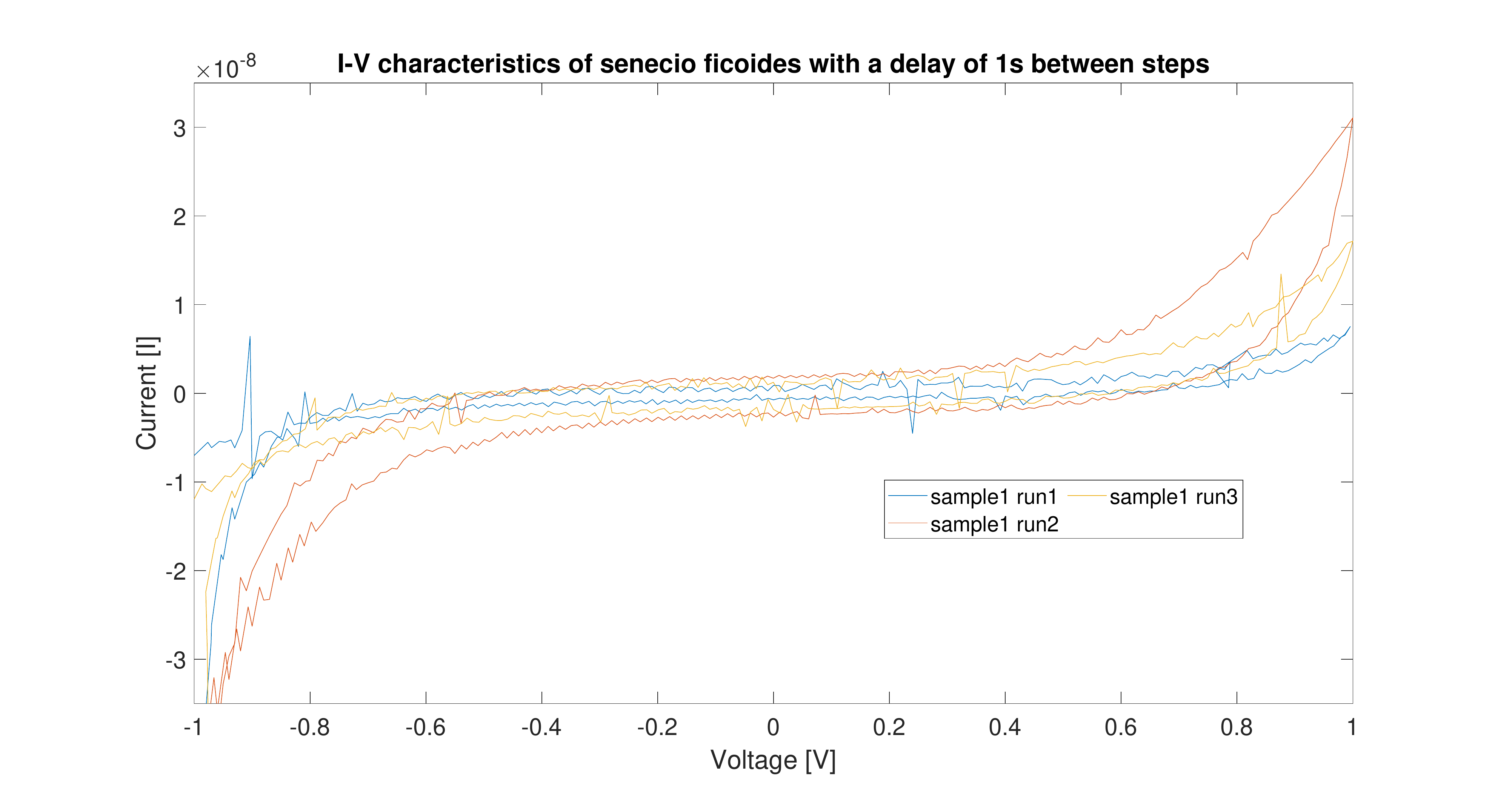}}
    \caption{Cyclic voltammetry (-1V to 1V) of senecio ficoides. (a) delay time between settings is 10ms, (b) delay time between settings is 100ms, (c) delay time between settings is 1000ms }
    \label{fig:senecio2Vpp}
\end{figure}

\end{document}